\documentstyle[sprocl,psfig]{article}
\def\zb{\bar{z}}

\def\be{\begin{equation}}
\def\ee{\end{equation}}
\def\bea{\begin{eqnarray}}
\def\eea{\end{eqnarray}}

\def\ap{{\alpha^\prime}}
\def\Tr{{\rm Tr}}
\def\ha{{\textstyle{1\over 2}}}
\def\H{{\bf H}}
\newcommand{\labell}[1]{\label{#1}}
\newcommand{\reef}[1]{(\ref{#1})}

\newcommand{\ennbee}[1]{\bigskip
\noindent\fbox{\noindent\parbox{4.65in}{{\it N.B.:} #1}}
\bigskip}

\newcommand{\insertion}[3]{\begin{figure}
\noindent\fbox{\parbox{4.6in}{\centerline{Insert #1:   {\bf #2}}
\smallskip 
#3}}
\end{figure}}

\newcommand{\subby}[1]{\bigskip
\leftline{$\bullet\quad{\mbox{\it #1}}$}\bigskip}

\def\IR{{\hbox{{\rm I}\kern-.2em\hbox{\rm R}}}}
\def\IB{{\hbox{{\rm I}\kern-.2em\hbox{\rm B}}}}
\def\IN{{\hbox{{\rm I}\kern-.2em\hbox{\rm N}}}}
\def\IC{\,\,{\hbox{{\rm I}\kern-.59em\hbox{\bf C}}}}
\def\IZ{{\hbox{{\rm Z}\kern-.4em\hbox{\rm Z}}}}
\def\IP{{\hbox{{\rm I}\kern-.2em\hbox{\rm P}}}}
\def\IH{{\hbox{{\rm I}\kern-.4em\hbox{\rm H}}}}
\def\ID{{\hbox{{\rm I}\kern-.2em\hbox{\rm D}}}}
\def\II{{\hbox{\rm I}\kern-.2em\hbox{\rm I}}}

\def\R{{\cal R}}

\begin{document}
\title{D--Brane Primer}
\author{Clifford V. Johnson\\\bigskip}
\address{Department of Mathematical Sciences\\
Science Laboratories, South Road\\
Durham    DH1 3LE\\
England, U.K.\\
\smallskip{\tt c.v.johnson@durham.ac.uk}}
%%%%%%%%%%%%%%%%%%%%%%%%%%%%%%%%
%
% You may repeat \author \address
% as often as necessary
%
%%%%%%%%%%%%%%%%%%%%%%%%%%%%%%%

\maketitle \abstracts{Following is a collection of lecture notes on
  D--branes, which may be used by the reader as preparation for
  applications to modern research applications such as: the AdS/CFT
  and other gauge theory/geometry correspondences, Matrix Theory and
  stringy non--commutative geometry, {\it etc}.  In attempting to be
  reasonably self--contained, the notes start from classical
  point--particles and develop the subject logically (but selectively)
  through classical strings, quantisation, D--branes, supergravity,
  superstrings, string duality, including many detailed applications.
  Selected focus topics feature D--branes as probes of both spacetime
  and gauge geometry, highlighting the role of world--volume curvature
  and gauge couplings, with some non--Abelian cases. Other
  advanced topics which are discussed are the (presently) novel tools
  of research such as fractional branes, the enhan\c con mechanism,
  D(ielectric)--branes and the emergence of the
  fuzzy/non--commutative sphere. (This is an expanded writeup of
  lectures given at ICTP, TASI, and BUSSTEPP.)}  \medskip
\noindent
%{\bf Contents}
\tableofcontents
\vfill\eject
\medskip

%\newpage
\bigskip
\bigskip

\section{Opening Remarks}
\label{remarks}
These lecture notes are supposed to represent, at least in part, the
introductory lectures on D--branes which I have given at a few schools
of one sort or another.  At some point last year, while preparing to
write some lecture notes in a publishable form, it occurred to me that
nobody really wanted another set of introductory ``D--notes'', (as I
like to call them).  After all, there have been many excellent ones,
dating as far back as 1996, not to mention at least two excellent text
books.~\cite{joebook,kiritsis}

Another problem which occurs is that well--meaning organisers want me
to give introductory lectures on D--branes, and assume that the
audience is going to ``pick up'' string theory along the way during
the lectures, but still seem quite keen that I get to the ``cool
stuff'' ---and usually in four or five lectures. Not being able to
bear the thought that I might be losing some of my audience, I thought
I'd write some notes for myself which try to take the ``informed
beginner'' from the very start (classical point particles), 
all the way to the modern applications (AdS/CFT, building a crystal
set, whatever), but making sure to stop to smell the flowers
along the way. It'll be  a bit more than four lectures, unfortunately, but
one should be able to pick and choose from the material.

This clearly calls for something somewhere between a serious text (for
which there is no need just yet) and another set of short lectures,
and here they are. I was hoping for them to be at least in part a sort
of useful toolbox, and not just a tour of what's happened or
happening. So as a result it reaches much further back than other
D--notes, and also necessarily a bit further forward, so as to make
contact with (and serve as a launching pad for) the other lecturers'
material at the school. Due to the remarkable activity in
the field, there is not a complete list of every paper written on
each topic. I am trying instead to supply a collection of notes that
can be worked through as preparatory material, so I list some of the
papers I found useful to this task, with an occasional partial attempt
at historical context. 

In terms of the later choice of topics, the notes hopefully fill in
some of the holes that other lecture notes on D--branes have left.
There is a rather detailed table of contents for aid in searching for
topics, and a list of some of the useful formulae that I (for one)
like to have to hand. (It's probably best just to tear those off and
throw the rest away.) There are lots of figures and (hopefully)
helpful insert boxes  to help the reader, especially in
the earlier parts.

I should mention that I think of this as a natural offspring of the
1996 project with Joe Polchinski and Shyamoli Chaudhuri,~\cite{dnotes}
and have inevitably borrowed many bits straight from there, and also
from Joe's excellent TASI notes from the same year.~\cite{joetasi}
Those form a sort of core which I have chipped away at, twisted,
stretched, embellished, and any other verb you can think of {\it
  except} ``improved'', simply because those lecture notes were trying
to do something quite different and still serve their purpose very
well.

Perhaps in conflict with some students' memory of the actual lectures,
the reader will not find any Star Wars references or jokes scattered
within. This is partly because I ran out of good Episode names, but
mostly because I actually saw last year's film. 

Quite seriously, I hope that the reader can use this collection of
notes as a means of pulling together lots of concepts that are in and
around string theory in a way that prepares them for actually doing
research in this wonderful and exciting subject. Since I wrote no
major problems to be done along with the material, let me end with a
few suggestions for some daily calisthenic exercises, if you will, of
a type which I have heard that all of the top researchers in the field
employ on a regular basis. They are listed on the next page. They
should be done first thing in the morning, or at least about the same
time each day. As experience is gained, you will find it fulfilling
to make up some of your own.

\bigskip
\bigskip

\noindent
I hope that you enjoy the notes. Look out for a fully hyperlinked web
version. \rightline{\it ---cvj}

\newpage

\centerline{\bf Daily Calisthenic Exercises~\footnote{I should stress
  that you do these exercises at your own risk. I cannot take any
  responsibility for injuries which result.}}

\bigskip
\bigskip
\bigskip

\begin{itemize}
\item{Wind an F--string around as large a circle as you can manage (or
    have room for in your office).  {\it Tip: If it is an open string,
      make sure that you firmly fix the ends on a handy D--brane before
      letting go, or you'll risk getting a nasty cut as it snaps
      back.}}
  
\item{Stretch a D--string between two parallel D--branes.  In the old
    days, we used to do this with F--strings, but that is now
    considered to be well beneath most young researchers' abilities.
    {\it Do {\bf not} be tempted to do this at strong coupling; all
      benefits of the exercise will be lost!}.}
  
\item{Try lifting a Coulomb branch from time to time, but be careful!
    This is one of the more advanced operations, and you should lift
    steadily to avoid any long term back pain.}
  
\item{As a nutritional supplement, try dissolving some D0--branes into
    your favourite drink, thereby giving both it and yourself a boost (in
    the M--direction).}
  
\item{I must admit that from time to time, as a treat, after such
    strenuous exercises I like to dry myself off with a slightly warm
    fuzzy sphere, which is surprisingly absorbent.}
\end{itemize}

\newpage

\section{String Worldsheet Perspective, Mostly}
This section will largely cover a lot of basic string theory material.
It may be skipped by many readers who want to go straight into the
properties of T--duality, {\it etc}, in the next section. There are
many issues which shall be covered only superficially here, and the
reader who wants to know more should consult introductory texts, some
of which are mentioned in the bibliography,
\cite{gsw,kiritsis,joebook} or the original references contained
within.
\subsection{Classical Point Particles}
\label{pparticles}
\begin{figure}[ht]
\centerline{\psfig{figure=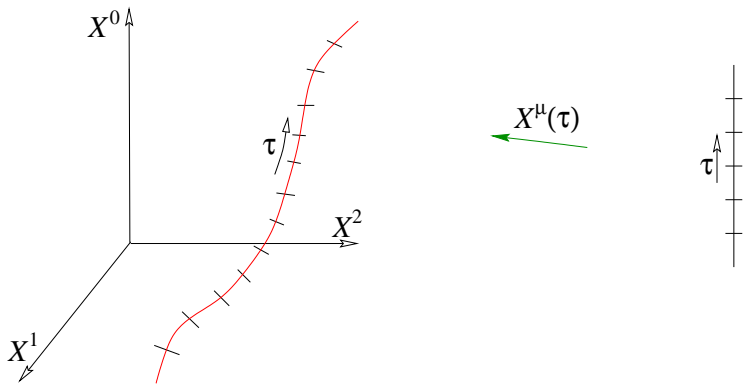,height=2.0in}}
\caption{A particle's world-line. The function $X^\mu(\tau)$ embeds the
world-line, parametrised by $\tau$, into spacetime, coordinatised by
$X^\mu$.}
\label{particlepic}
\end{figure}

{\subby{\it Writing an Action}}
Let us start by reminding ourselves about a description of a point
particle.  As a particle moves in the ``target spacetime'' (with
coordinates $(t\equiv X^0,X^1,\cdots,X^{D-1})$) and sweeps out a path
(see figure~\ref{particlepic}) in spacetime called a ``world--line'',
parametrised by $\tau$.

The infinitesimal path length swept out is:
\begin{equation}
d\ell=(-ds^2)^{1/2}=(-dX^\mu dX^\nu\eta_{\mu\nu})^{1/2}=(-dX^\mu
dX_\mu)^{1/2}\ ,
\end{equation}
and so the action is
\begin{equation}
S=-m\int d\ell=-m\int d\tau (-{\dot X}^\mu {\dot X}_\mu)^{1/2}\ ,
\end{equation}
where a dot denotes differentiation with respect to $\tau$.
Let us vary the action:
\begin{eqnarray}
\delta S=m\int d\tau (-{\dot X}^\mu {\dot X}_\mu)^{-1/2}{\dot X}^\nu 
\delta{\dot X}_\nu=m\int d\tau u^\nu\delta{\dot X}_\nu
=-m\int d\tau {\dot u}^\nu\delta X_\nu\ ,
\end{eqnarray}
where the last step used integration by parts, and
\begin{equation}
u^\nu\equiv (-{\dot X}^\mu {\dot X}_\mu)^{-1/2}{\dot X}^\nu\ .
\end{equation}
So for $\delta X$ arbitrary, we get ${\dot u}^\nu=0$, Newton's Law of
motion.

There is another action from which we can derive the same
physics. Consider the action
\begin{equation}
S^\prime={1\over2}\int 
d\tau\left(\eta^{-1}{\dot X}^\mu{\dot X}_\mu-\eta m^2\right)\ ,
\labell{sprime}
\end{equation}
for some independent function $\eta(\tau)$ defined on the world-line.

\bigskip
\noindent\fbox{\noindent\parbox{4.65in}{{\it N.B.:} In preparation for the
coming treatment of strings, think of the function $\eta$ as related
to the particle's ``world-line metric'', $\gamma_{\tau\tau}$, as
$\eta(\tau)=[-\gamma_{\tau\tau}(\tau)]^{1/2}$.  The function
$\gamma(\tau)$ ensures world-line reparametrisation invariance:
$$ds^2=\gamma(\tau)_{\tau\tau}d\tau d\tau=
\gamma^\prime(\tau)_{\tau^\prime\tau^\prime}d\tau^\prime d\tau^\prime\
.$$ }}
\bigskip

If we vary $S^\prime$ with respect to $\eta$:
\begin{equation}
\delta S^\prime={1\over2}\int d\tau\left[-\eta^{-2}{\dot X}^\mu{\dot
X}_\mu -m^2\right]\delta\eta\ .
\end{equation}
So for $\delta\eta$ arbitrary, we get an equation of motion
\begin{equation}
 \eta^2m^2+{\dot X}^\mu{\dot X}_\mu=0\ ,
\end{equation}
which we can solve with $\eta=m^{-1}(-{\dot X}^\mu{\dot
X}_\mu)^{1/2}$.  Upon substituting this into our expression
(\ref{sprime}) defining $S^\prime$, we get:
\begin{equation}
S^\prime=-{1\over2}\int d\tau \left\{m(-{\dot X}^\mu{\dot X}_\mu)^{1/2}+
(-{\dot X}^\mu{\dot X}_\mu)^{1/2}m^{-1}m\right\}=S\ ,
\end{equation}
showing that the two actions are equivalent.

Notice, however, that the action $S^\prime$ allows for a treatment of
the massless, $m=0$, case, in contrast to $S$. Another attractive
feature of $S^\prime$ is that it does not use the awkward square root
that $S$ does in order to compute the path length. The use of the
``auxiliary'' parameter $\eta$ allows us to get away from that. 

{\subby{\it Symmetries}}

There
are two notable symmetries of the action:

\begin{itemize}
\item{Spacetime Lorentz/Poincar\'e:
$$
X^\mu\to X^{\prime\mu}=\Lambda^\mu_{\phantom{\mu}\nu}X^\nu+A^\mu\ ,
$$ where $\Lambda$ is an $SO(1,3)$ Lorentz matrix and $A^\mu$ is an
arbitrary constant four--vector. This is a trivial global symmetry of
$S^\prime$ (and also $S$), following from the fact that we wrote them
in covariant form.}
\item{World line Reparametrisations:
\begin{eqnarray}
\delta X&=&\zeta(\tau){d X(\tau)\over d\tau}\nonumber\\
\delta\eta&=&{d\over d\tau}\left[\zeta(\tau)\eta(\tau)\right]\ ,
\nonumber
\end{eqnarray}
for some parameter $\zeta(\tau)$. This is a non--trivial local or
``gauge'' symmetry. This large extra symmetry on the world-line (and
its analogue when we come to study strings) is very useful. We can,
for example, use it to pick a nice gauge where we set
$\eta=m^{-1}$. This gives a nice simple action, resulting in a simple
expression for the conjugate momentum to $X^\mu$:
\begin{equation}
\Pi^\mu={\partial{\cal L}\over\partial{\dot X}^\mu}=m{\dot X}^\mu
\end{equation} 
We will use this much later.}
\end{itemize}

\subsection{Classical Bosonic Strings}
\label{bosonic}
\begin{figure}[ht]
\centerline{\psfig{figure=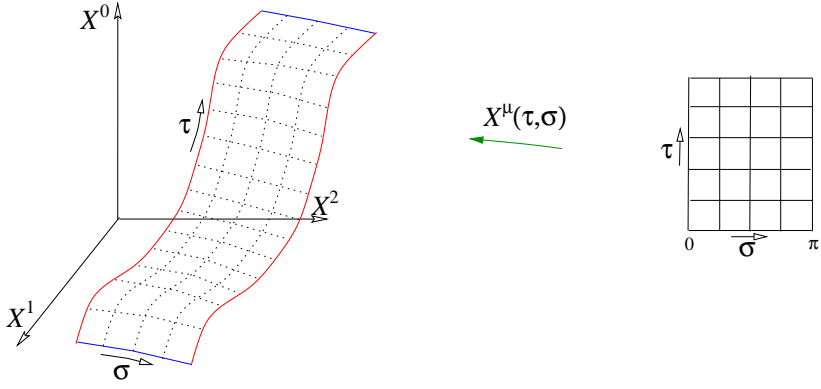,height=2.0in}}
\caption{A string's world-sheet. The function $X^\mu(\tau,\sigma)$
embeds the world-sheet, parametrised by $(\tau,\sigma)$, into spacetime,
coordinatized by $X^\mu$.}
\label{stringpic}
\end{figure}

Turning to strings, we parametrise the ``world-sheet'' which the
string sweeps out with coordinates $(\tau,\sigma)$. The latter is a
spatial coordinate, and for now, we take the string to be an open one,
with  $0\le\sigma\le\pi$ running from one end to the
other.  The string's evolution in spacetime is described by the
functions $X^\mu(\tau,\sigma)$, $\mu=0,\cdots,D-1$, giving the shape
of the string's world-sheet in target spacetime (see
figure~\ref{stringpic}).

There is an ``induced metric'' on the world-sheet given by
\begin{equation}
h_{ab}=\partial_aX^\mu\partial_bX^\nu\eta_{\mu\nu}\ ,
\end{equation}
with which we can perform meaningful measurements on the world-sheet as
an object embedded in spacetime.  Using this, we can define an action
analogous to the one we thought of first for the particle, by asking
that we extremize the area of the world-sheet:
\begin{equation}
S=-T\int d A=-T\int d\tau d\sigma\left(-{\rm det}h_{ab}\right)^{1/2}
\equiv\int d\tau d\sigma\,\, {{\cal L}}({\dot X},X^\prime;\sigma,\tau)\ .
\labell{ess}
\end{equation}
This is:
\begin{eqnarray}
S&=&-T\int d\tau d\sigma\left[\left({\partial
X^\mu\over\partial\sigma} {\partial X^\mu\over\partial\tau}\right)^2
-\left({\partial X^\mu\over\partial\sigma}\right)^2 \left({\partial
X^\mu\over\partial\tau}\right)^2 \right]^{1/2}\nonumber\\ &=&-T\int
d\tau d\sigma\left[ (X^\prime\cdot{\dot X})^2-X^{\prime2} {\dot
X}^2\right]^{1/2}\ ,
\end{eqnarray}
where $X^\prime$ means $\partial X/\partial\sigma$.
 Varying, we have generally:
\begin{eqnarray}
\delta S&=&\int d\tau d\sigma\left\{{\partial{\cal
L}\over\partial{\dot X}^\mu}\delta {\dot X}^\mu+{\partial{\cal
L}\over\partial X^{\prime\mu}}\delta X^{\prime\mu} \right\} \\ &=&\int
d\tau d\sigma\left\{-{\partial\over\partial\tau} {\partial{\cal
L}\over\partial{\dot X}^\mu}
-{\partial\over\partial\sigma}{\partial{\cal L}\over\partial
X^{\prime\mu}} \right\}\delta X^\mu+\int d\tau \left.
\left\{{\partial{\cal L}\over\partial X^{\prime\mu}} \delta
X^{\prime\mu}\right\}\right|_{\sigma=0}^{\sigma=\pi}\nonumber\ .
\end{eqnarray}
Asking this to be zero, we get:
\begin{eqnarray}
{\partial\over\partial\tau} {\partial{\cal L}\over\partial{\dot
X}^\mu} +{\partial\over\partial\sigma} {\partial{\cal L}\over\partial
X^{\prime\mu}}=0\quad{\rm and}\quad {\partial{\cal L}\over\partial
X^{\prime\mu}}=0\quad{\rm at}\quad\sigma=0,\pi\ ,
\end{eqnarray}
which are statements about the conjugate momenta:
\begin{eqnarray}
{\partial\over\partial\tau}
P^\mu_\tau
+{\partial\over\partial\sigma}P^\mu_\sigma=0
\quad{\rm and}\quad
P^\mu_\sigma=0\quad{\rm at}\quad\sigma=0,\pi\ .
\end{eqnarray}

Here, $P^\mu_\sigma$ is the momentum running along the string ({\it
i.e.,} in the $\sigma$ direction) while $P^\mu_\tau$ is the momentum
running transverse to it. The total spacetime momentum is given by
integrating up the infinitesimal (see figure 3):
\begin{equation}
dP^\mu=P^\mu_\tau d\sigma+P^\mu_\sigma d\tau\ .
\end{equation}
Actually, we can choose any slice of the world-sheet in order to
compute this momentum. A most convenient one is a slice $d\tau=0$,
revealing the string in its original paramaterisation: $P^\mu=\int
P^\mu_\tau d\sigma$, but any other slice will do.
\begin{figure}
\centerline{\psfig{figure=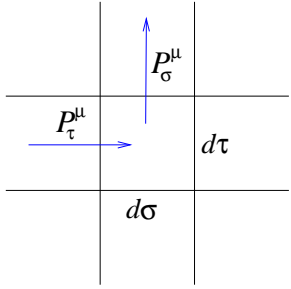,height=1.5in}}
\label{momentapic}
\caption{The infinitessimal momenta on the world sheet.}
\end{figure}
Similarly, one can define the angular momentum:
\begin{equation}
M^{\mu\nu}=\int(P^\mu_\tau X^\nu-P^\nu_\tau X^\mu)d\sigma\ .
\end{equation}
It is a simple exercise to work out the momenta for our particular
Lagrangian:
\begin{eqnarray}
P^\mu_\tau&=&T\,{{\dot X}^\mu X^{\prime2}-X^{\prime\mu}({\dot X}\cdot
X^\prime)\over \sqrt{({\dot X}\cdot X^\prime)^2-{\dot
X}^2X^{\prime2}}}\nonumber\\ P^\mu_\sigma&=&T\,{X^{\prime\mu}{\dot
X}^{2}-{\dot X}^\mu ({\dot X}\cdot X^\prime)\over \sqrt{({\dot X}\cdot
X^\prime)^2-{\dot X}^2X^{\prime2}}}\ .
\end{eqnarray}
It is interesting to compute the square of $P_\sigma^\mu$ using this,
and one finds that
\begin{equation}
P^2_\sigma\equiv P^\mu_\sigma P_{\mu\sigma}=-2T^2{\dot X}^2\ .
\end{equation}
This is our first (perhaps) non--intuitive classical result. We noticed that
$P_\sigma$ vanishes at the endpoints, in order to prevent momentum
from flowing off the ends of the string. The equation we just derived
implies that ${\dot X}^2=0$ at the endpoints, which is to say that
they move at the speed of light. This behaviour is a precursor of much
of what we will see in the quantum theory later.

\insertion{1}{$T$ is for Tension\label{insert1}}{As a first
  non--trivial example (and to learn that $T$, a mass per unit length,
  really is the string's tension) let us consider a closed string at
  rest lying in the $(X^1,X^2)$ plane.  We can make it by arranging
  that the $\sigma=0,\pi$ ends meet, that momentum flows across that
  join.  Such a configuration is:
\begin{eqnarray}
X^0&=&2R\tau;\nonumber\\
X^1&=&R\sin 2\sigma\nonumber\\
X^2&=&R\cos 2\sigma\ .\nonumber
\end{eqnarray}
It is worth taking the time to use this to show that one gets
$$
P^\mu_\tau=T\, (2R,0,0),\quad P^\mu_\sigma=T\, 
(0,2R\sin2\sigma,-2R\cos2\sigma)\ ,
$$
which is interesting, as a sketch shows:

%\begin{figure}[ht]
\centerline{\psfig{figure=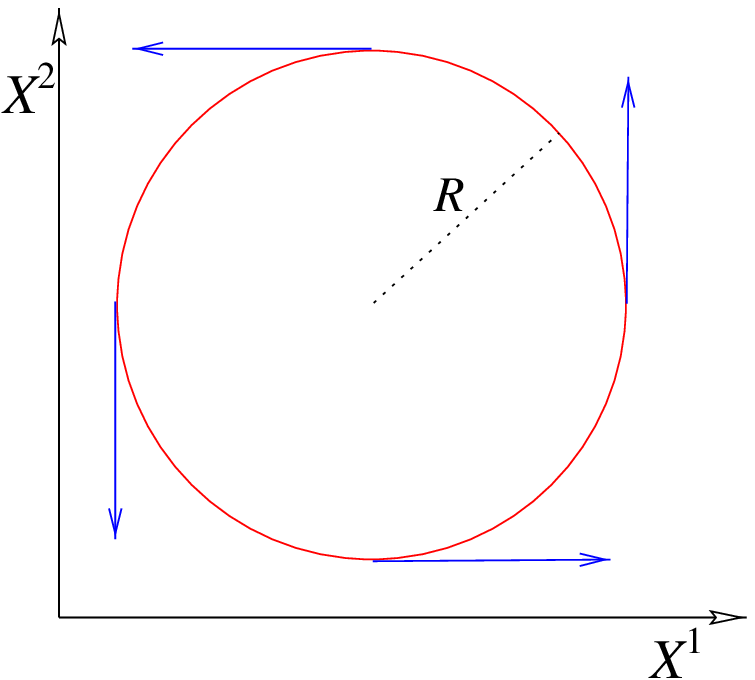,height=1.5in}}
%\end{figure}

The momentum is flowing around the string (which is lying in a circle
of radius~$R$).  The total momentum is
$$
P^\mu=\int_0^\pi d\sigma\, P_\tau^\mu\ .
$$
The only non--zero component is the mass--energy: 
$
M=2\pi RT={\rm length}{\times}T\ .
$}

Just like we did in the point particle case, we can introduce an
equivalent action which does not have the square root form that the
current one has. Once again, we do it by introducing a independent
metric, $\gamma_{ab}(\sigma,\tau)$, on the world-sheet, and write:
\begin{equation}
S^\prime=-{1\over4\pi\alpha^\prime}\int d\tau
d\sigma(-\gamma)^{1/2}\gamma^{ab} \partial_aX^\mu
\partial_bX^\nu\eta_{\mu\nu}= -{1\over4\pi\alpha^\prime}\int
d^2\!\sigma\, (-\gamma)^{1/2}\gamma^{ab} h_{ab}\ .
\labell{essprime}
\end{equation}
If we vary $\gamma$, we get
\begin{equation}
\delta S^\prime=-{1\over4\pi\alpha^\prime}\int
d^2\!\sigma\, \left\{-{1\over2}(-\gamma)^{1/2}\delta\gamma
\gamma^{ab}h_{ab}+(-\gamma)^{1/2}\delta\gamma^{ab}h_{ab}\right\}\ .
\end{equation}
Using the fact that $\delta\gamma=\gamma\gamma^{ab}\delta\gamma_{ab}=
-\gamma\gamma_{ab}\delta\gamma^{ab}$,
we get
\begin{equation}
\delta S^\prime=-{1\over4\pi\alpha^\prime}\int d^2\!\sigma\, 
(-\gamma)^{1/2}\delta
\gamma^{ab}\left\{h_{ab}-{1\over2}\gamma_{ab}\gamma^{cd}h_{cd}\right\}\
.
\end{equation}
Therefore we have
\begin{equation}
h_{ab}-{1\over2}\gamma_{ab}\gamma^{cd}h_{cd}=0\ ,
\end{equation}
from which we can derive
\begin{equation}
\gamma^{ab}h_{ab}=2(-h)^{1/2}(-\gamma)^{-1/2}\ ,
\labell{metricrelate}
\end{equation}
and so substituting into $S^\prime$, we recover (just as in the point
particle case) that it reduces to the Nambu--Goto action, $S$.

\insertion{2}{A Rotating Open String\label{insert2}}{As a second
  non--trivial example consider the following open string rotating at
  a constant angular velocity in the $(X^1,X^2)$ plane.  Such a
  configuration is:
\begin{eqnarray}
X^0&=&\tau;\nonumber\\
X^1&=&A\left(\sigma-{\pi\over2}\right)\cos \omega\tau\nonumber\\
X^2&=&A\left(\sigma-{\pi\over2}\right)\sin \omega\tau .\nonumber
\end{eqnarray}
This is what it looks like (the spinning string is shown in frozen
snapshots):

%\begin{figure}[ht]
\centerline{\psfig{figure=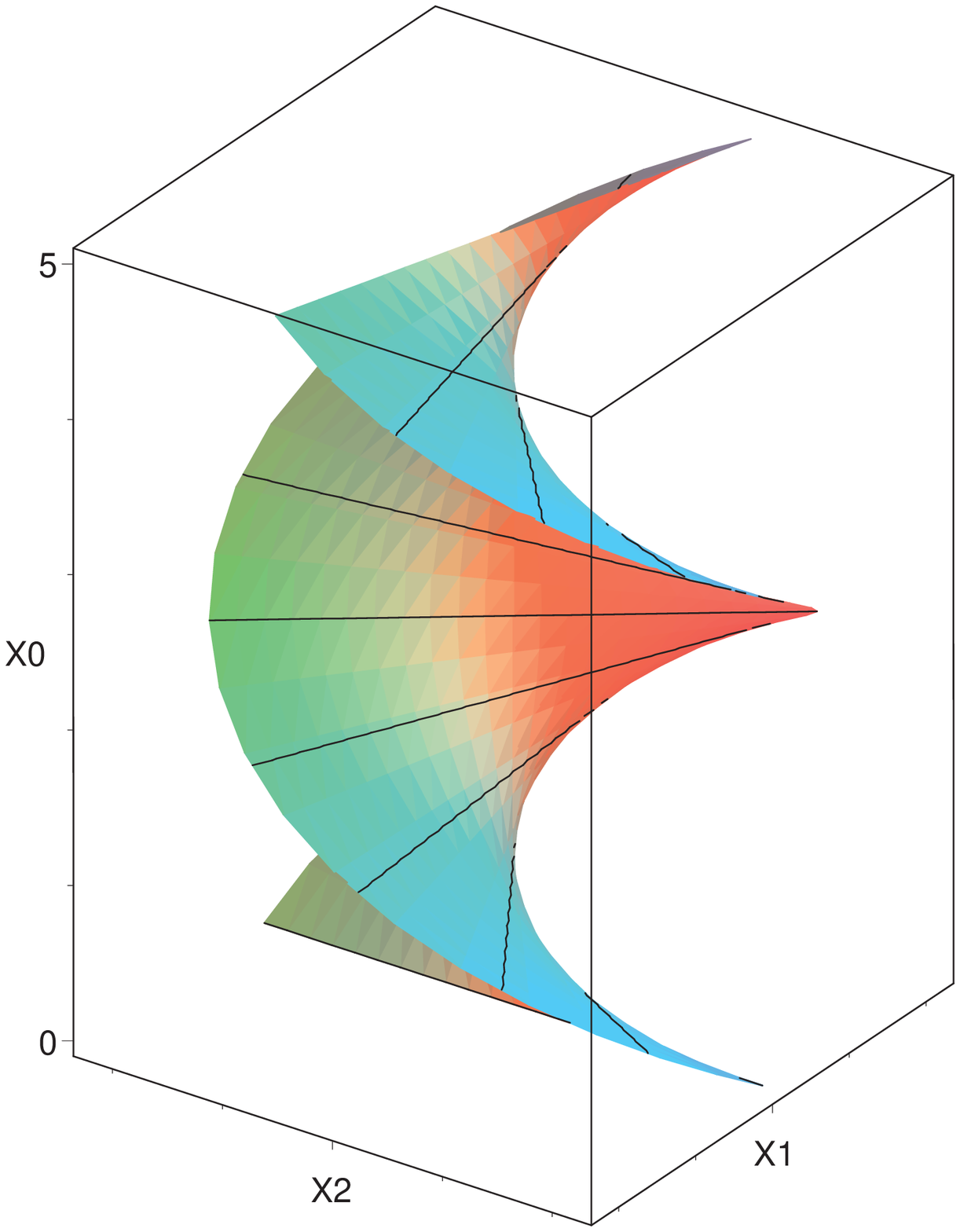,height=2.3in}}
%\end{figure}

It is again a worthwhile exercise to compute $P^\mu$, and also
$M^{\mu\nu}$.  With $J\equiv M^{12}$ and $M\equiv P^0$, some algebra
shows that
$$
{J\over M^2}={1\over2\pi T}=\alpha^\prime\ .
$$
This parameter, $\alpha^\prime$, is the slope of the celebrated
``Regge'' trajectories: the straight line plots of $J$ {\it vs.} $M^2$
seen in nuclear physics in the '60's. There remains the determination
of the intercept of this straight line graph with the $J$--axis. It
turns out to be 1 for the bosonic string as we shall see.}

{\subby{\it Symmetries}}

\noindent
Let us again study the symmetries of the action:
\begin{itemize}
\item{Spacetime Lorentz/Poincar\'e:
$$
X^\mu\to X^{\prime\mu}=\Lambda^\mu_{\phantom{\mu}\nu}X^\nu+A^\mu\ ,
$$ where $\Lambda$ is an $SO(1,3)$ Lorentz matrix and $A^\mu$ is an
arbitrary constant four--vector.  Just as before this is a trivial
global symmetry of $S^\prime$ (and also $S$), following from the fact
that we wrote them in covariant form.}
\item{Worldsheet Reparametrisations:
\begin{eqnarray}
\delta X^\mu&=&\zeta^a\partial_aX^\mu\nonumber\\
\delta\gamma^{ab}&=&\zeta^c\partial_c\gamma^{ab}-\partial_c\zeta^a\gamma^{cb}
-\partial_c\zeta^b\gamma^{ac}\ , \labell{reparams}
\end{eqnarray}
for two parameters $\zeta^a(\tau,\sigma)$. This is a non--trivial
local or ``gauge'' symmetry. This is a large extra symmetry on the
world-sheet of which we will make great use.}
\item{Weyl invariance:
\begin{eqnarray}
\gamma_{ab}\to\gamma^{\prime}_{ab}={\rm
e}^{2\omega}\gamma_{ab}\ ,
\labell{weyl}
\end{eqnarray}
 specified by a function $\omega(\tau,\sigma)$. This ability to do
local rescalings of the metric results from the fact that we did not
have to choose an overall scale when we chose $\gamma^{ab}$ to rewrite
$S$ in terms of $S^\prime$. This can be seen especially if we rewrite
the relation (\ref{metricrelate}) as
$(-h)^{-1/2}h_{ab}=(-\gamma)^{-1/2}\gamma_{ab}$.}

\end{itemize}

\ennbee{We note here for future use that there are just as many
parameters needed to specify the local symmetries (three) as there are
independent components of the world-sheet metric. This is very, very
useful, as we shall see.}

{\subby{\it String Equations of Motion}}

We can get equations of motion for the string by varying our action 
\reef{essprime} with respect to the $X^\mu$:
\begin{eqnarray}
\delta S^\prime&=&{1\over2\pi\alpha^\prime}\int d^2\!\sigma\, 
\partial_a\left\{
(-\gamma)^{1/2}\gamma^{ab}\partial_bX_\mu\right\}\delta
X^\mu\nonumber\\
&&\hskip2cm-{1\over2\pi\alpha^\prime}\int d\tau
\left. (-\gamma)^{1/2}\partial_\sigma X_\mu\delta
X^\mu\right|_{\sigma=0}^{\sigma=\pi}\ ,
\end{eqnarray}
which results in the equations of motion:
\begin{equation}
\partial_a\left(
(-\gamma)^{1/2}\gamma^{ab}\partial_b
X^\mu\right)\equiv(-\gamma)^{1/2}\nabla^2
X^\mu=0\ ,
\labell{motion}
\end{equation}
with {\it either}:
\begin{equation}
\left.\matrix{X^{\prime\mu}(\tau,0)=0\cr 
X^{\prime\mu}(\tau,\pi)=0}\right\}\qquad
\matrix{\mbox{Open String}\cr \mbox{(Neumann b.c.'s)}}
\labell{neumann}
\end{equation}
{\it or}:
\begin{equation}
\matrix{X^{\prime\mu}(\tau,0)=X^{\prime\mu}(\tau,\pi) \cr
X^{\mu}(\tau,0)=X^{\mu}(\tau,\pi)\cr\gamma_{ab}(\tau,0)=
\gamma_{ab}(\tau,\pi)}\Biggr\}\qquad
\matrix{\mbox{Closed String}\cr\mbox{(periodic b.c.'s)}}
\labell{periodic}
\end{equation}
We shall study the equation of motion \reef{motion} and the
accompanying boundary conditions a lot later. We are going to look at
the standard Neumann boundary conditions mostly, and then consider the
case of Dirichlet conditions later, when we uncover
D--branes,\cite{dbranesi,dbranesii,dbranesiii,dbranesiv,nairet,gv,buscher}
using T--duality.\cite{tdual} Notice that we have taken the liberty of
introducing closed strings by imposing periodicity (see also insert
1 (p.\pageref{insert1})).

{\subby{\it More terms}}

Thinking of this theory as a two--dimensional model ---consisting of
$D$ bosonic fields $X^\mu(\tau,\sigma)$ with an action given by
\reef{essprime}, it is natural to ask whether there are other terms
which we might want to add to the theory.

Given that we are treating the two dimensional metric $\gamma_{ab}$ as
a dynamical variable, two other terms spring effortlessly to mind,
from the analogy with General Relativity. One is the Einstein--Hilbert
action (supplemented with a boundary term):
\begin{equation}
\chi={1\over4\pi\alpha^\prime}
\int_{\cal M} d^2\!\sigma\, (-\gamma)^{1/2}R+{1\over
2\pi\alpha^\prime}\int_{\partial {\cal M}} ds K\ , \labell{einsteinhilbert}
\end{equation}
where $R$ is the two--dimensional Ricci scalar on the world-sheet
${\cal M}$, $K$ is the extrinsic curvature on the boundary $\partial
{\cal M}$ and the other is:
\begin{equation}
\Theta={1\over4\pi\alpha^\prime}\int_{\cal M} d^2\!\sigma\, (-\gamma)^{1/2}\ ,
\labell{cosmocon}
\end{equation}
which is the cosmological term. What is their role here? Well, under
a Weyl transformation \reef{weyl}, we see that $(-\gamma)^{1/2}\to
e^{2\omega}(-\gamma)^{1/2}$ and $R\to e^{-2\omega}
(R-2\nabla^2\omega)$, and so $\chi$ is invariant, (because $R$ changes
by a total derivative which is cancelled by the variation of $K$) but
$\Theta$ is not.

So we will include $\chi$, but not $\Theta$ in what follows. Now, the
full string action resembles two--dimensional gravity coupled to $D$
bosonic ``matter'' fields $X^\mu$, and the equations of motion are of
course:
\begin{equation}
R_{ab}-{1\over2}\gamma_{ab}R=T_{ab}\ .
\labell{einsteineq}
\end{equation}
The left hand side vanishes identically in two dimensions, and so
there is no dynamics associated to \reef{einsteinhilbert}. The
quantity $\chi$ depends only on the topology of the world
sheet  and so will only matter when comparing world sheets of
different topology. This will arise when we compare results from
different orders of string perturbation theory and when we consider
interactions.

How does this work? Well, let us sketch it here: Let us add our new
term to the action, and consider the string action to be (we will
denote it $S$ from now on), and dropping the prime:
\begin{equation}
S={1\over4\pi\alpha^\prime}\int_{\cal M}d^2\!\sigma\, 
g^{1/2}g^{ab}\partial_aX^\mu\partial_bX_\mu+\lambda
\left\{{1\over4\pi\alpha^\prime}
\int_{\cal M} d^2\!\sigma\, (g)^{1/2}R+{1\over
2\pi\alpha^\prime}\int_{\partial {\cal M}} ds K\right\}\ ,
\labell{eulerterm}
\end{equation}
where $\lambda$ is ---for now--- and arbitrary parameter which we have
not fixed to any particular value. 

\ennbee{It will turn out that $\lambda$ is not a free parameter. In
the full string theory, it has dynamical meaning, and will be equivalent
to the expectation value of one of the massless fields ---the
``dilaton''--- described by the string.}

Note that we have anticipated something that we will do later, which
is to work with Euclidean signature to make sense of the topological
statements to follow: $\gamma_{ab}$ with signature $(-+)$ has been
replaced by $g_{ab}$ with signature $(++)$.

\begin{figure}[ht]
\centerline{\psfig{figure=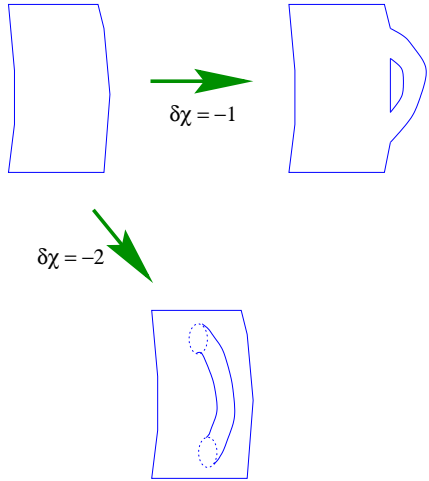,height=2.5in}}
\caption{Worldsheet topology change due to emission and reabsorption
of open and closed strings}
\label{split}
\end{figure}

So what will $\lambda$ do? Recall that it couples to Euler number, so
in the full path integral defining the string theory:
\begin{equation}
{\cal Z}=\int {\cal D}X{\cal D}g\,\, e^{-S}\ ,
\labell{pathintegral}
\end{equation}
resulting amplitudes will be weighted by a factor $e^{-\lambda\chi}$,
where $\chi=2-2h-b-c$. Here, $h,b,c$ are the numbers of handles,
boundaries and crosscaps, respectively, on the world sheet.  Consider
figure \ref{split}. An emission and reabsorption of an open string
results in a change $\delta\chi=-1$, while for a closed string it is
$\delta\chi=-2$. Therefore, relative to the tree level open string
diagram (disc topology), the amplitudes are weighted by $e^\lambda$
and $e^{2\lambda}$, respectively. The quantity $g_s\equiv e^\lambda$
therefore will be called the closed string coupling. Note that it is
the square of the open string coupling.

{\subby{\it The Stress Tensor}}

Let us also note that we can define a two--dimensional
energy--momentum tensor:
\begin{eqnarray}
T^{ab}(\tau,\sigma)\equiv-{4\pi\over\sqrt{-\gamma}}{\delta
S\over\delta\gamma_{ab}}
=-{1\over\alpha^\prime}\left\{\partial^aX_\mu\partial^b
X^\mu-{1\over2}\gamma^{ab}\gamma_{cd}\partial^cX_\mu\partial^d
X^\mu\right\}\
.
\end{eqnarray}
Notice that 
\begin{equation}
T^a_a\equiv\gamma_{ab}T^{ab}=0\ .
\labell{traceless}
\end{equation}
This is a consequence of Weyl symmetry.  Reparametrisation
invariance, $\delta_\gamma S^\prime=0$, translates here into (see
discussion after eqn.\reef{einsteineq})
\begin{equation}
T^{ab}=0\ .
\labell{vanish}
\end{equation}
These are the classical properties of the theory we have uncovered so
far. Later on, we shall attempt to ensure that they are true in the
quantum theory also, with interesting results.

{\subby{\it Gauge Fixing}}

\noindent
Now recall that we have three local or ``gauge'' symmetries of the
action:
\begin{eqnarray}
\mbox{2d
reparametrisations}:\,\,\sigma,\tau&\to&{\tilde\sigma}(\sigma,\tau),
{\tilde\tau}(\sigma,\tau) \nonumber\\
\mbox{Weyl}:\,\,\gamma_{ab}&\to&\exp({2\omega(\sigma,\tau)})
\gamma_{ab}\ .
\end{eqnarray}
The two dimensional metric $\gamma_{ab}$ is also specified by three
independent functions, as it is a symmetric $2\times2$ matrix. We may
therefore use the gauge symmetries (see \reef{reparams}, \reef{weyl})
to choose $\gamma_{ab}$ to be a particular form:
\begin{equation}
\gamma_{ab}=\eta_{ab}=\pmatrix{-1&0\cr\phantom{-}0&1}\ .
\labell{conformalgauge}
\end{equation}
In this ``conformal'' gauge, our $X^\mu$ equations of motion
\reef{motion} become:
\begin{equation}
\left({\partial^2\over\partial\sigma^2}-{\partial^2\over\partial\tau^2}
\right)X^\mu(\tau,\sigma)=0\ ,
\end{equation}
the two dimensional wave equation. As this is
$\partial_{\sigma^+}\partial_{\sigma^-} X^\mu=0$, we see that the full
solution to the equation of motion can be written in the form:
\begin{equation}
X^\mu(\sigma,\tau)=X_L^\mu(\sigma^+)+X^\mu_R(\sigma^-)\ ,
\labell{waves}
\end{equation}
where $\sigma^\pm\equiv\tau\pm\sigma$.

\ennbee{Write $\sigma^\pm=\tau\pm\sigma$. This gives metric
$ds^2=-d\tau^2+d\sigma^2\to -d\sigma^+d\sigma^-.$ So we have
$\eta_{-+}=\eta_{+-}=-1/2$, $\eta^{-+}=\eta^{+-}=-2$ and
$\eta_{++}=\eta_{--}=\eta^{++}=\eta^{--}=0$. Also,
$\partial_\tau=\partial_++\partial_-$ and
$\partial_\sigma=\partial_+-\partial_-$.}

Our constraints on the stress tensor become:
\begin{eqnarray}
T_{\tau\sigma}&=&T_{\sigma\tau}\equiv {1\over\alpha^\prime} {\dot X}^\mu
X^\prime_\mu=0\nonumber \\
T_{\sigma\sigma}&=&T_{\tau\tau}={1\over2\alpha^\prime}\left({\dot X}^\mu{\dot
X}_\mu+X^\prime\!{}^\mu X^\prime_\mu\right)=0\ ,
\end{eqnarray}
or
\begin{eqnarray}
T_{++}&=&{1\over2}(T_{\tau\tau}+T_{\tau\sigma})=
 {1\over\alpha^\prime}\partial_+X^\mu\partial_+X_\mu\equiv 
 {1\over\alpha^\prime}{\dot X}^2_L=0\nonumber\\
T_{--}&=&{1\over2}(T_{\tau\tau}-T_{\tau\sigma})=
 {1\over\alpha^\prime}\partial_-X^\mu\partial_-X_\mu\equiv  
{1\over\alpha^\prime}{\dot X}^2_R=0\ ,
\end{eqnarray}
and $T_{-+}$ and $T_{+-}$ are identically zero.

{\subby{\it The Mode Decomposition}}
Our equations of motion \reef{waves}, with our boundary conditions
\reef{neumann} and \reef{periodic} have the simple solutions:
\begin{equation}
X^\mu(\tau,\sigma)=x^\mu+2\alpha^\prime
p^\mu\tau+i(2\alpha^\prime)^{1/2}\sum_{n\neq0}{1\over n}\alpha_n^\mu
e^{-in\tau}\cos n\sigma\ ,
\labell{openmodes}
\end{equation}
for the open string
and 
\begin{eqnarray}
X^\mu(\tau,\sigma)&=&X^\mu_R(\sigma^-)+X_L^\mu(\sigma^+)\nonumber \\
X^\mu_R(\sigma^-)&=&{1\over2}x^\mu+\alpha^\prime
p^\mu\sigma^-+i\left({\alpha^\prime\over2}\right)^{1/2}
\sum_{n\neq0}{1\over n} \alpha_n^\mu e^{-2in\sigma^-}\nonumber\\
X^\mu_L(\sigma^+)&=&{1\over2}x^\mu+\alpha^\prime
p^\mu\sigma^++i\left({\alpha^\prime\over2}\right)^{1/2}
\sum_{n\neq0}{1\over n} {\tilde\alpha}_n^\mu e^{-2in\sigma^+}\ ,
\labell{closedmodes}
\end{eqnarray}
for the closed string, where, to ensure a real solution we impose
$\alpha_{-n}^\mu=(\alpha_n^\mu)^*$ and
${\tilde\alpha}_{-n}^\mu=({\tilde\alpha}_n^\mu)^*$. Note that $x^\mu$
and $p^\mu$ are the centre of mass position and momentum,
respectively. In each case, we can identify $p^\mu$ with the zero mode
of the expansion:
\begin{eqnarray}
\mbox{open
string:}\qquad\,\,\alpha_0^\mu&=&(2\alpha^\prime)^{1/2}p^\mu;\nonumber\\
\mbox{closed
string:}\qquad\,\,\alpha_0^\mu&=&\left({\alpha^\prime\over2}\right)^{1/2}
p^\mu\ .
\labell{zeromodes}
\end{eqnarray}

\ennbee{Notice that the mode expansion for the closed string
\reef{closedmodes} is simply that of a pair of independent left and
right moving travelling waves going around the string in opposite
directions. The open string expansion \reef{openmodes} on the other
hand, has a standing wave for its solution, representing the left and
right moving sector reflected into one another by the Neumann boundary
condition~\reef{neumann}.}

{\subby{\it A Residual Symmetry}}
Actually, we have not gauged away all of the local symmetry by
choosing the gauge \reef{conformalgauge}. We can do a left--right
decoupled change of variables:
\begin{equation}
\sigma^+\to f(\sigma^+)=\sigma^{\prime+};\,\, 
\sigma^-\to g(\sigma^-)=\sigma^{\prime-}\ .
\end{equation}
Then, as 
\begin{equation}
\gamma^\prime_{ab}=
{\partial\sigma^c\over\partial\sigma^{\prime a}}
{\partial\sigma^d\over\partial\sigma^{\prime b}}\gamma_{cd}\ ,
\end{equation}
we have 
\begin{equation}
\gamma^\prime_{+-}= \left({\partial
f(\sigma^+)\over\partial\sigma^{+}} {\partial
g(\sigma^-)\over\partial\sigma^{-}}\right)^{-1}\gamma_{+-}\ .
\end{equation}
However, we can undo this with a Weyl transformation of the form
\begin{equation}
\gamma^\prime_{+-}=\exp(2\omega_L(\sigma^+)
+2\omega_R(\sigma^-))\gamma_{+-}\ ,
\end{equation}
if $\exp(-2\omega_L(\sigma^+))=\partial_+f(\sigma^+)$ and
$\exp(-2\omega_R(\sigma^-))=\partial_-g(\sigma^-)$. So we still have a
residual ``conformal'' symmetry. As $f$ and $g$ are independent
arbitrary functions on the left and right, we have an infinite number
of conserved quantities on the left and right. This is because the
conservation equation $\nabla_aT^{ab}=0$, together with the result
$T_{+-}=T_{-+}=0$, turns into:
\begin{equation}
\partial_-T_{++}=0 \quad \mbox{and} \quad \partial_+T_{--}=0\ ,
\end{equation}
but since $\partial_-f=0=\partial_+g$, we have
\begin{equation}
\partial_-(f(\sigma^+)T_{++})=0\quad \mbox{and}\quad
\partial_+(g(\sigma^-)T_{--})=0\ ,
\end{equation}
resulting in an infinite number of conserved quantities.  The
fact that we have this infinite dimensional conformal symmetry is the
basis of some of the most powerful tools in the subject, for computing
in perturbative string theory.

{\subby{\it Hamiltonian Dynamics}}

Our Lagrangian density is 
\begin{equation}
{\cal L}=-{1\over4\pi\alpha^\prime}\left(\partial_\sigma
X^\mu\partial_\sigma X_\mu-\partial_\tau X^\mu\partial_\tau
X_\mu\right)\ ,
\end{equation}
from which we can derive that the conjugate momentum to $X^\mu$ is
\begin{equation}
\Pi^\mu={\delta{\cal L}\over\delta(\partial_\tau
X^\mu)}={1\over2\pi\alpha^\prime}{\dot X}^\mu\ .
\labell{conjugate}
\end{equation}
 So we have the equal time
Poisson brackets:
\begin{eqnarray}
&&\left[X^\mu(\sigma),\Pi^\nu(\sigma^\prime)\right]_{\rm
P.B.}=\eta^{\mu\nu}\delta(\sigma-\sigma^\prime)\ ,\\
&&\left[\Pi^\mu(\sigma),\Pi^\nu(\sigma^\prime)\right]_{\rm
P.B.}=0\ ,
\labell{poiss}
\end{eqnarray}
with the following results on the  oscillator modes:
\begin{eqnarray}
\left[\alpha_m^\mu,\alpha_n^\nu\right]_{\rm
P.B.}&=&\left[{\tilde\alpha}_m^\mu,{\tilde\alpha}_n^\nu\right]_{\rm
P.B.}=im\delta_{m+n}\eta^{\mu\nu}\nonumber\\
\left[p^\mu,x^\nu\right]_{\rm
P.B.}&=&\eta^{\mu\nu};\quad
\left[\alpha_m^\mu,{\tilde\alpha}_n^\nu\right]_{\rm
P.B.}=0\ .
\labell{poisson}
\end{eqnarray}
We can form the Hamiltonian density
\begin{eqnarray}
{\cal H}={\dot X}^\mu\Pi_\mu-{\cal
L}={1\over4\pi\alpha^\prime}\left(\partial_\sigma
X^\mu\partial_\sigma X_\mu+\partial_\tau X^\mu\partial_\tau
X_\mu\right)\ ,
\labell{destiny}
\end{eqnarray}
from which we can construct the Hamiltonian $H$ by integrating along
the length of the string.  This results in:
\begin{eqnarray}
H&=&\int_0^\pi d\sigma\, {\cal
H}(\sigma)={1\over2}\sum_{-\infty}^\infty
\alpha_{-n}\cdot\alpha_n\qquad\mbox{(open)}\nonumber\\
H&=&\int_0^{2\pi} d\sigma\, {\cal
H}(\sigma)={1\over2}\sum_{-\infty}^\infty
\left(\alpha_{-n}\cdot\alpha_n
+{\tilde\alpha}_{-n}\cdot{\tilde\alpha}_n\right)\qquad
\mbox{(closed)}\ .
\labell{hammy}
\end{eqnarray}
(We have used the notation
$\alpha_n\cdot\alpha_n\equiv\alpha^\mu_n\alpha_{n\mu}$)
The constraints $T_{++}=0=T_{--}$ on our energy--momentum tensor can
be expressed usefully in this language. We impose them mode by mode in
a Fourier expansion, defining:
\begin{eqnarray}
L_m={T\over2}\int_0^\pi
e^{-2im\sigma}T_{--}d\sigma={1\over2}\sum_{-\infty}^\infty
\alpha_{m-n}\cdot\alpha_n\ ,
\end{eqnarray}
and similarly for ${\bar L}_m$, using $T_{++}$. Using the Poisson
brackets \reef{poisson}, these can be shown to satisfy the ``Virasoro''
algebra:
\begin{eqnarray}
\left[L_m,L_n\right]_{\rm P.B.}&=&i(m-n)L_{m-n};\quad \left[{\bar
L}_m,{\bar L}_n\right]_{\rm P.B.}=i(m-n){\bar L}_{m-n};\nonumber\\
\left[{\bar
L}_m,L_n\right]_{\rm P.B.}&=&0\ .  \labell{virasoro}
\end{eqnarray}
Notice that there is a nice relation between the zero modes of our
expansion and the Hamiltonian:
\begin{eqnarray}
H=L_0\qquad\mbox{(open)};\quad H=L_0+{\bar L}_0\qquad\mbox{(closed)}\ .
\labell{hamilton}
\end{eqnarray}
So to impose our constraints, we can do it mode by mode and ask that
$L_m=0$ and ${\bar L}_m=0$, for all $m$. Looking at the zeroth
constraint results in something interesting. Note that
\begin{eqnarray}
L_0&=&{1\over2}\alpha_0^2
+2\times{1\over2}
\sum_{n=1}^\infty\alpha_{-n}\cdot\alpha_n+{D\over2}\sum_{n=1}^\infty
n\nonumber\\
&=&\alpha^\prime
p^\mu p_\mu+\sum_{n=1}^\infty\alpha_{-n}\cdot\alpha_n+{\rm const}\nonumber\\
&=&-\alpha^\prime M^2+\sum_{n=1}^\infty\alpha_{-n}\cdot\alpha_n+{\rm const}\ ,
\labell{expand}
\end{eqnarray}
where the constant is suspiciously infinite. We will ignore it for
now, and discuss it in the next section, where we study the quantum
theory.  Requiring $L_0$ to be zero ---diffeomorphism invariance---
results in a (spacetime) mass relation:
\begin{equation}
M^2= {1\over\alpha^\prime}\sum_{n=1}^\infty\alpha_{-n}
\cdot\alpha_n\qquad\mbox{(open)}\ ,
\labell{openmass}
\end{equation}
where we have used the zero mode relation \reef{zeromodes} for the
open string.  A similar exercise produces the mass relation for the
closed string:
\begin{equation}
M^2= {2\over\alpha^\prime}\sum_{n=1}^\infty\left(
\alpha_{-n} \cdot\alpha_n
+{\tilde\alpha}_{-n} \cdot{\tilde\alpha}_n\right) \qquad\mbox{(closed)}\ .
\labell{closedmass}
\end{equation}
These formulae \reef{openmass} and \reef{closedmass} give us the
result for the mass of a state in terms of how many oscillators are
excited on the string. The masses are set by the string tension
$T=(2\pi\alpha^\prime)^{-1}$, as they should be. Let us not dwell for
too long on these formulae however, as they are significantly modified
when we quantise the theory, since we have to understand the infinite
constant which we ignored.

\subsection{Quantised Bosonic Strings}
For our purposes, the simplest route to quantisation will be to
promote everything we met previously to operator statements, replacing
Poisson Brackets by commutators in the usual fashion:
$[\phantom{X},\phantom{X}]_{\rm P.B.}\to -i[\phantom{X},\phantom{X}]$.
This gives:
\begin{eqnarray}
\left[X^\mu(\tau,\sigma),
\Pi^\nu(\tau,\sigma^\prime)\right]
&=&i\eta^{\mu\nu}\delta(\sigma-\sigma^\prime)\ ;
\quad
\left[\Pi^\mu(\tau,\sigma),\Pi^\nu(\tau,\sigma^\prime)\right]=0
\nonumber\\
\left[\alpha_m^\mu,\alpha_n^\nu\right]&=&
\left[{\tilde\alpha}_m^\mu,{\tilde\alpha}_n^\nu\right]
\,=\,m\delta_{m+n}\eta^{\mu\nu}\nonumber\\
\left[x^\nu,p^\mu\right]&=&i\eta^{\mu\nu};\quad
\left[\alpha_m^\mu,{\tilde\alpha}_n^\nu\right]\,=\,0\ .  \labell{commute}
\end{eqnarray}

\ennbee{One of the first things that we ought to notice here is that
  $\sqrt{m}\alpha_{\pm m}^\mu$ are like creation and annihilation
  operators for the harmonic oscillator. There are actually $D$
  independent families of them ---one for each spacetime dimension---
 labelled by $\mu$.}

In the usual fashion, we will define our Fock space such that
$|0;k\!>$ is an eigenstate of $p^\mu$ with centre of mass momentum
$k^\mu$. This state is annihilated by $\alpha^\nu_m$.

What about our operators, the $L_m$? Well, with the usual ``normal
ordering'' prescription that all annihilators are to the right, the
$L_m$ are all fine when promoted to operators, except the Hamiltonian,
$L_0$. It needs more careful definition, since $\alpha^\mu_n$ and
$\alpha^\mu_{-n}$ {\it do not commute}. Indeed, as an operator, we
have that
\begin{equation}
L_0={1\over2}\alpha_0^2
+\sum_{n=1}^\infty\alpha_{-n}\cdot\alpha_n+\mbox{constant}\ ,
\end{equation}
where the apparently infinite constant is composed as  copy of the infinite
sum $(1/2)\sum_{n=1}^\infty n$ for each of the $D$ families of
oscillators.  As is of course to be anticipated, this
infinite constant can be regulated to give a finite answer,
corresponding to the total zero point energy of all of the harmonic
oscillators in the system.

\subby{The Constraints and Physical States}

For now, let us not worry about the value of the constant, and simply
impose our constraints on a state $|\phi\!>$ as:
\begin{equation}
(L_0-a)|\phi\!>=0;\qquad L_m|\phi\!>=0\quad \mbox{for } m>0\ ,
\labell{constraints}
\end{equation}
where our infinite constant is set by $a$, which is to be computed.
There is a reason why we have not also imposed this constraint for the
$L_{-m}$'s. This is because the Virasoro algebra \reef{virasoro} in
the quantum case is:
\begin{eqnarray}
\left[L_m,L_n\right]&=&(m-n)L_{m-n}+{D\over12}(m^3-m)\delta_{m+n};\quad
\left[{\bar L}_m,L_n\right]=0;\nonumber\\ 
\left[{\bar L}_m,{\bar L}_n\right]&=&(m-n){\bar
L}_{m-n}+{D\over12}(m^3-m)\delta_{m+n}\ ,
\labell{virasalgebra}
\end{eqnarray}
There is a central term in the algebra, which produces a non--zero
constant when $m=n$. Therefore, imposing both $L_m$ and $L_{-m}$ would
produce an inconsistency.

Note now that the first of our constraints \reef{constraints} produces
a modification to the mass formulae\footnote{This assumes that the
constant $a$ on each side are equal. At this stage, we have no other
choice. We have isomorphic copies of the same open string on the left
and the right, for which the values of $a$ are by definition the
same. When we have more than one consistent conformal field theory to
choose from, then we have the freedom to consider having
non--isomorphic sectors on the left and right. This is how the
heterotic string is made, for example.\cite{heterotic}}:
\begin{eqnarray}
M^2&=& {1\over\alpha^\prime}\left(\sum_{n=1}^\infty\alpha_{-n}
\cdot\alpha_n-a\right)\qquad\mbox{(open)}\nonumber\\
M^2&=&{2\over\alpha^\prime}\left(\sum_{n=1}^\infty\left( \alpha_{-n}
\cdot\alpha_n +{\tilde\alpha}_{-n} \cdot{\tilde\alpha}_n\right)
-2a\right) \qquad\mbox{(closed)}\ .  \labell{masses}
\end{eqnarray}
Notice that we can denote the (weighted) number of oscillators excited
as $N=\sum \alpha_{-n} \cdot\alpha_n$ $(=\sum nN_n)$ on the left and
${\bar N}=\sum {\tilde\alpha}_{-n} \cdot{\tilde\alpha}_n$ $(=\sum
n{\bar N}_n)$ on the right. $N_n$ and ${\bar N}_n$ are the true count,
on the left and right, of the number of copies of the oscillator
labelled by $n$ is present.

There is an extra condition in the closed string case. While
$L_0+{\bar L}_0$ generates time translations on the world sheet (being
the Hamiltonian), the combination $L_0-{\bar L}_0$ generates
translations in $\sigma$.  As there is no physical significance to
where on the string we are, the physics should be invariant under
translations in $\sigma$, and we should impose this as an operator
condition on our physical states:
\begin{equation}
(L_0-{\bar L}_0)|\phi>=0\ ,
\labell{levelmatch}
\end{equation}
which results in the ``level--matching'' condition $N={\bar N}$,
equating the number of oscillators excited on the left and the right.

In summary then, we have two copies of the open string on the left and
the right, in order to construct the closed string. The only extra
subtlety is that we should use the correct zero mode relation
\reef{zeromodes} and match the number of oscillators on each side
according to the level matching condition \reef{levelmatch}.

\subby{The Intercept and Critical Dimensions}

Let us consider the spectrum of states level by level, and uncover
some of the features, focusing on the open string sector.
Our first and simplest state is at level 0, {\it i.e.,} no
oscillators excited at all. There is just some centre of mass momentum
that it can have, which we shall denote as $k$. Let us write this
state as $|0;k\!>$. The first of our constraints \reef{constraints}
leads to an expression for the mass:
\begin{equation}
(L_0-a)|0;k\!>=0\qquad\Rightarrow\alpha^\prime k^2=a,\qquad\mbox{so
}M^2=-{a\over\alpha^\prime}\ .
\end{equation}
This state is a tachyonic state, having negative mass--squared.

The next simplest state is that with momentum $k$, and one oscillator
excited. We are also free to specify a polarisation vector
$\zeta^\mu$. We denote this state as
$|\zeta,k>\equiv(\zeta\cdot\alpha_{-1})|0;k\!>$; it starts out the
discussion with $D$ independent states. The first thing to observe is
the norm of this state:
\begin{eqnarray}
<\!\zeta;k||\zeta;k^\prime\!> &=&<\!0;k|\zeta^*\cdot\alpha_{1}
\zeta\cdot\alpha_{-1}|0;k^\prime\!>\nonumber\\
&=&\zeta^{*\mu}\zeta^\nu<\!0;k|\alpha^\mu_{1}
\cdot\alpha^\nu_{-1}|0;k^\prime\!> \nonumber\\
&=&\zeta\cdot\zeta<\!0;k|0;k^\prime\!>
=\zeta\cdot\zeta(2\pi)^D\delta^D(k-k^\prime)\ ,
\end{eqnarray}
where we have used the commutator \reef{commute} for the
oscillators. From this we see that the time-like component of $\zeta$
will produce a state with {\it negative norm}. Such states cannot be
made sense of in a unitary theory, and are often called\footnote{These
are not to be confused with the ghosts of the friendly variety
---Faddeev--Popov ghosts. These negative norm states are problematic
and need to be removed. 
%Perhaps they are better called the ``Phantom Menace''.
} 
``ghosts''.

Let us study the first constraint:
\begin{equation}
(L_0-a)|\zeta;k\!>=0\qquad\Rightarrow\quad\alpha^\prime k^2+1=a,\qquad
M^2={1-a\over\alpha^\prime}\ .
\end{equation}
The next constraint gives:
\begin{equation}
(L_1)|\zeta;k\!>=\sqrt{\alpha^\prime\over2}
k\cdot\alpha_1\zeta\cdot\alpha_{-1}|0;k>=0\qquad\Rightarrow,\qquad
k\cdot\zeta=0\ .
\end{equation}

Actually, at level 1, we can also make a special state of interest:
$|\psi\!>\equiv L_{-1}|0;k\!>$. This state has the special property that
it is orthogonal to any physical state, since
$<\!\phi|\psi\!>=<\!\psi|\phi\!>^*=<\!0;k|L_1|\phi\!>=0$. It also has
$L_1|\psi\!>=2L_0|0;k\!>=\alpha^\prime k^2|0;k>.$ This state is called a
``spurious'' state. 

So we note that there are three interesting cases for the level 1
physical state we have been considering:

\begin{enumerate}
\item{$a<1\Rightarrow M^2>0:$
\begin{itemize}
\item{momentum $k$ is timelike.}
\item{We can choose a frame where it is $(k,0,0,\ldots)$}
\item{Spurious state is not physical, since $k^2\neq0$.}
\item{$k\cdot\zeta=0$ removes the timelike polarisation. $D-1$ states left}
\end{itemize}}
\item{$a>1\Rightarrow M^2<0:$
\begin{itemize}
\item{momentum $k$ is spacelike.}
\item{We can choose a frame where it is $(0,k_1,k_2,\ldots)$}
\item{Spurious state is not physical, since $k^2\neq0$}
\item{$k\cdot\zeta=0$ removes a spacelike polarisation. $D-1$ states left, one which is
including ghosts and tachyons.}
\end{itemize}}
\item{$a=1\Rightarrow M^2=0:$
\begin{itemize}
\item{momentum $k$ is null.}
\item{We can choose a frame where it is $(k,k,0,\ldots)$}
\item{Spurious state is  physical {\it and} null, since $k^2=0$}
\item{$k\cdot\zeta=0$ and $k^2=0$ removes two polarisations; $D-2$
states left}
\end{itemize}}
\end{enumerate}

So if we choose case (3), we end up with the special situation that we
have a massless vector in the $D$ dimensional target spacetime. It
even has an associated gauge invariance: since the spurious state is
physical and null, and therefore we can add it to our physical state
with no physical consequences, defining an equivalence relation:
\begin{equation}
|\phi>\sim|\phi>+\lambda|\psi>\qquad\Rightarrow\qquad
\zeta^\mu\sim\zeta^\mu+\lambda k^\mu\ .  \labell{spurious}
\end{equation}
Case (1), while interesting, corresponds to a massive vector, where
the extra state plays the role of a longitudinal component. Case (2)
seems bad. We shall choose case (3), where $a=1$.

It is interesting to proceed to level two to construct physical and
spurious states, although we shall not do it here. The physical states
are massive string states. If we insert our level one choice $a=1$ and
see what the condition is for the maximal space spurious states to be
both physical and null, we find that there is a condition on the
spacetime dimension\footnote{We get a condition on the spacetime
dimension here because level 2 is the first time it can enter our
formulae for the norms of states, {\it via} the central term in the
the Virasoro algebra \reef{virasalgebra}.}: $D=26$.

In summary, we see that $a=1$, $D=26$ for the open bosonic string
gives a family of extra null states, giving something analogous to a
point of ``enhanced gauge symmetry'' in the space of possible string
theories. This is called a ``critical'' string theory, for many
reasons.  We have the 24 states of a massless vector we shall loosely
called the photon, $A_\mu$, since it has a $U(1)$ gauge invariance
\reef{spurious}.  There is a tachyon of $M^2=-1/\alpha^\prime$ in the
spectrum, which will not trouble us unduly.  We will actually remove
it in going to the superstring case. Tachyons will reappear from time
to time, representing situations where we have an unstable
configuration (as happens in field theory frequently).  Generally, it
seems that we should think of tachyons in the spectrum as pointing us
towards an instability, and in many cases, the source of the
instability is manifest. Indeed, this will be put to good use in
constructing stable non--BPS D--brane solitons in the lectures of John
Schwarz in this TASI school\cite{john}, following a recently developed
technique of Sen.~\cite{sentach} In these lectures, we will
try to map out the supersymmetric landscape, but ocassionally this
line of reasoning will appear.

Our analysis here extends to the closed string, since we can take two
copies of our result, use the appropriate zero mode relation
\reef{zeromodes}, and level matching. At level zero we get the closed
string tachyon which has $M^2=-4/\alpha^\prime$. At level zero we get
a tachyon with mass given by $M^2=-4/\alpha^\prime$, and at level 1
we get 24$^2$ massless states from
$\alpha^\mu_{-1}{\tilde\alpha}^\nu_{-1}|0;k\!>$. The traceless
symmetric part is the graviton, $G_{\mu\nu}$ and the antisymmetric
part, $B_{\mu\nu}$, is sometimes called the Kalb--Ramond field, and
the trace is is the dilaton, $\Phi$.

\subby{Had We Been More Careful}

A more careful treatment of our gauge fixing procedure
\reef{conformalgauge} would had seen us introduce Faddev--Popov
ghosts $(b,c)$ ({\it i.e.,} friendly ghosts) to ensure that we stay on
our chosen gauge slice in the full theory. Our resulting two
dimensional conformal field theory would have had an extra sector
coming from the $(b,c)$ ghosts.

The central term in the Virasoro algebra \reef{virasalgebra}
represents an anomaly in the transformation properties of the stress
tensor, spoiling its properties as a tensor under general coordinate
transformations. Generally:
\begin{equation}
 \left({\partial\sigma^{\prime+}\over\partial\sigma^+}\right)^2
T^\prime_{++}(\sigma^{\prime+})=
T_{++}(\sigma^+)-{{\rm c}\over12}
\left\{{2\partial_\sigma^3\sigma^\prime\partial_\sigma\sigma^\prime
-3\partial_\sigma^2\sigma^\prime\partial^2_\sigma\sigma^\prime\over
2\partial_\sigma\sigma^\prime\partial_\sigma\sigma^\prime
}\right\}\ ,
\end{equation}
where $\rm c$ is a number which depends upon the content of the theory. In
our case, we have $D$ bosons, which each contribute $1$ to c, for a
total anomaly of~$D$.

The ghosts do two crucial things: They contribute to the anomaly the
amount $-26$, and therefore we can retain all our favourite symmetries
for the dimension $D=26$. They also cancel the contributions to the
vacuum energy coming from the oscillators in the $\mu=0,1$ sector,
leaving $D-2$ transverse oscillators' contribution.

\begin{figure}[th]
\centerline{\psfig{figure=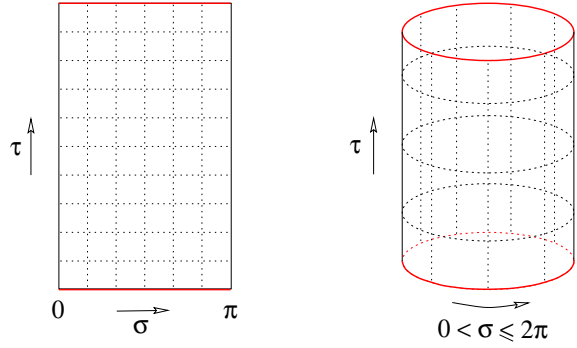,height=1.8in}}
\caption{String worldsheets as boxes upon which live two dimensional
conformal field theory.}
\label{boxes}
\end{figure}

The regulated value of $-a$ is the vacuum or ``zero point energy''
(z.p.e.) of the transverse modes of the theory. This zero point energy
is simply the Casimir energy arising from the fact that the two
dimensional field theory is in a box. The box is the infinite strip,
for the case of an open string, or the infinite cylinder, for the case
of the closed string (see figure \ref{boxes}).
A periodic (integer moded) boson such as the types we have here,
$X^\mu$, each contribute $-1/24$ to the vacuum energy (see insert 3
(p.\pageref{insert3})on a quick way to compute this). So we see that
in 26 dimensions, with only 24 contributions to count (see previous
paragraph), we get that $-a=24\times(-1/24)=-1$. (Notice that from
\reef{expand}, this implies that $\sum_{n=1}^\infty n=-1/12$, which is
in fact true in $\zeta$--function regularisation.)

Later, we shall have world sheet fermions $\psi^\mu$ as well, in the
supersymmetric theory. They each contribute $1/2$ to the anomaly.
World sheet superghosts will cancel the contributions from
$\psi^0,\psi^1$. Each anti--periodic fermion will give a
z.p.e. contribution of $-1/48$.

Generally, taking into account the possibility of both periodicities
for either bosons or fermions:
\begin{eqnarray}
\mbox{z.p.e.}&=&{1\over2}\omega\qquad\mbox{for boson};
\qquad-{1\over2}\omega\qquad\mbox{for fermion}\nonumber\\
\omega&=&{1\over24}-{1\over8}(2\theta-1)^2\qquad
\left\{\matrix{\theta&=&0&\quad\mbox{(integer modes)}\cr
\theta&=&{1\over2}&\quad\mbox{(half--integer modes)}}
\right.
\,\labell{zpe}
\end{eqnarray}
This is a formula which we shall use many times in what is to come.

\insertion{3}{Cylinders, Strips and the Complex
  Plane\label{insert3}}{As promised earlier, we will go from
  Lorentzian to Euclidean signature (making the action real) by
  sending $\tau\to i\tau$. Another thing we will often do is work on
  the complex plane, instead of the original world sheets we started
  with.

We go from one to the other using the exponential map, defining a
complex coordinate on the plane $z=e^{\tau-i\sigma}=e^w$. The closed
string's cylinder and the open string's strip (see figure
\ref{boxes}) map to the complex plane and the (upper) half complex
plane, respectively:
%\begin{figure}[ht]
\centerline{\psfig{figure=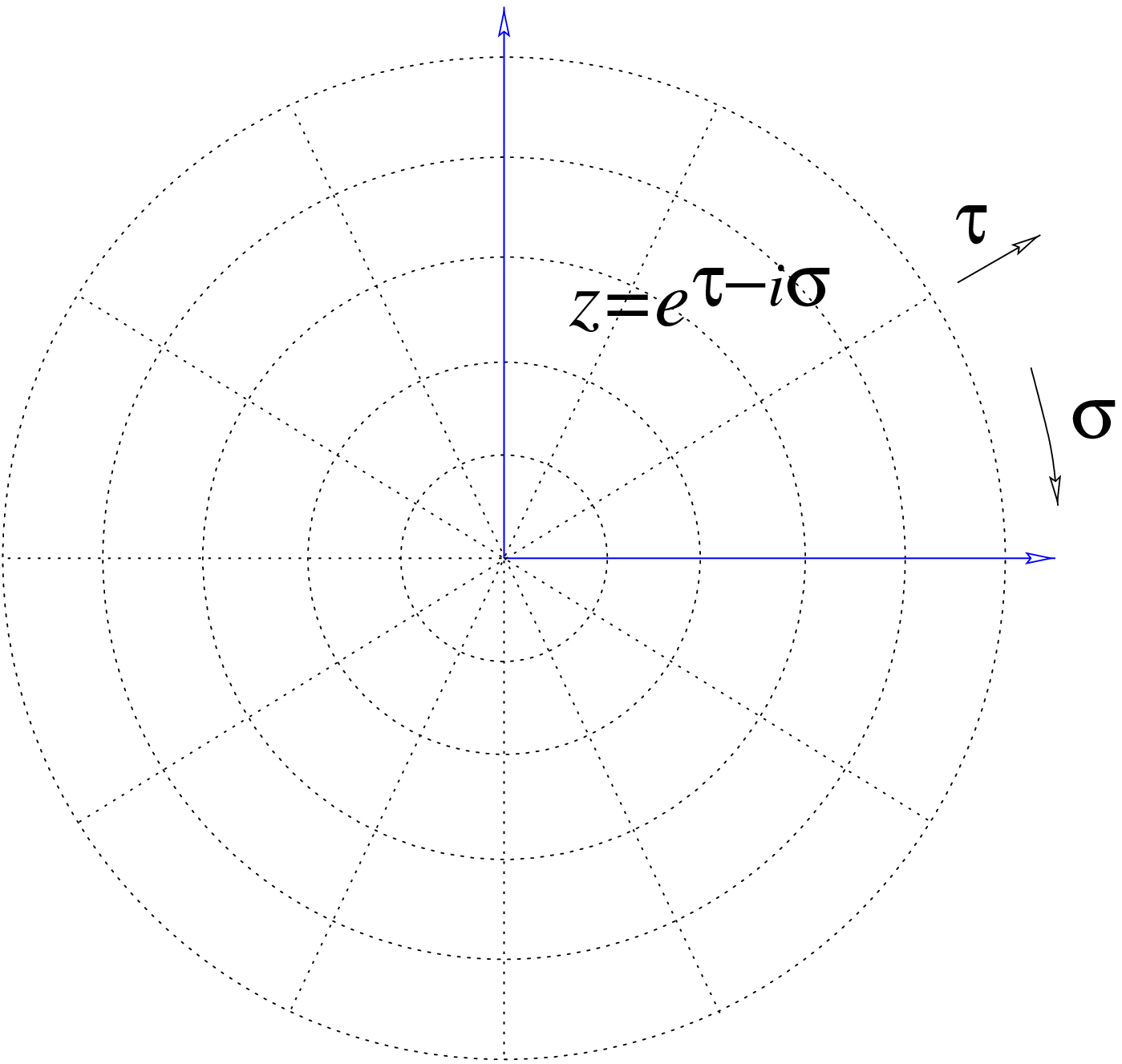,height=2.0in}}
%\end{figure}
Note that the Fourier expansions we have been working with to define
the modes become Laurent expansions on the complex plane, {\it e.g.}:
$$
T_{zz}(z)=\sum_{m=-\infty}^\infty{L_m\over z^{m+2}}\ .
$$
One of the most straightforward exercises is to compute the zero point
energy of the cylinder or strip (for a field of central charge $c$) by
starting with the fact that the plane has no Casimir energy. One
simply plugs the exponential change of coordinates $z=e^w$ into the
anomalous transformation for the energy momentum tensor and compute
the contribution to $T_{ww}$ starting with $T_{zz}$:
$$
T_{ww}=-z^2T_{zz}-{c\over 24}\ ,
$$
which results in the Fourier expansion on the cylinder, in terms of
the modes:
$$
T_{ww}(w)=-\sum_{m=-\infty}^\infty
\left(L_m-{c\over24}\delta_{m,0}\right)e^{i\sigma-\tau}\ .
$$}

\subby{States and Operators}

\begin{figure}
\centerline{\psfig{figure=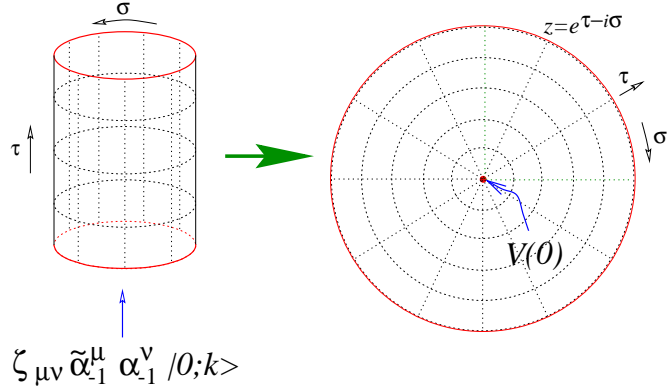,height=2.0in}}
\caption{The correspondence between states and operator insertions. A
closed string (graviton) state
$\zeta_{\mu\nu}\alpha^\mu_{-1}{\tilde\alpha}^\nu_{-1}|0;k\!>$ is set
up on the closed string at $\tau=-\infty$ and it propagates in.  This
is equivalent to inserting a graviton vertex operator
$V^{\mu\nu}(z)=\zeta_{\mu\nu}\partial_z X^\mu\partial_{\zb} X^\nu
:e^{ik\cdot X}:$ at $z=0$.}
\label{corresp}
\end{figure}

As we learned in insert 3, (p.\pageref{insert3}) we can work on the
complex plane with coordinate $z$. In these coordinates, our mode
expansions \reef{openmodes} and \reef{closedmodes} become:
\begin{equation}
X^\mu(z,\zb)=x^\mu-i\left({\alpha^\prime\over2}\right)^{1/2}
\alpha^\mu_0\ln z\zb+i\left({\alpha^\prime\over2}\right)^{1/2}
\sum_{n\neq0}{1\over n} \alpha_n^\mu \left(z^{-n}+\zb^{-n}\right)\ ,
\labell{opencomplexmodes}
\end{equation}
for the open string, and for the closed:
\begin{eqnarray}
X^\mu(z,\zb)&=&X^\mu_L(z)+X_R^\mu(\zb)\nonumber \\
X^\mu_L(z)&=&{1\over2}x^\mu-i\left({\alpha^\prime\over2}\right)^{1/2}
\alpha^\mu_0\ln z+i\left({\alpha^\prime\over2}\right)^{1/2}
\sum_{n\neq0}{1\over n} \alpha_n^\mu z^{-n}\nonumber\\
X^\mu_R(\zb)&=&{1\over2}x^\mu-i\left({\alpha^\prime\over2}\right)^{1/2}
{\tilde\alpha}^\mu_0\ln \zb+i\left({\alpha^\prime\over2}\right)^{1/2}
\sum_{n\neq0}{1\over n} {\tilde\alpha}_n^\mu \zb^{-n}\ ,
\labell{closedcomplexmodes}
\end{eqnarray}
where we have used the zero mode relations \reef{zeromodes}.
In fact, notice that:
\begin{eqnarray}
\partial_z X^\mu(z)&=&-i\left({\alpha^\prime\over2}\right)^{1/2}
\sum_{n} {\alpha}_n^\mu z^{-n-1}\nonumber\\ 
\partial_{\zb} X^\mu(\zb)&=&-i\left({\alpha^\prime\over2}\right)^{1/2}
\sum_{n} {\tilde\alpha}_n^\mu \zb^{-n-1}\ ,
\labell{fields}
\end{eqnarray}
and that we can invert these to get (for the closed string)
\begin{equation}
{\alpha}_{-n}^\mu=\left({2\over\alpha^\prime}\right)^{1/2}\oint
{dz\over2\pi}z^{-n}\partial_z X^\mu(z)\qquad
{\tilde\alpha}_{-n}^\mu=\left({2\over\alpha^\prime}\right)^{1/2}\oint
{dz\over2\pi}\zb^{-n}\partial_{\zb} X^\mu(z)\ ,
\labell{insertions}
\end{equation}
which are non--zero for $n\geq0$. This is suggestive: Equations
\reef{fields} define left--moving (holomorphic) and right--moving
(anti--holomorphic) fields.  We previously employed the objects on the
left in \reef{insertions} in making states by acting, {\it e.g.},
$\alpha^\mu_{-1}|0;k\!>$. The form of the right hand side suggests that this
is equivalent to performing a contour integral around an insertion of
a pointlike operator at the point $z$ in the complex plane (see figure
\ref{corresp}). For example, $\alpha_{-1}^\mu$ is related to the
residue $\partial_zX^\mu(0)$, while the $\alpha^\mu_{-m}$ correspond
to higher derivatives $\partial_z^m X^\mu(0)$. This is course makes
sense, as higher levels correspond to more oscillators excited on the
string, and hence higher frequency components, as measured by the
higher derivatives.

The state with no oscillators excited (the tachyon), but with some
momentum $k$, simply corresponds in this dictionary to the insertion
of:
\begin{equation}
|0;k>\qquad\Leftrightarrow\qquad 
\int d^2z  :e^{ik\cdot X}:
\end{equation}
This is reasonable, as it is the simplest form that allows the right
behaviour under translations: A translation by a constant vector,
$X^\mu\to X^\mu+T^\mu$, results in a multiplication of the operator
(and hence the state) by a phase $e^{ik\cdot T}$. The normal ordering
signs $::$ are there to remind that the expression means to expand and
keep all creation operators to the right, when expanding in terms of
the $\alpha_{\pm m}$'s.

The closed string level 1 vertex operator corresponds to the emission
or absorption of $G_{\mu\nu}$, $B_{\mu\nu}$ and $\Phi$:
\begin{equation}
\zeta_{\mu\nu}
\alpha_{-1}^\mu{\tilde\alpha}_{-1}^\nu|0;k\!>\qquad\Leftrightarrow\qquad
\int d^2z \,\,\zeta_{\mu\nu}\partial_z X^\mu\partial_{\zb} X^\nu
:e^{ik\cdot X}:
\end{equation}
where the symmetric part of $\zeta_{\mu\nu}$ is the graviton and the
antisymmetric part is the antisymmetric tensor.

More generally, in the full treatment of the string theory, where the
world sheet, ${\cal M}$, is not flat but curved, the vertex operator
is:
\begin{equation}
V={g_s\over\alpha^\prime}\int_{\cal M} d^2\!\sigma\,  \,g^{1/2}
\,\left\{(g^{ab}s_{\mu\nu}+i\epsilon^{ab} a_{\mu\nu})\partial_a
X^\mu\partial_b X^\nu e^{ik\cdot X}+\alpha^\prime\phi R e^{ik\cdot
X}\right\}\ , \labell{vertex}
\end{equation}
where $s$ is symmetric, $a$ is antisymmetric and $\phi$ is a constant,
and we have put in the closed string coupling to take into account the
fact that the world sheet topology changes when we emit or absorb a
closed string state. A linear combination of the trace of $s$ and
$\phi$ turn out to be the dilaton. The $\int d^2\!\sigma\, \, R$ coupling
is included as a possibility simply because it is allowed by Weyl and
reparametrisation invariance, as we stated earlier.

For the open string, the story is similar, but we get two copies of
the relations \reef{insertions} for the single set of modes
$\alpha_{-n}^\mu$ (recall that there are no ${\tilde\alpha}$'s). This
results in, for example the relation for the photon:
\begin{equation}
\zeta_\mu\alpha_{-1}^\mu|0;k\!>\qquad\Leftrightarrow\qquad \int dl
\,\,\zeta_\mu\partial_t X^\mu :e^{ik\cdot X}: \ ,
\labell{photon}
\end{equation}
where the integration is along the real line (the edge of the
half--plane, which corresponds to the vertical edges of the string
world-sheet on the left of figure~\ref{boxes}.  Also, $\partial_t$
means the derivative tangential to the boundary. The tachyon is simply
the boundary insertion of the momentum $:e^{ik\cdot X}:$ alone.

The fact that the photon is associated with the ends of the string is
a sort of quantum version of that which we saw in the classical analysis:
That the ends of the strings move with the speed of light.  Of course,
we see that there are other features which we did not see in the
classical analysis (like the tachyon), but our hard work in trying to
retain the classical symmetries after going to the quantum case has
paid off.

\subsection{Chan-Paton Factors}
\label{chanpat}
\begin{figure}[ht]
\centerline{\psfig{figure=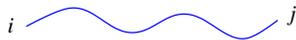,height=0.2in}}
\caption{An open string with Chan--Paton degrees of freedom.}
\label{chanpaton}
\end{figure}

\begin{figure}[ht]
\centerline{\psfig{figure=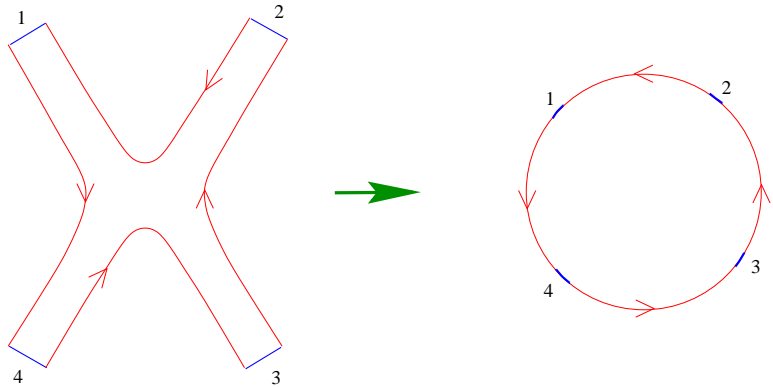,height=2in}}
\caption{ A four--point Scattering of open strings, and its 
conformally related disc amplitude.}
\label{fourpoint}
\end{figure}

While we are remarking upon the behaviour of the ends of the string,
let us endow them with a slightly more interesting property. We can
add non-dynamical degrees of freedom to the ends of the string without
spoiling spacetime Poincar\'e invariance or world--sheet conformal
invariance. These are called ``Chan--Paton'' degrees of freedom and by
declaring that their Hamiltonian is zero, we guarantee that they stay
in the state that we put them in.  In addition to the usual Fock space
labels we have been using for the state of the string, we ask that
each end be in a state $i$ or $j$ for $i,j$ from $1$ to $N$ (see
figure \ref{chanpaton}).
We use a family of $N\times N$ matrices, $\lambda^a_{ij}$, as a basis
into which to decompose a string wavefunction
\begin{equation}
|k;a\rangle = \sum^N_{i,j=1}|k,ij\rangle\lambda^a_{ij}.
\end{equation}
These wavefunctions are called ``Chan-Paton
factors''\cite{chan}. Similarly, all open string vertex operators
carry such factors.  For example, consider the tree--level (disc)
diagram for the interaction of four oriented open strings in
figure~\ref{fourpoint}.
As the Chan--Paton degrees of freedom are non-dynamical, the
right end of string \#1 must be in the same state as the left end of
string \#2, {\it etc.,} as we go around the edge of the disc. 
After summing over all the possible states involved in tying up the
ends, we are left with a trace of the product of Chan--Paton factors,
\begin{equation} 
\lambda^1_{ij}\lambda^2_{jk}\lambda^3_{kl} \lambda^4_{li} =
\Tr(\lambda^1\lambda^2\lambda^3\lambda^4). \label{thefactors} 
\end{equation}
All open string amplitudes will have a trace like this and
 are invariant under a global (on the world--sheet) $U(N)$:~\footnote
{The amplitudes are actually invariant under $GL(N)$, but this does not
leave the norms of states invariant.} 
\begin{equation}
\lambda^i \ \to\ U \lambda^i U^{-1},
\end{equation}
under which the endpoints transform as $\mathbf{N}$ and
$\bar{\mathbf{N}}$.  

Notice that the massless vector vertex operator
$V^{a\mu}=\lambda^a_{ij}\partial_t X^\mu\exp(ik\cdot X)$ transforms as
the adjoint under the $U(N)$ symmetry.  {\it This means that the global
symmetry of the world-sheet theory is promoted to a gauge symmetry in
spacetime.} It is a gauge symmetry because we can make a different
$U(N)$ rotation at separate points $X^\mu(\sigma,\tau)$ in spacetime.
 
\subsection{The Closed String Partition Function}
We have all of the ingredients we need to compute our first one--loop
diagram. It will be useful to do this as a warm up for more
complicated examples later, and in fact we will see structures in this
simple case which will persist throughout.

\begin{figure}[ht]
\centerline{\psfig{figure=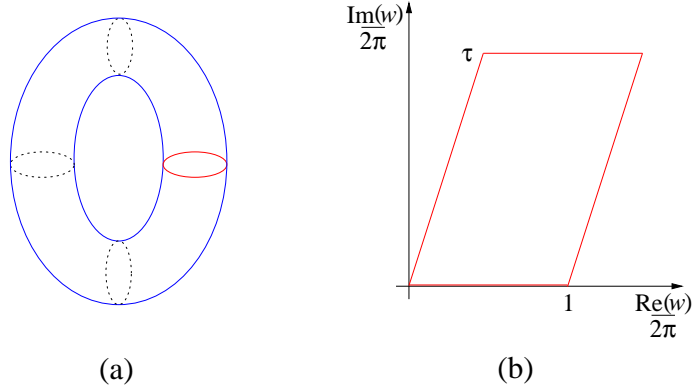,height=2.0in}}
\caption{{\it (a)} A closed string vacuum  diagram  
{\it (b)}. The flat torus and its
  complex structure.}
\label{torus}
\end{figure}

Consider the closed string diagram of figure \ref{torus}{\it (a)}.  This is
a vacuum diagram, since there are no external strings. This torus is
clearly a one loop diagram and in fact it is easily computed. It is
distinguished topologically by having two completely independent
one--cycles. To compute the path integral for this we are instructed,
as we have seen, to sum over all possible metrics representing all
possible surfaces, and hence all possible tori.

\begin{figure}[ht]
\centerline{\psfig{figure=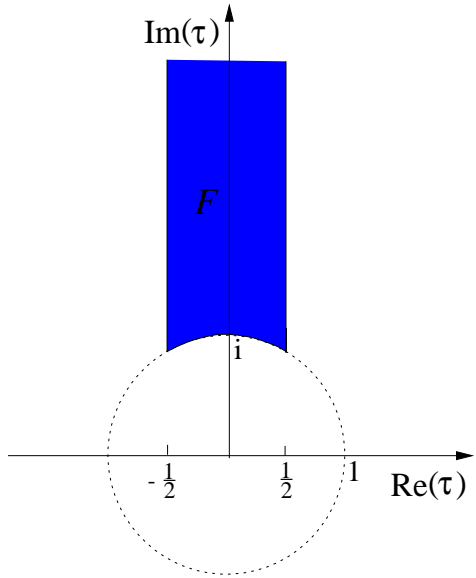,height=3.0in}}
\caption{The space of inequivalent tori is blue.}
\label{fundamental}
\end{figure}

Well, the torus is completely specified
by giving it a flat metric, and a complex structure, $\tau$, with
${\rm Im}\tau\geq0$. It can be described by the lattice given by
quotienting the complex $w$--plane by the equivalence relations
\begin{equation}
w\sim w+2\pi n\ ;\quad w\sim w+2\pi m\tau\ ,
\labell{equivalence}
\end{equation} for any integers $m$ and $n$,
as shown in figure \ref{torus}{\it (b)}. The two one--cycles can be
chosen to be horizontal and vertical. The complex number $\tau$
specifies the {\it shape} of a torus, which cannot be changed by
infinitesimal diffeomorphisms of the metric, and so we must sum
over all all of them. Actually, this naive reasoning will make us
overcount by a lot, since in fact there are a lot of $\tau$'s which
define the same torus.  For example, clearly for a torus with given
value of $\tau$, the torus with $\tau+1$ is the same torus, by the
equivalence relation \reef{equivalence}. The full family of
equivalent tori can be reached from any $\tau$ by the ``modular
transformations'':
\begin{eqnarray}
&T&: \quad\tau\to\tau+1\nonumber\\
&S&: \quad\tau\to -{1\over\tau}\ ,
\labell{modular}
\end{eqnarray}
which generate the group $SL(2,\IZ)$, which is represented here as the
group of $2\times 2$ unit determinant matrices with integer elements:
\begin{equation}
SL(2,\IZ):\quad \tau\to {a\tau+b\over c\tau+d}\ ;\quad\mbox{with}\quad
 \pmatrix{a&b\cr c&d}\ , \quad ad-bc=1\ .  \labell{essell}
\end{equation}
(It is worth noting that the map between tori defined by $S$ exchanges
the two one--cycles, therefore exchanging space and (Euclidean) time.)
The full family of inequivalent tori is given not by the
upper half plane $H_\perp$ ({\it i.e.,} $\tau$ such that ${\rm
  Im}\tau\geq0$) but the quotient of it by the equivalence relation
generated by the group of modular transformations. This is  ${\cal
  F}=H_\perp/PSL(2,\IZ)$, where the $P$ reminds us that we divide by
the extra $\IZ_2$ which swaps the sign on the defining $SL(2,\IZ)$
matrix, which clearly does not give a new torus. The commonly used
fundamental domain in the upper half plane corresponding to the
inequivalent tori is drawn in figure \ref{fundamental}. Any point
outside that can be mapped into it by a modular transformation.

The string propagation on our torus can be described as follows.
Imagine that the string is of length 1, and lies horizontally. Mark a
point on the string. Running time upwards, we see that the string
propagates for a time $t=2\pi{\rm Im}\tau\equiv 2\pi\tau_2$. One it
has got to the top of the diagram, we see that our marked point has
shifted rightwards by an amount $x=2\pi{\rm Re}\tau\equiv 2\pi\tau_1$.
We actually already have studied the operators which perform these two
operations. The operator for time translations is the Hamiltonian
\reef{hamilton}, $H=L_0+{\bar L}_0-({{\rm c}+{\bar {\rm c}}})/24$
while the operator for translations along the string is the momentum
$P=L_0-{\bar L}_0$ discussed above eqn.\reef{levelmatch}. Recall that
${\rm c}={\bar {\rm c}}{=}D{-}2{=}24$.  So our vacuum path
integral is
\begin{equation}
Z={\rm Tr}\left\{e^{-2\pi\tau_2 H}e^{2\pi {\rm i}\tau_1
P}\right\}={\rm Tr}q^{L_0-{c\over24}}{\bar q}^{{\bar L}_0-{{\bar
c}\over24}} \ .
\end{equation}
Here, $q\equiv e^{2\pi {\rm i}\tau}$, and the trace means a sum over
everything which is discrete and an integral over everything which is
continuous, which in this case, is simply $\tau$.  This is easily
evaluated, as the expressions for $L_0$ and ${\bar L}_0$ give a family
of simple geometric sums (see insert 4 (p.\pageref{insert4})), and the
result can be written as:
\begin{equation}
Z=\int_{\cal F} {d^2\tau\over \tau_2} Z(q)\ , \quad \mbox{where}
\end{equation}
\begin{equation}
Z(q)=|\tau_2|^{-12}(q{\bar
q})^{-1}\left|\prod_{n=1}^\infty\left(1-q^n\right)^{-24}\right|^2=
\left(\sqrt{\tau_2}\eta{\bar\eta}\right)^{-24}\ ,
\end{equation}
is the ``partition function'', with Dedekind's function
\begin{equation}
\eta(q)\equiv q^{1\over24}\prod_{n=1}^\infty\left( 1-q^n\right)\
;\quad \eta\left(-{1\over\tau}\right)=\sqrt{-{\rm i}\tau}\,\eta(\tau)\ .
\labell{dedekind}
\end{equation}

\insertion{4}{Partition Functions\label{insert4}}{It is not hard to do
  the sums. Let us look at one dimension, and so one family of
  oscillators $\alpha_{n}$. We need to consider
$$
\Tr\, q^{L_0}=\Tr\, q^{\sum_{n=0}^\infty \alpha_{-n}\alpha_{n}}\ . 
$$
We can see what the operator $q^{\sum_{n=0}^\infty
  \alpha_{-n}\alpha_{n}}$ means if we write it explicitly in a basis
of all possible multiparticle states of the form $\alpha_{-n}|0\!>$,
$(\alpha_{-n})^2|0\!>$, {\it etc.} :
$$
q^{\alpha_{-n}\alpha_{n}}=
\pmatrix{
1& & & & \cr
 &q^n & & & \cr
 & &q^{2n} & & \cr
 & & &q^{3n} & \cr
 & & & &\ddots \cr
}\ ,\nonumber
$$
and so clearly $\Tr q^{ \alpha_{-n}\alpha_{n}}=
\sum_{i=1}^\infty(q^n)^i=(1-q^n)^{-1}$, which is remarkably simple!
The final sum over all modes is trivial, since
$$\Tr\, q^{\sum_{n=0}^\infty \alpha_{-n}\alpha_{n}}=\prod_{n=0}^\infty
\Tr\, q^{ \alpha_{-n}\alpha_{n}}=\prod_{n=0}^\infty(1-q^n)^{-1} \ .
$$
We get a factor like this for all 24 dimensions, and we also get
contributions from both the left and right to give the result.

\medskip

Notice that if our modes were fermions, $\psi_{n}$, things would be
even simpler.  We would not be able to make multiparticle states
$(\psi_{-n})^2|0\!>$, (Pauli), and so we only have a $2{\times}2$
matrix of states to trace in this case, and so we simply get $$\Tr\, q^{
  \psi_{-n}\psi_{n}}=(1+q^n)\ .$$
Therefore the partition function is 
$$\Tr\, q^{\sum_{n=0}^\infty \psi_{-n}\psi_{n}}=\prod_{n=0}^\infty \Tr
q^{\psi_{-n}\psi_{n}}=\prod_{n=0}^\infty(1+q^n) \ .
$$
We will encounter such fermionic cases later.}

This is a pleasingly simple result.  One very interesting property it
has is that it is actually ``modular invariant''. It is invariant
under the $T$ transformation in \reef{equivalence}, since under
$\tau\to\tau+1$, we get that $Z(q)$ picks up a factor $\exp(2\pi{\rm
  i}(L_0-{\bar L}_0))$. This factor is precisely unity, as follows
from the level matching formula \reef{levelmatch}.  Invariance of
$Z(q)$ under the $S$ transformation $\tau\to-1/\tau$ follows from the
property mentioned in \reef{dedekind}, after a few steps of algebra,
and using the result $S:\,\,\tau_2\to\tau_2/|\tau|^2$.

Modular invariance of the
partition function is a crucial property. It means that we are
correctly integrating over all inequivalent tori, which is required of
us by diffeomorphism invariance of the original construction.
Furthermore, we are counting each torus only once, which is of course
important.

Note that $Z(q)$ really deserves the name ``partition function'' since
if it is expanded in powers of $q$ and $\bar q$, the powers in the
expansion ---after multiplication by $4/\alpha^\prime$--- refer to the
(mass)$^2$ level of excitations on the left and right, while the
coeeficient in the expansion gives the degeneracy at that level.  The
degeneracy is the number of partitions of the level number into
positive integers. For example, at level 3 this is 3, since we have
$\alpha_{-3},\alpha_{-1}\alpha_{-2}$, and
$\alpha_{-1}\alpha_{-1}\alpha_{-1}$.

The overall factor of $(q{\bar q})^{-1}$ sets the bottom of the tower
of masses. Note for example that at level zero we have the tachyon,
which appears only once, as it should, with $M^2=-4/\alpha^\prime$. At
level one, we have the massless states, with multiplicity $24^2$,
which is appropriate, since there are $24^2$ physical states in the
graviton multiplet $(G_{\mu\nu}, B_{\mu\nu}, \Phi)$.  Introducing a
common piece of terminology, a term $q^{w_1}{\bar q}^{w_2}$,
represents the appearance of a ``weight'' $(w_1,w_2)$ field in the 1+1
dimensional conformal field theory, denoting its left--moving and
right--moving weights or ``conformal dimensions''.

\subsection{Unoriented Strings}
\label{unory}
\subby{Unoriented Open Strings} There is an operation of world sheet
parity $\Omega$ which takes $\sigma \to \pi-\sigma$, on the open
string, and acts on $z=e^{\tau-i\sigma}$ as $z\leftrightarrow-\zb$.
In terms of the mode expansion \reef{opencomplexmodes}, $X^\mu(z,\zb)
\to X^\mu(-\zb,- z)$ yields
\begin{eqnarray}
x^\mu &\rightarrow& x^\mu \nonumber\\
p^\mu &\rightarrow& p^\mu \nonumber\\
\alpha^\mu_m &\rightarrow& (-1)^m\alpha^\mu_m\ .
\end{eqnarray}
This is a global symmetry of the open string theory and so, we can if
we wish also consider the theory that results when it is gauged, by
which we mean that only $\Omega$--invariant states are left in the
spectrum.  We must also consider the case when we take a string around
a closed loop, it is allowed to come back to itself only up to an over
all action of $\Omega$, which is to swap the ends. This means that we
must include unoriented worldsheets in our analysis. For open strings,
the case of the M\"{o}bius strip is a useful example to keep in mind.
It is on the same footing as the cylinder when we consider gauging
$\Omega$.  The string theories which result from gauging $\Omega$ are
understandably called ``unoriented string theories''.

\begin{figure}[ht]
\centerline{\psfig{figure=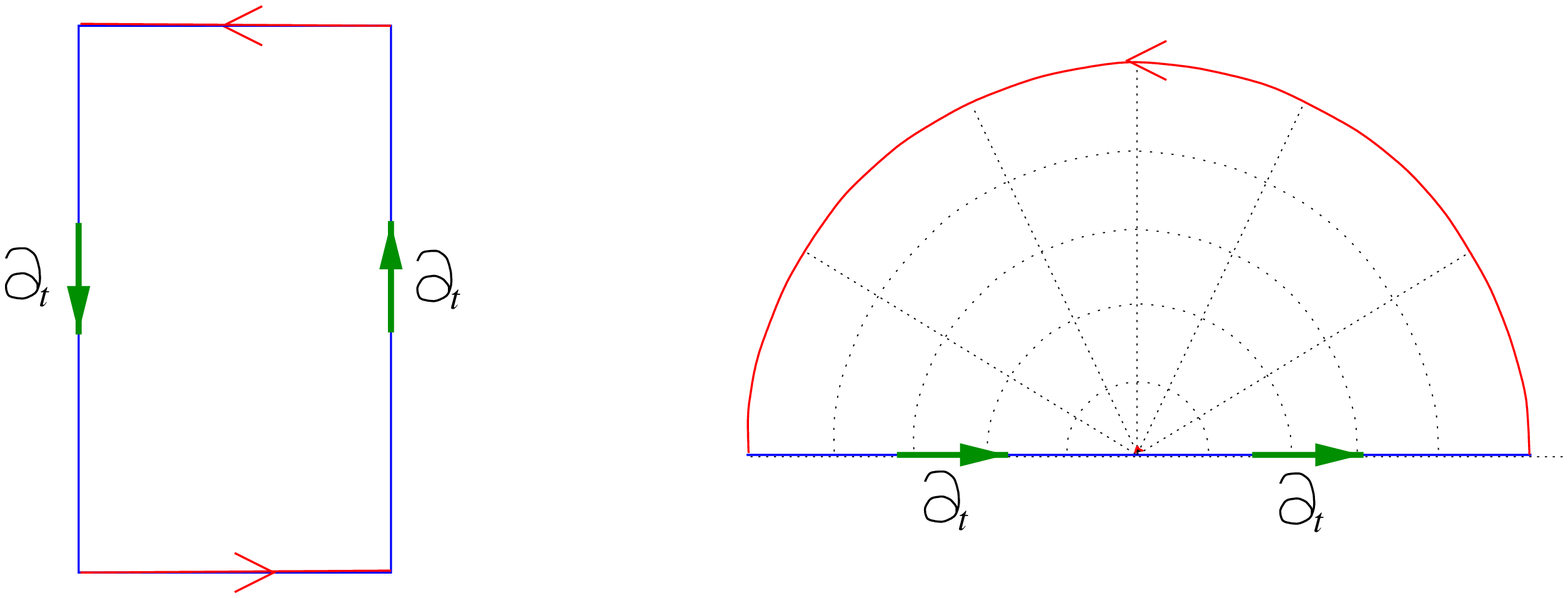,height=1.5in}}
\caption{The action of $\Omega$ on the photon vertex operator can be
deduced from seeing how exchanging the ends of the string changes the
sign of the tangent derivative, $\partial_t$.}
\label{omegaact}
\end{figure}

Let us see what becomes of the string spectrum when we perform this
projection.  The open string tachyon is even under $\Omega$ and so 
survives the projection. However, the photon, which has only one
oscillator acting, does not:
\begin{eqnarray}
\Omega|{ k}\rangle &=& +|{ k}\rangle \nonumber\\
\Omega\alpha^\mu_{-1}|{ k}\rangle &=&  -\alpha^\mu_{-1}|{ k}\rangle.
\label{photonout}
\end{eqnarray}
We have implicitly made a choice about the sign of $\Omega$ as it acts
on the vacuum The choice we have made in writing
eqn. (\ref{photonout}) corresponds to the symmetry of the vertex
operators~(\ref{photon}): the resulting minus sign comes from the
orientation reversal on the tangent derivative $\partial_t$ (see
figure~\ref{omegaact}).

Fortunately, we have endowed the string's ends with Chan--Paton
factors, and so there is some additional structure which can save the
photon.  While $\Omega$ reverses the Chan--Paton factors on the two
ends of the string, it can have some additional action:
\begin{equation}
\Omega\lambda_{ij}|{ k},ij \rangle\ \to\
\lambda^\prime_{ij} |{ k},ij \rangle, \quad
\lambda^\prime= M \lambda^{T} M^{-1}.
\end{equation}
This form of the action on the Chan--Paton factor follows from the
requirement that it be a symmetry of the  amplitudes which have
factors like those in eqn.~(\ref{thefactors}).

If we act twice with $\Omega$, this should square to the identity on
the fields, and leave only the action on the Chan--Paton degrees of
freedom. States should therefore be invariant under:
\begin{equation}
\lambda\to MM^{-T}\lambda M^TM^{-1}. \label{under}
\end{equation}
Now it should be clear that the $\lambda$ must span a complete set of
$N \times N$ matrices:  If strings with ends labelled $ik$ and
$jl$ are in the spectrum for {\it any} values of $k$ and $l$, then so
is the state $ij$.  This is because $jl$ implies $lj$ by CPT, and a
splitting--joining interaction in the middle gives $ik + lj \to ij +
lk$.  

Now equation~(\ref{under}) and Schur's lemma require $MM^{-T}$ to be
proportional to the identity, so $M$ is either symmetric or
antisymmetric.  This gives two distinct cases, modulo a choice of
basis. Denoting the $n\times n$ unit matrix as $I_n$, we have\,\cite{sms}
the symmetric case:
\begin{equation}
\qquad \ M=M^T=I_N
\end{equation}
In order for the photon $\lambda_{ij}\alpha^\mu_{-1}|{ k}\rangle$ to
be even under $\Omega$ and thus survive the projection, $\lambda$ must
be antisymmetric to cancel the minus sign from the transformation of
the oscillator state.  So $\lambda=-\lambda^T$, giving the gauge group
$SO(N)$.  For the antisymmetric case, we have:
\begin{equation}
M=-M^T=i\left[ \begin{array}{cc}
0&I_{N/2}\\ -I_{N/2}&0 \end{array}\right]
\end{equation}
For the photon to survive, $\lambda=-M\lambda^TM$, which is the
definition of the gauge group $USp(N)$. Here, we use the notation that
$USp(2)\equiv SU(2)$.  Elsewhere in the literature this group is often
denoted $Sp(N/2)$.

\subby{Unoriented Closed Strings} Turning to the closed string sector.
For closed strings, we see that the mode expansion
\reef{closedcomplexmodes} for $X^\mu(z,\zb)=X^\mu_L(z)+X^\mu_R(\zb)$
is invariant under a world--sheet parity symmetry $\sigma\to-\sigma$,
which is $z \to -\bar z$.  (We should note that this is a little
different from the choice of $\Omega$ we took for the open strings,
but more natural for this case. The two choices are related to other
by a shift of $\pi$.)  This natural action of $\Omega$ simply reverses
the left-- and right--moving oscillators:
\begin{equation}
\Omega\colon\qquad\alpha^\mu_n\leftrightarrow{\tilde\alpha}^\mu_n.
\end{equation}
Let us again gauge this symmetry, projecting out the states which are
odd under it.  Once again, since the tachyon contains no oscillators,
it is even and is in the projected spectrum. For the level 1
excitations:
\begin{equation}
\Omega\alpha^\mu_{-1}{\tilde\alpha}^\nu_{-1}|k\rangle= 
{\tilde\alpha}^\mu_{-1}\alpha^\nu_{-1}|k\rangle,
\end{equation}
and therefore it is only those states which are symmetric under
$\mu\leftrightarrow\nu$ ---the graviton and dilaton--- which survive
the projection. The antisymmetric tensor is projected out of the
theory.

\subby{Worldsheet Diagrams}

\begin{figure}[ht]
\centerline{\psfig{figure=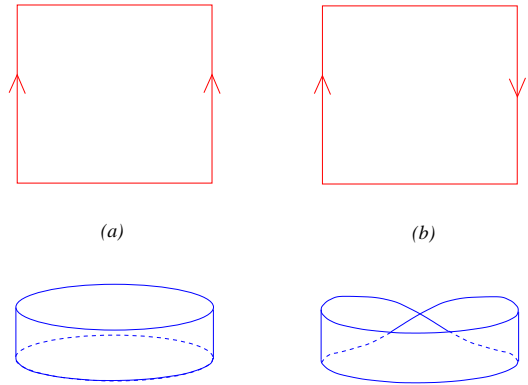,height=2.0in}}
\caption{{\it (a)} Constructing a cylinder or annulus by identifying a
pair of opposite edges of a rectangle.  {\it (b)} Constructing a
M\"obius strip by identifying after a twist.}
\label{mobius}
\end{figure}

As stated before, once we have gauged $\Omega$, we must allow for
unoriented worldsheets, and this gives us rather more types of string
worldsheet than we have studied so far.  Figure~\ref{mobius} depicts
the two types of one--loop diagram we must consider when computing
amplitudes for the open string. The annulus (or cylinder) is on the
left, and can be taken to represent an open string going around in a
loop. The M\"{o}bius strip on the right is an open string going around
a loop, but returning with the ends reversed. The two surfaces are
constructed by identifying a pair of opposite edges on a rectangle,
one with and the other without a twist.

\begin{figure}[ht]
\centerline{\psfig{figure=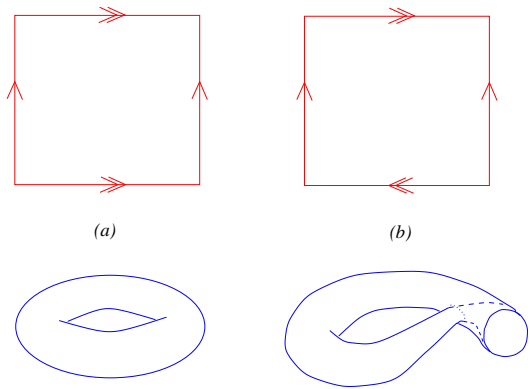,height=2.0in}}
\caption{{\it (a)} Constructing a torus by identifying opposite edges
of a rectangle.  {\it (b)} Constructing a Klein bottle by identifying
after a twist.}
\label{klein}
\end{figure}

Figure~\ref{klein} shows an example of two types of closed string
one--loop diagram we must consider. On the left is a torus, while on
the right is a Klein bottle, which is constructed in a similar way to
a torus save for a twist introduced when identifying a pair of edges.

In both the open and closed string cases, the two diagrams can be
thought of as descending from the oriented case after the insertion of
the normalised projection operator ${1\over2}\Tr(1+\Omega)$ into
one--loop amplitudes.

Similarly, the unoriented one-loop open string amplitude comes from
the annulus and M\"obius strip.  We will discuss these amplitudes in
more detail later.

The lowest order unoriented amplitude is the projective plane
$\mathbf{RP}^2$, which is a disk with opposite points identified.
\begin{figure}[ht]
\centerline{\psfig{figure=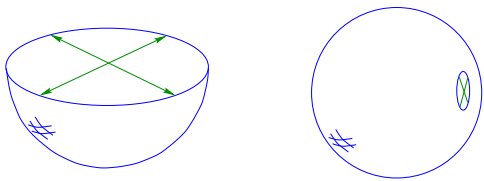,height=0.9in}}
\caption{Constructing the projective plane $\mathbf{RP}^2$ by
identifying opposite points on the disk.  This is equivalent to a
sphere with a crosscap insertion.}
\end{figure}
Shrinking the identified hole down, we recover the fact that
$\mathbf{RP}^2$ may be thought of as a sphere with a crosscap
inserted, where the crosscap is the result of shrinking the identified
hole.  Actually, a M\"obius strip can be thought of as a disc with a
crosscap inserted, and a Klein Bottle is a sphere with two crosscaps.
Since a sphere with a hole (one boundary) is the same as a disc, and a
sphere with one handle is a torus, we can classify all world sheet
diagrams in terms of the number of handles, boundaries and crosscaps
that they have.  Insert 5 (p.\pageref{insert5}) summaries all the
world sheet perturbation theory diagrams up to one loop.

\insertion{5}{World Sheet Perturbation Theory:
  Diagrammatics\label{insert5}}{It is worthwhile summarising all of
  the string theory diagrams up to one--loop in a table. Recall that
  each diagram is weighted by a factor $g_s^\chi=g_s^{2h-2+b+c}$ where
  $h,b,c$ are the numbers of handles, boundaries and crosscaps,
  respectively.

\bigskip
\bigskip
\begin{center}
\begin{tabular}[b]{|c|c|c|c|}
\hline
 & $g_s^{-2}$ & $g_s^{-1}$ & $g_s^0$ \\
\hline
$\matrix{\mbox{closed}\cr\mbox{oriented}}$&$\matrix{\cr\mbox{sphere $S^2$}
\cr\mbox{(plane)}\cr
\psfig{figure=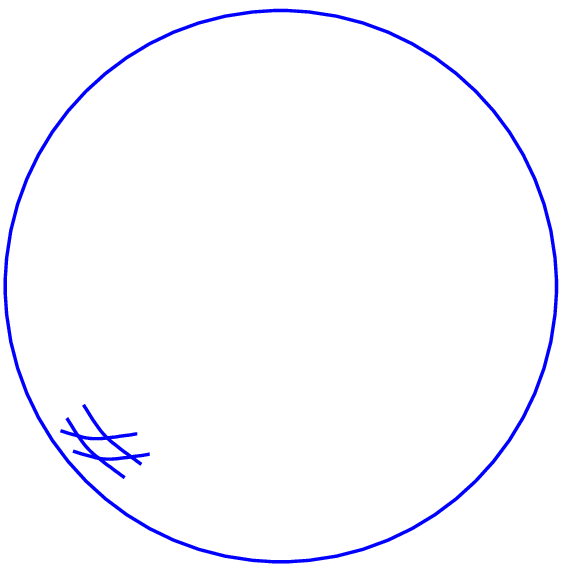,height=0.8in}
}$ 
& $\cdot$ &$\matrix{\cr\mbox{torus $T^2$}\cr\cr
\psfig{figure=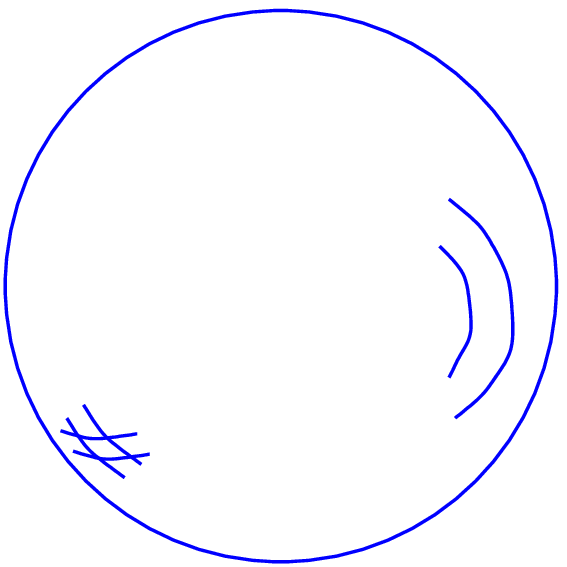,height=0.8in}
}$\\\hline
$\matrix{\mbox{open}\cr\mbox{oriented}}$&$\cdot$&$
\matrix{\cr\mbox{disc $D_2$}
\cr\mbox{(half--plane)}\cr
\psfig{figure=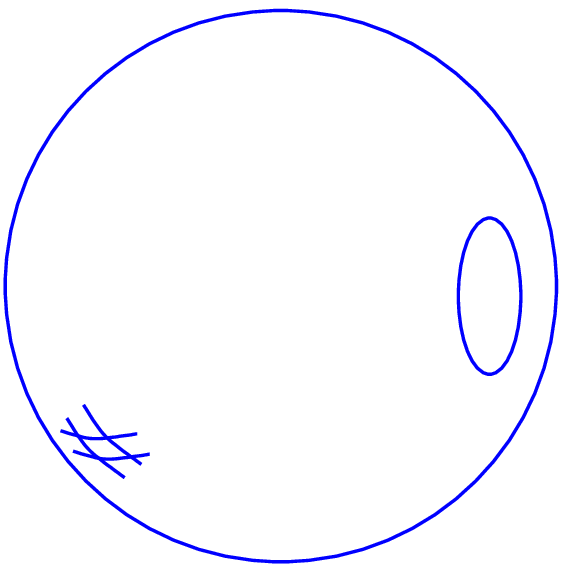,height=0.8in}
}$&$\matrix{\cr\mbox{cylinder $C_2$}
\cr\mbox{(annulus)}\cr
\psfig{figure=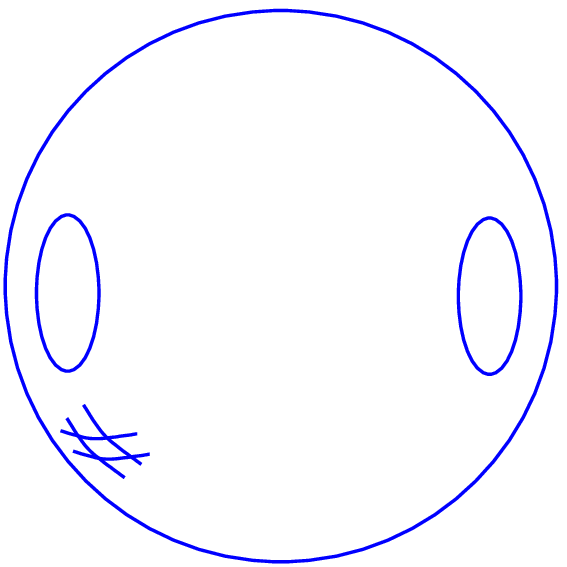,height=0.8in}
}$ \\
\hline
$\matrix{\mbox{closed}\cr\mbox{unoriented}}$&$\cdot$& 
$\matrix{\mbox{projective}\cr\mbox{plane }\mathbf{RP}^2\cr\cr
\psfig{figure=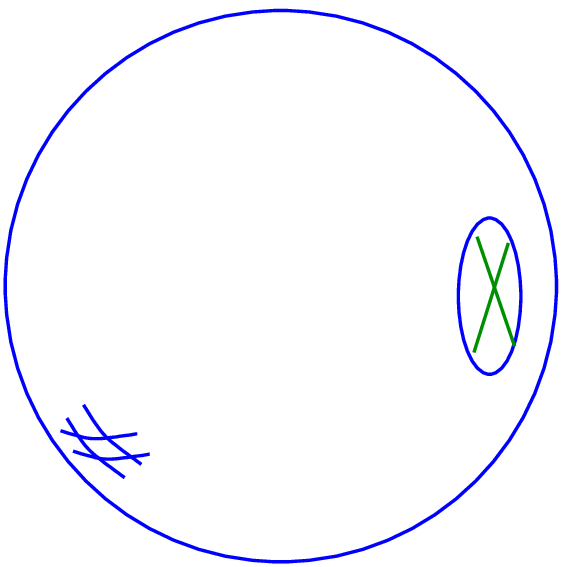,height=0.8in}
}$ & 
$\matrix{\cr\mbox{Klein Bottle KB}\cr\cr
\psfig{figure=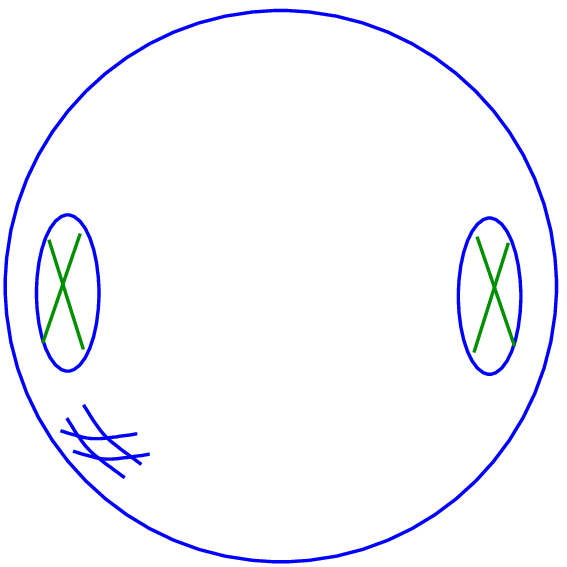,height=0.8in}
}$\\
\hline
$\matrix{\mbox{open}\cr\mbox{unoriented}}$&$\cdot$&$\cdot$& 
$\matrix{\cr\mbox{M\"obius
 Strip MS}\cr\cr
\psfig{figure=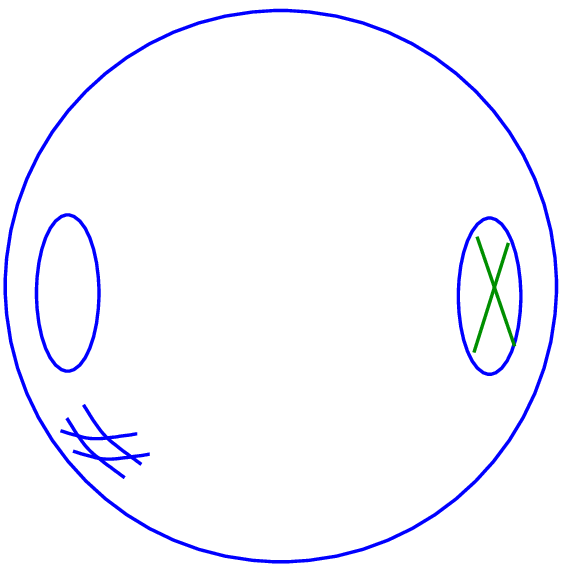,height=0.8in}
}$\\
\hline
\end{tabular}
\end{center}
}

\subsection{Strings in Curved Backgrounds}

So far, we have studied strings propagating in the (uncompactified)
target spacetime with metric $\eta_{\mu\nu}$. While this alone is
interesting, it is curved backgrounds of one sort or another which
will occupy much of this school, and so we ought to see how they fit
into the framework so far.

A natural generalisation of our action is simply to study the ``sigma
model'' action:
\begin{equation}
S_\sigma=-{1\over4\pi\alpha^\prime}\int d^2\!\sigma\, 
(-g)^{1/2}g^{ab} G_{\mu\nu}(X)\partial_aX^\mu
\partial_bX^\nu\ .
\labell{sigmamodel}
\end{equation}
Comparing this to what we had before \reef{essprime}, we see that from
the two dimensional point of view this still looks like a model of $D$
bosonic fields $X^\mu$, but with {\it field dependent} couplings given
by the non--trivial spacetime metric $G_{\mu\nu}(X)$. This is an
interesting action to study.

A first objection to this is that we seem to have cheated somewhat:
Strings are supposed to generate the graviton (and ultimately any
curved backgrounds) dynamically. Have we cheated by putting in such a
background by hand? Or a more careful, less confrontational question might
be: Is it consistent with the way strings generate the graviton to
introduce curved backgrounds in this way?

Well, let us see. Imagine, to start off, that the background metric is
only locally a small deviation from flat space:
$G_{\mu\nu}(X)=\eta_{\mu\nu}+h_{\mu\nu}(X)$, where $h$ is small.

Then, in conformal gauge, we can write in the Euclidean path integral
\reef{pathintegral}:
\begin{equation}
e^{-S_\sigma}=e^{-S}\left(1+{1\over4\pi\alpha^\prime}\int d^2z
h_{\mu\nu}(X)\partial_zX^\mu\partial_{\zb}X^\nu+\ldots\right)\ ,
\end{equation}
and we see that if $h_{\mu\nu}(X)\propto g_s\zeta_{\mu\nu}\exp(ik\cdot
X)$, where $\zeta$ is a symmetric polarisation matrix, we are simply
inserting a graviton emission vertex operator.  So we are indeed
consistent with that which we have already learned about how the
graviton arises in string theory. Furthermore, the insertion of the
full $G_{\mu\nu}(X)$ is equivalent in this language to inserting an
exponential of the graviton vertex operator, which is another way of
saying that a curved background is a ``coherent state'' of gravitons.

It is clear that we should generalise our success, by including sigma
model couplings which correspond to introducing background fields for
the antisymmetric tensor and the dilaton, mimicking \reef{vertex}:
\begin{equation}
S_\sigma={1\over4\pi\alpha^\prime}\int d^2\!\sigma\,  g^{1/2}
\,\left\{(g^{ab}G_{\mu\nu}(X)+i\epsilon^{ab} B_{\mu\nu}(X))\partial_a
X^\mu\partial_b X^\nu+\alpha^\prime\Phi R \right\}\ , \labell{curved}
\end{equation}
where $B_{\mu\nu}$ is the background antisymmetric tensor field and
$\Phi$ is the background value of the dilaton.  The next step is to do
a full analysis of this new action and ensure that in the quantum
theory, one has Weyl invariance, which amounts to the tracelessness of
the two dimensional stress tensor. Calculations (which we will not
discuss here) reveal that:~\cite{joebook,gsw}
\begin{equation}
T^a_{\phantom{a}a}=-{1\over2\alpha^\prime}\beta_{\mu\nu}^G
g^{ab}\partial_aX^\mu\partial_bX^\nu-{i\over2\alpha^\prime}
\beta_{\mu\nu}^B
\epsilon^{ab}\partial_aX^\mu\partial_bX^\nu-{1\over2}\beta^\Phi R\ .
\end{equation}
with
\begin{eqnarray}
\beta_{\mu\nu}^G
&=&\alpha^\prime\left(R_{\mu\nu}+2\nabla_\mu\nabla_\nu\Phi
-{1\over4}H_{\mu\kappa\sigma}H_\nu^{\phantom{\nu}\kappa\sigma}\right)
+O(\alpha^{\prime2}),\nonumber\\ 
\beta_{\mu\nu}^B
&=&\alpha^\prime\left(-{1\over2}\nabla^\kappa
H_{\kappa\mu\nu}+\nabla^\kappa \Phi H_{\kappa\mu\nu}
\right)+O(\alpha^{\prime2}),\nonumber\\
\beta^\Phi
&=&\alpha^\prime\left({D-26\over6\alpha^\prime}
-{1\over2}\nabla^2\Phi+\nabla_\kappa\Phi\nabla^\kappa \Phi 
-{1\over24}H_{\kappa\mu\nu}H^{\kappa\mu\nu}
\right)+O(\alpha^{\prime2})\ ,
\labell{eqnmtn}
\end{eqnarray}
with $H_{\mu\nu\kappa}\equiv\partial_{\mu}
B_{\nu\kappa}+\partial_{\nu} B_{\kappa\mu}+\partial_{\kappa}
B_{\mu\nu}.$ For Weyl invariance, we ask that each of these beta
functions for the sigma model couplings actually vanish. The
remarkable thing is that these resemble {\it spacetime field equations
  for the background fields.}  In fact, the field equations can be
derived from the following spacetime action:
\begin{eqnarray}
{\rm S}&=&{1\over2\kappa^2_0}\int
d^DX(-G)^{1/2}e^{-2\Phi}\left[R+4\nabla_\mu
\Phi\nabla^\mu\Phi-{1\over12}H_{\mu\nu\lambda}
H^{\mu\nu\lambda}\right.\nonumber\\
&&\hskip5cm
\left.-{2(D-26)\over3\alpha^\prime}+O(\alpha^\prime)\right]\ .
\labell{stringfrm}
\end{eqnarray}

\ennbee{Now we note something marvellous: $\Phi$ is a background field
  which appears in the closed string theory sigma model multiplied by
  the Euler density. So comparing to \reef{eulerterm} (and discussion
  following), we recover the remarkable fact that the string coupling
  $g_s$ is not fixed, but is in fact given by the value of one of the
  background fields in the theory: $g_s=e^{<\Phi>}$. So the only free
  parameter in the theory is the string tension.}  

\medskip
Turning to the open
string sector, we may also write the effective action which summarises
the leading order (in $\alpha^\prime$) open string physics at tree
level:
\begin{equation}
{\rm S}=-{{\rm C}\over 4}\int d^DX\, e^{-\Phi}\Tr
F_{\mu\nu}F^{\mu\nu}+O(\alpha^\prime)\ ,
\labell{yangmills}
\end{equation}
with ${\rm C}$ a dimensionful constant which we will fix later.  It is
of course of the form of the Yang--Mills action, where
$F_{\mu\nu}=\partial_\mu A_\nu-\partial_\nu A_\mu$. The field $A_\mu$
is coupled in sigma--model fashion to the boundary of the world sheet
by the boundary action:
\begin{equation}
\int_{\partial{\cal M}}d\tau\, A_\mu \partial_t X^\mu\ ,
\labell{abound}
\end{equation}
mimicking the form of the vertex operator \reef{photon}.

One should note the powers of $e^{\Phi}$ in the above actions. Recall
 that the expectation value of $e^\Phi$ sets the value of $g_s$. We
 see that the appearance of $\Phi$ in the actions are consistent with
 this, as we have $e^{-2\Phi}$ in front of all of the closed string
 parts, representing the sphere ($g_s^{-2}$) and $e^{-\Phi}$ for the
 open string, representing the disc ($g_s^{-1}$).

Notice that if we make the following redefinition of the background
fields:
\begin{equation}
{\tilde
G}_{\mu\nu}(X)=e^{2\Omega(X)}
G_{\mu\nu}=e^{4(\Phi_0-\Phi)/(D-2)}G_{\mu\nu}\ ,
\labell{framechange}
\end{equation}
and use the fact that the new Ricci scalar can be derived using:
\begin{equation}
{\tilde R}=e^{-2\Omega}\left[R-2(D-1)\nabla^2\Omega-(D-2)(D-1)\partial_\mu\Omega
\partial^\mu\Omega\right]\ ,
\end{equation}
The action \reef{stringfrm} becomes:
\begin{eqnarray}
{\rm S}&=&{1\over2\kappa^2}\int d^DX(-{\tilde G})^{1/2}\left[R-{4\over
D-2}\nabla_\mu
{\tilde\Phi}\nabla^\mu{\tilde\Phi}-{1\over12}e^{-8{\tilde\Phi}/(D-2)}
H_{\mu\nu\lambda}
H^{\mu\nu\lambda}\right.
\nonumber\\
&&\hskip5.5cm
\left.-{2(D-26)\over3\alpha^\prime}e^{4{\tilde\Phi}/(D-2)}
+O(\alpha^\prime)\right]\ ,\nonumber\\
\labell{einsteinfrm}
\end{eqnarray}
with ${\tilde\Phi}=\Phi-\Phi_0$, Looking at the part involving the
Ricci scalar, we see that we have the form of the standard
Einstein--Hilbert action ({\it i.e.,} we have removed the factor
involving the dilaton $\Phi$), with Newton's constant set by
\begin{equation}
\kappa\equiv\kappa_0e^{\Phi_0}=(8\pi G_N)^{1/2}\ .
\end{equation}
The standard terminology to note here is that the action
\reef{stringfrm} written in terms of the original fields is called the
``string frame'', while the action \reef{einsteinfrm} is referred to
as the ``Einstein frame'' action. It is in the latter frame that one
gives meaning to measuring quantities like gravitational mass--energy.
It is important to note the means to transform from the fields of one
to another, depending upon dimension \reef{framechange}.  See
also the supersymmetric cases much later in these notes.

\section{Target Spacetime Perspective, Mostly}
In this section we shall study T--duality.~\cite{tdual} This
is very dramatic symmetry of the theory of strings under a spacetime
transformation. It is a crucial consequence of the fact that strings
are extended objects.

\subsection{T--Duality for Closed Strings}
\label{tdualityclosed}
Let us start with closed strings, first focusing on the zero modes.
The mode expansion~(\ref{closedcomplexmodes}) can be written:
\begin{equation} 
X^\mu(z,\zb) = x^\mu + \tilde x^\mu -i\sqrt{{\alpha^\prime}\over2}
(\alpha^\mu_0+{\tilde\alpha}^\mu_0)\tau + \sqrt{{\alpha^\prime}\over2}
(\alpha^\mu_0-{\tilde\alpha}^\mu_0)\sigma + {\rm oscillators}.
\end{equation}
We have already identified  the spacetime momentum of the  string:
\begin{equation}
p^\mu = 
{1\over{\sqrt{2{\alpha^\prime}}}}(\alpha^\mu_0 + {\tilde\alpha}^\mu_0)\ .
\end{equation}
If we run around the string, {\it i.e.,} take $\sigma \to \sigma+2\pi$,
the oscillator term are periodic and we have 
\begin{equation}
X^\mu(z,\zb)\to X^\mu(z,\zb)
+2\pi\sqrt{\alpha^\prime\over2}(\alpha^\mu_0-{\tilde\alpha}^\mu_0)\ .
\end{equation}
So far, we have studied the situation of non--compact spatial
directions for which the embedding function $X^\mu(z,\zb)$ is
single--valued, and therefore the above change must be zero, giving 
\begin{equation}
\alpha^\mu_0={\tilde\alpha}^\mu_0
=\sqrt{{\alpha^\prime}\over2}p^\mu. \labell{boring}
\end{equation}
Momentum $P^\mu$ takes a continuum of values reflecting the fact that
the direction $X^\mu$ is non--compact.

Let us consider the case that we have a compact direction, say
$X^{25}$, of radius $R$. Our direction $X^{25}$ therefore has period
$2\pi R$. The momentum $p^{25}$ now takes the discrete values $n/R$,
for $n\in\IZ$.  Now, under $\sigma\sim\sigma+2\pi$, $X^{25}(z,\zb)$ is
not single valued, and can change by $2\pi wR$, for $w\in\IZ$.
Solving the two resulting equations gives:
\begin{eqnarray}
\alpha^{25}_0+{\tilde\alpha}^{25}_0 &=&
{2n\over R}\sqrt{{\alpha^\prime}\over2} \nonumber\\
\alpha^{25}_0-{\tilde\alpha}^{25}_0 &=& 
\sqrt{2 \over {\alpha^\prime}}wR
\end{eqnarray}
and so we have:
\begin{eqnarray}
\alpha^{25}_0 &=& \left({n\over R}+{wR\over{\alpha^\prime}}\right)
\sqrt{{\alpha^\prime}\over2}\equiv
P_L\sqrt{{\alpha^\prime}\over2} \nonumber\\
{\tilde\alpha}^{25}_0 &=&
\left({n\over R}-{wR\over{\alpha^\prime}}\right)\sqrt{{\alpha^\prime}\over2}
\equiv
P_R\sqrt{{\alpha^\prime}\over2}\ .
\labell{leftrightmomenta}
\end{eqnarray}

We can use this to compute the formula for the mass spectrum in the
remaining uncompactified 24+1 dimensions, using the fact that
$M^2=-p_\mu p^\mu$, where now $\mu=0,\ldots, 24$.
\begin{eqnarray}
M^2\ =\ -p^\mu p_\mu &=&
{2\over{\alpha^\prime}}(\alpha_0^{25})^2+{4\over{\alpha^\prime}}
({ N}-1) \nonumber\\
&=& {2\over{\alpha^\prime}}({\tilde\alpha}_0^{25})^2+{4\over{\alpha^\prime}}
( {\bar N}-1)\ ,
\end{eqnarray}
where $N, \bar N$ denote the total levels on the left-- and
right--moving sides, as before. These equations follow from the left
and right $L_0, {\bar L}_0$ constraints. Recall that the sum and
difference of these give the Hamiltonian and the level--matching
formulae. Here, they are modified, and a quick computation gives:
\begin{eqnarray}
&&M^2={n^2\over R^2}+{w^2 R^2\over
\alpha^{\prime2}}+{2\over\alpha^\prime} \left(N+{\tilde N}-2 \right)
\nonumber\\ &&nw+N-{\tilde N}=0 \ .
\labell{spectrum}
\end{eqnarray}
The key features here are that there are terms in addition to the
usual oscillator contributions. In the mass formula, there is a term
giving the contribution of the Kaluza--Klein tower of momentum states
for the string, and a term from the tower of winding states. This
latter term is a very stringy phenomenon. Notice that the level
matching term now also allows a mismatch between the number of left
and right oscillators excited, in the presence of discrete winding and
momenta.

In fact, notice that we can get our usual massless states by taking
\begin{equation}
n=w=0\ ;\qquad N={\bar N}=1\ .
\labell{kaluzaklein}
\end{equation}
If we write these states out(and the corresponding fields and
 vertex operators, for completeness), we have:
\medskip
\begin{center}
\begin{tabular}{|c|c|c|}
\hline
field&state&operator\\\hline
$G_{\mu\nu}$&
$(\alpha^\mu_{-1}{\tilde\alpha}^{\nu}_{-1}+
\alpha^{\nu}_{-1}{\tilde\alpha}^{\mu}_{-1})|0;k\!>$ &
$\partial X^\mu{\bar\partial}X^\nu+\partial X^\mu{\bar\partial}X^\nu$ \\
$B_{\mu\nu}$&
$(\alpha^\mu_{-1}{\tilde\alpha}^{\nu}_{-1}-
\alpha^{\nu}_{-1}{\tilde\alpha}^{\mu}_{-1})|0;k\!>$ & 
$\partial X^\mu{\bar\partial}X^\nu-\partial X^\mu{\bar\partial}X^\nu$ \\
$A_{\mu(R)}$&
$\alpha^\mu_{-1}
{\tilde\alpha}^{25}_{-1}|0;k\!>$ &
$\partial X^\mu{\bar\partial}X^{25}$ \\
$A_{\mu(L)}$&
${\tilde\alpha}^{\mu}_{-1}\alpha^{25}_{-1}|0;k\!>$ &
$\partial X^{25}{\bar\partial}X^{\mu}$ \\
$\phi\equiv G_{25,25}$&
$\alpha^{25}_{-1}{\tilde\alpha}^{25}_{-1}|0;k\!>$ &
$\partial X^{25}{\bar\partial}X^{25}$\\
\hline
\end{tabular}
\end{center}

\medskip
\noindent
where
$$
A_{\mu(R)}\equiv {1\over2}(G-B)_{\mu,25}\ ;\quad A_{\mu(L)}\equiv
{1\over2}(G+B)_{\mu,25}\ .
$$
(we have listed the zero momentum vertex operators for these states
also).

These 25 dimensional massless states are basically the components of
the graviton and antisymmetric tensor fields in 26 dimensions, now
relabelled. (There is also of course the dilaton $\Phi$, which we have
not listed.) There is a pair of gauge fields giving a
$U(1)_L{\times}U(1)_R$ gauge symmetry, and in addition a massless
scalar field $\phi$. Actually, $\phi$ is a massless scalar which can
have any background vacuum expectation value (vev), which in fact sets
the radius of the circle.  This is because the square root of the
metric component $G_{25,25}$ is indeed the measure of the radius of
the $X^{25}$ direction.

Let us now study the generic behaviour of the spectrum \reef{spectrum}
for different values of $R$.  For larger and larger $R$, momentum
states become lighter, and therefore it is less costly to excite them
in the spectrum. At the same time, winding states become heavier, and
are more costly. For smaller and smaller $R$, the reverse is true, and
it is gets cheaper to excite winding states and it is momentum states
which become more costly.

We can take this further: As $R \to \infty$, all of the winding states
{\it i.e.,} states with $w\neq 0$, become infinitely massive, while
the $w=0$ states with all values of $n$ go over to a continuum.  This
fits with what we expect intuitively, and we recover the fully
uncompactified result.

Consider instead the case $R\to 0$, where all of the momentum states
{\it i.e.,} states with $n \neq 0$, become infinitely massive.  If we
were studying field theory we would stop here, as this would be all
that would happen---the surviving fields would simply be independent
of the compact coordinate, and so we have performed a dimension
reduction.  In closed string theory things are quite different: the
pure winding states ({\it i.e.,} $n=0$, $w\neq0$, states) form a
continuum as $R\to 0$, following from our observation that it is very
cheap to wind around the small circle.  {\it Therefore, in the $R\to 0$
limit, an effective uncompactified dimension actually reappears!}

Notice that the formula \reef{spectrum} for the spectrum is invariant
under the exchange
\begin{equation}
n\leftrightarrow w\qquad\mbox{\rm and }\qquad R\leftrightarrow R'\equiv
{\alpha^\prime}/R\ .
\end{equation}
The string theory compactified on a circle of radius $R'$ (with
momenta and windings exchanged) is the ``T--dual'' theory, and the
process of going from one theory to the other will be referred to as
``T--dualising''.

The exchange takes (see \reef{leftrightmomenta})
\begin{equation}
\alpha^{25}_0 \rightarrow\alpha^{25}_0, \quad
{\tilde\alpha}_0^{25} \rightarrow -{\tilde\alpha}_0^{25} 
\labell{tparity}\ .
\end{equation}
The dual theories are identical in the fully interacting case as
well\cite{nairet}: If we write the radius $R$ theory in terms of
\begin{equation}
X^{\prime25}(z,\zb)=X^{25}(z)-X^{25}(\zb)\ . 
\labell{toneside}
\end{equation}
The energy--momentum tensor and other basic properties of the 
conformal field theory are invariant under this rewriting, and so are
therefore all of the correlation functions representing scattering
amplitudes, {\it etc.} The only
change, as follows from equation \reef{tparity}, is that the zero mode
spectrum in the new variable is that of the ${\alpha^\prime}/R$
theory.  These theories are physically identical;
T--duality, relating the $R$ and ${\alpha^\prime}/R$ theories, is an
exact symmetry of perturbative closed string theory.  The
transformation~(\ref{toneside}) can be regarded as a spacetime parity
transformation acting only on the right--moving (in the world sheet
sense) degrees of freedom.

\subsection{The Circle Partition Function}
\label{tdualing}
It is useful to consider the partition function to the theory on the
circle.  This is a computation as simple as the one we did for the
uncompactified theory earlier, since we have done the hard work in
working out $L_0$ and ${\bar L}_0$ for the circle
compactification. Each non--compact direction will contribute a factor
of $(\eta{\bar\eta})^{-1}$, as before, and the non--trivial part of
the final $\tau$--integrand, coming from the compact $X^{25}$
direction is:
\begin{equation}
Z(q,R)=(\eta{\bar\eta})^{-1}\sum_{n,w}q^{{\alpha^\prime\over4}P_L^2}
{\bar q}^{{\alpha^\prime\over4}P_R^2}\ ,
\labell{partfunR}
\end{equation}
where $P_{L,R}$ are given in \reef{leftrightmomenta}.  Our partition
function is manifestly T--dual, and is in fact also modular invariant:
Under $T$, it picks us a phase $\exp(\pi{\rm i}(P_L^2-P_R^2))$, which
is again unity, as follows from the second line in \reef{spectrum}:
$P_L^2-P_R^2=2nw$. Under $S$, the role of the time and space shifts as
we move on the torus are exchanged, and this in fact exchanges the
sums over momentum and winding. T--duality ensures that the
$S$--transformation properties of the exponential parts involving
$P_{L,R}$ are correct, while the rest is $S$ invariant as we have
already discussed. 

It is a useful exercise to expand this partition function out, after
combining it with the factors from the other non--compact dimensions
first, to see that at each level the mass (and level matching)
formulae \reef{spectrum} which we derived explicitly is recovered.

In fact, the modular invariance of this circle partition function is
part of a very important larger story. The left and right momenta
$P_{L,R}$ are components of a special two dimensional lattice,
$\Gamma_{1,1}$. There are two basis vectors $k=(1/R,1/R)$ and ${\hat
  k}=(R,-R)$. We make the lattice with arbitrary integer combinations
of these, $nk+w{\hat k}$, whose components are $(P_L,P_R)$. ({\it
  c.f.} \reef{leftrightmomenta}) If we define the dot products between
our basis vectors to be $k\cdot{\hat k}=2$ and $k\cdot k=0={\hat
  k}\cdot {\hat k}$, our lattice then has a Lorentzian signature, and
since $P_L^2-P_R^2=2nw\in 2\IZ$, it is called ``even''.  The ``dual''
lattice $\Gamma^*_{1,1}$ is the set of all vectors whose dot product
with $(P_L,P_R)$ gives an integer. In fact, our lattice is self--dual,
which is to say that $\Gamma_{1,1}=\Gamma^*_{1,1}$. It is the ``even''
quality which guarantees, invariance under $T$ as we have seen, while
it is the ``self--dual'' feature which ensures invariance under $S$.
In fact, $S$ is just a change of basis in the lattice, and the self
duality feature translates into the fact that the Jacobian for this is
unity.

The set of such lattices in this class is classified and is important
in string theory. An example is the lattice $\Gamma_{d,d}$ of left and
right momenta for strings compactified on a $d$ dimensional torus
$T^d$. There is a large space of inequivalent lattices of this type,
given by the shape of the torus (specified by background parameters in
the metric $G$) and the fluxes of the B--field through it. This
``moduli space'' of compactifications is isomorphic to 
\begin{equation}
{\cal M}={O(d,d)\over O(d)\times O(d)}\ ,
\end{equation}
In fact, the full set of T--duality transformations turns out to be
the {\it non--Abelian} $SO(d,d,\IZ)$, which is generated by the
T--dualities on all of the $d$ circles, linear redefinitions of the
axes, and discrete shifts of the B--field.

Two other examples are the lattices associated to the construction of
the modular invariant partition functions of the $E_8{\times}E_8$ and
$SO(32)$ heterotic strings.~\cite{heterotic}

\subsection{Self--Duality and Enhanced Gauge Symmetry}
\label{selfdual}
Given the relation we deduced between the spectra on radii $R$ and
$\alpha^\prime/R$, it is clear that there ought to be something
interesting about the theory at the radius $R=\sqrt{\alpha^\prime}$.
The theory should be self--dual, and this radius is the ``self--dual
radius''. There is something else special about this theory.

At this radius we have, using \reef{leftrightmomenta},
\begin{equation}
\alpha^{25}_0 = {(n+w)\over\sqrt{2}}\ ;\qquad
 {\tilde\alpha}^{25}_0={(n-w)\over\sqrt{2}}\ , 
\end{equation}
and so from the left and right we have:
\begin{eqnarray}
M^2\ =\ -p^\mu p_\mu &=&
{1\over{\alpha^\prime}}(n+w)^2+{4\over{\alpha^\prime}}
({ N}-1) \nonumber\\
&=& {2\over{\alpha^\prime}}(n-w)^2+{4\over{\alpha^\prime}}
( {\bar N}-1)\ .
\end{eqnarray}
So if we look at the massless spectrum, we have the conditions:
\begin{eqnarray}
(n+w)^2+4N=4\ ; \qquad (n-w)^2+4{\bar N}=4\ .
\end{eqnarray}
As before, we have the generic solutions $n=w=0$ with $N=1$ and ${\bar
N}=1$. These are the include the vectors of the $U(1){\times}U(1)$
gauge symmetry of the compactified theory.

Now however, we see that we have more solutions. In particular:
\begin{eqnarray}
n=-w=\pm1\ ,\quad N=1\ ,\,\,{\bar N}=0\ ;\qquad n=w=\pm1\ ,
\quad N=0\ ,\,\,{\bar N}=1\ .
\end{eqnarray}
The cases where the excited oscillators are in the non-compact
direction yield two pairs of massless vector fields. In fact, the
first pair go with the left $U(1)$ to make an $SU(2)$, while the
second pair go with the right $U(1)$ to make another $SU(2)$. Indeed,
they have the correct $\pm1$ charges under the Kaluza--Klein $U(1)$'s
in order to be the components of the W--bosons for the
$SU(2)_L{\times}SU(2)_R$ ``enhanced gauge symmetries''. The term is
appropriate since there is an extra gauge symmetry at this special
radius, given that new massless vectors appear there.

When the oscillators are in the compact direction, we get two pairs of
massless bosons. These go with the massless scalar $\phi$ to fill out
the massless adjoint Higgs field for each $SU(2)$. These are the
scalars whose vevs give the W--bosons their masses when we are away
from the special radius.

In fact, this special property of the string theory is succinctly
visible at all mass levels, by looking at the partition function
\reef{partfunR}. At the self dual radius, it can be rewritten as a sum
of squares of ``characters'' of the $su(2)$ affine Lie algrebra:
\begin{equation}
Z(q,R=\sqrt{\alpha^\prime})=\left|\chi_1(q)\right|^2+\left|\chi_2(q)
\right|^2\
,
\end{equation}
where
\begin{equation}
\chi_1(q)\equiv\eta^{-1}\sum_n q^{n^2}\ , \quad
\chi_2(q)\equiv\eta^{-1}\sum_n q^{(n+1/2)^2}
\end{equation}
It is amusing to expand these out (after putting in the other factors
of $(\eta{\bar\eta})^{-1}$ from the uncompactified directions) and
find the  massless states we discussed explicitly above.

%ref to put in\cite{dhs} 

In the language of two dimensional conformal field theory, there are
additional left-- and right--moving currents (fields with weights
(1,0) and (0,1)) present, whose vertex operators are exponentials. We
can construct the full set of vertex operators of the
$SU(2)_L{\times}SU(2)_R$  spacetime gauge symmetry:
\begin{eqnarray}
SU(2)_L\colon&&{\bar\partial}X^\mu\partial X^{25}(z),\ \ {\bar\partial}X^\mu
\exp(\pm
2iX^{25}(z)/\sqrt{{\alpha^\prime}}) \nonumber\\
SU(2)_R\colon&&\partial X^{\mu}\bar\partial X^{25}(z),\ \ \partial X^{\mu}
\exp(\pm
2iX^{25}(\zb)/\sqrt{{\alpha^\prime}})\ ,
\end{eqnarray}
corresponding to the massless vectors we constructed by hand above.

The vertex operator for the change of radius, $\partial X^{25}
{\bar\partial} X^{25}$, corresponding to the field $\phi$, transforms
as a $({\bf 3},{\bf 3})$ under $SU(2)_L{\times}SU(2)_R$, and therefore
a rotation by $\pi$ in one of the $SU(2)$'s transforms it into minus
itself.  The transformation $R \to {\alpha^\prime}/R$ is therefore the
${\bf Z}_2$ Weyl subgroup of the $SU(2) \times SU(2)$. Since
T--duality is part of the spacetime gauge theory, this is a clue that
it is an exact symmetry of the closed string theory, if we assume that
non--perturbative effects preserve the spacetime gauge symmetry.  We
shall see that this assumption seems to fit with non--perturbative
discoveries to be described later.

\subsection{T--duality in  Background Fields}

Notice that T--duality acts non-trivially on the dilaton, and therefore
modifies the string coupling:~\cite{gv,buscher}  After dimensional reduction on
the circle, the effective 25 dimensional string coupling read off from the
supergravity action is now $e^{\Phi} (2\pi R)^{-1/2}$.  T--Duality
requires this to be equal to $e^{\tilde\Phi}(2\pi R^\prime)^{-1/2}$,
the string coupling of the dual 25 dimensional theory, and therefore
\begin{equation}
e^{\tilde\Phi} = e^{\Phi} {{{\alpha^\prime}}^{1/2}\over R}\ ;\qquad 
{\tilde g}_s=g_s{{{\alpha^\prime}}^{1/2}\over R}\ . \labell{newg}
\end{equation}

This is just part of a larger statement about the T--duality
transformation properties of background fields in general. Starting
with background fields $G_{\mu\nu}$, $B_{\mu\nu}$ and $\Phi$, let us
first T--dualise in one direction, which we shall label $X^y$. In
other words, we mean that $X^y$ is a direction which is a circle of
radius $R$, and the dual circle $X^{\prime y}$ is a circle of radius
$R^\prime=\alpha^\prime/R$. The resulting background fields, ${\tilde
G}_{\mu\nu}$, ${\tilde B}_{\mu\nu}$ and $\tilde\Phi$, are given by:
\begin{eqnarray}
{\tilde G}_{yy}&=&{1\over G_{yy}}\ ;\qquad
e^{2{\tilde\Phi}}={e^{2{\Phi}}\over G_{yy}}\ ;\qquad {\tilde G}_{\mu
y}={B_{\mu y}\over G_{yy}}\ ; \qquad {\tilde B}_{\mu y}={G_{\mu y}\over
G_{yy}}\ , \nonumber\\ {\tilde G_{\mu\nu}}&=&G_{\mu\nu}-{G_{\mu y}G_{\nu
y}-B_{\mu y}B_{\nu y}\over G_{yy}}\ ,\nonumber\\ {\tilde
B_{\mu\nu}}&=&B_{\mu\nu}-{B_{\mu y}G_{\nu y}-G_{\mu y}B_{\nu y}\over
G_{yy}}\ . \labell{backgroundT}
\end{eqnarray}
Of course, we can T--dualise on many (say $d$) independent circles,
forming a torus $T^d$. It is not hard to deduce that one can
succinctly write the resulting T--dual background as follows. If we
define the $D{\times}D$ metric
\begin{equation}
E_{\mu\nu}=G_{\mu\nu}+B_{\mu\nu}\ , \labell{geebee}
\end{equation}
and if the circles are in the directions $X^i$, $i=1,\ldots,d$, with
the remaining directions labelled by $X^a$, then the dual fields are
given by
\begin{eqnarray}
{\tilde E}_{ij}&=&E^{ij}\ ;\qquad {\tilde E}_{aj}=E_{ak}E^{kj}\ ;\qquad
e^{2{\tilde\Phi}}=e^{2\Phi}\det(E^{ij})\ ,\nonumber\\ {\tilde
E}_{ab}&=&E_{ab}-E_{ai}E^{ij}E_{jb}\ , \labell{bigtdual}
\end{eqnarray}
where $E_{ik}E^{kj}=\delta_i^{\phantom{i}j}$ defines $E^{ij}$ as the
inverse of $E_{ij}$.  We will find this succinct form of the $O(d,d)$
T--duality transformation very useful later on.

\subsection{Another Special Radius: Bosonisation}
\label{bosonisation}
Before proceeding with T--duality discussion, let us pause for a
moment to remark upon something which will be useful later. In the
case that $R=\sqrt{(\alpha^\prime/2)}$, something remarkable
happens. The partition function is:
\begin{equation}
Z\left(q,R=\sqrt{\alpha^\prime\over2}\right)
=(\eta{\bar\eta})^{-1}\sum_{n,w}q^{{1\over2}{\left({n}+{w\over2}\right)^2}}
{\bar q}^{{1\over2}\left({n}-{w\over2}\right)^2}\ .
\labell{partfunRone}
\end{equation}
Note that the allowed momenta at this radius are
({\it c.f.} \reef{leftrightmomenta}):
\begin{eqnarray}
\alpha^{25}_0 &=& P_L\sqrt{{\alpha^\prime}\over2}
=\left({n}+{w\over2}\right) \nonumber\\ {\tilde\alpha}^{25}_0
&=&P_L\sqrt{{\alpha^\prime}\over2}=
\left({n}-{w\over2}\right) \ ,  \labell{momentaRone}
\end{eqnarray}
and so they span both integer and half--integer values. Now when $P_L$
is an integer, then so is $P_R$ and {\it vice--versa}, and so we have
two distinct sectors, integer and half--integer. In fact, we can
rewrite our partition function as a set of sums over these separate
sectors:
\begin{equation}
Z\left(R{=}\sqrt{\alpha^\prime\over2}\right)={1\over2}\left\{
\left|{1\over \eta}\sum_n q^{{1\over 2}n^2}   \right|^2+
\left|{1\over \eta}\sum_n (-1)^n q^{{1\over 2}n^2}   \right|^2+
\left|{1\over \eta}\sum_n q^{{1\over 2}\left(n+\ha\right)^2}   \right|^2
\right\}\ .
\end{equation}
The middle sum is rather like the first, except that there is a $-1$
whenever $n$ is odd. Taking the two sums together, it is just like we
have performed the sum (trace) over all the integer momenta, but
placed a projection onto even momenta, using the projector
\begin{equation}
P={1\over2}(1+(-1)^n)\ .
\labell{projector}
\end{equation}
In fact, an investigation will reveal that the third term can be
written with a partner just like it save for an insertion of $(-1)^n$
also, but that latter sum vanishes identically. This all has a
specific meaning which we will uncover shortly.

Notice that the partition function can be written in yet another nice
way, this time as
\begin{equation}
Z\left(R=\sqrt{\alpha^\prime\over2}\right)
={1\over 2}\left(|f^2_4(q)|^2+|f^2_3(q)|^2
+|f^2_2(q)|^2\right)\ ,
\labell{rewrite}
\end{equation}
where, for here and for future use, let us define
\begin{eqnarray}
f_1(q)\equiv\left[{\theta_1^\prime(0,\tau)\over2\pi\eta(\tau)}\right]^{1\over2}
&=& q^{1\over24}\prod_{n=1}^\infty(1-q^n) \equiv\eta(\tau)\nonumber\\
f_2(q)\equiv\left[{\theta_2(0,\tau)\over\eta(\tau)}\right]^{1\over2}
&=&\sqrt{2}q^{1\over24}\prod_{n=1}^\infty(1+q^n) \nonumber\\
f_3(q)\equiv\left[{\theta_3(0,\tau)\over\eta(\tau)}\right]^{1\over2}
&=&q^{-{1\over48}}\prod_{n=1}^\infty(1-q^{n-\ha}) \nonumber\\
f_4(q)\equiv\left[{\theta_4(0,\tau)\over\eta(\tau)}\right]^{1\over2}
&=&q^{-{1\over48}}\prod_{n=1}^\infty(1+q^{n-\ha}) \ ,
\labell{thethetas}
\end{eqnarray}
and note that 
\begin{eqnarray}
&&f_2\left(-{1\over\tau}\right)=f_4\left({\tau}\right)\ ;\quad 
f_3\left(-{1\over\tau}\right)=f_3\left({\tau}\right)\ ;\\
&&f_3\left(\tau+1\right)=f_4\left({\tau}\right)\ ;\quad
f_2\left(\tau+1\right)=f_2\left({\tau}\right)\ .
\labell{modularswop}
\end{eqnarray}

While the rewriting \reef{rewrite} might not look like much at first
glance, this is in fact the partition function of a single Dirac
fermion in two dimensions!: $ Z(R=\sqrt{\alpha^\prime/2})=Z_{\rm
  Dirac}.$ We have arrived at the result that a boson (at a special
radius) is in fact equivalent to a fermion. This is called
``Bosonisation'' or ``fermionisation'', depending upon one's
perspective. How can this possibly be true?

The action for a Dirac fermion,
$\Psi=(\Psi_L,\Psi_R)^T$ (which has two components in two dimensions)
is, in conformal gauge:
\begin{equation}
S_{\rm Dirac}={\rm i\over 2\pi}\int d^2\!\sigma\, \, {\bar
\Psi}\gamma^a\partial_a \Psi= {\rm i\over \pi}\int d^2\!\sigma\, \, {\bar
\Psi}_L{\bar \partial}\Psi_L-{\rm i\over \pi}\int d^2\!\sigma\, \, {\bar
\Psi}_R{ \partial}\Psi_R\ ,
\end{equation}
where we have used 
$$
\gamma^0={\rm i}\pmatrix{0&1\cr 1&0}\ ,\quad
\gamma^1={\rm i}\pmatrix{0&-1\cr 1&\phantom{-}0}\ . 
$$

Now, as a fermion goes around the cylinder $\sigma\to\sigma+2\pi$,
there are two types of boundary condition it can have: It can be
periodic, and hence have integer moding, in which case it is said to
be in the ``Ramond'' (R) sector. It can instead be antiperiodic, have
half integer moding, and is said to be in the ``Neveu--Schwarz'' (NS)
sector.

In fact, these two sectors in this theory map to the two sectors of
allowed momenta in the bosonic theory: integer momenta to NS and half
integer to R. The various parts of the partition function can be
picked out and identified in fermionic language. For example, the
contribution:
$$
\left|f^2_3(q)\right|^2\equiv
\left|q^{-{1\over24}}\right|^2\left|\prod_{n=1}^\infty
(1+q^{n-\ha})^2\right|^2\ ,
$$
looks very fermionic, (recall insert 4 (p.\pageref{insert4})) and
is in fact the trace over the contributions from the NS sector
fermions as they go around the torus. It is squared because there are
two components to the fermion, $\Psi$ and $\bar\Psi$. We have the
squared modulus beyond that since we have the contribution from the
left and the right.

The $f_4(q)$ contribution on the other hand, arises from the NS sector
with a $(-)^F$ inserted, where $F$ counts the number of fermions at
each level. The $f_2(q)$ contribution comes from the R sector, and
there is a vanishing contribution from the R sector with $(-1)^F$
inserted. We see that that the projector
\begin{equation}
P={1\over2}(1+(-1)^F)
\labell{projectortwo}
\end{equation}
is the fermionic version of the projector \reef{projector} we
identified previously. Notice that there is an extra factor of two in
front of the R sector contribution due to the definition of $f_2$.
This is because the R ground state is in fact degenerate. The modes
$\Psi_0$ and ${\bar \Psi}_0$ define two ground states which map into
one another. Denote the vacuum by $|s>$, where $s$ can take the values
$\pm\ha$. Then
\begin{eqnarray}
\matrix{\Psi_0|-\ha>=0\ ;& {\bar\Psi}_0|+\ha>=0\ ;\cr\cr
{\bar\Psi}_0|-\ha>=|+\ha>\ ;&\quad \Psi_0|+\ha>=|-\ha>\ ,} 
\end{eqnarray}
and $\Psi_0$ and ${\bar\Psi}_0$ therefore form a representation of the
two dimensional Clifford algebra. We will see this in more generality
later on.  In $D$ dimensions there are $D/2$ components, and the
degeneracy is $2^{D/2}$.

As a final check, we can see that the zero point energies work out
nicely too. The mnemonic (\ref{zpe}) gives us the zero point energy
for a fermion in the NS sector as $-1/48$, we multiply this by two since
there are two components and we see that that we recover the weight of
the ground state in the partition function. For the Ramond sector, the
zero point energy of a single fermion is $1/24$. After multiplying by
two, we see that this is again correctly obtained in our partition
function, since $-1/24+1/8=1/12$. It is awfully nice that the function
$f^2_2(q)$ has the extra factor of $2q^{1/8}$, just for this purpose.

This partition function is again modular invariant, as can be checked
using elementary properties of the $f$--functions \reef{modularswop}:
$f_2$ transforms into $f_4$ under the $S$ transformation, while under
T, $f_4$ transforms into $f_3$.

At the level of vertex operators, the correspondence between the
bosons and the fermions is given by:
\begin{equation}
\matrix{\Psi_L(z)=e^{i\beta X^{25}_L(z)}\ ;& {\bar \Psi}_L(z)=e^{-i\beta
  X^{25}_L(z)}\ ;\cr
\Psi_R({\bar
z})=e^{i\beta X^{25}_R({\bar z})}\ ;&
{\bar \Psi}_R({\bar z})=e^{-i\beta X^{25}_R({\bar z})}\ ,}
\labell{bosonize}
\end{equation}
where $\beta =\sqrt{2/\alpha^\prime}$. 
% ($X$ means $X^{25}$ in the he formulae above.) 
This makes sense, for the exponential factors define
fields single--valued under $X^{25}\to X^{25}+2\pi R$, at our special
radius $R=\sqrt{\alpha^\prime/2}$.  We also have
\begin{equation}
\Psi_L(z){\bar\Psi}_L(z)=\partial_z X^{25}\ ;\quad 
\Psi_R(\zb){\bar\Psi}_R(\zb)=\partial_{\zb} X^{25}\ ,  
\labell{fermionize}
\end{equation}
which shows how to combine two $(0,1/2)$ fields to make a $(0,1)$
field, with a similar structure on the left. Notice also that the
symmetry $X^{25}\to -X^{25}$ swaps $\Psi_{L(R)}$ and ${\bar
  \Psi}_{L(R)}$, a symmetry of interest in the next subsection.  Note
that while we encountered Bosonisation/fermionisation at a special
radius here, it works at other radii too. More generally, with care
taken to make sure that all of the content of the theory is consistent
(all physical operators are mutually local, {\it etc.,}), the
equivalence can be made precise at any radius. We shall briefly use
this fact in later sections, where it will be useful to write vertex
operators in various ways in the supersymmetric theories.

\subsection{String Theory on an Orbifold}
\label{orby}
There is a rather large class of string vacua, called
``orbifolds'',\cite{orbifoldrefs} with many applications in string
theory. We ought to study them, as many of the basic structures will
occur in their definition appear in more complicated examples later
on.

The circle $S^1$, parametrised by $X^{25}$ has the obvious $\IZ_2$
symmetry $R_{25}:X^{25}\to -X^{25}$. This symmetry extends to the full
spectrum of states and operators in the complete theory of the string
propagating on the circle. Some states are even under $R_{25}$, while
others are odd.  Just as we saw before in the case of $\Omega$, it
makes sense to ask whether we can define another theory from this one
by truncating the theory to the sector which is even. This would
define string theory propagating on the ``orbifold'' space
$S^1/\IZ_2$.

\begin{figure}[ht]
\centerline{\psfig{figure=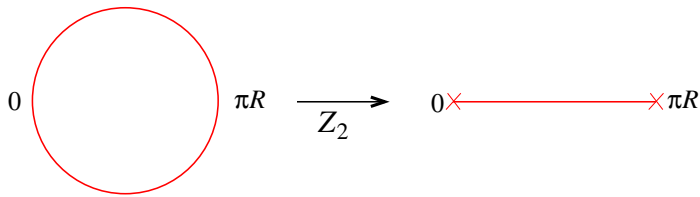,height=1.0in}}
\caption{A $\IZ_2$ orbifold of a circle, giving a line segment with
two fixed points.}
\label{orbifold}
\end{figure}

In defining this geometry, note that it is actually a line segment,
where the endpoints of the line are actually ``fixed points'' of the
$\IZ_2$ action. The point $X^{25}=0$ is clearly such a point and the
other is $X^{25}=\pi R\sim-\pi R$, where $R$ is the radius of the
original $S^1$. A picture of the orbifold space is given in figure
\ref{orbifold}.
In order to check whether string theory on this space is sensible, we
ought to compute the partition function for it. We can work this out
by simply inserting the projector
\begin{equation}
P={1\over2}(1+R_{25})\ ,
\end{equation}
which will have the desired effect of projecting out the $R_{25}$--odd
parts of the circle spectrum. So we expect to see two pieces to the
partition function: a part that is $\ha$ times $Z_{\rm circle}$, and
another part which is $Z_{\rm circle}$ with $R_{25}$ inserted.
Noting that the action of $R_{25}$ is 
\begin{eqnarray}
R_{25}:\left\{\matrix{\alpha_n^{25}\to -\alpha_n^{25}\cr
\tilde{\alpha}_n^{25}\to -{\tilde\alpha}_n^{25}}\right.\ ,
\end{eqnarray}
 the partition function is:
\begin{equation}
Z_{\rm orbifold}={1\over
2}\left[Z(R,\tau)+2\left(|f_2(q)|^{-2}+|f_3(q)|^{-2}
+|f_4(q)|^{-2}\right)\right]\ , \labell{partfunorb}
\end{equation}

The $f_2$ part is what one gets if one works out the projected piece,
but there are two extra terms. From where do they come? One way to see
that those extra pieces must be there is to realize that the first two
parts on their own cannot be modular invariant. The first part is of
course already modular invariant on its own, while the second part 
transforms \reef{modularswop} into $f_4$ under the $S$
transformation, so it has to be there too. Meanwhile, $f_4$ transforms
into $f_3$ under the $T$--transformation, and so that must be there
also, and so on.

While modular invariance is a requirement, as we saw, what is the
physical meaning of these two extra partition functions? What sectors
of the theory do they correspond to and how did we forget them?

The sectors we forgot are very stringy in origin, and arise in a
similar fashion to the way we saw windings appear in earlier sections.
There, the circle may be considered as a quotient of the real line
$\IR$ by a translation $X^{25}\to X^{25}+2\pi R$. There, we saw that
as we go around the string, $\sigma\to\sigma+2\pi$, the embedding map
$X^{25}(\sigma)$ is allowed to change by any amount of the lattice,
$2\pi Rw$. Here, the orbifold further imposes the equivalence
$X^{25}\sim -X^{25}$, and therefore, as we go around the string, we
ought to be allowed:
$$
X^{25}(\sigma+2\pi,\tau)=-X^{25}(\sigma,\tau)+2\pi w R\ ,
$$ for which the solution to the Laplace equation is:
\begin{equation}
X^{25}(z,{\bar z})=x^{25}+ {\rm
i}\sqrt{\alpha^\prime\over2}\sum_{n=-\infty}^\infty {1\over
\left(n+\ha\right)}
\left(\alpha^{25}_{n+\ha}z^{n+\ha}+{\widetilde\alpha}^{25}_{n+\ha}{\bar
z}^{n+\ha} \right)\ , \labell{twisted}
\end{equation}
with $x^{25}=0$ or $\pi R$, no zero mode $\alpha^{25}_0$ (hence no
momentum), and no winding: $w=0$. 

This is a configuration of the string allowed by our equations of
motion and boundary conditions and therefore has to be included in the
spectrum. We have two identical copies of these ``twisted sectors''
corresponding to strings trapped at $0$ and $\pi R$ in spacetime. They
are trapped, since $x^{25}$ is fixed and there is no momentum.

Notice that in this sector, where the boson $X^{25}(w,{\bar w})$ is
antiperiodic as one goes around the cylinder, there is a zero point
energy of $1/16$ from the twisted sector: it is a weight $(1/16,1/16)$
field, in terms of where it appears in the partition function.

Schematically therefore, the complete partition  function ought to be
\begin{equation}
Z_{\rm orb.}={\rm Tr}_{\rm untw'd}
\left( {(1+R_{25})\over2} q^{L_0-{1\over 24}}
{\bar q}^{{\bar L}_0-{1\over 24}}\right)+
{\rm Tr}_{\rm tw'd}\left(
{(1+R_{25})\over2} q^{L_0-{1\over 24}} {\bar q}^{{\bar L}_0-{1\over
24}}\right)
\end{equation}
to ensure modular invariance, and indeed, this is precisely what we
have in \reef{partfunorb}. The factor of two in front of the
twisted sector contribution is because there are two identical twisted
sectors, and we must sum over all sectors.

In fact, substituting in the expressions for the $f$--functions, one
can discover the weight $(1/16,1/16)$ twisted sector fields
contributing to the vacuum of the twisted sector. This simply comes
from the $q^{-1/48}$ factor in the definition of the
$f_{3,4}$--functions. They appear inversely, and for example on the
left, we have $1/48=-{{\rm c}/24}+1/16$, where ${\rm c}=1$.

Finally, notice that the contribution from the twisted sectors do not
depend upon the radius $R$. This fits with the fact that the twisted
sectors are trapped at the fixed points, and have no knowledge of the
extent of the circle. There are orbifolds which can be constructed to
have twisted sectors which are not trapped at fixed points. Their
contributions correspondingly do have knowledge of $R$. We will not
consider those here, in view of the rapidly disappearing available
space.

\subsection{T--Duality for Open Strings: D--branes}

Let us now consider the $R \to 0$ limit of the open string spectrum.
Open strings do not have a conserved winding around the periodic
dimension and so they have no quantum number comparable to $w$, so
something different must happen, as compared to the closed string
case. In fact, it is more like field theory: when $R \to 0$ the states
with nonzero momentum go to infinite mass, but there is no new
continuum of states coming from winding. So we are left with a a
theory in one dimension fewer.  A puzzle arises when one
remembers that theories with open strings  have closed strings
as well, so that in the $R\to 0$ limit the closed strings live in $D$
spacetime dimensions but the open strings only in $D-1$.

This is perfectly fine, though, since the interior of the open string
is indistinguishable from the closed string and so should still be
vibrating in $D$ dimensions.  The distinguished part of the open
string are the endpoints, and these are restricted to a $D-1$
dimensional hyperplane.

This is worth seeing in more detail. Write the open string mode expansion as
\begin{eqnarray}
 &&X^\mu(z,\zb)=X^\mu(z)+X^\mu(\zb)\ ,\quad{\rm where}\nonumber\\
&&\hskip1cm X^\mu(z)={x^\mu\over2}+{x^{\prime\mu}\over2}
-i\alpha^\prime p^\mu_0\ln z+i\left({\alpha^\prime\over2}\right)^{1/2}
\sum_{n\neq0}{1\over n} \alpha_n^\mu z^{-n}\ ,\nonumber\\
&&\hskip1cmX^\mu(\zb)={x^\mu\over2}-{x^{\prime\mu}\over2}
-i\alpha^\prime p^\mu_0\ln \zb+i\left({\alpha^\prime\over2}\right)^{1/2}
\sum_{n\neq0}{1\over n} \alpha_n^\mu\zb^{-n}\ ,
\labell{moreopenmodes}  
\end{eqnarray}
where $x^{\prime\mu}$ is an arbitrary number which cancels out when we
make the usual open string coordinate. Imagine that we place $X^{25}$ on a circle of radius $R$. The T--dual coordinate is
\begin{eqnarray}
&&X^{\prime\mu}(z,\zb)=X^\mu(z)-X^\mu(\zb)\nonumber\\
&&\hskip2cm =x^{\prime\mu}-i\alpha^\prime p^{25}\ln({z\over\zb})+ 
i(2\alpha^\prime)^{1/2}\sum_{n\neq0}{1\over n}\alpha^{25}_ne^{-in\tau}
\sin n\sigma\nonumber\\
&&\hskip2cm =x^{\prime\mu}+2\alpha^\prime p^{25}\sigma+ 
i(2\alpha^\prime)^{1/2}\sum_{n\neq0}{1\over n}\alpha^{25}_ne^{-in\tau}
\sin n\sigma\nonumber\\
&&\hskip2cm =x^{\prime\mu}+2\alpha^\prime {n\over R}\sigma+ 
i(2\alpha^\prime)^{1/2}\sum_{n\neq0}{1\over n}\alpha^{25}_ne^{-in\tau}
\sin n\sigma\ .
\labell{dualcoord}  
\end{eqnarray}

\begin{figure}[ht]
  \centerline{\psfig{figure=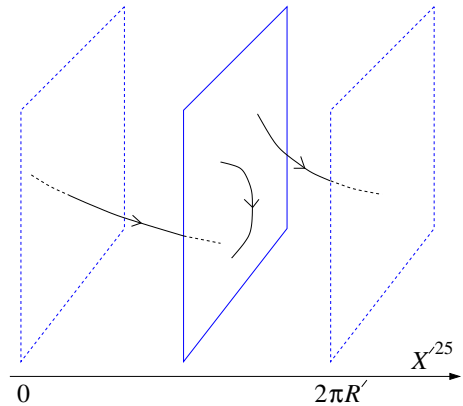,height=2.1in}}
\caption{Open strings with endpoints attached to a hyperplane.  The
dashed planes are periodically identified.  The strings shown have
winding numbers zero and one.}
\label{dbrane}
\end{figure}

Notice that there is no dependence on $\tau$ in the zero mode sector.
This is where momentum usually comes from in the mode expansion, and
so we have no momentum. In fact, since the oscillator terms vanish at
the endpoints $\sigma=0,\pi$, we see that {\it the endpoints do not
  move in the $X^{\prime 25}$ direction!}. Instead of the usual
Neumann boundary condition $\partial_n X\equiv \partial_\sigma X=0$,
we have $\partial_t X\equiv i\partial_\tau X=0$. More precisely, we
have the Dirichlet condition that the ends are at a fixed place:
\begin{eqnarray}
X'^{25}(\pi) - X'^{25}(0) % &=& \int_0^\pi d\sigma \partial_\sigma X'^{25}
%\ =\ i \int_0^\pi d\sigma \partial_\tau X^{25} \nonumber\\ &=& 
%2\pi {\alpha^\prime} p^{25}\ 
=\ \frac{2\pi {\alpha^\prime} n}{R}\ =\ 2\pi n R'. \labell{xchange}
\end{eqnarray}
In other words, the values of the coordinate $X'^{25}$ at the two ends
are  equal up to an integral multiple of the periodicity of the dual
dimension, corresponding to a string that winds as in
figure~\ref{dbrane}.

This picture is consistent with the fact that under T--duality, the
definition of the normal and tangential derivatives get exchanged:
\begin{eqnarray}
\partial_n X^{25}(z,\zb)&=&{\partial X^{25}(z)\over \partial z}
+{\partial X^{25}(\zb)\over \partial \zb}=\partial_t X^{\prime 25}(z,\zb)
\nonumber\\
\partial_t X^{25}(z,\zb)&=&{\partial X^{25}(z)\over \partial z}
-{\partial X^{25}(\zb)\over \partial \zb}=\partial_n X^{\prime 25}(z,\zb)\ .
\labell{derivs}
\end{eqnarray}

Notice that this all pertains to just the direction which we
T--dualised, $X^{25}$.  So the ends are still free to move in the
other $24$ spatial dimensions, which constitutes a hyperplane
called a ``D--brane''. There are 24 spatial directions, so we shall
denote it a D24--brane. 

\subby{Chan-Paton Factors and Wilson Lines} This picture becomes even
more rich when we include Chan--Paton factors.\cite{Djoe} Consider the
case of $U(N)$, the oriented open string. When we compactify the
$X^{25}$ direction, we can include a Wilson line $A_{25}={\rm
  diag}\{\theta_1,\theta_2,\ldots,\theta_N\}/2\pi R$, which
generically breaks $U(N) \to U(1)^N$. (See insert 6
(p.\pageref{insert6}) for a short discussion.) Locally this is pure
gauge,
\begin{equation}
A_{25} = -i\Lambda^{-1}\partial_{25}\Lambda,\qquad
\Lambda={\rm diag}\{e^{ i X^{25}\theta_1/2\pi R},e^{ i X^{25}\theta_2/2\pi
R}, \ldots , e^{ i X^{25}\theta_1/2\pi R} \}\ .
\end{equation} 
We can gauge $A_{25}$ away, but since the gauge transformation is not
periodic, the fields  pick up a phase
\begin{equation}
{\rm diag}\left\{ e^{-i\theta_1}, e^{-i\theta_2}, \ldots, e^{-i\theta_N}
\right\} \labell{wilphase}
\end{equation}
under $X^{25} \to X^{25} + 2\pi R$.  

\insertion{6}{Particles and Wilson Lines\label{insert6}}{The following
  illustrates an interesting gauge configuration which arises when
  spacetime has the non--trivial topology of a circle (with coordinate
  $X^{25}$) of radius $R$. Consider the case of $U(1)$. Let us make
  the following choice of constant background gauge potential:
\begin{equation}
A_{25}(X^\mu)=-{\theta\over2\pi R}=
-i\Lambda^{-1}{\partial\Lambda\over\partial X^{25}}\ ,
\nonumber
\end{equation}
where $ \Lambda(X^{25})=e^{-{i\theta X^{25}\over2\pi R}}\ .$
This is clearly pure gauge, but only locally. There still exists
non--trivial physics. Form the gauge invariant quantity (``Wilson
Line''):
\begin{equation}
W_q=\exp\left(iq\oint dX^{25} A_{25}\right)=e^{-iq\theta}\ .
\nonumber
\end{equation}
Where does this  observable show up? Imagine a point
particle of charge $q$ under the $U(1)$. Its action can be written
(see section 2) as:
\begin{equation}
S=\int d\tau\left\{{1\over2}{\dot X}^\mu{\dot
X}_\mu-iqA_\mu{\dot X}^\mu\right\}=\int d\tau {\cal L}\ .
\end{equation}
(The last term in the action is just $-iq\int A=-iq\int A_\mu dx^\mu$,
in the language of forms... this will be the natural coupling of a
world volume to an antisymmetric tensor, as we shall see.)
Recall that in the path integral we are computing $e^{-S}$. So if the
particle does a loop around $X^{25}$ circle, it will pick up a phase
factor of $W_q$.
Notice: the conjugate momentum to $X^\mu$ is 
$$\Pi^\mu=i{\partial {\cal L}\over\partial{\dot X}^\mu}=i{\dot X}^\mu\ ,$$
except for 
$$
\Pi^{25}=i{\dot X}^{25}-{q\theta\over2\pi R}={n\over R}
$$ where the last equality results from the fact that we are on a
circle.  Now we can of course gauge away $A$ with the choice
$\Lambda^{-1}$, but it will be the case that as we move around the
circle, {\it i.e.,} $X^{25}\to X^{25}+2\pi R$, the particle (and all
fields) of charge $q$ will pick up a phase $e^{iq\theta}$.  So the
canonical momentum is shifted to:
\begin{equation}
p^{25}={n\over R}+{q\theta\over2\pi R}\ .
\label{canonical}
\end{equation}}

\begin{figure}[ht]
\centerline{\psfig{figure=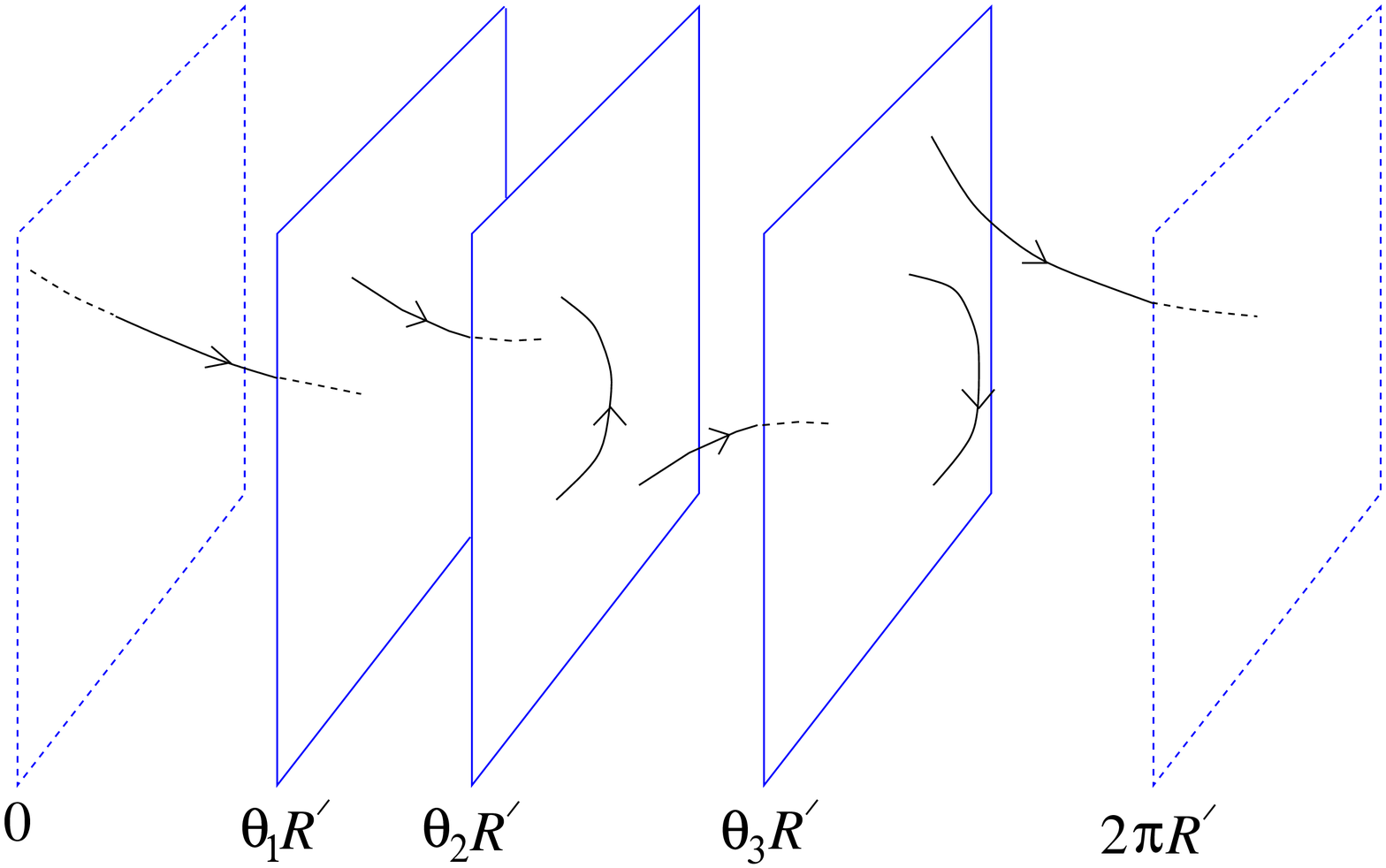,height=2.1in}}
\caption{Three D--branes  at different positions, with various strings
attached.}
\label{dbranes}
\end{figure}

What is the effect in the dual theory?  Due to the
phase~(\ref{wilphase}) the open string momenta are now fractional.  As
the momentum is dual to winding number, we conclude that the fields in
the dual description have fractional winding number, {\it i.e.,} their
endpoints are no longer on the same hyperplane.  Indeed, a string
whose endpoints are in the state $|ij\rangle$ picks up a phase
$e^{i(\theta_j - \theta_i)}$, so their momentum is $(2\pi n + \theta_j
- \theta_i)/2\pi R$.  Modifying the endpoint
calculation~(\ref{xchange}) then gives
\begin{equation}
X'^{25}(\pi) - X'^{25}(0) = (2\pi n + \theta_j - \theta_i) R'.
\end{equation}
In other words, up to an arbitrary additive constant, the endpoint in
state $i$ is at position
\begin{equation}
X'^{25}\ =\ \theta_i R' \ 
=\ 2\pi{\alpha^\prime} A_{25,ii}. \labell{positiondual}
\end{equation}  
We have in general $N$ hyperplanes at different positions as
depicted in figure~\ref{dbranes}.

\subsection{D--Brane Dynamics: Collective Co\"{o}rds and Gauge Theory}
\label{collective}
Clearly, the whole picture goes through if several coordinates
\begin{equation}
X^m=\{X^{25},X^{24},\ldots,X^{p+1}\}
\end{equation}
are periodic, and we rewrite the periodic dimensions in terms of the
dual coordinates.  The open string endpoints are then confined to $N$
$(p+1)$--dimensional hyperplanes, the D$(p+1)$--branes.  The Neumann
conditions on the world sheet, $\partial_nX^m(\sigma^1,\sigma^2)=0$,
have become Dirichlet conditions $\partial_t X^{\prime
  m}(\sigma^1,\sigma^2)=0$ for the dual coordinates.  In this
terminology, the original 26 dimensional open string theory theory
contains $N$ D25--branes. A 25--brane fills space, so the string
endpoint can be anywhere: it just corresponds to an ordinary
Chan--Paton factor.

It is natural to expect that the hyperplane is dynamical rather than
rigid.\cite{dbranesi} For one thing, this theory still has gravity, and it
is difficult to see how a perfectly rigid object could exist.  Rather,
we would expect that the hyperplanes can fluctuate in shape and
position as dynamical objects. We can see this by looking at the
massless spectrum of the theory, interpreted in the dual coordinates.

Taking for illustration the case where a single coordinate is
dualised, consider the mass spectrum.  The $D-1$ dimensional mass is
\begin{eqnarray}
M^2 &=& (p^{25})^2+{1\over{\alpha^\prime}}({N}-1) \nonumber\\
&=& \left( {[2\pi n+(\theta_i-\theta_j)]R^\prime\over2\pi{\alpha^\prime}}
\right)^2 +{1\over{\alpha^\prime}}(N-1).
\end{eqnarray}
Note that $[2\pi n+(\theta_i-\theta_j)]R^\prime$ is the minimum length
of a string winding between hyperplanes $i$ and $j$.  Massless states
arise generically only for non--winding ({\it i.e.,} $n=0$) open
strings whose end points are on the same hyperplane, since the string
tension contributes an energy to a stretched string.  We have
therefore the massless states (with their vertex operators):
\begin{eqnarray}
\alpha^{\mu}_{-1}|{ k};ii\rangle, && V = \partial_t X^\mu, \nonumber\\
\alpha^{m}_{-1}|{ k};ii\rangle, && V = \partial_t X^{25} = \partial_n
X'^{25}.
\end{eqnarray}
The first of these is a gauge field living on the D--brane, with $p+1$
components tangent to the hyperplane, $A^\mu(\xi^a)$,
$\mu,a=0,\ldots,p$. Here, $\xi^\mu=x^\mu$ are coordinates on the
D--branes' world--volume. The second was the gauge field in the
compact direction in the original theory.  In the dual theory it
becomes the transverse position of the D--brane (see
\reef{positiondual}). From the point of view of the world--volume, it
is a family of scalar fields, $\Phi^m(\xi^a)$,
($m=p+1,\ldots,D-1$) living there.

We saw this in equation (\ref{positiondual}) for a Wilson line, which
was a constant gauge potential.  Now imagine that, as genuine scalar
fields, the $\Phi^m$ vary as we move around on the world--volume of
the D--brane. This therefore embeds the brane into a variable place in
the transverse coordinates. This is simply describing a specific {\it
  shape} to the brane as it is embedded in spacetime. The
$\Phi^m(\xi^a)$ are exactly analogous to the embedding coordinate map
$X^\mu(\sigma,\tau)$ with which we described strings in the earlier
sections.

The values of the gauge field backgrounds describe the shape of the
branes as a soliton background, then. Meanwhile their quanta describe
fluctuations of that background.  This is the same phenomenon which we
found for our description of spacetime in string theory.  We started
with strings in a flat background and discover that a massless closed
string state corresponds to fluctuations of the geometry.  Here we
found first a flat hyperplane, and then discovered that a certain open
string state corresponds to fluctuations of its shape. Remarkably,
these open string states are simply gauge fields, and this is one of
the reasons for the great success of D--branes. There are other branes
in string theory (as we shall see) and they have other types of field
theory describing their collective dynamics. D--branes are special, in
that they have a beautiful description using gauge theory. Ultimately,
we can use the long experience of working with gauge theories to teach
us much about D--branes, and later, the geometry of D--branes and the
string theories in which they live can teach us a lot about gauge
theories.  This is the basis of the dialogue between gauge theory and
geometry which dominates the field at present.

It is interesting to look at the $U(N)$ symmetry breaking in the dual
picture where the brane can move transverse to their world--volumes.
When no D--branes coincide, there is just one massless vector each, or
$U(1)^N$ in all, the generic unbroken group.  If $k$ D--branes
coincide, there are new massless states because strings which are
stretched between these branes can achieve vanishing length.  Thus,
there are $k^2$ vectors, forming the adjoint of a $U(k)$ gauge
group.\cite{Djoe,edbound} This coincident position
corresponds to $\theta_1=\theta_2=\cdots=\theta_k$ for some subset of
the original $\{\theta\}$, so in the original theory the Wilson line
left a $U(k)$ subgroup unbroken. At the same time, there appears a set
of $k^2$ massless scalars: the $k$ positions are promoted to a matrix.
This is curious and hard to visualise, but plays an
important role in the dynamics of D--branes.\cite{edbound} We will
examine many consequences of this later in these notes.  Note that
if all $N$ branes are coincident, we recover the $U(N)$ gauge
symmetry.

While this picture seems a bit exotic, and will become more so in the
unoriented theory, the reader should note that all we have done is to
rewrite the original open string theory in terms of variables which
are more natural in the limit $R \ll \sqrt{{\alpha^\prime}}$.  Various
puzzling features of the small--radius limit become clear in the
T--dual picture.

Observe that, since T--duality interchanges Neumann and Dirichlet
boundary conditions, a further T--duality in a direction tangent to
a D$p$--brane reduces it to a D$(p-1)$--brane, while a T--duality in a
direction orthogonal turns it into a D$(p+1)$--brane.

\subsection{T--Duality and Orientifolds.}
\label{orientplanes}
The $R \to 0$ limit of an unoriented theory also leads to a new
extended object.  Recall that the effect of T--duality can also be
understood as a one--sided parity transformation.  For closed strings,
the original coordinate is $X^m(z,\zb)=X^m(z)+X^m(\zb)$. We have
already discussed how to project string theory with these coordinates
by $\Omega$.  The dual coordinate is $X^{\prime
  m}(z,\zb)=X^m(z)-X^m(\zb)$.  The action of world sheet parity
reversal is to exchange $X^\mu(z)$ and $X^\mu(\zb)$. This gives for
the dual coordinate:
\begin{equation}
X^{\prime m}(z,\zb) \leftrightarrow -X^{\prime m}(\zb,z)\ .
\end{equation}
{\it This is the product of a world--sheet and a spacetime parity
  operation.}  In the unoriented theory, strings are invariant under
the action of $\Omega$, while in the dual coordinate the theory is
invariant under the product of world--sheet parity and a spacetime
parity.  This generalisation of the usual unoriented theory is known
as an ``orientifold'', a term which mixes the term ``orbifold'' with
orientation reversal.

Imagine that we have separated the string wavefunction into its
internal part and its dependence on the centre of mass, $x^m$.
Furthermore, take the internal wavefunction to be an eigenstate of
$\Omega$.  The projection then determines the string wavefunction at
$-x^m$ to be the same as at $x^m$, up to a sign.  In practice, the
various components of the metric and antisymmetric tensor satisfy {\it
  e.g.,}
\begin{eqnarray}
G_{\mu\nu}(x^\mu,-x^m) = G_{\mu\nu}(x^\mu,x^m),&&
B_{\mu\nu}(x^\mu,-x^m) = -B_{\mu\nu}(x^\mu,x^m), \nonumber\\ G_{\mu
n}(x^\mu,-x^m) = -G_{\mu n}(x^\mu,x^m),&& B_{\mu n}(x^\mu,-x^m) =
B_{\mu n}(x^\mu,x^m), \nonumber\\ G_{mn}(x^\mu,-x^m) =
G_{mn}(x^\mu,x^m),&& B_{mn}(x^\mu,-x^m) = -B_{mn}(x^\mu,x^m)\ . 
\labell{horrid}
\end{eqnarray}
In other words, when we have $k$ compact directions, the T--dual
spacetime is the torus $T^{25-k}$ modded by a $\IZ_2$ reflection in
the compact directions.  So we are instructed to perform an orbifold
construction, modified by the extra sign.  In the case of a single
periodic dimension, for example, the dual spacetime is the line
segment $0 \leq x^{25} \leq \pi R^\prime$. The reader should remind
themselves of the orbifold construction in section \ref{orby}. At the
ends of the interval, there are fixed ``points'' , which are in fact
spatially 24--dimensional planes. Looking at the projections
\reef{horrid} in this case, we see that on these fixed planes, the
projection is just like we did for the $\Omega$--projection of the
25+1 dimensional theory in section \ref{unory}: The theory is
unoriented there, and half the states are removed. These orientifold
fixed planes are called ``O--planes'' for short. For this case,
we have two O24--planes.  (For $k$ directions we have $2^k$
O$(25-k)$--planes arranged on the vertices of a hypercube.) In
particular, we can usefully think of the original case of $k=0$ as
being on an O25--plane.

While the theory is unoriented on the O--plane, away from the
orientifold fixed planes, the local physics is that of the {\it
  oriented} string theory.  The projection relates the physics of a
string at some point $x^m$ to the string at the image point $-x^m$.

In string perturbation theory, orientifold planes are not dynamical.
Unlike the case of D--branes, there are no string modes tied to the
orientifold plane to represent fluctuations in its shape. Our
heuristic argument in the previous subsection that gravitational
fluctuations force a D--brane to move dynamically does not apply to
the orientifold fixed plane.  This is because the
identifications~(\ref{horrid}) become {\it boundary conditions} at the
fixed plane, such that the incident and reflected gravitational waves
cancel.  For the D--brane, the reflected wave is higher order in the
string coupling.

The orientifold construction was discovered via
T--duality\,\cite{dbranesi} and independently from other points of
view.\cite{opentwist,dbranesiii} One can of course consider more
general orientifolds which are not simply T--duals of toroidal
compactifications. The idea is simply to combine a group of discrete
symmetries with $\Omega$ such that the resulting group of operations
(the ``orientifold group'', $G_{\Omega}$) is itself a symmetry of some
string theory. One then has the right to ask what the nature of the
projected theory obtained by dividing by $G_{\Omega}$ might be. This
is a fruitful way of construction interesting and useful string
vacua.~\cite{atishreview} We shall have more to say about this later,
since in superstring theory we shall find that O--planes, like
D--branes , are sources of various closed string sector fields.
Therefore there will be additional consistency conditions to be
satisfied in constructing an orientifold, amounting to making sure
that the field equations are satisfied.

So far our discussion of orientifolds was just for the closed string
sector. Let us see how things are changed in the presence of open
strings. In fact, the situation is similar.  Again, let us focus for
simplicity on a single compact dimension.  Again there is one
orientifold fixed plane at $0$ and another at $\pi R^\prime$.
Introducing $SO(N)$ Chan--Paton factors, a Wilson line can be brought
to the form
\begin{equation}
{\rm diag}
\{\theta_1,-\theta_1,\theta_2,-\theta_2,\cdots,
\theta_{N/2},-\theta_{N/2}\}.
\end{equation}
Thus in the dual picture there are ${1\over 2}N$ D--branes on the line
segment $0\leq X^{\prime25}<\pi R^\prime$, and ${1\over 2}N$ at their
image points under the orientifold identification. 

\begin{figure}[ht]
\centerline{\psfig{figure=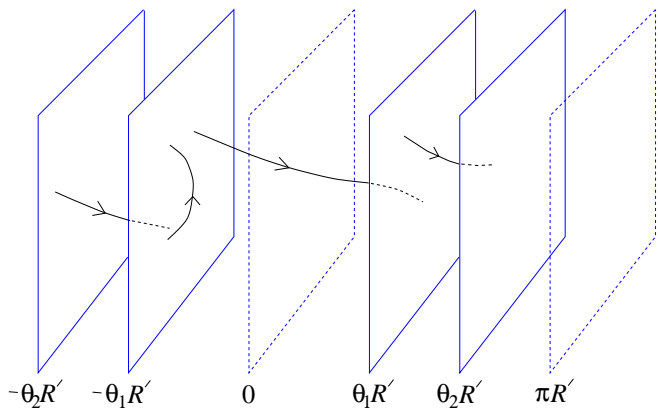,height=2.1in}}
\caption{Orientifold planes at $0$ and $\pi R'$. There are D--branes at
  $\theta_1 R'$ and $\theta_2 R'$, and their images at -$\theta_1 R'$
  and $-\theta_2 R'$.  $\Omega$ acts on any string by a combination of
  a spacetime reflection through the planes and reversing the
  orientation arrow.}
\label{orient}
\end{figure}

Strings can stretch between D--branes and their images as shown in
figure~\ref{orient}.
The generic gauge group is $U(1)^{N/2}$, where all branes are
separated.  As in the oriented case, if $m$ D--branes are coincident
there is a $U(m)$ gauge group.  However, now if the $m$ D--branes in
addition lie at one of the fixed planes, then strings stretching
between one of these branes and one of the image branes also become
massless and we have the right spectrum of additional states to fill
out $SO(2m)$. The maximal $SO(N)$ is restored if all of the branes are
coincident at a single orientifold plane.  Note that this maximally
symmetric case is asymmetric between the two fixed planes.  Similar
considerations apply to $USp(N)$. As we saw before, the difference
between the appearance of the two groups is in a sign on the matrix
$M$ as it acts on the string wavefunction. Later, we shall see that
this sign is correlated with the sign of the charge and tension of
the orientifold plane.

We should emphasise that there are $\frac{1}{2}N$ dynamical D--branes
but an $N$--valued Chan--Paton index.  An interesting case is when $k +
\frac{1}{2}$ D--branes lie on a fixed plane, which makes sense because
the number $2k + 1$ of indices is integer.  A brane plus image can
move away from the fixed plane, but the number of branes remaining is
always half--integer. This anticipates a discussion which we shall
have about fractional branes much later, in section \ref{fractions}
 even outside the context of
orientifolds.

\subsection{The D--Brane Tension}
\label{tense}
The D--brane is a dynamical object, and as such, feels the force of
gravity.  The tension of the brane controls its response to outside
influences which try to make it change its shape, absorb energy, {\it
  etc.}, just as we saw for the tension of a string. If we introduce
coordinates $\xi^a$, $a = 0, \ldots p$ on the brane, we can begin to
write an action for the dynamics of the brane in terms of fields
living on the world--volume in much the same way that we did for the
string, in terms of fields living on the world--sheet.  As we
discussed before, the fields on the brane are the embedding
$X^\mu(\xi)$ and the gauge field $A_a(\xi)$. We shall ignore the
latter for now and concentrate just on the embedding part, which is
enough to get the tension right.  The action is (again by direct
analogy to the particle and string case)
\begin{equation}
S_p = -T_p \int d^{p+1}\xi\, e^{-\Phi} 
\det\!^{1/2} G_{ab}\ , \labell{actone}
\end{equation}
where $G_{ab}$ is the induced metric on the brane, otherwise known as
the ``pull--back'' of the spacetime metric $G_{\mu\nu}$ to the brane:
\begin{equation}
G_{ab}\equiv{\partial X^\mu\over
\partial \xi^a}{\partial X^\nu\over\partial\xi^b}
 G_{\mu\nu} \ .
\labell{induced}
\end{equation} $T_p$ is the tension of the D$p$--brane.
The dilaton dependence $e^{-\Phi} = g_s^{-1}$ arises because this is
an open string tree level action.

Before computing the tension, we should note that we can get a
recursion relation for it from T--duality\cite{ght,shanta} The mass
of a D$p$--brane wrapped around a $p$--torus  $T^p$ is
\begin{equation}
T_p e^{-\Phi} \prod_{i=1}^p (2\pi R_i)\ .\labell{wrapmass}
\end{equation}  T--dualising on the single direction 
$X^p$ and recalling the transformation~(\ref{newg}) of the dilaton, we
can rewrite the mass~(\ref{wrapmass}) in the dual variables:
\begin{equation}
T_p (2\pi\sqrt{{{\alpha^\prime}}}) e^{-\Phi'} \prod_{i=1}^{p-1} (2\pi R_i) \ =\
T_{p-1} e^{-\Phi'} \prod_{i=1}^{p-1} (2\pi R_i)\ .
\end{equation}
Hence,
\begin{equation}
T_p = T_{p-1}/2\pi\sqrt{{{\alpha^\prime}}}\qquad\Rightarrow\qquad
T_p=T_{p'}(2\pi\sqrt{\alpha^\prime})^{p'-p}\ . \labell{trec}
\end{equation} 
Where we performed the duality  recursively to deduce the general relation.

\begin{figure}[ht]
\centerline{\psfig{figure=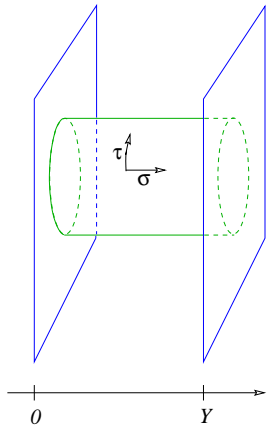,height=2.2in}}
\caption{Exchange of a closed string between two D--branes. 
This is equivalent to a vacuum loop of an open string with one end on each
D--brane.}
\label{tiefighter}
\end{figure}

Let us now compute the  D--brane tension $T_{p}$. As noted above, it
is proportional to $g_s^{-1}$.  One could calculate it from the
gravitational coupling to the D--brane, given by the disk with a
graviton vertex operator.  However, it is much easier to obtain the
absolute normalisation as follows.  Consider two parallel 
D$p$--branes at positions $X^{\prime\mu}=0$ and
$X^{\prime\mu}=Y^\mu$. These two objects can feel each other`s
presence by exchanging closed strings as shown in figure~\ref{tiefighter}.
This string graph is an annulus, with no vertex operators.  It is
therefore as easily calculated as our closed string one loop
amplitudes done earlier.  Extracting the poles from graviton and
dilaton exchange then give the coupling $T_p$ of closed string states
to the D--brane.

Let us parametrise the string world--sheet as
$(\sigma^2=\tau,\sigma^1=\sigma)$ where now $\tau$ is periodic and
runs from $0$ to $2\pi t$, and $\sigma$ runs (as usual) from $0$ to
$\pi$.  This vacuum graph (a cylinder) has the single modulus $t$,
running from $0$ to $\infty$.  If we slice horizontally, so that
$\sigma^2=\tau$ is world--sheet time, we get an open string going in a
loop.  If we instead slice vertically, so that $\sigma$ is time, we
see a single closed string propagating in the tree channel.  (The
world--line of the open string boundary can be regarded as a vertex
connecting the vacuum to the single closed string, {\it i.e.,} a
one--point closed string vertex,~\cite{chans,rrex}
which is a useful picture in the ``boundary state'' formalism, which
we will not use here.)

This diagram will occur explicitly again in these lectures. It also
appears implicitly in many other modern aspects in this series of
lectures in this school and beyond: String theory produces many
examples where one--loop gauge/field theory results (open strings) are
related to tree level geometrical/gravity results. This is all
organised by diagrams of this form, and is the basis of much of the
gauge theory/geometry correspondences to be discussed.

Let us consider the limit $t \to 0$ of the loop amplitude. This is the
ultra--violet limit for the open string channel, since the circle of
the loop is small.  However, this limit is correctly interpreted as an
{\it infrared} limit of the closed string. (This is one of the
earliest ``dualities'' of string theory, discussed even before it was
known to be a theory of strings.) Time--slicing vertically shows that
the $t \to 0$ limit is dominated by the lowest lying modes in the
closed string spectrum.  This all fits with the idea that there are no
``ultraviolet limits'' of the moduli space which could give rise to
high energy divergences.  They can always be related to amplitudes
which have a handle pinching off. This physics is controlled by the
lightest states, or the long distance physics. (This relationship is
responsible for the various ``UV/IR'' relations which are a popular
feature of current research.)

One--loop vacuum amplitudes are given by the Coleman--Weinberg formula,
which can be thought of as the sum of the zero point energies of all
the modes: (see insert 7 (p.\pageref{insert7}))~\cite{cw}
\begin{equation}
{\cal A} = V_{p+1}\int \frac{d^{p+1}k}{(2\pi)^{p+1}}\int_0^\infty {dt
\over 2t} \sum_{I} e^{-2\pi {\alpha^\prime} t (k^2 + M_I^2)} .
\end{equation}
Here the sum $I$ is over the physical spectrum of the string,
{\it i.e.,} the transverse spectrum, and the momentum $k$ is in the
$p+1$ extended directions of the D--brane world--sheet.

\insertion{7}{Vacuum Energy\label{insert7}}{The Coleman--Weinberg
  formula evaluates the one--loop vacuum amplitude, which is simply
  the logarithm of the partition function ${\cal A}=Z_{\rm vac}$ for
  the complete theory:
  $$
  \ln\left(Z_{\rm
      vac}\right)=-{1\over2}\Tr\ln\left(\Box^2+M^2\right)=-{V_D\over2}\int{d^Dk\over(2\pi)^D}
  \ln\left(k^2+M^2\right)\ .
$$
But since we can write
$$
-{1\over2}\ln(k^2+M^2)=\int_0^\infty {dt\over2t}e^{-{(k^2+M^2)}t/2}\ ,
$$
we have
$$
{\cal A}=V_D\int {d^Dk\over(2\pi)^D}\int_0^\infty {dt\over
  2t}e^{-{(k^2+M^2)}t/2}\ .
$$
Recall finally that $(k^2+M^2)/2$ is just the Hamiltonian, $H$,
which in our case is just $L_0/\alpha^\prime$ (see \reef{hamilton}).}

\insertion{8}{Translating Closed to Open\label{insert8}}{ Compare our
  open string appearance of $f_1(q)$, for $q=e^{-2\pi t}$ with the
  expressions for $f_1(q)$, ($q=e^{2\pi\tau}$) defined in our closed
  string discussion in \reef{thethetas}. Here the argument is real.
  The translation between definitions is done by setting $t=-{\rm
    Im}\, \tau$. From the modular transformations \reef{modularswop},
  we can deduce some useful asymptotia.  While the asymptotics as
  $t\to\infty$ are obvious, we can get the $t\to 0$ asymptotics using
  \reef{modularswop}
\begin{equation}
f_{1}(e^{-{\pi}/{s}}) = \sqrt{s}\,f_{1}(e^{-\pi s}),\quad
f_{3}(e^{-{\pi}/{s}}) = f_{3}(e^{-\pi s}),\quad
f_{2}(e^{-{\pi}/{s}}) = f_{4}(e^{-\pi s}) . 
\end{equation}}

The mass
spectrum is given by
\begin{equation}
M^2={1\over{\alpha^\prime}}\left(\sum_{n=1}^\infty
{\alpha}^i_{-n}\alpha_n^i-1\right)+{Y \cdot Y\over4\pi^2
\alpha^{\prime 2}}
\end{equation}
where $Y^m$ is the separation of the D--branes.  The sums over the
oscillator modes work just like the computations we did before (see
insert 4 (p.\pageref{insert4})), giving
\begin{equation}
{\cal A}=2V_{p+1} \int_0^\infty {dt\over 2t} (8\pi^2 {\alpha^\prime}
t)^{-{(p+1)\over2}} e^{-Y \cdot Y t/ 2\pi{\alpha^\prime}} f_1(q)^{-24}\ .
\end{equation}
Here $q=e^{-2\pi t}$, and the overall factor of 2 is from exchanging
the two ends of the string. (See insert 8 (p.\pageref{insert8}) for
news of $f_1(q)$)

In the present case, (using the asymptotics derived in insert 8)
\begin{equation}
{\cal A}= 2 V_{p+1} \int_0^\infty {dt\over 2t} (8\pi^2 {\alpha^\prime}
t)^{-{(p+1)\over2}} e^{-Y \cdot Y t/ 2\pi{\alpha^\prime}} t^{12} \left( e^{2\pi/t}
+ 24 + \ldots \right).  \labell{cpoles}
\end{equation}
The leading divergence is from the tachyon and is an uninteresting
bosonic artifact.  The massless pole, from the second term, is
\begin{eqnarray}
{\cal A} &\sim& V_{p+1}{24\over 2^{12}} (4\pi^2{\alpha^\prime})^{11-p}
\pi^{(p-23)/2} \Gamma((23 - p)/2) |Y|^{p-23} \nonumber\\
& =& V_{p+1} {24 \pi\over 2^{10}} (4\pi^2{\alpha^\prime})^{11-p}
G_{25-p}(Y^2)
\end{eqnarray}
where $G_d(Y^2)$ is the massless scalar Green's function in $d$
dimensions.

We can  compare this with a field theory calculation, the exchange of
graviton plus dilaton between a pair of D--branes.  The propagator is
from the bulk action~(\ref{stringfrm}) and the couplings are from the
D--brane action~(\ref{actone}).  This is a bit of effort because the
graviton and dilaton mix, but in the end one finds
\begin{equation}
{\cal A} \sim {D-2 \over 4} V_{p+1} T_p^2 \kappa_0^2
 G_{25-p}(Y^2) \labell{fieldpole}
\end{equation}
and so
\begin{equation}
T_p = {\sqrt{\pi}\over 16 \kappa_0} (4\pi^2{\alpha^\prime})^{(11-p)/2}\ .
\labell{bosonict}
\end{equation}
This agrees with the recursion relation~(\ref{trec}).  The actual
D--brane tension includes a factor of the string coupling  from the
action~(\ref{actone}),
\begin{equation}
\tau_p = {\sqrt{\pi}\over 16 \kappa} (4\pi^2{\alpha^\prime})^{(11-p)/2}\ 
\end{equation}
where $\kappa = \kappa_0 g_s$, and we shall use $\tau$ this to denote
the tension when we include the string coupling henceforth, and
reserve $T$ for situations where the string coupling is included in
the background field $e^{-\Phi}$.( This will be less confusing than it
sounds, since it will always be clear from the context which we mean.)

Notice then that the tension $\tau_p$ of a D$p$--brane is of order
$g_s^{-1}$. This follows from the fact that the diagram connecting the
brane to the closed string sector is a disc diagram, and insert 5
(p.\pageref{insert5}) shows reminds us that this is of order
$g_s^{-1}$. An immediate consequence of this is that they will produce
non--perturbative effects of order $\exp(-1/g_s)$ in string theory,
since their action is of the same order as their mass.

Formula \reef{bosonict} will not concern us much beyond these
sections, since we will derive a new one for the superstring case
later.

The asymptotics~(\ref{cpoles}) can be interpreted in terms
of a sum over closed string states exchanged between the two D--branes.
One can write the cylinder path integral in Hilbert space formalism
treating $\sigma_1$ rather than $\sigma_2$ as time.  It then has the
form
\begin{equation}
\langle B | e^{-(L_0 + \tilde L_0)\pi/t} | B \rangle
\end{equation}
where the {\it boundary state} $|B\rangle$ is the closed string state
created by the boundary loop.  We will not have time to develop this
formalism but it is useful in finding the couplings between closed and
open strings.\cite{chans,rrex}

\subsection{The Orientifold Tension}

\begin{figure}[ht]
\centerline{\psfig{figure=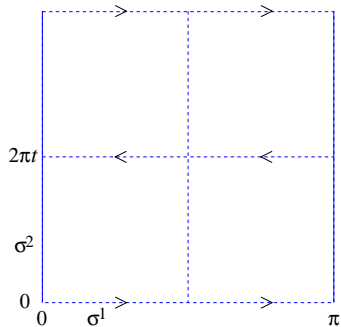,height=1.7in}}
\caption{Two copies of the fundamental region for the M\"obius strip.}
\label{mobiusfund}
\end{figure}

\begin{figure}[ht]
\centerline{\psfig{figure=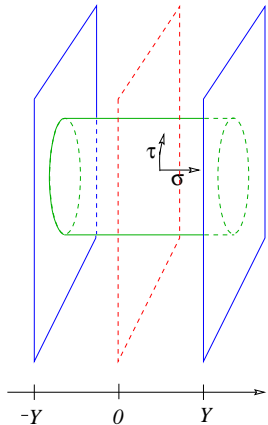,height=2.2in}}
\caption{The M\"obius strip as the exchange of closed strings between
a brane and its mirror image. The dotted plane is the orientifold plane.}
\label{mirrortie}
\end{figure}

The O--plane, like the D--brane, couples to the dilaton and metric.
The amplitude is the same as in the previous section, but with
$\mathbf{RP}^2$ in place of the disk; {\it i.e.,} a crosscap replaces
the boundary loop.  The orientifold identifies $X^m$ with $-X^m$ at
the opposite point on the crosscap, so the crosscap is localised near
one of the orientifold fixed planes.  Again the easiest way to
calculate this is {\it via} vacuum graphs, the cylinder with one or
both boundary loops replaced by crosscaps.  These give the M\"obius
strip and Klein bottle, respectively.  To understand this, consider
figure~\ref{mobiusfund}, which shows two copies of the fundamental
region for the M\"obius strip.
The lower half is identified with the reflection of the upper, and the
edges $\sigma^1 = 0, \pi$ are boundaries.  Taking the lower half as
the fundamental region gives the familiar representation of the
M\"obius strip as a strip of length $2\pi t$, with ends twisted and
glued.  Taking instead the left half of the figure, the line $\sigma^1
= 0$ is a boundary loop while the line $\sigma^1 = \pi/2$ is
identified with itself under a shift $\sigma^2 \to \sigma^2 + 2\pi t$
plus reflection of $\sigma^1$: it is a crosscap.  The same
construction applies to the Klein bottle, with the right and left
edges now identified. Another way to think of the M\"obius strip
amplitude we are going to compute  here is as representing the
exchange of a closed string between a D---brane and its mirror image,
as shown in figure~\ref{mirrortie}. The identification with a twist is
performed on the two D--branes, turning the cylinder into a M\"obius
strip.

The M\"obius strip is given by the
vacuum amplitude
\begin{equation}
{\cal A}_{\rm M} = V_{p+1}\int \frac{d^{p+1}k}{(2\pi)^{p+1}} \int_0^\infty
{dt \over 2t} \sum_{I} {\Omega_i\over 2} e^{-2\pi {\alpha^\prime} t (p^2 + M_I^2)} ,
\end{equation}
where $\Omega_I$ is the $\Omega$ eigenvalue of state $i$.  The
oscillator contribution to $\Omega_I$ is $(-1)^N$ from
eq.~(\ref{photonout}).~\footnote {In the directions orthogonal to the
  brane and orientifold there are two additional signs in $\Omega_I$
  which cancel: the world-sheet parity contributes an extra minus sign
  in the directions with Dirichlet boundary conditions (this is
  evident from the mode expansions we shall list later, in
  equations~(\ref{modexps})), and the spacetime reflection an
  additional sign.}  For the $SO(N)$ open string the Chan--Paton
factors have ${1\over 2}N(N+1)$ even states and ${1\over 2}N(N-1)$ odd
for a net of $+N$. For $USp(N)$ these numbers are reversed for a net
of $-N$.  Focus on a D--brane and its image, which correspondingly
contribute $\pm 2$.  The diagonal elements, which contribute to the
trace, are those where one end is on the D--brane and one on its image.
The total separation is then $Y^m = 2X^m$.  Then,
\begin{eqnarray}
{\cal A}_{\rm M} &=& 
\pm V_{p+1} \int_0^\infty {dt\over 2t} (8\pi^2 {\alpha^\prime}
t)^{-{(p+1)\over2}} e^{-2X \cdot X t/ \pi{\alpha^\prime}}\times \nonumber\\
&&\qquad\qquad\qquad\qquad\qquad
\times
\left[ q^{-2} \prod_{k=1}^\infty (1+q^{4k-2})^{-24} (1-q^{4k})^{-24} \right]
\end{eqnarray}
The factor in braces  is
\begin{eqnarray}
f_3(q^2)^{-24} f_1(q^2)^{-24} &=& (2t)^{12} 
f_3(e^{-\pi/2t})^{-24} f_1(e^{-\pi/2t})^{-24} \nonumber\\
&=& (2t)^{12} \left(e^{\pi/2t} - 24 + \ldots \right)\ .
\end{eqnarray}
One thus finds a pole
\begin{equation}
\mp 2^{p - 12} V_{p+1} {3 \pi\over 2^{6}}  (4\pi^2{\alpha^\prime})^{11-p}
G_{25-p}(X^2)\ .  \labell{orten}
\end{equation}
This is to be compared with the field theory result ${D-2\over2}
V_{p+1} T_p T'_p \kappa_0^2 G_{25-p}(Y^2)$, where $T'_p$ is the
O--plane tension.  A factor of 2 as compared to the earlier field
theory calculation~(\ref{fieldpole}) comes because the spacetime
boundary forces all the flux in one direction.  Thus the O--plane
and D--brane tensions are related
\begin{equation}
\tau'_p = \mp 2^{p - 13} \tau_p.
\end{equation}
A similar calculation with the Klein bottle gives a result proportional
to $\tau_p^{\prime 2}$.

Noting that there are $2^{25 - p}$ O--planes, the total O--plane
source is $\mp 2^{12} \tau_p$.  The total source must vanish because
the volume is finite and there is no place for flux to go.  Thus there
are $2^{(D-2)/2}=2^{12}$ D--branes (times two for the images) and the
group is $SO(2^{13}) = SO(2^{D/2})$.\cite{sobig} For this group the
``tadpoles'' associated with the dilaton and graviton, representing
violations of the field equations, cancel at order $g_s^{-1}$.  This
has no special significance in the bosonic string, as the one loop
$g_s^0$ tadpoles are nonzero and imaginary due to the tachyon
instability, but similar considerations will give a restriction on
anomaly free Chan--Paton gauge groups in the superstring.

\section{Worldvolume Actions I:  Dirac--Born--Infeld}
\label{Dactionone}
In the previous section, we wrote only part of the world--volume
action for the D--branes: that involving the embedding fields
$X^m(\xi)=2\pi\alpha^\prime \Phi^m(\xi)$ ($m=p+1,\ldots,D-1$) and
their coupling given by the induced metric on the worldvolume. We
should expect, however, to have to take into account new couplings for
the $X^m(\xi)$ as a result of other background spacetime fields like
the antisymmetric tensor $B_{\mu\nu}$, which must again appear as an
induced tensor $B_{ab}$ on the worldvolume, {\it via} a formula like
\reef{induced}. (Recall that $a=0,\ldots,p$). Furthermore, we should
also write the action for the collective modes which we uncovered in
section~\ref{collective}, the world--volume gauge fields $A^a(\xi)$.

The first thing to notice that that there is a restriction due to
spacetime gauge symmetry on the precise combination of $B_{ab}$ and
$A^a$ which can appear in the action.  The combination $B_{ab} +
2\pi{\alpha^\prime} F_{ab}$ can be understood as follows. In the
world--sheet sigma model action of the string, we have the usual
closed string term \reef{curved} for $B$ and the boundary action
\reef{abound} for $A$. So the fields appear in the combination:
\begin{equation}
{1\over2\pi\alpha^\prime} \int_{\cal M} B + \int_{\partial\cal M} A\ .
\end{equation} We have 
written everything in terms of differential forms, since $B$ and $A$
are antisymmetric. For example $\int A\equiv \int A_a d\xi^a$.  

This action is invariant under the spacetime gauge transformation
$\delta A = d\lambda$. However, the spacetime gauge transformation
$\delta B = d\zeta$ will give a surface term which must be cancelled
with the following gauge transformation of $A$: $\delta A =
-\zeta/2\pi\alpha^\prime$.  So the combination $B + 2\pi\alpha^\prime
F$, where $F=dA$ is invariant under both symmetries; This is the
combination of $A$ and $B$ which must appear in the action in order
for spacetime gauge invariance to be preserved.

\subsection{Tilted D--Branes}
\label{tiltaway}
There are many ways to deduce the world--volume action. One way is to
simply redo the computation for Weyl invariance of the complete sigma
model, including the boundary terms, which will result in the
$p+1$--dimensional equations of motion for the worldvolume fields
$G_{ab}$ $B_{ab}$ and $A_a$. One can then deduce the
$p+1$--dimensional worldvolume action from which those equations of
motion may be derived. We will comment on this below.

Another way is to use T--duality to build the action piece by
piece. For the purposes of these lectures and the various
applications, this second way is perhaps more instructive.
 
Consider\cite{bachas,bachasreview,tdbi} a D2--brane extended in the
$X^1$ and $X^2$ directions, and let there be a constant gauge field
$F_{12}$. (We leave the other dimensions unspecified, so the brane
could be larger by having extent in other directions. This will not
affect our discussion.) We can choose a gauge in which $A_2 =X^1
F_{12}$.  Now consider T--dualising along the 2-direction. The
relation~(\ref{positiondual}) between the potential and coordinate
gives
\begin{equation}
X'^2 = 2\pi{\alpha^\prime} X^1 F_{12}\ ,
\end{equation}
This says that the resulting D1--brane is tilted at an angle
$\theta=\tan^{-1}(2\pi\alpha^\prime F_{12})$ to the $X^2$--axis!  This
gives a geometric factor in the D1--brane world--volume action,
\begin{equation}
S\sim\int_{\rm D1} ds=\int dX^1\, \sqrt{1 + (\partial_1 X'^2)^2} = \int dX^1\, \sqrt{1 + (2\pi
{\alpha^\prime} F_{12})^2}\ . \labell{tilt}
\end{equation}
We can always boost the D--brane to be aligned with the coordinate axes
and then rotate to bring $F_{\mu\nu}$ to block-diagonal form, and in
this way we can reduce the problem to a product of factors like
\reef{tilt} giving a determinant:
\begin{equation}
S\sim\int d^{D}X\,\det\!^{1/2}(\eta_{\mu\nu}+2\pi\alpha^\prime
F_{\mu\nu})\ .  \labell{borninfeld}
\end{equation}
This
is the Born--Infeld action\cite{frad}.

In fact, this is the complete action (in a particular ``static'' gauge
which we will discuss later) for a space--filling D25--brane in flat
space, and with the dilaton and antisymmetric tensor field set to
zero.  In the language of section~2.7, Weyl invariance of the open
string sigma model \reef{abound} amounts to the following open string
analogue of \reef{eqnmtn} for the open string sector:
\begin{equation}
\beta^A_\mu=\alpha^\prime\left({1\over 1-(2\pi\alpha^\prime
F)^2}\right)^{\nu\lambda}\partial_{(\nu}F_{\lambda)\mu}=0\ ,
\end{equation}
these equations of motion follow from the action. In fact, in contrast
to the Maxwell action written previously \reef{yangmills}, and the
closed string action \reef{stringfrm}, this action is true to all
orders in $\alpha^\prime$, although only for slowly varying field
strengths; there are corrections from derivatives of
$F_{\mu\nu}$.~\cite{openfolk}

\subsection{The Dirac--Born--Infeld Action}
We can uncover a lot of the rest of the action by simply dimensionally
reducing.  Starting with \reef{borninfeld}, where
$F_{\mu\nu}=\partial_\mu A_\nu-\partial_\nu A_\mu$ as usual (we will
treat the non--Abelian case later) let us assume that $D-p-1$ spatial
coordinates are very small circles, small enough that we can neglect
all derivatives with respect to those directions, labelled $X^m$,
$m=p+1,\ldots,D-1$. (The uncompactified coordinates will be labelled
$X^a$, $a=0,\ldots,p$.) In this case, the matrix whose determinant
appears in \reef{borninfeld} is:
\begin{equation}
\pmatrix{N&-A^T\cr A&M}\ ,
\end{equation}
where
\begin{equation}
N=\eta_{ab}+2\pi\alpha^\prime F_{ab}\ ;\qquad M=\delta_{mn}\ ;\qquad
A=2\pi\alpha^\prime \partial_a A_m\ .
\end{equation}
Using the fact that its determinant can be written as
$|M||N+A^TM^{-1}A|$, our action becomes~\cite{gibbonsT}
\begin{equation}
S\sim -\int d^{p+1}X\,\det\!^{1/2}(\eta_{ab}+ \partial_aX^m\partial_bX_m
+2\pi\alpha^\prime F_{ab})\ ,
\labell{staticdiracborninfeld}
\end{equation}
up to a numerical factor (coming from the volume of the torus we
reduced on.  Once again, we used the T--duality rules
\reef{positiondual} to replace the gauge fields in the T--dual
directions by coordinates: $2\pi\alpha^\prime A_m=X^m$.

This is (nearly) the action for a D$p$--brane and we have uncovered
how to write the action for the collective coordinates $X^m$
representing the fluctuations of the brane transverse to the
world--volume. There now remains only the issue of putting in the case
of non--trivial metric, $B_{\mu\nu}$ and dilaton. This is easy to
guess given that which we have encountered already:
\begin{equation}
S_p=-T_p\int d^{p+1}\xi\,e^{-\Phi}\det\!^{1/2}(G_{ab}+B_{ab}
+2\pi\alpha^\prime
F_{ab})\ .  \labell{diracborninfeld}
\end{equation}
This is the Dirac--Born--Infeld Lagrangian, for arbitrary background
fields. The factor of the dilaton is again a result of the fact that
all of this physics arises at open string tree level, hence the factor
$g_s^{-1}$, and the $B_{ab}$ is in the right place because of
spacetime gauge invariance. $T_p$ and $G_{ab}$ are in the right place
to match onto the discussion we had when we computed the tension.
Instead of using T--duality, we could have also deduced this action by
a generalisation of the sigma model methods described earlier, and in
fact this is how it was first derived in this context.~\cite{leigh}

We have re--introduced independent coordinates $\xi^a$ on the
world--volume. Note that the actions
\reef{tilt},\reef{staticdiracborninfeld} were written using a choice
where we aligned the world--volume with the first $p+1$ spacetime
coordinates as $\xi^a=X^a$, leaving the $D-p-1$ transverse coordinates
called $X^m$. We can always do this using worldvolume and spacetime
diffeomorphism invariance. This choice is called the ``static gauge'',
and we shall use it quite a bit in these notes.  Writing this out (for
vanishing dilaton) using the formula \reef{induced} for the induced
metric, for the case of $G_{\mu\nu}=\eta_{\mu\nu}$ we see that we get
the action \reef{staticdiracborninfeld}.

\subsection{The Action of T--Duality}
It is amusing~\cite{tdbi,robdielectric} to note that our full action
obeys (as it should) the rules of T--duality which we already wrote
down for our background fields.  The action for the D$p$--brane is
built out of the determinant $|E_{ab}+2\pi\alpha^\prime F_{ab}|$,
where the $(a,b =0,\ldots p)$ indices on $E_{ab}$ mean that we have
performed the pullback of $E_{\mu\nu}$ (defined in \reef{geebee})  to
the worldvolume.  This matrix becomes, if we
T--dualise on $n$ directions labelled by $X^i$ and use the rules we
wrote in \reef{bigtdual}:
\begin{equation}
\left| \matrix{E_{ab}-E_{ai}E^{ij}E_{jb}+2\pi\alpha^\prime F_{ab}&
-E_{ak}E^{kj}-2\pi\alpha^\prime \partial_aX^i\cr
E^{ik}E_{kb}+2\pi\alpha^\prime \partial_bX^i& E^{ij}}\right|\ ,
\end{equation}
which has determinant $|E^{ij}||E_{ab}+2\pi\alpha^\prime F_{ab}|$.  In
forming the square root, we get again the determinant needed for the
definition of a T--dual DBI action, as the extra determinant
$|E^{ij}|$ precisely cancels the determinant factor picked up by the
dilaton under T--duality. (Recall,  $E^{ij}$ is the inverse of $E_{ij}$.)

Furthermore, the tension $T_{p'}$ comes out correctly, because there
is a factor of $\Pi_i^n (2\pi R_i)$ from integrating over the torus
directions, and a factor $\Pi_i^n (R_i/\sqrt{\alpha^\prime})$ from
converting the factor $e^{-\Phi}$, (see \reef{newg}), which fits
nicely with the recursion formula \reef{trec} relating $T_p$ and
$T_{p^\prime}$.

The above was done as though the directions on which we dualised were
all Neumann or all Dirichlet. Clearly, we can also extrapolate to the
more general case.

\subsection{Non--Abelian Extensions}
\label{nonabelstuff}
For $N$ D--branes the story is more complicated.  The various fields
on the brane representing the collective motions, $A_a$ and $X^m$,
become matrices valued in the adjoint.  In the Abelian case, the
various spacetime background fields (here denoted $F_\mu$ for the sake
of argument) which can appear on the worldvolume typically depend on
the transverse coordinates $X^m$ in some (possibly) non--trivial way.
In the non--Abelian case, with $N$ D--branes, the transverse
coordinates are really $N\times N$ matrices,
$2\pi\alpha^\prime\Phi^m$, since they are T--dual to non--Abelian
gauge fields as we learned in previous sections, and so inherit the
behaviour of gauge fields (see eqn.\reef{positiondual}). We write them
as $\Phi^m=X^m/(2\pi\alpha^\prime)$.  So not only should the
background fields $F_\mu$ depend on the Abelian part, but they ought
to possibly depend (implicitly or explicitly) on the full non--Abelian
part as $F(\Phi)_\mu$ in the action.

Furthermore, in \reef{diracborninfeld} we have used the partial
derivatives $\partial_a X^\mu$ to pull back spacetime indices $\mu$ to
the worldvolume indices, $a$, {\it e.g.,} $F_a=F_\mu\partial_a X^\mu
$, and so on. To make this gauge covariant in the non--Abelian case,
we should pull back with the covariant derivative: $F_a=F_\mu{\cal
  D}_a X^\mu =F_\mu({\partial}_a X^\mu+[A_a,X^\mu])$.

With the introduction of non--Abelian quantities in all of these
places, we need to consider just how to perform a trace, in order to
get a gauge invariant quantity to use for the action.  Starting with
the fully Neumann case \reef{borninfeld}, a first guess is that things
generalise by performing a trace (in the fundamental of $U(N)$) of the
square rooted expression. The meaning of the $\Tr$ needs to be stated,
It is proposed that is means the ``symmetric'' trace, denoted ``STr''
which is to say that one symmetrises over gauge indices, consequently
ignoring all commutators of the field strengths encountered while
expanding the action.~\cite{tseyt} (This suggestion is consistent with
various studies of scattering amplitudes and also the BPS nature of
various non--Abelian soliton solutions. There is still apparently some
ambiguity in the definition which results in problems beyond fifth
order in the field
strength.~\cite{malcolm,dominic,myersscatter,otherthoughts,tseytlinreview})

Once we have this action, we can then again use T--duality to deduce
the form for the lower dimensional, D$p$--brane actions. The point is
that we can reproduce the steps of the previous analysis, but keeping
commutator terms such as $[A_a,\Phi^m]$ and $[\Phi^m,\Phi^n]$.  We
will not reproduce those steps here, as they are similar in spirit to
that which we have already done (for a complete discussion, please
consult some of the
literature.~\cite{tseyt,malcolm,tseytlinreview,robdielectric}). The
resulting action is:
\begin{eqnarray}
S_p&=&-T_p\int
d^{p+1}\xi\,e^{-\Phi}{\cal L}\ ,\quad \mbox{where}\nonumber\\
{\cal L}&=&{\rm STr}\left\{
%\sqrt
{\det\!^{1/2}[E_{ab}+E_{ai}(Q^{-1}-\delta)^{ij}
E_{jb}+2\pi\alpha^\prime
F_{ab}]\det\!^{1/2}[Q^i_{\phantom{i}j}]}\right\}\ ,
\labell{nonabel}
\end{eqnarray}
where $Q^i_{\phantom{i}j}=\delta^i_{\phantom{i}j}
+i2\pi\alpha^\prime[\Phi^i,\Phi^k]E_{kj}$, and we have raised indices
with $E^{ij}$.

\subsection{Yang--Mills Theory}
\label{yangmillsstuff}
In fact, we are now in a position to compute the constant $C$ in
eqn.\reef{yangmills}, by considering $N$ D25--branes, which is the same
as an ordinary (fully Neumann) $N$-valued Chan--Paton factor.
Expanding the D25--brane Lagrangian~(\ref{borninfeld}) to second order
in the gauge field, we get
\begin{equation}
- {T_{25} 
\over 4} (2\pi{\alpha^\prime})^2 e^{-\Phi}\Tr F_{\mu\nu}F^{\mu\nu},
\end{equation}
with the trace in the fundamental representation of $U(N)$.  This
gives the precise numerical relation between the open and closed
string couplings.\cite{occoup}

Actually, with Dirichlet and Neumann directions, performing the same
expansion, and in addition noting that
\begin{equation}
\det[Q^i_{\phantom{i}j}]=1-{(2\pi\alpha^\prime)^2\over
4}[\Phi^i,\Phi^j][\Phi^i,\Phi^j]+\cdots\ ,
\end{equation}
one can write the leading order action \reef{nonabel} as
\begin{eqnarray}
S_p = -{T_p(2\pi\alpha^\prime)^2\over
4} \int d^{p+1}\xi\, e^{-\Phi} \Tr \left[F_{ab}F^{ab}
+2{\cal D}_a\Phi^i{\cal D}_a\Phi^i+[\Phi^i, \Phi^j]^2\right]
\biggr\}\ .
\labell{expanddn}
\end{eqnarray}

This is the dimensional reduction of the $D$--dimensional Yang--Mills
term, displaying the non--trivial commutator for the adjoint
scalars. This is an important term in many modern applications, as we
shall see. Note that the $(p+1)$--dimensional Yang--Mills coupling for
the theory on the branes is 
\begin{equation} 
g^2_{{\rm YM},p}=g_s
T_p^{-1}(2\pi\alpha^\prime)^{-2}\ .
\labell{yangmillsbos}
\end{equation}
This is worth noting. With the superstring value of $T_p$ which we
will compute later, it is used in many applications to give the correct
relation between gauge theory couplings and string quantities.

\subsection{BIons, BPS Saturation and Fundamental Strings}
\label{BIonsection}
We can of course treat the Dirac--Born--Infeld action as an
interesting theory in its own right, and seek for interesting
solutions of it. These solutions will have both a $(p+1)$--dimensional
interpretation and a $D$--dimensional
one.~\cite{gibbonsT,juancurt,morebions,roberto,gauntlett}

We shall not dwell on this in great detail, but include a brief
discussion here to illustrate an important point, and refer to the
literature for more complete
discussions. More details will
appear when we get to the supersymmetric case. One can derive an
expression for the energy density contained in the fields on the
world--volume:~\cite{gauntlett}
\begin{equation}
{\cal E}^2=E^aE^bF_{ca}F_{cb}+E^aE^bG_{ab}+{\rm
det}(G+2\pi\alpha^{\prime}F)\ ,
\end{equation}
where here the matrix $F_{ab}$ contains only the magnetic components
({\it i.e.}  no time derivatives) and $E^a$ are the electric
components, subject to the Gauss Law constraint ${\vec\nabla}\cdot
{\vec E}=0$.  Also, as before
\begin{equation}
G_{ab}=\eta_{ab}+\partial_a X^m\partial_b X^m\ ,\qquad m=p+1,\ldots ,D-1\ .
\end{equation}

Let us consider the case where we have no magnetic components and only
one of the transverse fields, say $X^{25}$, switched on. In this case,
we have
\begin{equation}
{\cal E}^2=(1\pm {\vec E}\cdot {\vec\nabla}X^{25})^2+({\vec
E}\mp{\vec\nabla}X^{25})^2\ ,
\end{equation}
and so we see that we have the Bogomol'nyi condition
\begin{equation}
{\cal E}\geq |{\vec E}\cdot{\vec\nabla}X^{25}|+1\ .
\end{equation}
This condition is saturated if ${\vec E}=\pm {\vec \nabla}X^{25}$. In
such a case, we have
\begin{equation}
\nabla^2 X^{25}=0\qquad\Rightarrow\qquad X^{25}={c_p\over r^{p-2}}\ ,
\end{equation} 
a harmonic solution, where $c_p$ is a constant to be determined.
The total energy (beyond that of the brane itself) is, integrating
over the world--volume:
\begin{eqnarray}
E_{\rm tot}&=&\lim_{\epsilon\to\infty}T_p\int_\epsilon^\infty r^{p-1}
drd\Omega_{p-1}({\vec\nabla}X^{25})^2 =\lim_{\epsilon\to\infty}T_p
{c_p^2(p-2)\Omega_{p-1}\over \epsilon^{p-2}}\nonumber\\
&=&\lim_{\epsilon\to\infty}T_pc_p(p-2)\Omega_{p-1}X^{25}(\epsilon)\ ,
\end{eqnarray}
where $\Omega_{p-1}$ is the volume of the sphere $S^{p-1}$ surrounding
our point charge source, and we have cut off the divergent integral by
integrating down to $r=\epsilon$. (We will save the case of $p=1$ for
later.~\cite{dasguptamukhi,gauntlett}) Now we can choose a value of
the electric flux such that
% is 
% \begin{equation}
% T_p\int_{S^{p-1}}(2\pi\alpha^\prime)F_{0r}=1\ ,
% \end{equation}
we get $(p-2)c_p\Omega_{p-1}T_p=(2\pi\alpha^\prime)^{-1}$.
~\footnote{In the supersymmetric case, this has a physical meaning,
  since overall consistency of the D--brane charges set a minimum
  electric flux. Here, it is a little more arbitrary, and so we choose
  a value by hand to make the point we wish to illustrate.} Putting
this into our equation for the total energy, we see that the
(divergent) energy of our configuration is:
\begin{equation}
E_{\rm tot}={1\over 2\pi\alpha^\prime}X^{25}(\epsilon)\ .
\end{equation}

\begin{figure}[ht]
\centerline{\psfig{figure=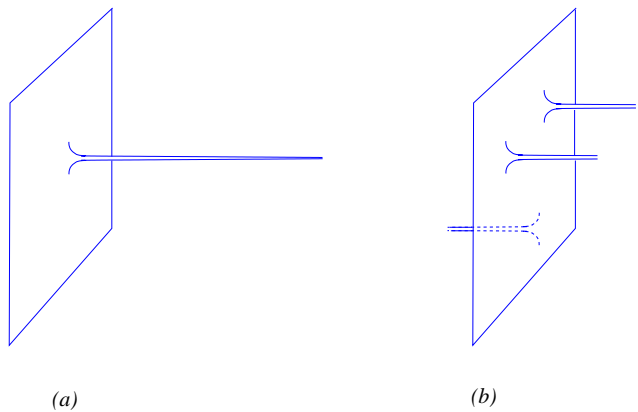,height=2.1in}}
\caption{The $D$ dimensional interpretation of the BIon solution. $(a)$ It is
  an infinitely long spike representing a fundamental string ending on
  the D--brane. $(b)$ BIons are BPS and therefore can be added
  together at no cost to make a multi--BIon solution.}
\label{bions}
\end{figure}

What does this mean? Well, recall that $X^{25}(\xi)$ gives the
transverse position of the brane in the $X^{25}$ direction. So we see
that the brane has grown a semi--infinite spike at $r=0$, and the base
of this spike is our point charge.  The interpretation of the
divergent energy is simply the (infinite) length of the spike
multiplied by a mass per unit length. But this mass per unit length is
precisely the fundamental string tension $T=(2\pi\alpha^\prime)^{-1}$!
In other words, the spike solution is the fundamental string stretched
perpendicular to the brane and ending on it, forming a point electric
charge, known as a ``BIon''. See figure \ref{bions}$(a)$. In fact, a
general BIon includes the non--linear corrections to this spike solution,
which we have neglected here, having only written the linearised
solution.

It is a worthwhile computation to show that if test source with the
same charges is placed on the brane, there is no force of attraction
or repulsion between it and the source just constructed, as would
happen with pure Maxwell charges. This is because our sources have in
addition to electric charge, some scalar ($X^{25}$) charge, which can
also be attractive or repulsive. In fact, the scalar charges are such
that the force due to electromagnetic charges is cancelled by the
force of the scalar charge, another characteristic property of these
solutions, which are said to be ``Bogomol'nyi--Prasad--Sommerfield''
(BPS)--saturated.\cite{bogo,onemonopole} We shall encounter solutions
with this sort of behaviour a number of times in what is to follow.

Because of this property, the solution is easily generalised to
include any number of BIons, at arbitrary positions, with positive and
negative charges. The two choices of  charge simply represents strings either
leaving from, or arriving on the brane. See figure \ref{bions}$(b)$.

\section{Superstrings and D--Branes}
\subsection{Open Superstrings: First Look}
\label{superlook}
Let us go back to the beginning, almost. We can generalise the bosonic
string action we had earlier to include fermions. In conformal gauge
it is:
\begin{equation}
S = {1\over4\pi} \int_{\cal M} d^2\!\sigma\,   
\left\{ {1\over\alpha^\prime} \partial X^{\mu} \bar{\partial} X_{\mu}
     + \psi^{\mu} \bar{\partial} \psi_{\mu} + 
       \tilde{\psi}^{\mu}\partial \tilde{\psi}_{\mu} \right\}
\end{equation}
where the open string world--sheet is the strip $0 < \sigma < \pi$, $-
\infty <\tau< \infty$.
%  We will use $\sigma^1=\sigma$ and
% $\sigma^2=\tau$ for some of what follows.

\ennbee{Actually, $\alpha^\prime$ is the loop expansion parameter
analogous to $\hbar$ on worldsheet. It is therefore natural for the
fermions' kinetic terms to be normalised in this way.} 

We get a
modification to the energy--momentum tensor from before (which we now
denote as $T_B$, since it is the bosonic part):
\begin{equation}
T_B(z)=-{1\over\alpha^\prime}\partial X^\mu\partial 
X_\mu-{1\over2}\psi^\mu\partial \psi_\mu\ ,
\end{equation}
which is now accompanied by a fermionic energy--momentum tensor:
\begin{equation}
T_F(z)=i{2\over\alpha^\prime}
\psi^\mu\partial X_\mu\ .
\end{equation}
This enlarges our theory somewhat, while much of the logic of what we
did in the purely bosonic story survives intact here. Now, one
extremely important feature which we encountered in section
\ref{bosonisation} is the fact that the equations of motion admit two
possible boundary conditions on the world--sheet fermions consistent
with Lorentz invariance. These are denoted the ``Ramond'' (R) and the
``Neveu--Schwarz'' (NS) sectors:
\begin{eqnarray}
{\rm R\colon} && \psi^{\mu}(0, \tau) =\tilde{\psi^{\mu}}
(0,\tau) 
 \qquad\psi^{\mu}(\pi ,\tau )= \tilde{\psi^{\mu}}(\pi,\tau)
\nonumber\\ 
{\rm NS\colon} && \psi^{\mu}(0, \tau) =-\tilde{\psi^{\mu}}
(0,\tau) 
\qquad\psi^{\mu}(\pi ,\tau )= \tilde{\psi^{\mu}}(\pi,\tau)
\end{eqnarray}
We are free to choose the boundary condition at, for example the
$\sigma = \pi$ end, in order to have a $+$ sign, by redefinition of
$\tilde\psi$.  The boundary conditions and equations of motion are
 summarised by the ``doubling trick'': Take just
left--moving (analytic) fields $\psi^\mu$ on the range $0$ to $2\pi$
and define $\tilde\psi^\mu(\sigma,\tau)$ to be $\psi^\mu(2\pi
- \sigma,\tau)$.  These left--moving fields are periodic in the
Ramond (R) sector and antiperiodic in the Neveu-Schwarz (NS).

On the complex $z$--plane, the NS sector fermions are half--integer moded
while the R sector ones are integer, and we have:
\begin{equation}
\psi^\mu(z)=\sum_{r}{\psi^\mu_r\over z^{r+1/2}}\ ,
\,\,\mbox{where}\,\, {r\in\IZ}\,\mbox{or}\,\,{r\in\IZ+\ha} 
\end{equation}
and canonical quantisation gives 
\begin{equation}
\{\psi_r^\mu,\psi^\nu_s\}=
\{{\tilde\psi}_r^\mu,{\tilde\psi}^\nu_s\}=\eta^{\mu\nu}\delta_{r+s}\ .
\end{equation}
Similarly we have 
\begin{eqnarray}
T_B(z)&=&\sum_{m=-\infty}^\infty{L_m\over z^{m+2}}\quad\mbox{as
before, and}\nonumber\\ T_F(z)&=&\sum_{r}{G_r\over z^{r+3/2}}\ ,
\,\,\mbox{where}\,\,
{r\in\IZ}\,\,\mbox{(R)}\,\,\,\mbox{or}\,\,\,{\IZ+\ha}\,\,\mbox{(NS)}
\end{eqnarray}

Correspondingly, the Virasoro algebra is enlarged, 
with the non--zero (anti) commutators
being
\begin{eqnarray}
[L_m,L_n]&=&(m-n)L_{m+n} +{{\rm c}\over12}(m^3-m)\delta_{m+n}\nonumber\\
\{G_r,G_s\}&=&2L_{r+s}+{{\rm c}\over12}(4r^2-1)\delta_{r+s}\nonumber\\
\left[L_m,G_r\right]&=&{1\over 2}(m-2r)G_{m+r} \ , \\
\end{eqnarray}
with
\begin{eqnarray}
L_m&=&{1\over2}\sum_m:\alpha_{m-n}\cdot\alpha_{m}:
+{1\over4}\sum_r (2r-m):\psi_{m-r}\cdot\psi_{r}:+a\delta_{m,0}\nonumber\\
G_r&=&\sum_n\alpha_{n}\cdot\psi_{r-n}\ .
\end{eqnarray}
In the above, $\rm c$ is the total contribution to the conformal
anomaly, which is $D+D/2$, where $D$ is from the $D$ bosons while
$D/2$ is from the $D$ fermions.

The values of $D$ and $a$ are again determined by any of the methods
mentioned in the discussion of the bosonic string. For the
superstring, it turns out that $D=10$ and $a=0$ for the R sector and
$a=-1/2$ for the NS sector. This comes about because the contributions
from the $X^0$ and $X^1$ directions are cancelled by the Faddev--Popov
ghosts as before, and the contributions from the $\psi^0$ and $\psi^1$
oscillators are cancelled by the superghosts.  Then, the computation
uses the mnemonic/formula given in \reef{zpe}.
\begin{eqnarray}
\mbox{NS sector:}\quad
\mbox{zpe}&=&8\left(-{1\over24}\right)+8\left(-{1\over48}\right)
=-{1\over2}\ ,\nonumber\\ \mbox{R sector:}\quad
\mbox{zpe}&=&8\left(-{1\over24}\right)+8\left({1\over24}\right) =0\ .
\end{eqnarray} 
As before, there is a physical state condition imposed by annihilating
with the positive modes of the (super) Virasoro generators:
\begin{eqnarray}
G_r|\phi\rangle=0\ ,\,\,\, r>0\ ;\quad L_n|\phi\rangle=0\ ,\,\, n>0\ ;
\quad (L_0-a)|\phi\rangle=0\ .
\labell{physical}
\end{eqnarray}
The $L_0$ constraint leads to a mass formula:
\begin{equation}
M^2={1\over\alpha^\prime}\left(\sum_{n,r}
 \alpha_{-n}\cdot\alpha_n+r\psi_{-r}\cdot\psi_r 
-a  \right)\ .
\end{equation}
In the NS sector the ground state is a Lorentz singlet and is assigned
odd fermion number, {\it i.e.,} under the operator $(-1)^F$, it has
eigenvalue~$-1$.  
%his assignment is necessary in order for
%$(-1)^F$ to be multiplicatively conserved.~\footnote{In the `$-1$
%picture'\,\cite{fms} the matter part of the ground state vertex
%operator is the identity but the ghost part has odd fermion number.
%In the `0 picture' this is reversed.\label{picfoot}} 

In order to achieve spacetime supersymmetry, the spectrum is projected
on to states with even fermion number. This is called the ``GSO
projection'',\cite{gso} and for our purposes, it is enough to simply
state that this obtains spacetime supersymmetry, as we will show at
the massless level. A more complete treatment ---which gets it right
for all mass levels--- is contained in the full superconformal field
theory. The GSO projection there is a statement about locality with
the gravitino vertex operator.

%MARKER

Since the open string tachyon clearly has $(-1)^F=-1$, it is removed
from the spectrum by GSO. This is our first achievement, and justifies
our earlier practice of ignoring the tachyons appearance in the
bosonic spectrum in what has gone before. Fro what we will do for the
rest of the these notes, the tachyon will largely remain in the
wings, but it (and other tachyons) do have a role to play, since they
are often a signal that the vacuum wants to move to a (perhaps) more
interesting place. We will see this in a couple of places before the
end.  (See John Schwarz's discussion of the construction of non--BPS
D--branes, in this school.~\cite{john})

Massless particle states in ten dimensions are classified by their
$SO(8)$ representation under Lorentz rotations, that leave the
momentum invariant: $SO(8)$ is the ``Little group'' of $SO(1,9)$.  The
lowest lying surviving states in the NS sector are the eight transverse
polarisations of the massless open string photon, $A^{\mu}$, made by 
exciting the $\psi$ oscillators:
\begin{equation}\psi^{\mu}_{-1/2} |{ k}\rangle, \qquad M^2=0\ .
\end{equation} 
These states clearly form the vector of $SO(8)$. They have $(-)^F=1$
and so survive GSO.

In the R sector the ground state energy always vanishes because the
world--sheet bosons and their superconformal partners have the same
moding.  The Ramond vacuum has a 32--fold degeneracy, since the
$\psi^{\mu}_0$ take ground states into ground states. The ground
states form a representation of the ten dimensional Dirac matrix
algebra
\begin{equation}
 \{\psi^{\mu}_0 , \psi^{\nu}_0 \} = \eta^{\mu \nu}\ .
\end{equation}
(Note the similarity with the standard $\Gamma$--matrix algebra, 
$\{\Gamma^\mu,\Gamma^\nu \}=2\eta^{\mu\nu}$. 
We see that $\psi^\mu_0\equiv\Gamma^\mu/\sqrt{2}$.)

For this representation, it is useful to choose  this basis:
\begin{eqnarray}
d^{\pm}_i &=& {{1}\over{\sqrt 2}}\left ( \psi^{2i}_0\pm i \psi^{2i+1}_0\right
) \qquad i=1,\cdots,4 \nonumber\\ 
 d^{\pm}_0 &=& {{1}\over{\sqrt 2}}\left ( \psi^{1}_0 \mp \psi^{0}_0\right )\ .
\labell{nicebasis}
\end{eqnarray}
In this basis, the Clifford algebra takes the form
\begin{equation}
\{ d^{+}_i, d^{-}_j \}=\delta_{ij}\ .
\end{equation}
The $d^{\pm}_i$, $i = 0, \cdots, 4$ act as raising and lowering
operators, generating the $2^{10/2}=32$ Ramond ground states.  Denote these
states 
\begin{equation} 
|s_0,s_1,s_2,s_3,s_4 \rangle = |{\bf s}\rangle
\labell{Rstates}
\end{equation}
where each of the $s_i$ takes the values  $\pm\ha$, and where
\begin{equation}
d^{-}_{i} | -\ha , -\ha , -\ha , -\ha , -\ha \rangle = 0  
\end{equation}
while $d^{+}_i$ raises $s_i$ from $-\ha$ to $\ha$.  This notation has physical
meaning:  The fermionic part of the
ten--dimensional Lorentz generators is 
\begin{equation}
 S^{\mu \nu} = - {{i}\over{2}} \sum_{r \in {\bf Z }+\kappa}
[\psi^{\mu}_{-r},\psi^{\nu }_r]\ .
\labell{Rlorentz}
\end{equation}
The states \reef{Rstates} above are eigenstates of $S_0 = iS^{01}$,
$S_i = S^{2i,2i+1}$, with $s_i$ the corresponding eigenvalues.  Since
by construction the Lorentz generators \reef{Rlorentz} always flip an
even number of $s_i$, the Dirac representation ${\bf 32}$ decomposes
into a ${\bf 16}$ with an even number of $-\ha$'s and ${\bf 16}'$ with
an odd number.

The physical state conditions \reef{physical}, on these ground states,
reduce to $G_0=(2\alpha^\prime)^{1/2}p_{\mu}\psi^{\mu}_0$. (Note that
$G_0^2\sim L_0$.) Let us pick the (massless) frame $p^0 = p^1$.  This
becomes
\begin{equation}
G_0%=\alpha^{\prime 1/2}p_1\left(\Gamma_0+\Gamma_1\right)
=\alpha^{\prime 1/2}p_1\Gamma_0\left(1-\Gamma_0\Gamma_1\right)
=2\alpha^{\prime 1/2}p_1\Gamma_0\left(\ha-S_0\right)\ ,
\end{equation}which means that $s_0 = \ha$, giving a
sixteen--fold degeneracy for the {\it physical} Ramond vacuum.  This
is a representation of $SO(8)$ which decomposes into ${\bf 8_s}$ with
an even number of $-\ha$'s and ${\bf 8_c}$ with an odd number.  One is
in the $\bf 16$ and the $\bf 16'$, but the two choices, $\bf 16$ or
$\bf 16'$, are physically equivalent, differing only by a spacetime
parity redefinition, which would therefore swap the $\bf 8_s$ and the
$\bf 8_c$.

%It is useful to think of the
%GSO projection in terms of locality of the OPE of a general vertex
%operator with the gravitino vertex operator.  Suppose we take a
%projection which includes the operator $e^{-\phi/2 +
%i(H_0+H_1+H_2+H_3+H_4)/2}$, where the $H_i$ are the bosonization of
%$\psi^\mu$.\cite{fms} {\bf DISCUSS BOSONIZATION SOMEWHERE} 
%In the NS sector this has a branch cut with the
%ground state vertex operator $e^{-\phi}$, accounting for the sign
%discussed in footnote~{\it\ref{picfoot}} for the $-1$ picture vertex
%operator.  
In the R sector 
%the ghost plus longitudinal part of the OPE
%is local, so we have 
the GSO projection amounts to requiring 
\begin{equation}
 \sum_{i=1}^4 s_i = 0 \pmod 2,
\end{equation}
picking out the ${\bf 8_s}$. Of course, it is just a convention that
we associated an even number of $\ha$'s with the $\bf 8_s$; a
physically equivalent discussion with things the other way around
would have resulted in $\bf 8_c$. The difference between these two is
only meaningful when they are both present, and at this stage we only
have one copy, so either is as good as the other.

The ground state spectrum is then ${\bf 8}_v \oplus {\bf 8}_s$, a
vector multiplet of $D=10$, $N=1$ spacetime supersymmetry.  Including
Chan-Paton factors gives again a $U(N)$ gauge theory in the oriented
theory and $SO(N)$ or $USp(N)$ in the unoriented. This completes our
tree--level construction of the open superstring theory.

\subsection{Closed Superstrings: Type II}
Just as we saw before, the closed string spectrum is the product of
two copies of the open string spectrum, with right-- and left--moving
levels matched.  In the open string the two choices for the GSO
projection were equivalent, but in the closed string there are two
inequivalent choices, since we have to pick two copies to make a close string.

Taking the same projection on both sides gives the ``type IIB'' case,
 while taking them opposite gives ``type IIA''. These lead to the
 massless sectors
\begin{eqnarray}
{\rm Type ~ IIA\colon} && ({\bf 8_v}\oplus{\bf 8_s}) \otimes
   ({\bf 8_v}\oplus{\bf 8_{c}}) \nonumber\\
{\rm Type ~ IIB\colon} && ({\bf 8_v}\oplus{\bf 8_s}) \otimes
   ({\bf 8_v}\oplus{\bf 8_{s}}) \ .
\end{eqnarray}
  
Let us expand out these products to see the resulting Lorentz
($SO(8)$) content.  In the NS--NS sector, this is
\begin{equation}
{\bf 8_v} \otimes {\bf 8_v} = \Phi \oplus B_{\mu\nu} \oplus G_{\mu\nu}
={\bf 1} \oplus {\bf 28} \oplus {\bf 35} .
\end{equation}
In the R--R sector, the IIA and IIB spectra are respectively
\begin{eqnarray}
{\bf 8_s} \otimes {\bf 8_c} &=& [1] \oplus [3] = {\bf 8_v} \oplus
{\bf 56_t} \nonumber\\
{\bf 8_s} \otimes {\bf 8_s} &=& [0] \oplus [2] \oplus [4]_+
= {\bf 1} \oplus {\bf 28}  \oplus {\bf 35}_+ .
\end{eqnarray}
Here $[n]$ denotes the $n$-times antisymmetrised representation of
$SO(8)$, and $[4]_+$ is self--dual.  Note that the representations
$[n]$ and $[8-n]$ are the same, as they are related by contraction
with the 8-dimensional $\epsilon$--tensor.  The NS-NS and R-R spectra
together form the bosonic components of $D=10$ IIA (nonchiral) and IIB
(chiral) supergravity respectively; We will write their effective
actions shortly.

\insertion{9}{Forms and Branes\label{insert9}}{It is useful to
  emphasise and summarise here, for later use, the structure of the
  bosonic content
  of the two theories. \\

Common to both type IIA and IIB are the NS--NS sector fields 
$$\Phi\ , G_{\mu\nu}\ , B_{\mu\nu}\ .$$
The latter is a rank two
antisymmetric tensor potential, and we have seen that the fundamental
closed string couples to it electrically by the coupling
$$\nu_1\int_{{\cal M}_2} B_{(2)}\ ,$$
where
$\nu_1=(2\pi\alpha^\prime)^{-1}$, ${\cal M}_2$ is the world sheet,
with coordinates $\xi^a$, $a=1,2$.  $B_{(2)}=B_{ab}d\xi^a d\xi^b$, and
$B_{ab}$ is the pullback of
$B_{\mu\nu}$ {\it via} \reef{induced}.\\

By ten dimensional Hodge duality, we can also construct a six form
potential $B_{(6)}$, by the relation $dB_{(6)}=*dB_{(2)}$. There is a
natural electric coupling $\nu_5\int_{{\cal M}_6} B_{(6)}$, to the
world--volume ${\cal M}_6$ of a five dimensional extended object. This
NS--NS charged object, which is commonly called the ``NS5--brane'' is
the magnetic dual of the fundamental string.~\cite{fivebranes,chstwo}
It is in fact, in the ten dimensional sense, the monopole of the
$U(1)$ associated to
$B_{(2)}$.\\

The string theory has other potentials, from the R--R sector: 
\begin{eqnarray}
&&{\rm type~IIA}:\quad C_{(1)}\ , C_{(3)}\ , C_{(5)}\ ,
C_{(7)}\nonumber \\ &&{\rm type~IIB}:\quad C_{(0)}\ , C_{(2)}\ ,
C_{(4)}\ , C_{(6)}\ , C_{(8)} \nonumber
\end{eqnarray}
where in each case the last two are Hodge 
duals of the first two, and $C_{(4)}$ is self dual. 
(A  $p$--form potential and a rank $q$--form
potential are Hodge dual to one another in $D$ dimensions if $p+q=D-2$.)\\

As we shall discuss at length later, we expect that there should be
$p$--dimensional extended sources which couple to all of these {\it via} an
electric coupling of the form:
$$
Q_p\int_{{\cal M}_{p+1}}C_{(p+1)}
$$
to their $p+1$--dimensional world volumes ${\cal M}_{p+1}$. {\it
Continued....}}

\insertion{9}{\it Continued....}{One of the most striking and far
  reaching results of modern string theory is the fact that the most
  basic such R--R sources are the superstrings' D--brane solutions,
  and furthermore that their charges $\mu_p$ are the smallest allowed
  by consistency (see \ref{diracconsist}), suggesting that they are
  the basic sources from which all R--R charged objects may be
  constructed, at least in principle, and
  often in practice. \\

So we see that type~IIA contains a D0--brane and its magnetic dual, a
D6--brane, and a D2--brane and its magnetic cousin, a D4--brane. The
last even brane is a ten--dimensional domain wall type solution, the
D8--brane, which as we shall later see pertains to the type~IA or
type~I$^\prime$ theory.\\

Meanwhile, type~IIB has a string--like D1--brane, which is dual to a
D5--brane. There is a self--dual D3--brane, and there is an instanton
which is the D(-1)--brane, and its Hodge dual, the D7--brane.  To
complete the list of odd branes, we note that there is a spacetime
filling D9--brane which pertains to the type~IB or type~I theory.}

In the NS--R and R--NS sectors are
the products
\begin{eqnarray}
{\bf 8_v} \otimes {\bf 8_c} &=& 
{\bf 8_s}\oplus{\bf 56_c} \nonumber\\
{\bf 8_v} \otimes {\bf 8_s} &=&
{\bf 8_c}\oplus{\bf 56_s}.
\end{eqnarray}
The $\bf 56_{s,c}$ are gravitinos. Their vertex operators are made
roughly by tensoring a NS field $\psi^\mu$ with a vertex operator
${\cal V}_\alpha=e^{-\varphi/2}{\bf S}_\alpha$, where the latter is
a``spin field'', made by bosonising the $d_i$'s of equation
\reef{nicebasis} and building:
\begin{equation}
{\bf S}=\exp{\left[i\sum_{i=0}^4 s_iH^i\right]}\ ;
\quad d_i=e^{\pm {\rm i}H^i}\ .
\end{equation} 
(The factor $e^{-\varphi/2}$ are the bosonisation of the
Faddev--Popov ghosts, about which we will have nothing more to say
here.)  The resulting full gravitino vertex operators, which correctly
have one vector and one spinor index, are two fields of weight $(0,1)$
and $(1,0)$, respectively, depending upon whether $\psi^\mu$ comes
from the left or right. These are therefore holomorphic and
antiholomorphic world--sheet currents, and the symmetry associated to
them in spacetime is the supersymmetry.  In the IIA theory the two
gravitinos (and supercharges) have opposite chirality, and in the IIB
the same.

Let us develop further the vertex operators for the R--R
states.~\footnote{The reader should consult a more advanced
  text\cite{joebook} for details.} This will involve a product of spin
fields,\cite{fms} one from the left and one from the right.  These
again decompose into antisymmetric tensors, now of $SO(9,1)$: \be V =
{\cal V}_\alpha{\cal V}_\beta ( \Gamma^{[\mu_1} \cdots
\Gamma^{\mu_n]}C)_{\alpha\beta} G_{[\mu_1 \cdots \mu_n]}(X)
\label{rrver}
\end{equation}
with $C$ the charge conjugation matrix.  In the IIA theory the product
is ${\bf 16} \otimes {\bf 16'}$ giving even $n$ (with $n \cong 10-n$)
and in the IIB theory it is ${\bf 16} \otimes {\bf 16}$ giving odd
$n$.  As in the bosonic case, the classical equations of motion
follow from the physical state conditions, which at the massless level
reduce to $G_0 \cdot V = \tilde{G}_0 \cdot V = 0.$ The relevant part
of $G_0$ is just $p_\mu \psi^\mu_0$ and similarly for $\tilde G_0$.
The $p_\mu$ acts by differentiation on $G$, while $\psi_0^\mu$ acts on
the spin fields as it does on the corresponding ground states: as
multiplication by $\Gamma^\mu$.  Noting the identity \be \Gamma^\nu
\Gamma^{[\mu_1} \cdots \Gamma^{\mu_n]} = \Gamma^{[\nu} \cdots
\Gamma^{\mu_n]} + \left( \delta^{\nu\mu_1} \Gamma^{[\mu_2} \cdots
\Gamma^{\mu_n]} + {\rm perms} \right) \label{gamma}
\end{equation}
and similarly for right multiplication, the physical state conditions
become
\be
dG=0 \qquad\qquad d{}^* G = 0.
\end{equation}
These are the Bianchi identity and field equation for an antisymmetric
tensor field strength.  This is in accord with the representations found:
in the IIA theory we have odd--rank tensors of $SO(8)$ but even-rank
tensors of $SO(9,1)$ (and reversed in the IIB), the extra index being
contracted with the momentum to form the field strength.
It also follows that R-R amplitudes involving elementary strings vanish
at zero momentum, so strings do not carry R-R charges.

As an aside, when the dilaton background is nontrivial, the Ramond
generators have a term $\phi_{,\mu} \partial\psi^\mu$, and the Bianchi
identity and field strength pick up terms proportional to
$d\phi \wedge G$ and $d\phi \wedge {}^*G$.  The Bianchi identity is
nonstandard, so $G$ is not of the form $dC$.  Defining $G' = e^{-\phi} G$
removes the extra term from both the Bianchi identity and field strength.
The field $G'$ is thus decoupled from the dilaton.
In terms of the action, the fields $G$ in the vertex operators appear
with the usual closed string $e^{-2\phi}$ but with non-standard dilaton
gradient terms.  The fields we are calling $G'$ (which in fact are the
usual fields used in the literature) have a dilaton-independent action.

\subsection{Open Superstrings: Second Look --- Type~I from Type IIB}
As we saw in the bosonic case, we can construct an unoriented theory
by projecting onto states invariant under world sheet parity,
$\Omega$. In order to get a consistent theory, we must of course
project a theory which is invariant under $\Omega$ to start
with. Since the left and right moving sectors have the same GSO
projection for type~IIB, it is invariant under $\Omega$, so we can
again form an unoriented theory by gauging. We cannot gauge
$\Omega$ in type ~IIA to get a consistent theory, but see later.

Projecting onto $\Omega=+1$ interchanges left--moving and right--moving
oscillators and so one linear combination of the R-NS and NS-R
gravitinos survives, so there can be only one supersymmetry surviving.
In the NS--NS sector, the dilaton and graviton are symmetric under
$\Omega$ and survive, while the antisymmetric tensor is odd and is
projected out. In the R--R sector, by counting we can see that the
$\bf 1$ and ${\bf 35}_+$ are in the symmetric product of ${\bf 8_s}
\otimes {\bf 8_s}$ while the $\bf 28$ is in the antisymmetric.  The
R--R state is the product of right-- and left--moving fermions, so
there is an extra minus in the exchange. Therefore it is the $\bf 28$
that survives.  The bosonic massless sector is thus ${\bf 1} \oplus
{\bf 28} \oplus {\bf 35}$, and together with the surviving gravitino,
this give us the $D=10$ $N=1$ supergravity multiplet.  

Sadly, this supergravity is in fact anomalous, and requires an
additional sector to cancel the anomaly.\cite{gsdiv} This sector turns
out to be $N=1$ supersymmetric Yang--Mills theory, with gauge group
$SO(32)$ or $E_8{\times}E_8$. Happily, we already know at least one
place to find the first choice: We can use the low--energy (massless)
sector of $SO(32)$ unoriented open superstring theory. This fits
nicely, since as we have seen before, at one loop open strings couple
to closed strings.

In the language we learned in section \ref{orientplanes}, we put a single
(spacetime--filling) O9--plane into type~IIB theory, making the type~IIB
theory into the unoriented $N=1$ closed string theory. This is
anomalous, but we can cancel the resulting anomalies by adding 16
D9--branes. There is a general principle here: such an inconsistency
in the low energy theory should be related to some inconsistency in
the full string theory, and we will discuss this later once we have
uncovered the roles of D--branes in superstring theory a bit more,
using T--duality.

We have just constructed our first (and in fact, the simplest) example
of an ``orientifolding'' of a superstring theory to get another. More
complicated orientifolds may be constructed by gauging combinations of
$\Omega$ with other discrete symmetries of a given string theory which
form an ``orientifold group'' $G_\Omega$ under which the theory is
invariant.~\cite{atishreview} Generically, there will be the
requirement to cancel anomalies by the addition of open string sectors
({\it i.e.}  D--branes), which results in consistent new string theory
with some spacetime gauge group carried by the D--branes. In fact,
these projections give rise to gauge groups containing any of $U(n)$,
$USp(n)$ factors, and not just $SO(n)$ sectors.

\subsection{The 10 Dimensional Supergravities}
\label{supergravity}
Just as we saw in the case of the bosonic string, we can truncate
consistently to focus on the massless sector of the string theories,
by focusing on low energy limit $\alpha^\prime\to 0$. Also as before,
the dynamics can be summarised in terms of a low energy effective
(field theory) action for these fields, commonly referred to as
``supergravity''.

The bosonic part of the low energy action for the type IIA string theory
in ten dimensions may be written ({\it c.f.}~(\ref{stringfrm}))
as (the wedge product is understood):~\cite{sugraref,joebook,gsw}
\begin{eqnarray}
&&S_{\rm IIA} = {1\over2\kappa_0^2}\int\!d^{10}\!x (-G)^{1/2} \left\{
e^{-2 \Phi}\left[R + 4 ( \nabla \phi)^2 -{1 \over
12}(H^{(3)})^2\right] \right.\nonumber\\ &&\qquad\quad\left.- {1 \over
4} (G^{(2)})^2- {1 \over
48}(G^{(4)})^2\right\}-{1\over4\kappa_0^2}\int B^{(2)}
dC^{(3)}dC^{(3)}\ .  \labell{Aaction}
\end{eqnarray}
As before $G_{\mu\nu}$ is the metric in string frame, $\Phi$ is the dilaton,
$H^{(3)}=dB^{(2)}$ is the field strength of the NS--NS two form, 
while the Ramond-Ramond field strengths are
$G^{(2)}=dC^{(1)}$ and $G^{(4)}=dC^{(3)}+H^{(3)}\wedge C^{(1)}$.

For the bosonic part in the case of type~IIB, we have:
\begin{eqnarray}
&&S_{\rm IIB} = {1\over2\kappa_0^2}\int\!d^{10}\!x (-G)^{1/2}
\left\{e^{-2 \phi}\left[R + 4( \nabla \phi)^2
-{1 \over 12} (H^{(3)})^2\right] 
\right.
\nonumber\\
&&\left.\qquad\qquad\qquad
- {1 \over 12} (G^{(3)} + C^{(0)} H^{(3)})^2 
-{1\over2}(d C^{(0)})^2-{1\over 480}(G^{(5)})^2\right\}\nonumber\\
&&\qquad\qquad\qquad
+{1\over4\kappa_0^2}\int 
\left(C^{(4)}+{1\over2}B^{(2)}\,C^{(2)}\right)\,G^{(3)}\,H^{(3)}\ .
\labell{Baction}
\end{eqnarray}
Now, $G^{(3)}=dC^{(2)}$ and $G^{(5)}=dC^{(4)}+H^{(3)}C^{(2)}$ are R--R
field strengths, and $C^{(0)}$ is the RR scalar.  (Note that we have
canonical normalisations for the kinetic terms of forms: there is a
prefactor of the inverse of $-2\times p!$ for a $p$--form field
strength.)  There is a small complication due to the fact that we
require the R--R four form $C^{(4)}$ to be self dual, or we will have
too many degrees of freedom. We write the action here and remind
ourselves to always impose the self duality constraint
$F^{(5)}=\,^*\!F^{(5)}$ by hand in the equations of
motion.%\cite{boon}.

%$$
%(C^{(0)},C^{(1)},C^{(2)},C^{(3)},C^{(4)}) =
%(\chi=A^{(0)},-A^{(1)},A^{(2)},A^{(3)},A^{(4)}-{1\over2}B\,A^{(2)})\ .
%$$

Equation \reef{framechange} tells us  that in ten dimensions,
we must use:
\begin{equation}
{\widetilde G}_{\mu\nu}=e^{(\Phi_0-\Phi)/2}G_{\mu\nu}\ .
\labell{einst}
\end{equation} to convert these actions to the
Einstein frame. As before, (see discussion below \reef{einsteinfrm})
 Newton's constant will be set by
\begin{equation}
2\kappa^2\equiv 2\kappa_0^2 g_s^2
=16\pi G_N
= (2\pi)^7\alpha^{\prime 4}g_s^2\ ,
\labell{nice}
\end{equation}
where the latter equality can be established by direct computation. We
will see that it gives a very natural normalisation for the masses and
charges of the various branes in the theory.  Also $g_s$ is set by the
asymptotic value of the dilaton at infinity: $g_s\equiv e^{\Phi_0}.$

Those were the actions for the ten dimensional supergravities with
thirty--two supercharges. Let us consider those with sixteen
supercharges. For the bosonic part of type~I, we can construct it by
dropping the fields which are odd under $\Omega$ and then adding the
gauge sector, plus a number of cross terms which result from cancelling
anomalies (see later):
\begin{eqnarray}
&&S_{\rm I} = {1\over2\kappa_0^2}\int\!d^{10}\!x (-G)^{1/2}
\Biggl\{e^{-2 \Phi}\left[R + 4( \nabla \phi)^2\right] \nonumber\\
&&\hskip4cm - {1 \over 12} ({\widetilde
G}^{(3)})^2-{\alpha^\prime\over8}e^{-\Phi}\Tr (F^{(2)})^2\Biggr\}\ .
\labell{Iaction}
\end{eqnarray}
Here
\begin{equation}
{\widetilde G}^{(3)}=dC^{(2)}-{\alpha^\prime\over4}\left[\omega_{\rm
3Y}(A)-\omega_{\rm 3L}(\Omega)\right]\ ,
\end{equation}
where the Chern--Simons three form is (with a similar expression for
$\omega_{\rm 3L}$ in terms of the spin connection $\Omega$):
\begin{equation}
\omega_{\rm 3Y}(A)\equiv \Tr\left(A\wedge dA+{2\over3}A\wedge A\wedge
A\right)\ ,\, \,\mbox{with}\,\,\, d\omega_{\rm 3Y}=\Tr F\wedge F\ .
\end{equation}

As a curiosity which will be meaningful later, notice that a simple
redefinition of fields:
\begin{eqnarray}
G_{\mu\nu}(\mbox {type I})&=&e^{-\Phi}G_{\mu\nu}
(\mbox{heterotic})\nonumber\\ \Phi(\mbox {type
I})&=&-\Phi(\mbox{heterotic})\nonumber\\ {\widetilde G}^{(3)}(\mbox
{type I})&=&{\widetilde H}^{(3)}(\mbox{heterotic})\nonumber\\
A_{\mu}(\mbox {type I})&=&A_\mu(\mbox{heterotic})\ ,
\labell{redefinition}
\end{eqnarray}
takes one from the type I Lagrangian to:
\begin{eqnarray}
S_{\rm H} = {1\over2\kappa_0^2}\int\!d^{10}\!x (-G)^{1/2}
e^{-2 \Phi}\Biggl\{ R + 4( \nabla \phi)^2
%\nonumber\\&&\hskip4cm
- {1 \over 12} ({\widetilde H}^{(3)})^2-{\alpha^\prime\over 8}\Tr
(F^{(2)})^2\Biggr\}\ , \labell{Haction}
\end{eqnarray}
where (renaming $C^{(2)}\to B^{(2)}$)
\begin{equation}
{\widetilde H}^{(3)}=dB^{(2)}-{\alpha^\prime\over4}\left[\omega_{\rm
3Y}(A)-\omega_{\rm 3L}(\Omega)\right]\ .
\end{equation}
This will turn out to be the low energy effective Lagrangian of a pair
of closely related closed string theories known as ``heterotic''
string theories,\cite{heterotic} which we have not yet explicitly
encountered in our development so far. (In \reef{Haction},
$\alpha^\prime$ is measured in heterotic units of length.)

We can immediately see two things about these
theories: The first is that $B_{\mu\nu}$ and $A_\mu$ are actually
closed string fields from the NS--NS sector, as can be deduced from
the power of the dilaton which appears, showing that all terms arise
from closed string tree level. The second deduction is that since from
eqn.\reef{redefinition} the dilaton relations tell us that
$g_s(\mbox{type I})=g_s^{-1}({\rm heterotic})$, we will be {\it forced
  to consider these theories when we study the type~I string in the
  limit of infinite coupling}.

\subsection{The K3 Manifold from a Superstring Orbifold}
\label{k3orbifold}
Before we go further, let us briefly revisit the idea of strings
propagating on an orbifold, and take it a bit further. Imagine that we
compactify one of our closed string theories on the four torus, $T^4$.
Let us take the simple case where there the torus is simply the
product of four circles, $S^1$, each with radius $R$. This simply asks
that the four directions (say) $x^6,x^7,x^8$ and $x^9$ are periodic
with period $2\pi R$. This does not not affect any of our discussion
of supercharges, {\it etc}, and we simply have a six dimensional
theory with the same amount of supersymmetry as the ten dimensional
theory which we started with. It is ${\cal N}=4$ in six dimensions. As
discussed in section \ref{tdualing}, there is a large $O(4,4,\IZ)$
pattern of T--duality groups available to us, and all the the
associated enhanced gauge symmetries present at special radii.

Let us proceed further and orbifold the theory by the $\IZ_2$ group
which has the action
\begin{equation}
{\bf R}:\quad x^6,x^7,x^8,x^9\to -x^6,-x^7,-x^8,-x^9\ ,
\labell{zeetwo}
\end{equation}
which is clearly a good symmetry to divide by.

We can construct the resulting six dimensional spectrum 
by first working out (say) the left---moving spectrum, 
seeing how it transforms under ${\bf R}$ and 
then tensoring with another copy from the right
in order to construct the closed string spectrum.

Let us now introduce a bit of notation which will be useful in the
future. Use the label $x^m$, $m=6,7,8,9$ for the orbifolded
directions, and use $x^\mu$, $\mu=0,\ldots,5$, for the remaining.  Let
us also note that the ten dimensional Lorentz group is decomposed as
$$
SO(1,9)\supset SO(1,5)\times SO(4)\ .
$$
We shall label the transformation properties of our massless states
in the theory under the $SU(2)\times SU(2)=SO(4)$ Little group.  Just
as we did before, it will be useful in the Ramond sector to choose a
labelling of the states which refers to the rotations in the planes
$(x^0,x^1)$, $(x^2,x^3)$, {\it etc.,} as eigenstates $s_0,s_1...s_4$
of the operator $S^{01}, S^{23}$, etc, (see \reef{Rstates} and
\reef{Rlorentz} and surrounding discussion).

With this in mind, we can list the states on the left which survive
the GSO projection:
\bigskip
\begin{center}
\begin{tabular}{|c|c|c|c|}
\hline
sector&state&${\bf R}$ charge&$SO(4)$ charge\\\hline
NS$^{\phantom\ha}$&
$\psi^\mu_{-\ha}|0;k\!> $ &
 $+$&$\bf (2,2)$\\
{}&
$\psi^m_{-\ha}|0;k\!>$ & $-$&$\bf 4(1,1)$\\\hline
R$^{\phantom\ha}$$_{\phantom{\ha}}$&
$|s_1s_2s_3s_4\!>$;\, $s_1=+s_2,\, s_3=-s_4$ & $+$&$\bf 2(2,1)$\\
{}$_{\phantom{\ha}}$&
$|s_1s_2s_3s_4\!>$;\, $s_1=-s_2,\, s_3=+s_4$ & $-$&$\bf 2(1,2)$\\
\hline
\end{tabular}
\end{center}
\bigskip
Crucially, we should also examine the ``twisted sectors'' which will
arise, in order to make sure that we get a modular invariant theory.
The big difference here is that in the twisted sector, the moding of
the fields in the $x^m$ directions is shifted. For example, the bosons
are now half--integer moded.  We have to recompute the zero point
energies in each sector in order to see how to get massless states
(see \reef{zpe}):
\begin{eqnarray}
\mbox{NS sector zpe:}\,
&&4\left(-{1\over24}\right)+4\left(-{1\over48}\right)
+4\left({1\over48}\right)+4\left({1\over24}\right)
=0\ ,\nonumber\\ 
\mbox{R sector zpe:}\,
&&4\left(-{1\over24}\right)+4\left({1\over24}\right)
+4\left({1\over48}\right)+4\left(-{1\over48}\right) =0\ .
\labell{energies}
\end{eqnarray} 
This is amusing, both the Ramond and NS sectors have zero vacuum
energy, and so the integer moded sectors will give us degenerate
vacua.  We see that it is only states $|s_1s_2\!>$ which contribute
from the R--sector (since they are half integer moded in the $x^m$
directions) and the NS sector, since it is integer moded in the $x^m$
directions, has states $|s_3s_4\!>$. (It is worth seeing in
\reef{energies} how we achieved this ability to make a massless field
in this case. The single twisted sector ground state in the bosonic
orbifold theory with energy $1/48$, was multiplied by 4 since there are
four such orbifolded directions. Combining this with the contribution
from the four unorbifolded directions produced just the energy needed
to cancel the contribution from the fermions.)

The states and their charges are therefore (after imposing GSO):
\bigskip
\begin{center}
\begin{tabular}{|c|c|c|c|}
\hline
sector&state&${\bf R}$ charge&$SO(4)$ charge\\\hline
NS$^{\phantom{\ha}}$&
$|s_3s_4\!>$;\,  $s_3=-s_4$ & $+$&$\bf 2(1,1)$\\
R$^{\phantom{\ha}}$&
$|s_1s_2>$;\, $s_1=-s_2$ & $-$&$\bf (1,2)$\\
\hline
\end{tabular}
\end{center}
\bigskip
Now we are ready to tensor. Recall that we could have taken the
opposite GSO choice here to get a left moving with the identical
spectrum, but with the swap ${\bf (1,2)}\leftrightarrow{\bf (2,1)}$.
Again we have two choices: Tensor together two identical GSO
choices, or two opposite. In fact, since six dimensional
supersymmetries are chiral, and the orbifold will keep only two of the
four we started with, we can write these choices as $(0,2)$ or $(1,1)$
supersymmetry, resulting from type~IIB or~IIA on K3.  Let us write
the result for the bosonic spectra:
\bigskip
\begin{center}
\begin{tabular}{|c|c|}
\hline
sector&$SO(4)$ charge\\\hline
NS--NS$_{\phantom{\ha}}$&
$\matrix{\bf (3,3)+(1,3)+(3,1)+(1,1)\cr 
10{\bf (1,1)}+6{\bf (1,1)}}$\\\hline
R--R (IIB)&$\matrix{3{\bf (3,1)}+4{\bf(1,1)}\cr
3{\bf (1,3)}+4{\bf(1,1)}}$\\\hline
R--R (IIA)&$\matrix{4{\bf(2,2)}\cr
4{\bf(2,2)}}$\\\hline
\end{tabular}
\end{center}
\bigskip
and for the twisted sector we have:
\bigskip
\begin{center}
\begin{tabular}{|c|c|}
\hline
sector&$SO(4)$ charge\\\hline
NS--NS$_{\phantom{\ha}}$&
$3{\bf (1,1)}+{\bf (1,1)}$\\\hline
R--R (IIB) &$ {\bf (1,3)+(1,1)}$\\\hline
R--R (IIA) &$ {\bf(2,2)}$\\
\hline
\end{tabular}
\end{center}
\bigskip
Recall now that we have two twisted sectors for each orbifolded circle,
and hence there are 16 twisted sectors in all, for $T^4/\IZ_2$. Therefore, to
make the complete model, we must take sixteen  copies of  the content
of the twisted sector table above.

Now let identify the various pieces of the spectrum. The gravity
multiplet $G_{\mu\nu}+B_{\mu\nu}+\Phi$ is in fact the first line of
our untwisted sector table, coming from the NS--NS sector, as
expected. The field $B$ can be seen to be broken into its self--dual
and anti--self--dual parts $B^+_{\mu\nu}$ and $B^-_{\mu\nu}$,
transforming as ${\bf(1,3)}$ and ${\bf(3,1)}$.  There are sixteen other
scalar fields, (${\bf(1,1)}$), from the untwisted NS--NS sector. The
twisted sector NS--NS sector has 4$\times$16 scalars. Not including
the dilaton, there are 80 scalars in total from the NS--NS sector.

Turning to the R--R sectors, we must consider the cases of IIA and IIB
separately.  For type~IIA, there are 8 one--forms (vectors,
${\bf(2,2)}$) from the untwisted sector and 16 from the twisted,
giving a total of 24 vectors.  For type~IIB, the untwisted R--R sector
contains three self--dual and three anti--self--dual tensors, while
there are an additional 16 self--dual tensors ${\bf(1,3)}$. We
therefore have 19 self--dual $C^+_{\mu\nu}$ and 3 anti--self--dual
$C^-_{\mu\nu}$. There are also eight scalars from the untwisted R--R
sector and 16 scalars from the twisted R--R sector. In fact, including
the dilaton, there are 105 scalars in total for the type~IIB case.

Quite remarkably, there is a geometrical interpretation of all of this
data in terms of compactifying type~II string theory on a smooth
manifold. The manifold is K3. It is a four dimensional manifold
containing 22 independent two--cycles, which are topologically
two--spheres more properly described as the complex surface $\IP^1$,
in this context. Correspondingly the space of two forms which can be
integrated over these two cycles is 22 dimensional. So we can choose a
basis for this space. Nineteen of them are self--dual and three of them
are anti--self--dual, in fact. The space of metrics on K3 is in fact
parametrised by 58 numbers.

In compactifying the type~II superstrings on K3, the ten dimensional
gravity multiplet and the other R--R fields gives rise to six
dimensional fields by direct dimensional reduction, while the
components of the fields in the K3 give other fields.  The six
dimensional gravity multiplet arises by direct reduction form the
NS--NS sector, while 58 scalars arise, parametrising the 58
dimensional space of K3 metrics which the internal parts of the
metric, $G_{mn}$, can choose. Correspondingly, there are 22 scalars
arising from the 19+3 ways of placing the internal components of the
antisymmetric tensor, $B_{mn}$ on the manifold. A commonly used
terminology is that the form has been ``wrapped'' on the 22
two--cycles to give 22 scalars.

In the R--R sector of type~IIB, there is one scalar in ten dimensions,
which directly reduces to a scalar in six. There is a two--form, which
produces 22 scalars, in the same way as the NS--NS two form did. The
self--dual four form can be integrated over the 22 two cycles to give
22 two forms in six dimensions, 19 of them self--dual and 3
anti--self--dual. Finally, there is an extra scalar from wrapping the
four form entirely on K3.  This is precisely the spectrum of fields
which we computed directly in the type~IIB orbifold.

Alternatively, while the NS--NS sector of type~IIA gives rise to the
same fields as before, there is in the R--R sector a one form, three
form and five form. The one form directly reduces to a one form in six
dimensions. The three form gives rise to 22 one forms in six
dimensions while the five form gives rise to a single one form.  We
therefore have 24 one forms (generically carrying a $U(1)$ gauge
symmetry) in six dimensions.  This also completes the smooth
description of the type~IIA on K3 spectrum, which we computed
directly in the orbifold limit. We shall have more to say about this
spectrum later.

The connection between the orbifold and the smooth K3 manifold is as
follows.\cite{page,walton,billo,seibergk3,aspin} K3 does indeed have a
geometrical limit which is $T^4/\IZ_2$, and it can be arrived at by
tuning enough parameters, which corresponds here to choosing the vev's
of the various scalar fields.  Starting with the $T^4/\IZ_2$, there
are 16 fixed points which look locally like $\IR^4/\IZ^2$, a singular
point of infinite curvature. It is easy to see where the 58 geometric
parameters of the K3 metric come from in this case. Ten of them are
just the symmetric $G_{mn}$ constant components, on the internal
directions. This is enough to specify a torus $T^4$, since the
hypercube of the lattice in $\IR^4$ is specified by the ten angles
between its unit vectors, ${\bf e}^m\cdot{\bf e}^n$. Meanwhile each of
the 16 fixed points has 3 scalars associated to its metric geometry.
(The remaining fixed point NS--NS scalar in the table is from the
field B, about which we will have more to say later.)

The three metric scalars can be tuned to resolve or ``blow up'' the
fixed point, and smooth it out into the $\IP^1$ which we mentioned
earlier.  (This accounts for 16 of the two--cycles. The other six
correspond to the six $\IZ_2$ invariant forms $dX^m\wedge dX^n$ on the
four--torus.)  The smooth space has a known metric, the
``Eguchi--Hanson'' metric,\cite{eguchihanson} which is {\it locally}
asymptotic to $\IR^4$ (like the singular space) but with a global
$\IZ_2$ identification. Its metric is:
\begin{equation}
ds^2=\left(1-\left({a\over r}\right)^4\right)^{-1}dr^2 + 
r^2\left(1-\left({a\over r}\right)^4\right)
(d\psi+\cos\theta d\phi)^2+r^2(d\theta^2+\sin^2\theta d\phi^2)\ ,
\labell{eguchi}
\end{equation}
where $\theta,\phi$ and $\psi$ are $S^3$ Euler angles.  The point
$r=a$ is an example of a ``bolt'' singularity. Near there, the space
is topolgically $\IR^2_{r\psi}\times S^2_{\theta\phi}$, with the $S^2$
of radius $a$, and the singularity is a coordinate one provided $\psi$
has period~$2\pi$. (See insert 10, (p.\pageref{insert10}).)  Since on
$S^3$, $\psi$ would have period $4\pi$, the space at infinity is
$S^3/\IZ_2$, just like an $\IR^4/\IZ_2$ fixed point. For small enough
$a$, the Eguchi--Hanson space can be neatly slotted into the space
left after cutting out the neighbourhood of the fixed point. The bolt
is in fact the $\IP^1$ of the blowup mentioned earlier. The parameter
$a$ controls the size of the $\IP^1$, while the other two parameters
correspond to how the $\IR^2$ (say) is oriented in $\IR^4$.

\insertion{10}{A Closer Look at the Eguchi--Hanson
  Space; The ``Bolt''\label{insert10}}{Let us establish some of the properties
  claimed in the main body of the text, while uncovering a useful
  technique. First, introduce some handy notation for later: The
  $SU(2)_L$ invariant  one--forms are:
\begin{eqnarray}
&&\sigma_1=-\sin\psi d\theta+\cos\psi\sin\theta d\phi\ ;\nonumber\\
&&\sigma_2=\cos\psi d\theta+\sin\psi\sin\theta d\phi\ ;\nonumber\\
&&\sigma_3=d\psi +\cos\theta d\phi\ ,
\labell{oneforms}
\end{eqnarray}
($0<\theta<\pi$, $0<\phi<2\pi$, $0<\psi<4\pi$ are the $S^3$ Euler
angles), which satisfy
$$
d\sigma_i={1\over2}\epsilon_{ijk}\sigma_j\wedge\sigma_k\ .\quad{\rm Also,}\quad
\sigma_1^2+\sigma_2^2\equiv d\Omega_2^2\ ,
$$
is the round $S^2$ metric. (The $SU(2)_R$ invariant choice comes from
$\psi\leftrightarrow\phi$.)\\

Now we can write the metric in the
manifestly $SU(2)$ invariant form:
$$
ds^2=\left(1-\left({a\over r}\right)^4\right)^{-1}dr^2 +
r^2\left(1-\left({a\over r}\right)^4\right)\sigma_3^2
+r^2(\sigma_1^2+\sigma_2^2)\ .
$$
The $S^3$'s in the metric are the natural 3D ``orbits'' of the
$SU(2)$
action. The $S^2$ of $(\theta,\phi)$ is a special 2D ``invariant
submanifold''.\\

To examine the potential singularity at $r=a$, look {\it near} $r=a$.
Choose, if you will, $r=a+\varepsilon$ for small $\varepsilon$, and:
$$
ds^2={a\over4\varepsilon}
\left[d\varepsilon^2+16\varepsilon^2(d\psi+\cos\theta
  d\phi)^2\right]+(a^2+2a\varepsilon)d\Omega_2^2\ ,
$$
which as $\varepsilon\to0$ is obviously topologically looking
locally like $\IR^2_{\varepsilon,\psi}\times S^2_{\theta,\phi}$, where
the $S^2$ is of radius $a$. (Globally, there is a fibred structure due
to the $d\psi d\phi$ cross term.)  Incidentally, this is the quickest
way to see that the Euler number of the space has to be equal to that
of an $S^2$, which is 2. {\it Continued...}  } \insertion{10}{\it
Continued...}{ Now, the point is that $r=a$ is a true singularity for
arbitrary choices of periodicity $\Delta\psi$ of $\psi$, since there
is a conical deficit angle in the plane. In other words, we have to
ensure that as we get to the origin of the plane, $\varepsilon=0$, the
$\psi$--circles have circumference $2\pi$, no more or less.
Infinitesimmally, we make those measures with the metric, and so the
condition is:
$$
2\pi=\lim_{\varepsilon\to0}\left({d(2\sqrt{a}\varepsilon^{1/2})\Delta\psi\over
d\varepsilon \sqrt{(a/4)}\varepsilon^{-1/2}}\right)\ ,
$$
which gives $\Delta\Psi=2\pi$. So in fact, we must spoil our $S^3$
which was a nice orbit of the $SU(2)$ isometry, by performing an
$\IZ_2$ identification on $\psi$, giving it half its usual period. In
this way, the ``bolt'' singularity $r=a$ is just a harmless artefact
of coordinates.~\cite{nuts,egh} Also, we are left with an
$SO(3)=SU(2)/\IZ_2$ isometry of the metric. The space at infinity is
$S^3/\IZ_2$. }

The Eguchi--Hanson space is the simplest example of an
``Asymptotically Locally Euclidean'' (ALE) space, which K3 can always
be tuned to resemble locally. These spaces are
classified~\cite{hitchinpoly} according to their identification at
infinity, which can be any discrete subgroup, $\Gamma$,
~\cite{kleinian} of the $SU(2)$ which acts on the $S^3$ at infinity,
to give $S^3/\Gamma$. These subgroups have been classified by
McKay,\cite{mckay} and have an A--D--E classification.  The metrics on
the A--series are known explicitly as the Gibbons--Hawking
metrics,\cite{gibbonshawking} which we shall display later, and
Eguchi--Hanson is in fact the simplest of this series, corresponding
to $A_1$.\cite{prasad} We shall later use a D--brane as a probe of
string theory on a $\IR^4/\IZ_2$ orbifold, an example which will show
that the string theory correctly recovers all of the metric data
\reef{eguchi} of these fixed points, and not just the algebraic data
we have seen here.

For completeness, let us compute one more thing about K3 using this
description. The Euler characteristic, in this situation, can be
written in two ways~\cite{egh}
\begin{eqnarray}
\chi(K3)&=&{1\over32\pi^2}\int_{K3} \sqrt{g}\left(R_{abcd} R^{abcd}-
4R_{ab}R^{ab}+R^2\right)\nonumber\\
&=&
{1\over32\pi^2}\int_{K3} \sqrt{g}\epsilon_{abcd}R^{ab} R^{cd}\nonumber\\
&=&-{1\over16\pi^2}\int_{K3} {\rm Tr}R\wedge R=24\ .
\labell{eulerK3}
\end{eqnarray} 
Even though no explicit metric for K3 has been written, we can compute
$\chi$ as follows.\cite{walton,egh} If we take a manifold $M$, divide
by some group $G$, remove some fixed point set $F$, and add in some
set of new manifolds $N$, the Euler characteristic of the new manifold
is $\chi=(\chi(M)-\chi(F))/|G|+\chi(N)$. Here, $G={\bf R}\equiv\IZ_2$,
and the Euler characteristic of the Eguchi--Hanson space is equal to
$2$, from insert 10 (p.\pageref{insert10}). That of a point is 1, and
of the torus is zero.  We therefore get $\chi(K3)=-16/2+32=24$, which
will be of considerable use later on.

So we have constructed the consistent, supersymmetric string
propagation on the K3 manifold, using orbifold
techniques. We shall use this manifold to illustrate a number of
beautiful properties of D--branes and string theory in the rest of
these lectures. See also Paul Aspinwall's lectures in this school for more
applications of such manifolds to the subject of
duality.~\cite{aspinlects}

We should mention in passing that it is possible to construct a whole
new class of string ``compactification'' vacua by including D--branes
in the spectrum in such a way that their contribution to spacetime
anomalies, {\it etc}, combines with that of the pure geometry in a way
that is crucial to the consistency of the model. This gives the idea
of a ``D--manifold''~\cite{dmanifold,engineers}, which we will not
review here in detail. The analogue of the orbifold method for making
these supersymmetric vacua is the generalised ``orientifold''
construction already mentioned. There are constructions of ``K3
Orientifolds'' which follow the ideas presented in this section,
combined with D--brane orbifold techniques to be developed in later
sections.~\cite{GP,sagnotti,ericme,atishk3,atishreview,moreori,berk}
Six dimensional supersymmetric D--manifolds constructed as
orientifolds have been constructed.~\cite{somemore} Also, there have
been important studies of the strong coupling nature of orientifold
vacua,~\cite{nonpert} making connections to
``F--theory'',~\cite{ftheory}, some beautiful geometric technology for
studying type~IIB string vacua with variable coupling $g_s$, which
unfortunately we do not have time or space to review here.  There are
also pure conformal field theory techniques for constructing
D--manifolds, which are not pure orbifolds of the type considered
here.~\cite{cftbranes}

\subsection{T--Duality of Type II Superstrings}
\label{tdualc}
T--duality on the closed oriented Type II theories has a somewhat more
interesting effect than in the bosonic case.~\cite{dhs,dbranesi}
Consider compactifying a single coordinate $X^9$, of radius $R$.  In
the $R\to \infty$ limit the momenta are $p^9_R = p^9_L$, while in the
$R \to 0$ limit $p^9_R = -p^9_L$.  Both theories are $SO(9,1)$
invariant but under {\it different} $SO(9,1)$'s.  T--duality, as a
right--handed parity transformation (see \reef{tparity}), reverses the
sign of the right--moving $X^9(\zb)$; therefore by superconformal
invariance it does so on $\tilde\psi^9(\zb)$.  Separate the Lorentz
generators into their left-and right--moving parts $M^{\mu\nu} +
\widetilde M^{\mu\nu}$. Duality reverses all terms in $\widetilde
M^{\mu 9}$, so the $\mu 9$ Lorentz generators of the T--dual theory
are $M^{\mu 9} - \widetilde M^{\mu 9}$. In particular this reverses
the sign of the helicity $\tilde s_4$ and so switches the chirality on
the right--moving side.  If one starts in the IIA theory, with
opposite chiralities, the $R\to 0$ theory has the same chirality on
both sides and is the $R\to\infty$ limit of the IIB theory, and {\it
  vice--versa}. In short, T--duality, as a one--sided spacetime parity
operation, reverses the relative chiralities of the right-- and
left--moving ground states.  The same is true if one dualises on any
odd number of dimensions, whilst dualising on an even number returns
the original Type II theory.

Since the IIA and IIB theories have different R--R fields, T$_9$
duality must transform one set into the other.  The action of duality
on the spin fields is of the form
\begin{equation}
S_{\alpha} (z) \to S_{\alpha} (z),\qquad \tilde{S}_{\alpha} (\bar{z})
\to P_9 \tilde{S}_{\alpha} (\bar{z})
\end{equation}
for some matrix $P_9$, the parity transformation (9-reflection) on the
spinors.  In order for this to be consistent with the action
$\tilde\psi^9 \to -\tilde\psi^9$, $P_9$ must anticommute with
$\Gamma^9$ and commute with the remaining $\Gamma^\mu$.  Thus $P_9 =
\Gamma^9\Gamma^{11}$ (the phase of $P_9$ is determined, up to sign, by
hermiticity of the spin field).  Now consider the effect on the R-R
vertex operators~(\ref{rrver}).  The $\Gamma^{11}$ just contributes a
sign, because the spin fields have definite chirality.  Then by the
$\Gamma$-matrix identity~(\ref{gamma}), the effect is to add a 9-index
to $G$ if none is present, or to remove one if it is.  The effect on
the potential $C$ ($G = dC$) is the same.  Take as an example the Type
IIA vector $C_\mu$.  The component $C_9$ maps to the IIB scalar $C$,
while the $\mu\neq 9$ components map to $C_{\mu 9}$.  The remaining
components of $C_{\mu\nu}$ come from $C_{\mu \nu 9}$, and so on.

Of course, these relations should be translated into rules for
T--dualising the spacetime fields in the supergravity actions
\reef{Aaction} and \reef{Baction}. The NS--NS sector fields'
transformations are the same as those shown in equations
\reef{backgroundT},\reef{bigtdual}, while for the R--R potentials:
\cite{tdualstuff}
\begin{eqnarray}
\tilde{C}^{(n)}_{\mu\cdots\nu\alpha y}&=&
C^{(n-1)}_{\mu\cdots\nu\alpha}-(n-1)
{C^{(n-1)}_{[\mu\cdots\nu| y}G_{|\alpha]y}\over G_{yy}}
\labell{moretdual}\\
\tilde{C}^{(n)}_{\mu\cdots\nu\alpha\beta}&=&
C^{(n+1)}_{\mu\cdots\nu\alpha\beta y}
+nC^{(n-1)}_{[\mu\cdots\nu\alpha}B_{\beta]y}
+n(n-1){C^{(n-1)}_{[\mu\cdots\nu|y}B_{|\alpha|y}G_{|\beta]y}\over
G_{yy}}
\nonumber
\end{eqnarray}

\subsection{T--Duality of Type I Superstrings}

Just as in the case of the bosonic string, the action of T--duality in
the open and unoriented open superstring theory produces D--branes and
orientifold planes.  Having done it once, (say on $X^9$ with radius
$R$), we get a T$_9$--dual theory on the line interval $S^1/\IZ_2$,
where $\IZ_2$ acts as the reflection $X^9\to-X^9$. The $S^1$ has
radius $R^\prime=\alpha^\prime/R)$. There are 16 D8--branes and their
mirror images (coming from the 16 D9--branes), together with two
orientifold O8--planes located at $X^9=0,\pi R^\prime$. This is called
the ``Type~I$^\prime$'' theory (and sometimes the ``Type~IA'' theory),
about which we will have more to say later as well.

Starting with the type~IIB theory, we can carry this out $n$ times on
$n$ directions, giving us 16 D$(9-n)$ and their mirror images through
$2^n$ O$(9-n)$--planes arranged on the hypercube of fixed points of
$T^n/\IZ_2$, where the $\IZ_2$ acts as a reflection in the $n$
directions. If $n$ is odd, we are in type~IIA string theory, while we
are back in type~IIB otherwise.

Let us focus here on a single D--brane, taking a limit in which the
other D--branes and the O--planes are distant and can be ignored.
Away from the D--brane, only closed strings propagate.  The local
physics is that of the Type II theory, with two gravitinos.  This is
true even  though we began with the unoriented Type I theory which
has only a single gravitino.  The point is that the closed string
begins with two gravitinos, one with the spacetime supersymmetry on
the right--moving side of the world--sheet and one on the left.  The
orientation projection of the Type I theory leaves one linear
combination of these.  But in the T--dual theory, the orientation
projection does not constrain the local state of the string, but
relates it to the state of the (distant) image gravitino.  Locally
there are two independent gravitinos, with equal chiralities if $n$,
(the number of dimensions on which we dualised) is even and opposite
if $n$ is odd.

This is all summarised nicely by saying that while the type I string
theory comes from projecting the type~IIB theory by $\Omega$, the
T--dual string theories come from projecting type~II string theory
compactified on the torus $T^n$ by $\Omega \prod_m [R_m(-1)^F]$, where
the product over $m$ is over all the $n$ directions, and $R_m$ is a
reflection in the $m$th direction. This is indeed a symmetry of the
theory and hence a good symmetry with which to project.  So we have
that T--duality takes the orientifold groups into one another:
\begin{equation} 
\{\Omega\}\leftrightarrow \{1, \Omega \textstyle{\prod_m} [R_m(-1)^F]\}\ .
\end{equation}
This is a rather trivial example of an orientifold group, since it
takes type~II strings on the torus $T^n$ and simply gives a theory
which is simply related to type~I string theory on $T^n$ by $n$
T--dualities. Nevertheless, it is illustrative of the general
constructions of orientifold backgrounds made by using more
complicated orientifold groups. This is a useful piece of technology
for constructing string backgrounds with interesting gauge groups, with
fewer symmetries, as a starting point for phenomenological
applications.

\subsection{D--Branes as BPS Solitons}
\label{bpsstates}
While there is type~II string theory in the bulk, ({\it i.e.,} away
from the branes and orientifolds), notice that the open string
boundary conditions are invariant under only one supersymmetry.  In
the original Type I theory, the left--moving world--sheet current for
spacetime supersymmetry $j_\alpha(z)$ flows into the boundary and the
right--moving current $\tilde j_\alpha(\bar z)$ flows out, so only the
total charge $Q_\alpha + \tilde Q_\alpha$ of the left- and
right-movers is conserved.  Under T--duality this becomes
\begin{equation}
 Q_\alpha + \left({\textstyle \prod_m} P_m\right) \tilde
Q_\alpha\ ,
\end{equation}
where the product of reflections $P_m$ runs over all the dualised
dimensions, that is, over all directions orthogonal to the D--brane.
Closed strings couple to open, so the general amplitude has only one
linearly realized supersymmetry.  That is, the vacuum without D--branes
is invariant under $N=2$ supersymmetry, but the state containing the
D--brane is invariant under only $N=1$: {\it it is a BPS
  state.}\cite{gojoe,gdinst}

BPS states must carry conserved charges.  In the present case there is
only one set of charges with the correct Lorentz properties, namely
the antisymmetric R-R charges.  The world volume of a $p$--brane
naturally couples to a ($p + 1$)--form potential $C_{(p+1)}$, which has
a ($p + 2$)-form field strength $G_{(p+2)}$.  This identification can
also be made from the $g_s^{-1}$ behaviour of the D--brane tension: this
is the behaviour of an R--R soliton.\cite{blackp,coni},
as will be developed further later.

The IIA theory has D$p$--branes for $p = 0$, 2, 4, 6, and 8.  The
vertex operators~(\ref{rrver}) describe field strengths of all even
ranks from 0 to 10.  By a $\Gamma$-matrix identity the $n$-form and
$(10-n)$-form field strengths are Hodge dual to one another, so a
$p$--brane and $(6-p)$--brane are sources for the same field, but one
`magnetic' and one `electric.'  The field equation for the 10-form
field strength allows no propagating states, but the field can still
have a physically significant energy
density~\cite{gojoe,romans,joeandy}.

The IIB theory has D$p$--branes for $p = -1$, 1, 3, 5, 7, and 9.  The
vertex operators~(\ref{rrver}) describe field strengths of all odd
ranks from 1 to 9, appropriate to couple to all but the 9--brane.  The
9--brane does couple to a nontrivial {\it potential,} as we will see
below.

A $(-1)$--brane is a Dirichlet instanton, defined by Dirichlet
conditions in the time direction as well as all spatial
directions.\cite{parton} Of course, it is not clear that T--duality
in the time direction has any meaning, but one can argue for the
presence of $(-1)$--branes as follows.  Given $0$--branes in the IIA
theory, there should be virtual $0$--brane world--lines that wind in a
purely spatial direction.  Such world--lines are required by quantum
mechanics, but note that they are essentially instantons, being
localised in time.  A T--duality in the winding direction then gives
a $(-1)$--brane.  One of the first clues to the relevance of
D--branes,\cite{Djoe} was the observation that D--instantons, having
action $g_s^{-1}$, would contribute effects of order $e^{-1/g_s}$ as
expected from the behaviour of large orders of string perturbation
theory.\cite{shenk1}

The D--brane, unlike the fundamental string, carries R--R charge. This
is consistent with the fact that they are BPS states, and so there
must be a conserved charge. A more careful argument, involving the
R--R vertex operators, can be used to show that they {\it must} couple
thus, and furthermore that fundamental strings cannot carry R--R
charges.

\subsection{The D--Brane Charge and Tension}
The bosonic discussion of section \ref{Dactionone} will supply us with
the worldvolume action \reef{diracborninfeld} for the bosonic
excitations of the D--branes in this supersymmetric context.  Now that
we have seen that D$p$--branes are BPS states, and couple to R--R
sector $(p+1)$--form potential, we ought to compute their charges and
new values for the tensions. 

Focusing on the R--R sector for now, supplementing the spacetime
supergravity action with the D--brane action we must have at least
(recall that the dilaton will not appear here, and also that we cannot
write this for $p=3$):
\begin{equation}
S=-{1\over2\kappa_0^2}\int
G_{(p+2)}{}^*G_{(p+2)}+\mu_p\int_{{\cal M}_{p+1}} C_{(p+1)},
\labell{RRpart}
\end{equation}
where $\mu_p$ is the charge of the D$p$--brane under the
($p{+}1$)--form $C_{(p+1)}$.  ${\cal M}_{p+1}$ is the world--volume of
the D$p$--brane.

Now the same vacuum cylinder diagram as in the bosonic string, as we
did in section \ref{tense}. With the fermionic sectors, our trace must
include a sum over the NS and R sectors, and furthermore must include
the GSO projection onto even fermion number. Formally, therefore, the
amplitude looks like:~\cite{gojoe}
\begin{equation}
{\cal A}=\int_0^\infty {dt\over 2t} {\rm Tr}_{\rm
NS+R}\left\{{1+(-1)^F\over2}e^{-2\pi t L_0} \right\}\ .
\end{equation}
Performing  the traces over the
open superstring spectrum gives
\begin{eqnarray}
{\cal A} &=& 2V_{p+1} \int {dt\over 2t}\, 
(8\pi^2 {\alpha^\prime} t)^{-(p+1)/2}
e^{- t{Y^2\over2\pi \alpha'}} 
\nonumber\\
&& \qquad\qquad\qquad\qquad\qquad 2^{-1} f_1^{-8}(q)\left\{ -f_2(q)^{8}
+ f_3(q)^8 - f_4(q)^8 \right\},
\end{eqnarray}
where again $q =e^{-2\pi t}$.  The three terms in the braces come from
the open string R sector with ${1\over2}$ in the trace, from the NS
sector with ${1\over2}$ in the trace, and the NS sector with
${1\over2} (-1)^F$ in the trace; the R sector with ${1\over2} (-1)^F$
gives no net contribution.  In fact, these three terms sum to zero by
Jacobi's {\it ``aequatio identico satis abstrusa''}, as they ought to since
the open string spectrum is supersymmetric, and we are computing a
vacuum diagram.  

What does this result mean? Recall that this vacuum diagram also
represents the exchange of closed strings between two identical
branes. the result ${\cal A}=0$ is simply a restatement of the fact
that D--branes are BPS states: The net forces from the NS--NS and R--R
exchanges cancel.  ${\cal A}=0$ has a useful structure, nonetheless,
and we can learn more by identifying the separate NS--NS and R--R
pieces.  This is easy, if we look at the diagram afresh in terms of
closed string: In the terms with $(-1)^F$, the world--sheet fermions
are {\it periodic} around the cylinder thus correspond to R--R
exchange. Meanwhile the terms without $(-1)^F$ have {\it
anti--periodic} fermions and are therefore NS--NS exchange.

Obtaining the $t\to 0$ behaviour as before (use the limits in insert 8
(p.\pageref{insert8})) gives
\begin{eqnarray}
{\cal A}_{\rm NS}\ =\ - {\cal A}_{\rm R} &\sim & {1\over2} V_{p+1}
 \int{dt\over t} (2\pi t)^{-(p+1)/2} (t/2\pi\alpha')^4 e^{-
 t{Y^2\over8\pi^2 \alpha'^2}} \nonumber\\ &=& V_{p+1} 2\pi
 (4\pi^2\alpha')^{3-p} G_{9-p}(Y^2).
\end{eqnarray} 
Comparing with field theory calculations gives\,\cite{gojoe}
\begin{equation}
2\kappa_0^2\mu_p^2 = 2 \kappa^2 \tau_p^2 = 2\pi
(4\pi^2\alpha')^{3-p}. \label{dcharge}
\end{equation}
Finally, using the explicit expression \reef{nice} for $\kappa$ in
 terms of string theory quantities, we get an extremely simple form
 for the charge:
\begin{equation}
\mu_p=(2\pi)^{-p}\alpha^{\prime -{(p+1)\over2}}\
,\quad\mbox{and}\quad \tau_p=g_s^{-1}\mu_p\ .
\labell{thecharge}
\end{equation}
(For consistency with the discussion in the bosonic case, we shall
still use the symbol $T_p$ to mean $\tau_p g_s$, in situations where we
write the action with the dilaton present. It will be understood then
that $e^{-\Phi}$ contains the required factor of $g_s^{-1}$.) 

It is worth updating our bosonic formula \reef{yangmillsbos}
for the coupling of the Yang--Mills theory which appears on the
world--volume of D$p$--branes with our superstring result above, to give:
\begin{equation}
g^2_{{\rm YM},p}=\tau_p^{-1}(2\pi\alpha^\prime)^{-2}=(2\pi)^{p-2}
\alpha^{\prime(p-3)/2}\ ,
\labell{yangmillscoupling}
\end{equation}
a formula we will use a lot in what is to follow.

Note that our formula for the tension \reef{thecharge} gives for the
D1--brane
\begin{equation}
\tau_1={1\over 2\pi\alpha^\prime g_s}\ ,
\end{equation}
which sets the ratios of the tension of the fundamental string,
$\tau^{\rm F}_1\equiv T=(2\pi\alpha^\prime)^{-1}$ , and the D--string
to be simply the string coupling $g_s$. This is a very elegant
normalisation is is extremely natural.

D--branes that are not parallel feel a net force since the
cancellation is no longer exact.  In the extreme case, where one of
the D--branes is rotated by $\pi$, the coupling to the dilaton and
graviton is unchanged but the coupling to the R--R tensor is reversed
in sign. So the two terms in the cylinder amplitude add, instead of
cancelling, and Jacobi cannot help us. The result is:
\begin{equation}
{\cal A}=V_{p+1}\int {dt\over t}(2\pi t)^{-(p+1)/2}e^{-{t}
(Y^2-2\pi\alpha^\prime)/8\pi^2\alpha^{\prime2}} f(t)
\end{equation}
where $f(t)$ approaches zero as $t\to0$. Differentiating this with
respect to $Y$ to extract the force per unit world--volume, we get 
\begin{equation}
F(Y)=Y\int {dt\over t}(2\pi t)^{-(p+3)/2} e^{-{t}
(Y^2-2\pi\alpha^\prime)/
8\pi^2\alpha^{\prime2}} f(t)\ .
\end{equation}
The point to notice here is that the force diverges as
$Y^2\to2\pi\alpha^\prime$. This is significant. One would expect a
divergence, of course, since the two oppositely charged objects are on
their way to annihilating.~\cite{tomlen} The interesting feature it
that the divergence begins when their separation is of order the
string length. This is where the physics of light fundamental strings
stretching between the two branes begins to take over. Notice that the
argument of the exponential is $tU^2$, where $U=Y/(2\alpha^\prime)$ is
the energy of the lightest open string connecting the branes. A scale
like $U$ will appear again, as it is a useful guide to new variables
to D--brane physics at ``substringy''
distances\cite{shenk2,short,shorty} in the limit where $\alpha^\prime$
and $Y$ go to zero.

Orientifold planes also break half the supersymmetry and are R-R
and NS-NS sources.  In the original Type I theory the orientation
projection keeps only the linear combination $Q_\alpha + \tilde
Q_\alpha$.  In the T--dualised theory this becomes $Q_\alpha +
(\prod_m P_m) \tilde Q_\alpha$ just as for the D--branes.  The force
between an orientifold plane and a D--brane can be obtained from the
M\"obius strip as in the bosonic case; again the total is zero and can
be separated into NS-NS and R-R exchanges.  The result is similar to
the bosonic result~(\ref{orten}),
\begin{equation}
\mu'_p = \mp 2^{p - 5} 
\mu_p, \qquad \tau'_p = \mp 2^{p - 5} \tau_p \
.
\labell{orrycharge}
\end{equation}
Since there are $2^{9-p}$ orientifold planes, the total O--plane
charge is $\mp 16 \mu_p$, and the total fixed-plane tension is $\mp 16
\tau_p$.

A nonzero total tension represents a source for the graviton and
dilaton. By the Fischler--Susskind mechanism\cite{fsuss}, at order
$g_s$ those background fields become become time dependent as in a
consistent way. A non--zero total R--R source is more serious, since
this would mean that the field equations are inconsistent (there are
uncancelled tadpoles): There is a violation of Gauss' Law, as R--R
flux lines have no place to go in the compact space $T^{9-p}$.  So our
result tells us that on $T^{9-p}$, we need exactly 16 D--branes, with
the $SO$ projection, in order to cancel the R--R $G_{(p+2)}$ form
charge. This gives the T--dual of $SO(32)$, completing our simple
orientifold story.

The spacetime anomalies for $G \neq SO(32)$ are thus accompanied by a
divergence\cite{gsdiv} in the full string theory, as promised, with
inconsistent field equations in the R--R sector: As in field theory,
the anomaly is related to the ultraviolet limit of a (open string)
loop graph.  But this ultraviolet limit of the annulus/cylinder ($t\to
\infty$) is in fact the infrared limit of the closed string tree
graph, and the anomaly comes from this infrared divergence.  From the
world--sheet point of view, as we have seen in the bosonic case,
inconsistency of the field equations indicates that there is a
conformal anomaly that cannot be cancelled.  The prototype of  this
is the original $D=10$ Type I theory.\cite{rrex} The $N$ D9--branes and 
single O9--plane couple to an R--R
10-form,
\begin{equation}
 (32 \mp N) {\mu_{10}\over 2} \int A_{10} ,
\end{equation}
and the field equation from varying~$A_{10}$ is just $G =
SO(32)$.

\subsection{Dirac Charge Quantisation}
\label{diracconsist}
We are of course studying a quantum theory, and so the presence of
both magnetic and electric sources of various potentials in the theory
should give some cause for concern. We should check that the values of
the charges are consistent with the appropriate generalisation
of~\cite{dirac} the Dirac quantisation condition.  The field strengths
to which a D$p$--brane and D$(6 - p)$--brane couple are dual to one
another, $G_{(p+2)} = *G_{(8-p)}$.

We can integrate the field strength $*G_{(p+2)}$ on
an ($8 - p$)-sphere surrounding a D$p$--brane, and using the
action~(\ref{RRpart}), we find  a total flux $\Phi= \mu_p$.  We can
write $*G_{(p+2)} = G_{(8-p)} = d C_{(7-p)}$ everywhere except on a
Dirac ``string'' (it is really a sheet), 
at the end of which lives the D$(6-p)$ ``monopole''.
Then
\begin{equation}
\Phi= {1\over 2\kappa_0^2}\int_{S_{8-p}} *G_{(p+2)}
={1\over 2\kappa_0^2}\int_{S_{7-p}} C_{(7-p)}\ .
\end{equation}
where we perform the last integral on a small sphere surrounding the Dirac 
string.  A ($6 - p$)--brane circling the string picks up a phase
$e^{i \mu_{6 - p}\Phi}$.  The condition that the string be invisible is
\begin{equation}
\mu_{6 - p} \Phi = {1\over 2\kappa_0^2}\mu_{6 - p} \mu_p = 2\pi n.
\end{equation}
{\it The D--branes' charges~(\ref{dcharge}) satisfy this with the minimum
quantum $n=1$.}

While this argument does not apply directly to the case $p=3$, as the
self--dual 5--form field strength has no covariant action, the result
follows by T--duality.  A topological derivation of the D--brane
charge has been given. There are mathematical structures with deep
roots, {\it e.g.} ``K--theory'', which seem to capture the physics of
the R--R charges in string theory, and this is a subject of exciting
research.~\cite{ghs,ktheory} The lectures of John Schwarz in this
school develop some of the techniques of Sen~\cite{sentach} which are
relevant to constructing branes from the K--theory point of
view.~\cite{john,olsen}

\section{Worldvolume  Actions II: Curvature Couplings}

\subsection{Tilted D--Branes and Branes within Branes}
There are additional terms in the action \reef{RRpart} which we just
wrote down, involving the D--brane gauge field.  Again these can be
determined from T--duality.  Consider, as an example, a D1--brane in
the 1--2 plane.  The action is 
\begin{equation}
 \mu_1\int dx^0\, dx^1\, \left(
  C_{01} + \partial_1 X^2 C_{02}\right)\ .
\end{equation}
Under a T--duality in the $x^2$--direction this becomes 
\begin{equation}
 \mu_2\int
dx^0dx^1dx^2\, \left( C_{012} + 2\pi{\alpha^\prime} F_{12} C_0
\right)\ . \labell{rrf}
\end{equation}
We have used the T--transformation of the $C$ fields as discussed in
section \ref{tdualc}, and also the recursion relation
\reef{trec} between D--brane tensions.
  
This has an interesting interpretation. As we saw before in section
\ref{tiltaway}, a D$p$--brane tilted at an angle $\theta$ is
equivalent to a D$(p+1)$--brane with a constant gauge field of
strength $F=(1/2\pi\alpha^\prime)\tan\theta$. Now we see that there is
additional structure: the flux of the gauge field couples to the R--R
potential $C^{(p)}$. In other words, the flux acts as a source for a
D$(p-1)$--brane living in the worldvolume of the D$(p+1)$--brane. In
fact, given that the flux comes from an integral over the whole
world--volume, we cannot localise the smaller brane at a particular
place in the world--volume: it is ``smeared'' or ``dissolved'' in the
world--volume.

In fact, we shall see when we come to study supersymmetric
combinations of D--branes that supersymmetry requires the D0--brane to
be completely smeared inside the D2--brane. It is clear here how it
manages this, by being simply T--dual to a tilted D1--brane. We shall
see many consequences of this later.

\subsection{Branes Within Branes: Anomalous Gauge Couplings}
\label{anomalousF}
The T--duality argument of the previous section is easily generalised,
with the Chern--Simons like result~\cite{rract,douginst} 
\begin{equation}
\mu_p\int_{{\cal M}_{p+1}}
\left[\textstyle{\sum}_p C_{(p+1)}\right] \wedge{\rm
Tr}\,e^{2\pi{\alpha^\prime} F + B} \ , \labell{csact}
\end{equation}
(We have included non--trivial $B$ on the basis of the argument given
at the beginning of section \ref{Dactionone}.) So far, the gauge trace
has the obvious meaning. We note that there is the possibility that in
the full non--Abelian situation, the $C$ can depend on {\it
  non--commuting} transverse fields $X^i$, and so we need something
more general.  We will return to this later. The expansion of the
integrand~(\ref{csact}) involves forms of various rank; the notation
means that the integral picks out precisely the terms that are
proportional to the volume form of the D$p$--brane.

Looking at the first non--trivial term in the expansion of the
exponential in the action we see that there is the term that we
studied above corresponding to the dissolution of a D$(p-2)$--brane
into the sub 2--plane in the D$p$--brane's world volume formed by the
axes $X^i$ and $X^j$, if field strength components $F_{ij}$ are turned
on.

At the next order, we have  a term which is quadratic in $F$:
\begin{equation}
\mu_p {(2\pi\alpha^\prime)^2\over2}\int C_{(p-3)}\wedge{\rm
Tr}F\wedge F={\mu_{p-4}\over8\pi^2} \int C_{(p-3)}\wedge{\rm
Tr}F\wedge F\ .
\labell{quadratic}
\end{equation} We have used the fact that $\mu_{p-4}/\mu_p=
(2\pi\sqrt{\alpha^\prime})^4.$ Interestingly, we see that if we excite
an instanton configuration on a 4 dimensional sub--space of the
D$p$--brane's worldvolume, it is equivalent to precisely one unit of
D$(p-4)$--brane charge! In fact, this term is already recognisable
from the study of consistency of the type~I string theory in ten
dimensions from just field theory considerations.  There is a modified
3--form field strength, ${\widetilde G}_{(3)}$, which is
\begin{equation}
{\widetilde G}_{(3)}=dC_{(2)}-{\alpha^\prime\over4}
\left[\omega_{3Y}-\omega_{3L}\right]\ ,
\end{equation}
with action
\begin{equation}
S=-{1\over4\kappa^2}\int{\widetilde G}_{(3)} \wedge^*{\widetilde G}_{(3)}\ .
\end{equation}
Since $d\omega_{3Y}={\rm Tr}(F\wedge F)$  and 
$d\omega_{3L}={\rm Tr}(R\wedge R)$,  
this gives, after integrating by parts
\begin{equation}
{\alpha^\prime\over8\kappa^2}\int C_{(6)}\wedge\left({\rm Tr}F\wedge F
-{\rm Tr}R\wedge R\right)\ .
\labell{coupled}
\end{equation}
An evaluation of the coefficient of the quadratic term in $F$ shows
that it is precisely that in \reef{quadratic}, for $p=9$.
Furthermore, the Green--Schwarz anomaly cancellation mechanism
\cite{gsdiv} requires a term
\begin{equation}
C_{(2)}\wedge X_8\ ,
\labell{greenschwarz}
\end{equation}
where
\begin{equation}
X_8={1\over(2\pi)^4}\left({1\over 48}{\rm Tr}F^4
-{1\over192}{\rm Tr}F^2{\rm Tr}R^2\right)+{1\over128}p_1^2(R)
-{1\over96}p_2(R)\ ,
\end{equation}
the pure gauge part of which can again be found by expanding \reef{csact}
to quartic order.  The terms involving curvature will be shown to
arise in the next section, where the $p_i$, the Pontryagin classes,
will be defined very shortly.

Since we see that the gauge couplings are correct, giving the correct
results known from ten dimensional string theory, we ought to take
seriously the implications of the terms involving
curvature. It is clear that there must be curvature terms in the
action for the D$p$--branes and O$p$-planes also.

\subsection{Branes Within Branes: Anomalous ``Curvature'' Couplings}

There are indeed curvature terms of the sort which we deduced in the
previous subsection, from knowledge of the anomaly in string theory.
Their presence may be deduced in many other ways, for example using
string duality. A more straightforward way is to generalise the type
of anomaly arguments used for the ten dimensional type~I string
supergravity+Yang--Mills case to include not just D9--branes, but all
branes, treating them as surfaces upon which anomalous theories
reside.~\cite{inflow,ghs} A topological argument can be applied to
constrain the form of the couplings required on the world--volumes in
order to make the bulk+brane theory consistent. We will not review the
details of the argument here, but merely quote the
result:~\cite{bsv,ghs}
\begin{equation}
\mu_p\int_{{\cal M}_{p+1}}\sum_i C_{(i)}\left[ e^{2\pi\alpha^\prime
 F+B} \right] \sqrt{{\hat {\cal A}}(4\pi^2\alpha^\prime R)}\ ,
\labell{curvy}
\end{equation}
where the ``A--roof'' or ``Dirac'' genus has its square root defined as:
\begin{equation}
\sqrt{{\hat {\cal
A}}(R)}=1-{p_1(R)\over48}+p_1^2(R){7\over11520}-{p_2(R)\over2880}+\cdots
\end{equation}
The $p_i(R)$'s are the $i$th Pontryagin class. For example, 
\begin{eqnarray}
p_1(R)&=&-{1\over8\pi^2}{\rm Tr} R\wedge R\nonumber\\
p_2(R)&=&
{1\over(2\pi)^4}\left(-{1\over4}{\rm Tr}R^4+{1\over8}({\rm Tr}R^2)^2\right)
\ ,
\end{eqnarray}
Expanding, we have 
\begin{equation}
-{\mu_p(4\pi^2\alpha^\prime)^2\over 48}
\int_{{\cal M}_{p+1}}C_{(p-3)}\wedge p_1(R)\ .
\end{equation}
So we see another way to get a D$(p-4)$--brane: wrap the brane on a
four dimensional surface of non--zero $p_1(R)$. Indeed, as we saw in
equation \reef{eulerK3}, the K3 surface has $p_1=2\chi=48$, and so
wrapping a D$p$--brane on K3 gives D$(p-4)$--brane charge of
$-1$!~\cite{bsv} We will return to this later.

In fact, we can see that this is not the whole story. We can not
reproduce the correct coefficient of the curvature terms in
\reef{coupled} from the anomalous couplings on the D9--branes alone.
Happily, there is a nice resolution to this problem,~\cite{sunilone}
which is found with the O9--plane. It is present since the type~I
string theory is an orientifold of the Type~IIB theory. An O9--brane
does not have open strings ending on it, as we have seen, and
therefore there are no gauge fields on their world--volume. This fits
with the fact that we already have the correct gauge couplings of
type~I.  As they are objects with finite tension, however, they are
natural candidates to have curvature couplings. To get the coupling in
\reef{coupled} right, it can be seen that there must be a coupling
($p=9$):
\begin{equation}
{(\pi^2\alpha^\prime)^2
{\tilde\mu}_p\over48\pi^2}\int C_{(p-3)}\wedge {\rm Tr}R\wedge R
\end{equation} 
on its worldvolume, (${\tilde\mu}_p=-2^{5-p}\mu_p$), and we have
written the general expression at this order for all negative
O$p$--planes.  This (for $p=9$) combined with the contribution from
the sixteen D9--branes, gives the correct total curvature coupling.

In general, the couplings for this class of O$p$--planes may be
deduced from anomaly--inflow type arguments,~\cite{mukhioplane} as was
the case for the D$p$--branes, and a general formula written in terms
of a index, just like the D--brane case. The answer is:~\cite{hirzy}
\begin{equation}
{\tilde\mu}_p\int_{{\cal M}_{p+1}}\sum_i C_{(i)} \sqrt{{\hat {\cal
L}}(\pi^2\alpha^\prime R)}\ ,
\end{equation}
where the ``Hirzebruch'' polynomial, $\hat {\cal L}$, has its square
root defined as:
\begin{equation}
\sqrt{\hat{\cal
L}(R)}=1+{p_1(R)\over6}-p_1^2(R){1\over90}+p_2(R){7\over90}+\cdots
\end{equation}
In fact, the fourth order terms also give us the correct couplings to
complete the $C_{(2)}\wedge X_8$ needed for consistency.
There are more general  types of O--plane in string theory than the type
we have considered here, and for which curvature couplings have been
derived.~\cite{otheroplanei,mukhisurya,otheroplaneii}

\subsection{Further Non--Abelian Extensions}
\label{nonablestuff}
One can use T--duality to go a bit further and deduce the non--Abelian
form of the action, being mindful of the sort of complications
mentioned at the beginning of section \reef{nonabelstuff}.  In the
absence of curvature terms~\footnote{An important issue is the nature
  of the coupling of curvature and R--R potentials in such
  non--Abelian situations. Given the enhan\c con phenomenon discussed
  later on, it is clear that there are such effective couplings.} it
turns out to be:~\cite{robdielectric,taylorvan}
\begin{equation}
\mu_p\int_{p-\rm brane} {\rm Tr}\left(\left[e^{2\pi\alpha^\prime {\bf
i}_\Phi{\bf i}_\Phi}\textstyle{\sum}_p C_{(p+1)} \right]
e^{2\pi{\alpha^\prime} F + B}\right) \ . \labell{nonabeliancs}
\end{equation}

Here, we ascribe the same meaning to the gauge trace as we did
previously (see section \reef{nonabelstuff}). The meaning of ${\bf i}_X$ is
as the ``interior product'' in the direction given by the vector $\Phi^i$,
which produces a form of one degree fewer in rank. For example, on a
two form $C_{(2)}(\Phi)=(1/2)C_{ij}(\Phi)dX^idX^j$, we have
\begin{equation}
{\bf i}_\Phi C_{(2)}= \Phi^iC_{ij}(\Phi)dX^j\ ;\quad {\bf i}_\Phi{\bf i}_\Phi
C_{(2)}(\Phi)=\Phi^j\Phi^iC_{ij}(\Phi) ={1\over2}[\Phi^i,\Phi^j]C_{ij}(\Phi)\ ,
\end{equation} 
where we see that the result of acting twice is non--vanishing when we
allow for the non--abelian case, with $C$ having a nontrivial
dependence on $\Phi$.  We shall see this action work for us to produce
interesting physics later.

\subsection{Even More Curvature Couplings}
We deduced curvature couplings to the R--R potentials a few
subsections ago. In particular, such couplings induce the charge of
lower $p$ branes by wrapping larger branes on topologically
non--trivial surfaces.  

In fact, as we saw before, if we wrap a D$p$--brane on K3, there is
induced precisely $-1$ units of charge of a D$(p-4)$--brane. This
means that the charge of the effective $(p-4)$--dimensional object is
\begin{equation}
\mu=\mu_pV_{\rm K3}-\mu_{p-4}\ ,
\labell{wrapcharge}
\end{equation}
where $V_{\rm K3}$ is the volume of the K3. However, we can go further
and notice that since this is a BPS object of the six dimensional
${\cal N}=2$ string theory obtained by compactifying on K3, we
should expect that it has a tension which is
\begin{equation}
\tau=\tau_pV_{\rm K3}-\tau_{p-4}=g^{-1}_s\mu\ .
\labell{wraptension}
\end{equation}
If this is indeed so, then there must be a means by which the
curvature of K3 induces a shift in the tension in the world--volume
action. Since the part of the action which refers to the tension is
the Dirac--Born--Infeld action, we deduce that there must be a set of
curvature couplings for that part of the action as well.\cite{sunilone}
Some of them are given by the following:~\cite{sunilone,bbg}
\begin{eqnarray}
&&S=-\tau_p\int d^{p+1}\xi \,\, e^{-\Phi}{\rm det}^{1/2}(G_{ab}+{\cal F}_{ab})
\Biggl(1-{1\over768\pi^2}\times\nonumber\\
&&\hskip1cm \left(\R_{abcd}\R^{abcd}
-\R_{\alpha\beta ab}R^{\alpha\beta ab}
+2{\hat \R}_{\alpha\beta}{\hat \R}^{\alpha\beta}
-2{\hat \R}_{ab}{\hat \R}^{ab}
\right)
+O(\alpha^{\prime 4})\Biggr)\ ,\nonumber\\
\end{eqnarray}
where ${\cal R}_{abcd}=(4\pi^2\alpha^\prime) R_{abcd}$, {\it etc.},
and $a,b,c,d$ are the usual tangent space indices running along the
brane's world volume, while $\alpha,\beta$ are normal indices, running
transverse to the world--volume.  

Some explanation is needed. Recall: the embedding of the brane into
$D$--dimensional spacetime is achieved with the functions
$X^\mu(\xi^a)$, ($a=0,\ldots,p; \mu=0,\ldots,D-1$) and the pullback of
a spacetime field $F_\mu$ is performed by soaking up spacetime indices
$\mu$ with the local ``tangent frame'' vectors $\partial_a X^\mu$, to
give $F_a=F_\mu \partial_a X^\mu$. There is another frame, the
``normal frame'', with basis vectors $\zeta^\mu_\alpha$,
($\alpha=p+1,\ldots,D-1$). Orthonormality gives
$\zeta^\mu_\alpha\zeta^\nu_\beta G_{\mu\nu}=\delta_{\alpha\beta}$ and
also we have $\zeta^\mu_\alpha\partial_a X^\nu G_{\mu\nu}=0$. 

We can pull back the spacetime Riemann tensor
$R_{\mu\nu\kappa\lambda}$ in a number of ways, using these different
frames, as can be seen in the action. $\hat R$ with two indices are
objects which were constructed by contraction of the {\it
  pulled--back} fields. They are {\sl not} the pull back of the bulk
Ricci tensor, which vanishes at this order of string perturbation
theory anyway.

In fact, for the case of K3, it is Ricci flat and everything with
normal space indices vanishes and so we get only $R_{abcd}R^{abcd}$
appearing, which alone computes the result \reef{eulerK3} for us, and
so after integrating over K3, the action becomes:
\begin{equation}
S=-\int d^{p-3}\xi\, \,
 e^{-\Phi}\left[\tau_pV_{\rm K3}-\tau_{p-4}\right]{\rm det}^{1/2}(G_{ab}+
{\cal F}_{ab})\ ,
\labell{wrapped}
\end{equation}
where again we have used the recursion relation between the D--brane
tensions. So we see that we have correctly reproduced the shift in the
tension that we expected on general grounds for the effective
D$(p-4)$--brane. We will use this action later.

\section{The D$p$--D$p'$ System}
\label{peepee}
Simple T--duality gives parallel D--branes all with the same dimension
but  we can consider more general configurations.  In this
section we consider two D--branes, D$p$ and D$p^\prime$, each parallel
to the coordinate axes. (We can of course have D--branes at
angles,~\cite{bdl} but we will not consider this here.) An open string
can have both ends on the same D--brane or one on each.  The $p-p$ and
$p'-p'$ spectra are the same as before, but the $p-p'$ strings are
new.  Since we are taking the D--branes to be parallel to the
coordinate axes, there are four possible sets of boundary conditions
for each spatial coordinate $X^i$ of the open string, namely NN
(Neumann at both ends), DD, ND, and DN.  What really will matter is
the number $\nu$ of ND plus DN coordinates.  A T--duality can switch
NN and DD, or ND and DN, but $\nu$ is invariant.  Of course $\nu$ is
even because we only have $p$ even or $p$ odd in a given theory.

The respective mode expansions are 
\begin{eqnarray}
{\rm NN\colon}&& X^\mu(z,\zb) = x^\mu - i\ap p^\mu
\ln(z\zb) + i\sqrt{\ap\over2} \sum_{m\neq0} {\alpha_m^\mu\over
m}(z^{-m}+\zb^{-m}), \nonumber\\
{\rm DN, ND\colon}&& X^\mu(z,\zb) = i\sqrt{\ap\over2} \sum_{r \in\IZ +1/2}
{\alpha_r^\mu\over r}(z^{-r}\pm\zb^{-r}),
\label{modexps}\\
{\rm DD\colon}&& X^\mu(z,\zb) = -i \frac{\delta X^\mu}{2\pi} \ln(z/\zb)
+ i\sqrt{\ap\over2} \sum_{m\neq0}
{\alpha_m^\mu\over m}(z^{-m}-\zb^{-m})\ . \nonumber
\end{eqnarray}
In particular, the DN and ND coordinates have half--integer moding.
The fermions have the same moding in the Ramond sector (by definition)
and opposite in the Neveu--Schwarz sector.  The string zero point
energy is 0 in the R sector as always, and using \reef{zpe} we get:
\begin{equation}
(8-\nu)\left(-{1\over 24} - {1\over 48}\right) 
+ \nu \left({1\over 48} + {1\over 24}\right) = -{1\over 2} + {\nu\over 8}
\label{nszpe}
\end{equation}
in the NS sector.

The oscillators can raise the level in half--integer units, so only
for $\nu$ a multiple of 4 is degeneracy between the R and NS sectors
possible.  Indeed, it is in this case that the D$p$--D$p'$ system is
supersymmetric.  We can see this directly.  As discussed in
sections~\ref{tdualc} and~\ref{bpsstates}, a D--brane leaves unbroken
the supersymmetries
\begin{equation}
Q_\alpha+ P {\tilde Q}_\alpha\ , \label{unb1}
\end{equation}
where $P$ acts as a reflection in the direction transverse to the
D--brane.  With a second D--brane, the only unbroken supersymmetries
will be those that are also of the form
\begin{equation}
Q_\alpha+ P' {\tilde Q}_\alpha = Q_\alpha+ P (P^{-1}P') {\tilde Q}_\alpha\ .
\end{equation}
with $P'$ the reflection transverse to the second D--brane.  Then the
unbroken supersymmetries correspond to the $+1$ eigenvalues of
$P^{-1}P'$.  In DD and NN directions this is trivial, while in DN and
ND directions it is a net parity transformation.  Since the number
$\nu$ of such dimensions is even, we can pair them as we did in
section \ref{superlook}, and write $P^{-1}P'$ as a product of rotations
by $\pi$,
\begin{equation}
e^{i\pi (J_1 + \ldots + J_{\nu/2}) }\ .
\end{equation}
In a spinor representation, each $e^{i\pi J}$ has eigenvalues $\pm
{i}$, so there will be unbroken supersymmetry only if $\nu$ is a
multiple of 4 as found above.~\footnote{We will see that
there are supersymmetric {\it bound states} when $\nu = 2$.}

For example, Type~I theory, besides the D9--branes, will have
D1--branes and D5--branes.  This is consistent with the fact that the
only R-R field strengths are the three-form and its Hodge--dual
seven--form.  The D5--brane is required to have two Chan--Paton
degrees of freedom (which can be thought of as images under $\Omega$)
and so an $SU(2)$ gauge group.\cite{edsmall,GP}

When $\nu = 0$, $P^{-1}P' = 1$ identically and there is a full
ten-dimensional spinor of supersymmetries.  This is the same as for
the original Type~I theory, to which it is T--dual.  In $D=4$ units,
this is ${\cal N}=4$, or sixteen supercharges.  For $\nu=4$ or $\nu=8$
there is $D=4$ ${\cal N}=2$ supersymmetry.

Let us now study the spectrum for $\nu = 4$, saving $\nu = 8$ for
later. Sometimes it is useful to draw a quick table showing where the
branes are located. Here is one for the (9,5) system, where the
D5--brane is pointlike in the $x^6,x^7,x^8,x^9$ directions and the
D9--brane is (of course) extended everywhere:
\bigskip
\begin{center}
\begin{tabular}{|c|c|c|c|c|c|c|c|c|c|c|}
\hline
&$x^0$&$x^1$&$x^2$&$x^3$&$x^4$&$x^5$&$x^6$&$x^7$&$x^8$&$x^9$\\\hline
D9&$-$&$-$&$-$&$-$&$-$&$-$&$-$&$-$&$-$&$-$\\\hline
D5&$-$&$-$&$-$&$-$&$-$&$-$&$\bullet$&$\bullet$&$\bullet$&$\bullet$
\\
\hline
\end{tabular}
\end{center}
\bigskip
A dash under $x^i$ means that the brane is extended in that direction,
while a dot means that it is pointlike there.

Continuing with our analysis, we see that the NS zero--point energy is
zero.  There are four periodic world--sheet fermions $\psi^i$, namely
those in the ND directions.  The four zero modes generate $2^{4/2}$ or
four ground states, of which two survive the GSO projection.  In the R
sector the zero--point energy is also zero; there are four periodic
transverse $\psi$, from the NN and DD directions not counting the
directions $\mu=0,1$.  Again these generate four ground states of
which two survive the GSO projection.  The full content of the
$p$--$p'$ system is then is half of an $N=2$ hypermultiplet.  The
other half comes from the $p'$--$p$ states, obtained from the
orientation reversed strings: these are distinct because for $\nu \neq
0$ the ends are always on different D--branes.

Let us write the action for the bosonic $p-p'$ fields $\chi^A$,
starting with $(p,p') = (9,5)$.  Here $A$ is a doublet index under the
$SU(2)_R$ of the $N=2$ algebra.  The field $\chi^A$ has charges
$(+1,-1)$ under the $U(1) \times U(1)$ gauge theories on the branes,
since one end leaves, and the other arrives.  The minimally coupled
action is then
\begin{equation}
\int d^6\xi\, \left( \sum_{a=0}^5 \left| (\partial_a + iA_a - i A'_a) \chi
\right|^2 + 
\biggr(\frac{1}{4g_{{\rm YM},p}^2} + \frac{1}{4g_{{\rm YM},p'}^2} \biggl)
\sum_{I=1}^3 (\chi^{\dagger}\tau^I\chi)^2 \right)\ ,
\label{59act}
\end{equation}
with $A_a$ and $A_a'$ the brane gauge fields, $g_{{\rm YM},p}$ and
$g_{{\rm YM},p'}$ the effective Yang--Mills couplings
\reef{yangmillscoupling}, and $\tau^I$ the Pauli matrices.  The
second term is from the $N=2$ D--terms for the two gauge fields. It
can also be written as a commutator $\Tr\, [\phi^i,\phi^j]^2$ for
appropriately chosen fields $\phi^i$, showing that its form is
controlled by the dimensional reduction of an $F^2$ pure Yang--Mills
term. See section \ref{aleprobe} for more on this.

The integral is
over the 5--brane world-volume, which lies in the 9--brane
world-volume.  Under T--dualities in any of the ND directions, one
obtains $(p,p') = (8,6)$, $(7,7)$, $(6,8)$, or $(5,9)$, but the
intersection of the branes remains $(5+1)$-dimensional and the
$p$--$p'$ strings live on the intersection with action~(\ref{59act}).
In the present case the $D$-term is nonvanishing only for $\chi^A =
0$, though more generally (say when there are several coincident $p$
and $p'$--branes), there will be additional massless charged fields
and flat directions arise.

Under T--dualities in $r$ NN directions, one obtains $(p,p') = (9-r,5-r)$. 
The action becomes 
\begin{eqnarray}
&&\int d^{6-r}\xi\, \left( \sum_{a=0}^{5-r} \left| (\partial_a + iA_a
- i A'_a) \chi \right|^2\ + \frac{ \chi^{\dagger}\chi
}{(2\pi{\alpha^\prime})^2 } \sum_{a=6-r}^{5} ( X_a - X'_a )^2
\right.\nonumber\\ &&\qquad\qquad\qquad\qquad\qquad+\
\left.\biggl(\frac{1}{4g_{{\rm YM},p}^2} + \frac{1}{4g_{{\rm YM}p'}^2} \biggr)
\sum_{i=1}^3 (\chi^{\dagger}\tau^I\chi)^2 \right)\ .
\label{xyact}
\end{eqnarray}
The second term, proportional to the separation of the branes, is from
the tension of the stretched string.

\subsection{The BPS Bound}

The ten dimensional ${\cal N}=2$ supersymmetry algebra (in a Majorana
basis) is
\begin{eqnarray}
&& \{ Q_\alpha, Q_\beta \}\ 
=\ 2(\Gamma^0 \Gamma^\mu)_{\alpha\beta} ( P_\mu + Q^{\rm NS}_\mu/2\pi\ap )
\nonumber\\ 
&& \{ \tilde Q_\alpha, \tilde Q_\beta\}\ =\ 2(\Gamma^0
\Gamma^\mu)_{\alpha\beta} ( P_\mu - Q^{\rm NS}_\mu/2\pi\ap )
\nonumber\\ 
&& \{ Q_\alpha, \tilde Q_\beta \}\ =\ 2 \sum_p {\tau_p\over p!} (\Gamma^0
\Gamma^{m_1} \ldots \Gamma^{m_p})_{\alpha\beta} Q^{\rm R}_{m_1\ldots m_p} \ .
\labell{central}
\end{eqnarray}
Here $Q^{\rm NS}$ is the charge to which the NS--NS two--form couples,
it is essentially the winding of a fundamental string stretched along
${\cal M}_1$:
\begin{equation}
Q^{\rm NS}_\mu\equiv  {Q^{\rm NS}\over v_1}\int_{{\cal M}_1} dX^\mu 
\ ,\quad {\rm with}
\quad Q^{\rm NS}={1\over {\rm Vol}\, S^7}\int_{S^7} e^{-2\Phi}\, {}^*H^{(3)}
\end{equation}
and the charge $Q^{\rm NS}$ is normalised to one per unit spatial
world--volume, $v_1=L$, the length of the string.  It is obtained by
integrating over the $S^7$ which surrounds the string.  The $Q^{\rm
  R}$ are the R--R charges, defined as a generalisation of winding on
the space ${\cal M}_p$:
\begin{equation}
Q^{\rm R}_{\mu_1\dots\mu_p}\equiv  {Q^{\rm R}_p\over v_p}\int_{{\cal M}_p} 
dX^{\mu_1}\wedge\cdots dX^{\mu_p} \ ,\quad {\rm with}
\quad Q_p^{\rm R}={1\over {\rm Vol}\, S^{8-p}}\int_{S^{8-p}}\!\!
 {}^*G^{(p+2)}\ .
\end{equation}

The sum in \reef{central} runs over all orderings of indices, and we
divide by $p!$ Of course, $p$ is even for IIA or odd for IIB.  The
R--R charges appear in the product of the right- and left--moving
supersymmetries, since the corresponding vertex operators are a
product of spin fields, while the NS-NS charges appear in right--right
and left--left combinations of supercharges.

As an example of how this all works, consider an object of length $L$,
with the charges of $p$ fundamental strings (``F--strings'', for
short) and $q$ D1--branes (``D--strings) in the IIB theory, at rest
and aligned along the direction $X^1$.  The anticommutator implies
\begin{equation} 
{1\over2} 
\left\{ \left[ \begin{array}{c} Q_\alpha \\ \tilde Q_\alpha \end{array}
\right] , \left[ Q_\beta\ \tilde Q_\beta \right] \right\}
= \left[ \begin{array}{cc} 1&0 \\ 0&1 \end{array} \right] M
\delta_{\alpha\beta} +
\left[ \begin{array}{cc} p&q/g_s \\ q/g_s&-p \end{array} \right]
{{L (\Gamma^0 \Gamma^1)}_{\alpha\beta}\over{2\pi{\alpha^\prime}}}\ .
\end{equation} The eigenvalues of $\Gamma^0
\Gamma^1$ are $\pm 1$ so those of the right--hand side are $M \pm L
(p^2 + q^2/g^2)^{1/2}/2\pi\ap$.  The left side is a positive matrix,
and so we get the ``BPS bound'' on the tension\,\cite{schwarz}
\begin{equation}
\frac{M}{L} \geq 
\frac{\sqrt{p^2 + q^2/g^2_s}}{2\pi{\alpha^\prime}}\equiv \tau_{p,q}
\ . \label{fdbps}
\end{equation}
Quite pleasingly, this is saturated by the fundamental string, $(p,q)
= (1,0)$, and by the D--string, $(p,q) = (0,1)$.

It is not too hard to extend this to a system with the quantum numbers
of Dirichlet $p$ and $p'$ branes. The result for $\nu$ a multiple of~4
is
\begin{equation}
M \geq \tau_p v_p + \tau_{p'} v_{p'} \label{marg}
\end{equation}
and for $\nu$ even but not a multiple of~4 it is\,~\footnote{The
  difference between the two cases comes from the relative sign of
  $\Gamma^M (\Gamma^{M'})^T$ and $\Gamma^{M'} (\Gamma^{M})^T$. }
\begin{equation}
M \geq \sqrt{\tau^2_p v^2_p + \tau^2_{p'} v^2_{p'} }\ .
\label{deep}
\end{equation}
The branes are wrapped
on tori of volumes $v_p$ and $v'_p$ in order to make the masses finite.

The results~(\ref{marg}) and (\ref{deep}) are consistent with the earlier
results on supersymmetry breaking.  For $\nu$ a multiple of 4, a
separated $p$--brane and $p'$--brane do indeed saturate the
bound~(\ref{marg}).  For $\nu$ not a multiple of four, they do not saturate
the bound~(\ref{deep}) and cannot be supersymmetric.

\subsection{FD Bound States}
\label{bounds}
\begin{figure}
\centerline{\psfig{figure=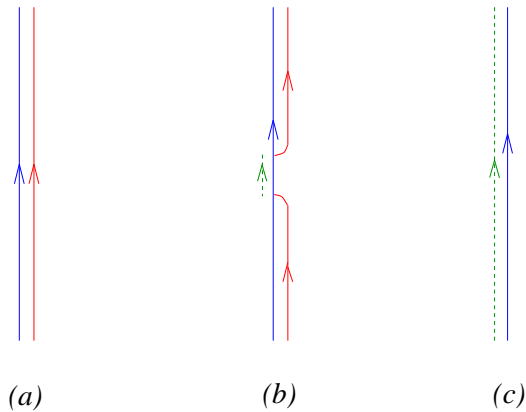,height=2.1in}}
\caption{{\it (a)} A parallel D--string and F--string, which is not 
  supersymmetric.  {\it (b)} The F--string breaks, its ends attaching
  to the D-string, resulting in {\it (c)} the final supersymmetric
  state, a D--string with flux.}
\label{boundfd}
\end{figure}

Consider a parallel D--string and F--string lying along $X^1$.  The
total tension
\begin{equation}
\tau_{D1} + \tau_{F1} = \frac{g_s^{-1} + 1}{ 2\pi{\alpha^\prime}}
\end{equation} 
exceeds the BPS bound~(\ref{fdbps}) and so this configuration is not
supersymmetric.  However, it can lower its energy\cite{edbound} as shown in
figure~\ref{boundfd}.  
The F--string breaks, its endpoints attached to the D--string.  The
endpoints can then move off to infinity, leaving only the D--string
behind.  Of course, the D--string must now carry the charge of the
F-string as well.  This comes about because the F--string endpoints are
charged under the D--string gauge field, so a flux runs between them;
this flux remains at the end.  
% Varying the NS-NS $B_ {\mu\nu}$ field in
% the D--brane action~ (\ref{dact}), one sees that it has a source
% proportional to the  invariant flux $F_{ab} + B_{ab} / 2\pi\ap$.  
Thus the final D--string carries both the NS--NS and R--R two--form
charges.  The flux is of order $g_s$, its energy density is of
order~$g_s$, and so the final tension is $(g_s^{-1} +
O(g_s))/2\pi{\alpha^\prime}$.  This is below the tension of the
separated strings and of the same form as the BPS bound~(\ref{fdbps})
for a $(1,1)$ string.  A more detailed calculation shows that the
final tension saturates the bound,~\cite{rract} so the state is
supersymmetric.  In effect, the F--string has dissolved into the D--string,
leaving flux behind.

We can see quite readily that this is a supersymmetric situation using
T--duality. We can choose a gauge in which the electric flux is
$F_{01}={\dot A}_1$. T--dualizing along the $x^1$ direction, we ought
to get a D0--brane, which we do, except that it is moving with
constant velocity, since we get ${\dot X}^1=2\pi\alpha^\prime {\dot
  A}_1$. This clearly has the same supersymmetry as a stationary
D0--brane, having been simply boosted.

To calculate the number of BPS states we should put the strings in a
box of length $L$ to make the spectrum discrete.  For the $(1,0)$
F-string, the usual quantisation of the ground state gives eight
bosonic and eight fermionic states moving in each direction for $16^2
= 256$ in all.  This is the ultrashort representation of
supersymmetry: half the 32 generators annihilate the BPS state and the
other half generate $2^8 = 256$ states.  The same is true of the
$(0,1)$ D-string and the $(1,1)$ bound state just found, as will be
clear from the later duality discussion of the D--string.

It is worth noting that the $(1,0)$ F--string leaves unbroken half the
supersymmetry and the $(0,1)$ D--string leaves unbroken a different
half of the supersymmetry.  The $(1,1)$ bound state leaves unbroken
not the intersection of the two (which is empty), but yet a different
half.  The unbroken symmetries are linear combinations of the unbroken
and broken supersymmetries of the D--string.

All the above extends immediately to $p$ F--strings and one D--string,
forming a supersymmetric $(p,1)$ bound state.  The more general case
of $p$ F-strings and $q$ D--strings is more complicated.  The gauge
dynamics are now non--Abelian, the interactions are strong in the
infrared, and no explicit solution is known.  When $p$ and $q$ have a
common factor, the BPS bound makes any bound state only neutrally
stable against falling apart into subsystems.  To avoid this
complication let $p$ and $q$ be relatively prime, so any
supersymmetric state is discretely below the continuum of separated
states.  This allows the Hamiltonian to be deformed to a simpler
supersymmetric Hamiltonian whose supersymmetric states can be
determined explicitly, and again there is one ultrashort
representation, $256$ states.  The details, which are quite intricate,
are left to the reader to consult in the
literature~\cite{edbound,joebook}.

\subsection{The Three--String Junction}
\label{threestring}

\begin{figure}[ht]
\centerline{\psfig{figure=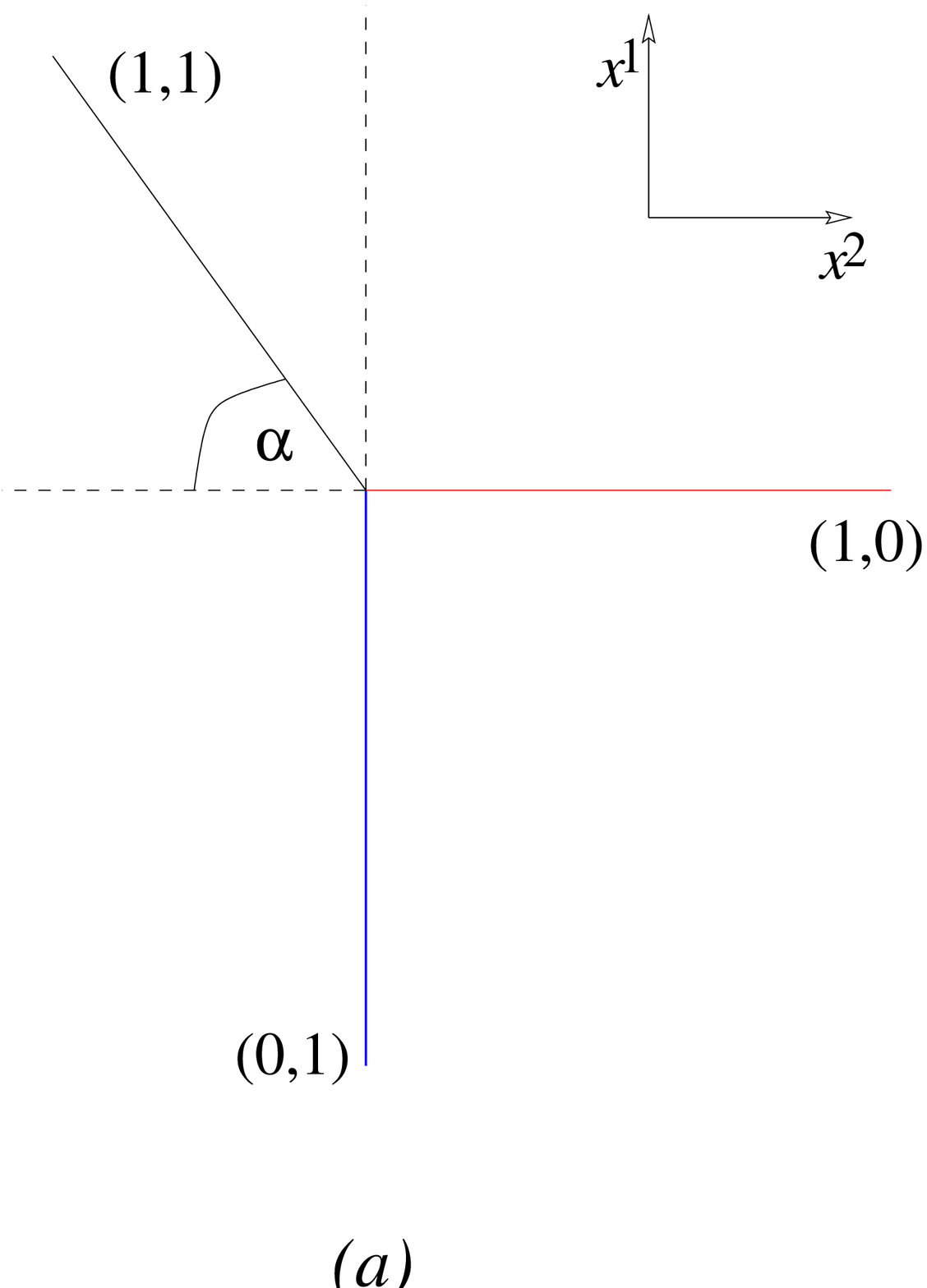,height=2.2in}\hskip1cm
\psfig{figure=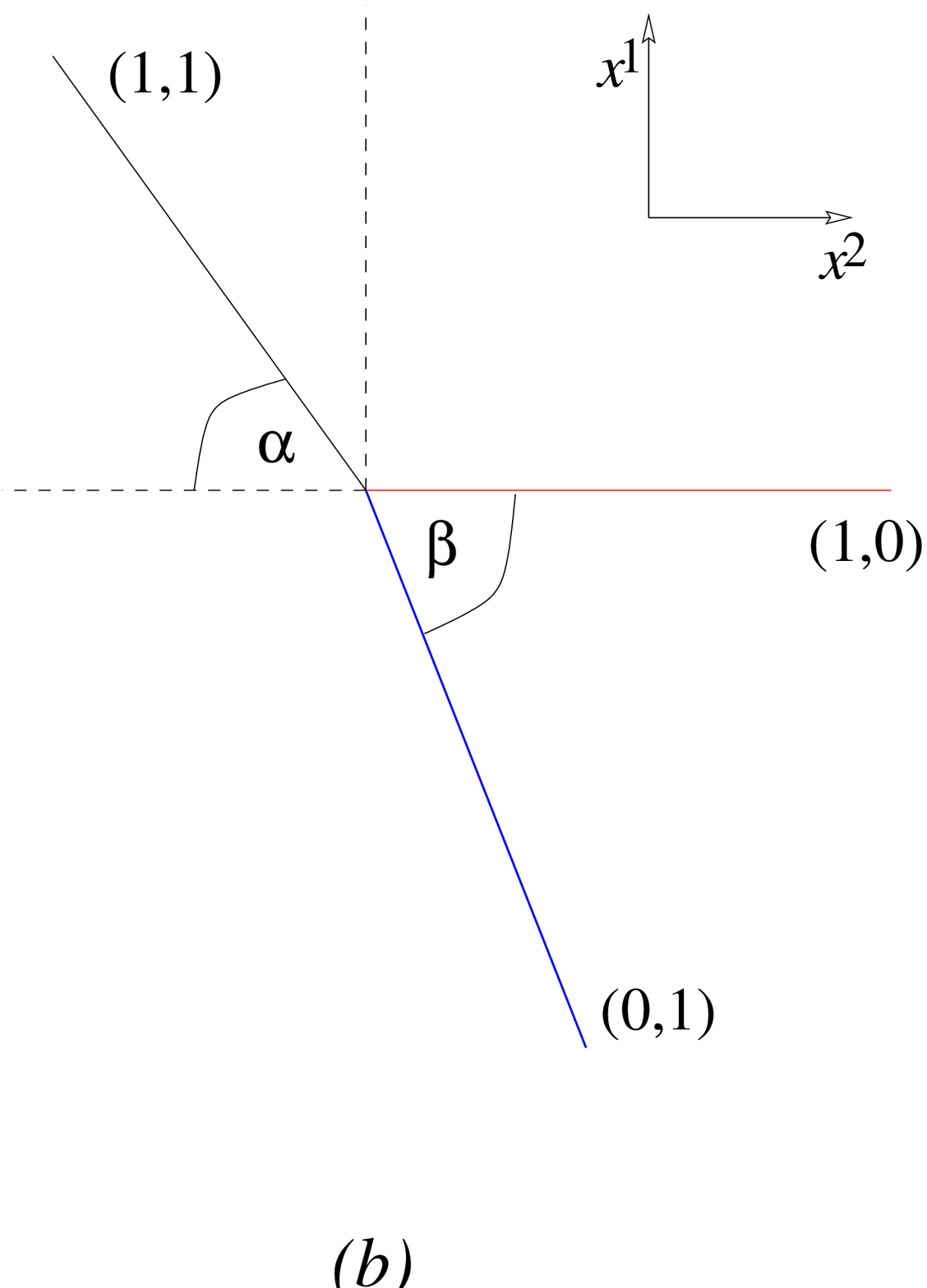,height=2.2in}}
\caption{ {\it (a)} When an F--string ends on a D--string it causes it
  to  bend at an angle set by the string coupling. On the other side
  of the junction is a (1,1) string. This is in fact a BPS state.
  {\it (b)} Switching on some amount of the R--R scalar can vary the
  other angle, as shown.}
\label{balancefig}
\end{figure}

Let us consider further the BPS saturated formula derived and studied
in the two 
previous subsections, and write it as follows:
\begin{equation}
\tau_{p,q}=\sqrt{(p\tau_{1,0})^2+(q\tau_{0,1})^2}\ .
\labell{writeit}
\end{equation}
An obvious solution to this is 
\begin{equation}
\tau_{p,q}\sin\alpha=q\tau_{0,1}\ ,\quad
\tau_{p,q}\cos\alpha=p\tau_{1,0}\ .
\labell{balance}
\end{equation}
with $\tan\alpha=q/(pg_s)$.  Recall that these are tensions of
strings, and therefore we can interpret the equations \reef{balance}
as balance conditions for the components of forces! In fact, it is the
required balance for three strings,~\cite{threestringstuff,schwarzlect}
and we draw the case of $p=q=1$ in figure \ref{balancefig}.
 
Is this at all consistent with what we already know? The answer is
yes. An F--string is allowed to end on a D--string by definition, and
a (1,1) string is produced, due to flux conservation, as we discussed
above.  The issue here is just how we see that there is bending. The
first thing to notice is that the angle $\alpha$ goes to $\pi/2$ in
the limit of zero string coupling, and so the D--string appears in
that case to run straight. This had better be true, since it is then
clear that we simply were allowed to ignore the bending in our
previous weakly coupled string analysis. (This study of the bending of
branes beyond zero coupling has important consequences for the study
of one--loop gauge theory data.~\cite{edbend} We shall study some of
this later on.)

Parenthetically, it is nice to see that in the limit of infinite
string coupling, $\alpha$ goes to $0$. The diagram is better
interpreted as a D--string ending on an F--string with no resulting
bending. This fits nicely with the fact that the D-- and F--strings
exchange roles under the strong/weak coupling duality (``S--duality'') of
the type IIB string theory.

When we wrote the linearized BIon equations in section
\ref{BIonsection}, we ignored the 1+1 dimensional case. Let us now
include that part of the story here as a 1+1 dimensional gauge theory
discussion. There is a flux $F_{01}$ on the world--volume, and the end
of the F--string is an electric source. Given that there is only one
spatial dimension, the F--string creates a discontinuity on the flux,
such that {\it e.g}:~\cite{dasguptamukhi,gauntlett}
\begin{equation}
F_{01}=\left\{\matrix{g_s\ ,&x_1>0\cr 0\ ,& x_1<0} \right.\ ,
\labell{discont}
\end{equation}
so we can choose a gauge such that
\begin{equation}
A_0=\left\{\matrix{g_sx^1\ ,&x_1>0\cr 0\ ,& x_1<0} \right.\ .
\labell{discontin}
\end{equation}
Just as in section \ref{BIonsection}, this is BPS is one of the eight
scalars $\Phi^m$ is also switched on so that
\begin{equation}
\Phi^2(x^1)=A_0\ .
\end{equation} 
How do we interpret this?  Since $(2\pi\alpha^\prime)\Phi^2$
represents the $x^2$ position of the D--string, we see that for
$x^1<0$ the D--string is lying along the $x^1$ axis, while for
$x^1>0$, it lies on a line forming an angle $\tan^{-1}(1/g_s)$ with
the $x^1$, axis. 

Recall the T$_1$--dual picture we mentioned in the previous section,
where we saw that the flux on the D--string (making the (1,1) string)
is equivalent to a D0--brane moving with velocity
$(2\pi\alpha^\prime)F_{01}$. Now we see that the D0--brane loses its
velocity at $x^1=0$. This is fine, since the apparent impulse is
accounted for by the momentum carried by the F--string in the T--dual
picture. (One has to tilt the diagram in order to T--dualize along the
(1,1) string in order to see that there is F--string momentum.)

Since as we have seen many times that the presence of flux on the
world--volume of a D$p$--brane is equivalent to having a dissolved
D$(p-2)$--brane, {\it i.e.,} non--zero $C_{(p-1)}$ source, we can
modify the flux on the $x^1<0$ part of the string this way by turning
on the R--R scalar $C_0$. This means that $\Phi^2(x^1)$ will be linear
there too, and so the angle $\beta$ between the D-- and F--strings can
be varied too (see figure \ref{balancefig}{\it (b)}). It is
interesting to derive the balance conditions from this, and then
convert it into a modified tension formula, but we will not do that
here.~\cite{dasguptamukhi}

It is not hard to imagine that given the presence we have already
deduced of a general $(p,q)$ string in the theory that there are
three--string junctions to be made out of any three strings such that
the $(p,q)$--charges add up correctly, giving a condition on the
angles at which they can meet. This is harder to do in the full
non--Abelian gauge theories on their world--volumes, but in fact a
complete formula can be derived using the underlying $SL(2,\IZ)$
symmetry of the type~IIB string theory. We will have more to say about
this symmetry later.

General three--string junctions have been shown to be important in a
number of applications, and there is a large literature on the subject
which we are unfortunately not able to review here.

\subsection{$0$--$p$ Bound States}
\label{boundstates}

Bound states of $p$--branes and $p'$--branes have many applications.
Some of them will appear in our later lectures, and so it is worth
listing some of the results here.  Here we focus on $p' = 0$, since we
can always reach it from a general $(p,p^\prime)$ using T--duality.

\subby{0--0 bound states:}

The BPS bound for the quantum numbers of $n$ 0--branes is $n \tau_0$,
so any bound state will be at the edge of the continuum.  What we
would like to know is if there is actually a true bound state wave
function, {\it i.e.,} a wavefunction which is normalisable. To make
the bound state counting well defined, compactify one direction and
give the system momentum $m/R$ with $m$ and $n$ relatively
prime.~\cite{senbound} The bound state now lies discretely below the
continuum, because the momentum cannot be shared evenly among unbound
subsystems.

This bound state problem is T--dual to the one just considered.
Taking the T--dual, the $n$ D0--branes become D1--branes, while the
momentum becomes winding number, corresponding to $m$ F-strings.
There is therefore one ultrashort multiplet of supersymmetric states
when $m$ and $n$ are relatively prime.~\cite{senbound} This bound
state should still be present back in infinite volume, since one can
take $R$ to be large compared to the size of the bound state. There is
a danger that the size of the wavefunction we have just implicitly
found might simply grow with $R$ such that as $R\to \infty$ it becomes
non--normalisable again. More careful analysis is needed to show this.
It is sufficient to say here that the bound states for arbitrary
numbers of D0--branes are needed for the consistency of string
duality, so this is an important problem. Some strong arguments have
been presented in the literature, ($n=2$ is proven) but the general
case is not yet proven. We give an embarrasingly incomplete list of
papers in this topic in references.~\cite{bounding}

\subby{0--2 bound states:}

Now the BPS bound~(\ref{deep}) puts any bound state discretely below
the continuum.  One can see a hint of a bound state forming by
noticing that for a coincident D0--brane and D2--brane the NS 0--2
string has a negative zero--point energy~(\ref{nszpe}) and so a
tachyon (which survives the GSO projection), indicating instability
towards something.  In fact the bound state (one short representation)
is easily described: the D0--brane dissolves in the D2--brane, leaving
flux, as we have seen numerous times.  The brane R--R
action~(\ref{csact}) contains the coupling $C_{(1)} F$, so with the
flux the D2--brane also carries the D0--brane charge.~\cite{towndf} There
is also one short multiplet for $n$ D0--branes.  This same bound state
is always present when $\nu = 2$.

\subby{ 0--4 bound states:} 

The BPS bound~(\ref{marg}) makes any bound state marginally stable, so
the problem is made well--defined as in the 0--0 case by compactifying
and adding momentum.~\cite{senbound2} The interactions in the
action~(\ref{xyact}) are relevant in the infrared so this is again a
hard problem, but as before it can be deformed into a solvable
supersymmetric system.  Again there is one multiplet of bound
states.~\cite{senbound2} Now, though, the bound state is invariant only
under $\frac{1}{4}$ of the original supersymmetry, the intersection of
the supersymmetries of the D0--brane and of the D4--brane.  The bound
states then lie in a short (but not ultrashort) multiplet of $2^{12}$
states.

For $2$ D0--branes and one D4--brane, one gets the correct count as
follows.~\cite{vafa1} Think of the case that the volume of the
D4--brane is large.  The $16$ supersymmetries broken by the D4--brane
generate $256$ states that are delocalized on the D4--brane.  The 8
supersymmetries unbroken by the D4--brane and broken by the D0--brane
generate $16$ states (half bosonic and half fermionic), localized on
the D0--brane. The total number is the product $2^{12}$.  Now count
the number of ways two D0--branes can be put into their $16$ states on
the D4--brane: there are $8$ states with both D0--branes in the same
(bosonic) state and $\ha 16\cdot 15$ states with the D--branes in
different states, for a total of $8\cdot 16$ states.  But in addition,
the two--branes can bind, and there are again $16$ states where the
bound state binds to the D4--brane.  The total, tensoring again with
the D4--brane ground states, is $9 \cdot 16 \cdot 256$.

For $n$ D0--branes and one D4--brane, the degeneracy $D_n$ is given by the
generating functional\,~\cite{vafa1}
\begin{equation}
\sum_{n=0}^\infty q^n D_n\ =\ 256\ \prod_{k=1}^\infty \left( \frac{1 + q^k}{1
- q^k}\right)^8\ ,  \label{degen}
\end{equation}
where the term $k$ in the product comes from bound states of $k$
D0--branes then bound to the D4--brane. A recent paper discussing the
D0--D4 bound state, with more references, can be found in the
references.~\cite{morebound}

\subby{ 0--6 bound states:} 

The relevant bound is~(\ref{deep}) and again any bound state would be
below the continuum.  The NS zero-point energy for 0--6 strings is
positive, so there is no sign of decay.  One can give D0--brane charge
to the D6--brane by turning on flux, but there is no way to do this and
saturate the BPS bound.  So it appears that there are {\it no}
supersymmetric bound states.  Incidentally, and unlike the 0--2 case,
the 0--6 interaction is repulsive, both at short distance
and at long.~\cite{joebook}

\subby{ 0--8 bound states:} 

The case of the D8--brane is special, since it is rather big. It is a
domain wall, since there is only one spatial dimension transverse to
it. In fact, the D8--brane on its own is not really a consistent
object. Trying to put it into type~IIA runs into trouble, since the
string coupling blows up a finite distance from it on either side
because of the nature of its coupling to the dilaton.  To stop this
happening, one has to introduce a pair of O8--planes, one on each
side, since they (for SO groups) have negative charge ($-8$ times that
of the D8--brane) and can soak up the dilaton. We therefore should
have 16 D8--branes for consistency, and so we end up in the
type~I$\prime$ theory, the T--dual of Type~I. The bound state problem
is now quite different, and certain details of it pertain to the
strong coupling limit of certain string theories, and their
``matrix''~\cite{matrixT} formulation.~\cite{eightplane,matrixhet} We shall
revisit this in section \ref{heterstrong}.

\section{D--Branes, Strong Coupling, and String Duality}

One of the most striking results of the middle '90's was the
realization that all of the string theories are in fact dual to one
another at strong coupling.~\cite{hullt,town,wit}~\footnote{There are
  excellent reviews in the literature, some of which are listed in the
  bibliography.~\cite{duality,schwarzlect,townreview}} This also
brought eleven dimensional supergravity in the picture and started the
search for M--theory, the dynamical theory within which all of those
theories would fit as various effective descriptions of perturbative
limits.

All of this is referred to as the ``Second Superstring Revolution''.
Every revolution is supposed to have a hero or heroes.  We shall
consider branes to be cast in that particular role, since they (and
D--branes especially) supplied the truly damning evidence of the
strong coupling fate of the various strign theories.

We shall discuss aspects of this in the present section. We simply
study the properties of various D--branes in the various string
theories, and then trust to that fact that as they are BPS states,
many of these properties will survive at strong coupling.

\subsection{D1--Brane Collective Dynamics}
Let us first study the D1--brane. This will be appropriate to the
study of type~IIB and the type~I string by $\Omega$--projection.  Its
collective dynamics as a BPS soliton moving in flat ten dimensions is
captured by the 1+1 dimensional world--volume theory, with 16 or 8
supercharges, depending upon the theory we are in. (See figure
\ref{dstrings}{\it (a)}.)

It is worth first setting up a notation and examining the global
symmetries. Let us put the D1--brane to lie along the $x^1$ direction,
as we will do many times in what is to come. 
This arrangement of branes breaks the Lorentz group up as follows:

\begin{equation} 
 SO(1,9) \supset SO(1,1)_{01}\times SO(8)_{2-9}\ ,
\labell{lorentzi}
\end{equation}
Accordingly, the supercharges decompose under \reef{lorentzi}  as
\begin{equation}
{\bf 16}={\bf 8}_++{\bf 8}_-
\labell{susy}
\end{equation}
where $\pm$ subscripts
denote a chirality with respect to $SO(1,1)$.

For the 1--1 strings, there are 8 Dirichlet--Dirichlet (DD)
directions, the Neveu--Schwarz (NS) sector has zero point energy
$-1/2$. The massless excitations form vectors and scalars in the 1+1
dimensional model. For the vectors, the Neumann--Neumann (NN)
directions give a gauge field $A^{\mu}$.
Now, the gauge field has no local
dynamics, so the only contentful bosonic excitations are the transverse
fluctuations. These  come from the 8 Dirichlet--Dirichlet
(DD) directions $x^m$, $m=2,\cdots,9$, and are
\begin{equation}
\phi^m(x^0,x^1): \quad\lambda_\phi\psi^{m}_{-{1\over2}} |0>\ .
\end{equation} 

The fermionic states $\xi$ from the Ramond (R) sector
(with zero point energy~0, as always) are built on the vacua
formed by the zero modes $\psi_0^i, i{=}0,\ldots,9$. This gives
the initial ${\bf 16}$. The
GSO projection acts on the vacuum in this sector as:
\begin{equation}
(-1)^F=e^{i\pi(S_0+S_1+S_2+S_3+S_4)}\ .
%\Gamma^0\Gamma^1\ldots\Gamma^9\ .
\labell{gso}   
\end{equation}
A left or right--moving state obeys
$\Gamma^0\Gamma^1\xi_{\pm}=\pm\xi_\pm$, and so the projection onto
$(-1)^F\xi{=}\xi$ says that left and right moving states are odd and
(respectively) even under $\Gamma^2\ldots\Gamma^9$, which is to say
that they are either in the ${\bf 8_s}$ or the ${\bf 8_c}$.  So we see
that the GSO projection simply correlates world sheet chirality with
spacetime chirality: $\xi_-$ is in the ${\bf 8}_c$ of $SO(8)$
and~$\xi_+$ is in the~${\bf 8}_s$.

\subsection{Type IIB/Type IIB Duality}
So we have seen that for a D1--brane in type~IIB string theory, the
right--moving spinors are in the ${\bf 8_s}$ of $SO(8)$, and the
left-moving spinors in the ${\bf 8_c}$.  These are the same as the
fluctuations of a fundamental IIB string, in static
gauge.~\cite{edbound} There, the supersymmetries $Q_\alpha$ and
$\tilde Q_\alpha$ have the same chirality.  Half of each spinor
annihilates the F-string and the other half generates fluctuations.
Since the supersymmetries have the same $SO(9,1)$ chirality, the
$SO(8)$ chirality is correlated with the direction of motion.

So far we have been using the string metric. We can switch to the
Einstein metric, $g_{\mu\nu}^{\rm (E)} = e^{-\Phi/2} g_{\mu\nu}^{\rm
  (S)}$, since in this case gravitational action has no dependence on
the dilaton, and so it is an invariant under duality.  The tensions in
this frame are:
\begin{eqnarray}
\mbox{F--string:} && g^{1/2}_s / 2\pi{\alpha^\prime} \nonumber\\
\mbox{D--string:} && g^{-1/2}_s / 2\pi{\alpha^\prime}\ .
\end{eqnarray}
Since these are BPS states, we are able to trust these formulae at
arbitrary values of $g_s$.

Let us see what interpretation we can make of these formulae: At weak
coupling the D--string is heavy and the F--string tension is the
lightest scale in the theory.  At strong coupling, however, the
D--string is the lightest object in the theory, (A dimensional
argument shows that the lowest--dimensional branes have the lowest
scale.~\cite{hullscale}) and it is natural to believe that the theory
can be reinterpreted as a theory of weakly coupled D--strings, with $g_s'
= g_s^{-1}$.  One cannot prove this without a non--perturbative definition
of the theory, but quantising the light D--string implies a large
number of the states that would be found in the dual theory, and
 self--duality of the IIB theory seems by far the
simplest interpretation---given that physics below the Planck energy
is described by some specific string theory, it seems likely that
there is a unique extension to higher energies.  This agrees with the
duality deduced from the low energy action and other
considerations.~\cite{hullt,wit,fstring} In particular, the
NS--NS and R--R two-form potentials, to which the D-- and F--strings
respectively couple, are interchanged by this duality. 

This duality also explains our remark about the strong and weak
coupling limits of the three string junction depicted in figure
\ref{balancefig}.  The roles of the D-- and F--strings are swapped in
the $g_s\to0,\infty$ limits, which fits with the two limiting values
$\alpha\to\pi/2,0$.

The full duality group of the $D=10$ Type IIB theory is expected to be
$SL(2,\IZ)$.~\cite{hullt,wit} This relates the
fundamental string not only to the R--R string but to a whole set of
strings with the quantum numbers of $p$ F--strings and $q$ D--strings
for $p$ and $q$ relatively prime.~\cite{schwarz} The bound states found
in section~\ref{bounds} are just what is required for $SL(2,\IZ)$
duality.~\cite{edbound} As the coupling and the R--R scalar are varied,
each of these strings becomes light at the appropriate point in moduli
space.

\subsection{Type I/Heterotic}
\label{hetdual}
Let us now consider the D1--brane in the Type I theory. We must modify
our previous analysis in two ways. First, we must project onto
$\Omega$--even states.

As in section 2.6, the $U(1)$ gauge field $A$ is in fact projected
out, since $\partial_t$ is odd under $\Omega$. The normal derivative
$\partial_n$, is even under $\Omega$, and hence the $\Phi^m$
survive. Turning to the fermions, we see that $\Omega$ acts as
$e^{i\pi(S_1+S_2+S_3+S_4)}$
%$\Gamma^2\ldots\Gamma^9$, 
and so the left--moving ${\bf 8_c}$ is projected out and the
right-moving ${\bf 8_s}$ survives.

\begin{figure}[ht]
\centerline{\psfig{figure=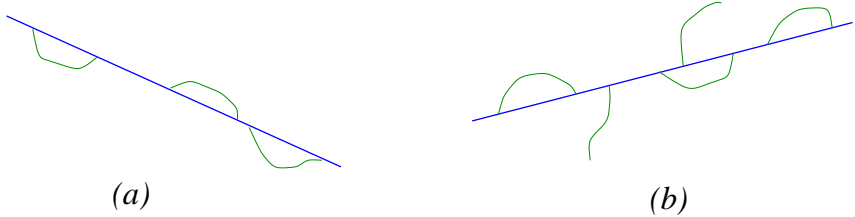,height=1.1in}}
\caption{D1--branes {\it (a)} 
  in Type IIB theory its fluctuations are described by 1--1 strings.
  {\it (b)} in Type~I string theory, there are additional
  contributions from 1--9 strings.}
\label{dstrings}
\end{figure}

Recall that D9--branes must be introduced after doing the $\Omega$
projection of the type~IIB string theory. These are the $SO(32)$
Chan--Paton factors.  This means that we must also include the massless
fluctuations due to strings with one end on the D1-brane and the other
on a D9--brane. (See figure \ref{dstrings}{\it (b)}) The zero point
energy in the NS sector for these states is $1/2$, and so there is way
to make a massless state. The R sector has zero point energy zero, as
usual, and the ground states come from excitations in the $x^0,x^1$
direction, since it is in the NN sector that the modes are integer.
The GSO projection $(-)^F=\Gamma^0\Gamma^1$ will project out one of
these, $\lambda_-$, while the right moving one will remain.  The
$\Omega$ projection simply relates 1--9 strings to 9--1 strings, and
so places no constraint on them. Finally, we should note that the 1--9
strings, as they have one end on a D9--brane, transform as vectors of
$SO(32)$.

Now, by the argument that we saw in the case of the type~IIB string,
we should deduce that this string becomes a light fundamental string
in some dual string theory at strong coupling. In these notes we have
not seen such a string before. It has $(0,8$ world sheet
supersymmetry, and a left moving family of 32 fermions transforming as
the {\bf 32} of $SO(32)$.

Happily, there is precisely one such string theory in ten dimensions
with this property. It is a closed string theory called the $SO(32)$
``heterotic'' string.~\cite{heterotic} There is in fact another ten
dimensional heterotic string, with gauge group $E_8\times E_8$. It has
a storng coupling limit we will examine shortly. Upon compactifying on
a circle, the two heterotic string theories are perturbatively related
by T--duality.~\cite{narain,ginsp}

We have obtained the $SO(32)$ string  here with the
spacetime supersymmetry realized in Green--Schwarz form and with a
left--moving ``current algebra'' in fermionic form~\cite{polwit}, which
realises a spacetime $SO(32)$ gauge symmetry. In fact, recall that we
had already deduced that such a string theory might exist, by looking
at the supergravity sector in section~\ref{supergravity}. This is just
how type~I/heterotic duality was deduced
first~\cite{wit}$^{\!,\,}$~\cite{fstring} and then D--brane
constructions were used to test it more sharply~\cite{polwit}.

Actually, heterotic string experts know that the fermionic $SO(32)$
current algebra requires a GSO projection.  By considering a closed
D1--brane we see that the $\Omega$ projection removes the $U(1)$ gauge
field, but in fact allows a discrete gauge symmetry: a holonomy $\pm
1$ around the D1--brane.  This discrete gauge symmetry is the GSO
projection, and we should sum over all consistent possibilities.  The
heterotic strings have spinor representations of $SO(32)$, and we need
to be able to make them in the Type~I theory, in order for duality to
be correct.  In the {R} sector of the discrete D1--brane gauge theory,
the 1--9 strings are periodic.  The zero modes of the fields $\Psi^i$,
representing the massless 1--9 strings, satisfy the Clifford algebra
\begin{equation}
\{ \Psi^i_0 , \Psi^j_0 \} = \delta^{ij}, \qquad i,j= 1,
\cdots , 32 .
\end{equation}
The quantisation now proceeds just as for the fundamental heterotic
string, giving spinors ${\bf 2^{31}} + \overline{\bf 2^{31}}$, one of
which is removed by the discrete gauge symmetry.

\insertion{11}{Dual Branes from 10D String--String
  Duality\label{insert11}}{There is an instructive way to see how the
  D--string tension turns into that of an F--string. In terms of
  supergravity fields, part of the duality transformation involves
$$
G_{\mu\nu}\to e^{-{\tilde\Phi}}{\tilde G}_{\mu\nu} \ ,\quad
  \Phi\to-{\tilde\Phi}\ ,
$$
  where the quantities on the right, with
  tildes, are in the dual theory. This means that in addition to
  $g_s={\tilde g}_s^{-1}$, for the relation of the string coupling to
  the dual string coupling, there is also a redefinition of the string
  length, {\it via}
$$
\alpha^\prime={\tilde g}_s{\tilde\alpha}^\prime \ ,
$$
which is the same as 
$$
\alpha^\prime g_s={\tilde\alpha}^\prime \ .
$$
Starting with the D--string tension, these relations give:
$$
\tau_1={1\over 2\pi\alpha^\prime g_s}\to {1\over
  2\pi{\tilde\alpha}^\prime}=
\tau^{\rm F}_1\ ,
$$
precisely the tension of the fundamental string in the dual string
theory, measured in the
correct units of length.\\

One might understandably ask the question about the fate of other
D--branes. For the type~IIB's D3--brane:
$$
\tau_3={1\over (2\pi)^3\alpha^{\prime2}g_s}\to{1\over (2\pi)^3
{\tilde\alpha}^{\prime2}{\tilde g}_s}=\tau_3\ ,
$$
showing that the dual object is again a D3--brane. For the D5--brane,
in either type~IIB or type~I theory:
$$
\tau_5={1\over (2\pi)^5\alpha^{\prime3}g_s}\to{1\over (2\pi)^5
{\tilde\alpha}^{\prime3}{\tilde g}^2_s}=\tau^{\rm F}_5\ ,
$$
This is the tension of a fivebrane which is {\it not} a D5--brane.
This is intersting, since for both dualities, the R--R 2--form
$C^{(2)}$ is exchanged for the NS--NS 2--form $B^{(2)}$, and so this
fivebrane is magnetically charged under the latter. It is in fact that
magnetic dual of the fundamental string. Its $g_s^{-2}$ behaviour
identifies it as a soliton of the NS--NS fields $(G,B,\Phi)$.
{\it  Continued...}}
\insertion{11}{{\it Continued...}}{So we
conclude that there exists in both the type~IIB and $SO(32)$ heterotic
theories such a brane, and in fact such a brane can be constrcuted
directly as a soliton solution. They should be called ``F5--branes'',
but this name never stuck. They go by various names like
``NS5--brane'' or ``solitonic fivebrane'', and so on.  As they are
constructed completely out of closed string fields, T--duality along a
direction parallel to the brane does not change its dimensionality, as
would happen for a D--brane. We conclude therefore that they also exist in the
T--dual Type~IIA and $E_8\times E_8$ string theories.  For the
heterotic cases, the soliton solution also involves a background gauge
field, which is in fact an instanton. We shall deduce this from
duality later also, when we uncover more properties of branes within
branes.\\

One last feature worth mentioning is the worldvolume theory describing
the low energy collective motions of the branes. This can be worked
out directly, and string duality is consistent with the answers: From
the duality, we can immediately deduce that the type~IIB's NS5--brane
must have a vector multiplet, just like the D5--brane. There is
non--chiral (1,1) six dimensional supersymmetry on the worldvolume.
Just like with D5--branes, there is enhanced gauge symmetry when many
coincide.~\cite{edcomm} For the type~IIA NS5--brane, things are
different. A duality argument can be used to show that the brane
actually carries a two--form potential, and so there is a six
dimensional tensor multiplet on the brane. There is a chiral (0,2)
supersymmetry on the brane. The gauge symmetry associated to this
multiplet is also enhanced when many branes coincide.

That there is either a (1,1) vector multiplet or a (0,2) tensor
multiplet was first uncovered by direct analysis of the collective
dynamics of the NS5--branes as solitons in the full type~II theories.
~\cite{callan} }

\subsection{Type IIA/M--Theory}

In the IIA theory, the D0--brane has a mass $\tau_0 = \ap^{-1/2}g_s$
in the string metric.  As $g_s \to \infty$, this mass is the lightest
scale in the theory.  In addition, we have seen in
section~\ref{boundstates} that $n$ D0--branes have a single
supersymmetric bound state with mass $n\tau_0$.  This evenly spaced
tower of states is characteristic of the appearance of an additional
dimension, where the momentum (Kaluza--Klein) states have masses $n/R$
and form a continuum is $R \to \infty$. Here,
$R=\alpha^{\prime1/2}g_s$, so weak coupling is small $R$ and the
theory is effectively ten dimensional, while strong coupling is large
$R$, and the theory is eleven dimensional. We saw such Kaluza--Klein
behaviour in section~\ref{tdualityclosed}. The charge of the $n$th
Kaluza--klein particle corresponds to $n$ units of momentum $1/R$ in
the hidden dimension. In this case, this $U(1)$ is the R--R one form
of type~IIA, and so we interpret D0--brane charge as eleven
dimensional momentum.  In this way, we are led to consider eleven
dimensional supergravity as the strong coupling limit of the type~IIA
string. This is only for {\it low energy}, of course, and the issue of
the complete description of the short distance physics at strong
coupling to complete the ``M--theory'', is yet to be settled. It
cannot be simply eleven dimensional supergravity, since that theory
(like all purely field theories of gravity) is ill--defined at short
distances.  The most widely examined proposal for the structure of the
short distance physics is ``Matrix Theory''~~\cite{matrixT}, although
there are other interesting proposals.~\cite{otherM}

It is worth noting that the existence of the bound states and of the
eleventh dimension was inferred even before the significance of
D--branes was understood, because they are required by
lower--dimensional
``U--dualities''.~\cite{town,wit}

To relate the coupling to the size of the eleventh dimension we need
to compare the respective Einstein--Hilbert actions,~\cite{wit}
\begin{equation}
\frac{1}{2\kappa_0^2 g_s^2} \int d^{10} x\, \sqrt{-G_{\rm s}} R_{\rm s}
= \frac{2\pi R}{2\kappa_{11}^2} \int d^{10} x\, \sqrt{-G_{11}} R_{11}\
.
\end{equation}
The string and M theory metrics are equal up to a rescaling,
\begin{equation}
G_{{\rm s}\mu\nu} = \zeta^2 G_{{\rm 11}\mu\nu}  \label{mmet}
\end{equation}
and so $\zeta^8 = 2\pi R \kappa_0^2 g_s^2 / \kappa_{11}^2$.  The
respective masses are related $nR^{-1} = m_{11} = \zeta m_{\rm s} = n
\zeta \tau_0$ or $R = \ap^{1/2} g_s/\zeta$.  Combining these with the
result~(\ref{nice}) for $\kappa_0$, we obtain
\begin{equation}
\zeta = g_s^{1/3} \left[2^{7/9} \pi^{8/9} \ap \kappa_{11}^{-2/9} \right]
\end{equation}
and 
\begin{equation}
R = g_s^{2/3} \left[2^{-7/9} \pi^{-8/9} \kappa_{11}^{2/9} \right]\ .
\label{mrad}
\end{equation}
In order to emphasise the basic structure we hide in braces numerical
factors and factors of $\kappa_{11}$ and $\ap$.  The latter factors
are determined by dimensional analysis, with $\kappa_{11}$ having
units of (M theory length$^{9/2}$) and $\ap$ (string theory
length$^2$).  We are free to set $\zeta = 1$, using the same metric
and units in M--theory as in string theory.  In this case
\begin{equation}
\kappa_{11}^2 = g_s^3 \left[2^7 \pi^8 \ap^{9/2} \right]\ .
\end{equation}
The reason for not always doing so is that when we have a series of
dualities, as below, there will be different string metrics.

For completeness, let us note that if we define Newton's constant as
before {\it via} $2\kappa_{11}^2=16\pi G_N^{11}$, then we have:
\begin{eqnarray}
\kappa_{11}^2=2^7\pi^8\ell_p^9\ ;\quad
\ell_p=g_s^{1/3}\sqrt{\alpha^\prime}\ ;\quad G_N^{11}=16\pi^7\ell_p^9
\ .\labell{elevenrelation}
\end{eqnarray}

It is interesting to track the eleven--dimensional origin of the
various branes of the IIA theory.~\cite{towndf,schwarzlect,duality} The
D0--branes are, as we have just seen, Kaluza-Klein states. The
F1--branes, the IIA strings themselves, are wrapped membranes
(``M2--branes'') of M--theory.~\cite{mtwo} The D2--branes are
membranes transverse to the eleventh dimension $X^{10}$.  The
D4--branes are M--theory fivebranes (``M5--branes'')~\cite{mfive}
wrapped on $X^{10}$, while the NS (symmetric) 5--branes are M5--branes
transverse to $X^{10}$.  The D6--branes, being the magnetic duals of
the D0--branes, are Kaluza--Klein monopoles~\cite{kkmono} (we shall see this
directly later in section \ref{d2probemagic}).  As mentioned before
the D8--branes have a more complicated fate.  To recap, the point is
that the D8--branes cause the dilaton to diverge within a finite
distance,~\cite{polwit} and must therefore be a finite distance from
an orientifold plane, which is essentially a boundary of spacetime as
we saw in section \ref{orientplanes}.  As the coupling grows, the
distance to the divergence and the boundary necessarily shrinks, so
that they disappear into it in the strong coupling limit: they become
part of the gauge dynamics of the nine--dimensional boundary of
M--theory,~\cite{horwit} used to make the $E_8\times E_8$ heterotic
string, to be discussed in more detail below. This raises the issue of
the strong coupling limit of orientifolds in general. There are
various results in the literature, but since the issue are
complicated, and because the techniques used are largely strongly
coupled field theory deductions, which take us well beyond the scope
of these lectures, we will have to refer the reader to the
literature.~\cite{seibergprobe,senhance,SWtwo,strongorient}

One can see further indication of the eleventh dimension in the
dynamics of the D2--brane.  In $2+1$ dimensions, the vector field on
the brane is dual to a scalar, through Hodge duality of the field
strength, $*F_2 = d\phi$.  This scalar is the eleventh embedding
dimension.~\cite{duffeleven,schmidhuber,towndf} Carrying out the
duality in detail, the D2--brane action is found to have a hidden
eleven--dimensional Lorentz invariance. We shall see this feature in
certain probe computations later on in section \ref{d2probemagic}.

\subsection{$E_8 \times E_8$ Heterotic String/M--Theory on {\cal I}}
\label{heterstrong}

We have deduced the duals of four of the five ten dimensional string
theories.  Let us study the final one, the $E_8 \times E_8$ heterotic
string, which is T--dual to the $SO(32)$ string.~\cite{narain,ginsp}
Compactify on a large radius $R$ and turn on a Wilson line which
breaks $E_8\times E_8$ to $SO(16) \times SO(16)$.  This is T--dual to
the $SO(32)$ heterotic string, again with a Wilson line breaking the
group to $SO(16) \times SO(16)$.  The couplings and radii are related
\begin{eqnarray}
R' &=& R^{-1} \left[ {\alpha^\prime} \right] , \nonumber\\
g_s' &=& g_s R^{-1} \left[ {\ap}^{1/2} \right] \ .
\end{eqnarray}
Now use Type~I - heterotic duality to write this as a Type~I theory
with\,~\cite{wit}
\begin{eqnarray}
R_{\rm I} &=& g_s'^{-1/2} R'\ =\ g_s^{-1/2} R^{-1/2} \left[
{\alpha^\prime}^{3/4} \right], \nonumber\\ g_{s,{\rm I}} &=& g_s'^{-1}\ =\
g_s^{-1} R \left[ {\ap}^{-1/2} \right]\ .
\end{eqnarray}
The radius is very small, so it is useful to make another T--duality,
to the `Type I$'$' theory.  The compact dimension is then a segment of
length $\pi R_{\rm I'}$ with eight D 8--branes at each end, and
\begin{eqnarray}
R_{\rm I'} &=& R_{\rm I}^{-1} \left[ {\alpha^\prime} \right] \ =\
g_s^{1/2} R^{1/2} \left[ {\alpha^\prime}^{1/4} \right], \nonumber\\
g_{s,{\rm I'}} &=& g_{s,{\rm I}} R_{\rm I}^{-1} \left[ 2^{-1/2} {\ap}^{1/2}
\right]\ =\ g_s^{-1/2} R^{3/2} \left[ 2^{-1/2} {\alpha^\prime}^{-3/4}
\right]\ .
\end{eqnarray}
Now take $R \to \infty$ to recover the original ten-dimensional theory
(in particular the Wilson line is irrelevant and the original
$E_8\times E_8$ restored).  Both the radius and the coupling of the
Type I$'$ theory become large.  The physics between the ends of the
segment is given locally by the IIA string, and so the strongly
coupled limit is eleven dimensional.  Taking into account the
transformations~(\ref{mmet}), (\ref{mrad}), the radii of the two
compact dimensions in M--theory units are
\begin{eqnarray}
R_9 &=& \zeta_{\rm I'}^{-1} R_{\rm I'}\ =\ g_s^{2/3} \left[2^{-11/18}
\pi^{-8/9} \kappa_{11}^{2/9} \right]\label{r910} \\ R_{10} &=& g_{s,{\rm
I'}}^{2/3} \left[2^{-7/9} \pi^{-8/9} \kappa_{11}^{2/9} \right] \ =\
g_s^{-1/3} R \left[2^{-10/9} \pi^{-8/9} \ap^{-1/2} \kappa_{11}^{2/9}
\right]\ .  \nonumber
\end{eqnarray}
As $R\to \infty$, $R_{10} \to\infty$ also, while $R_9$ remains fixed
and (for $g$ large) large compared to the Planck scale.  Thus, in the
strongly coupled limit of the ten-dimensional $E_8\times E_8$
heterotic string an eleventh dimension again appears, a segment of
length $R_9$, with one $E_8$ factor on each endpoint.~\cite{horwit}

\subsection{U--Duality}

An interesting feature of string duality is the enlargement of the
duality group under further toroidal compactification.  There is a lot
to cover, and it is somewhat orthogonal to most of what we want to do
for the rest of the notes, so we will err on the side of brevity (for
a change). The example of the Type~II string on a five--torus $T^5$ is
useful, since it is the setting for the simplest black hole state
counting, which Amanda Peet will cover in her lectures in this
school.~\cite{peettasi}

Let us first count the gauge fields. This can be worked out simply by
counting the number of ways of wrapping the metric and the various
p--form potentials in the theory on the five circles of the $T^5$ to
give a one--form in the remaining five non--compact directions. From
the NS--NS sector there are 5 Kaluza--Klein gauge bosons and 5 gauge
bosons from the antisymmetric tensor.  There are 16 gauge bosons from
the dimensional reduction of the various R--R forms: The breakdown is
10+5+1 from the forms $C^{(3)}$, $C^{(5)}$ and $C^{(1)}$,
respectively.  Finally, in five dimensions, one can form a two form
field strength from the Hodge dual $*H$ of the 3--form field strength
of the NS--NS $B_{\mu\nu}$, thus defining another gauge field.

Let us see how T--duality acts on these.  The T--duality is $SO(5,5;\IZ)$,
as discussed in section \ref{tdualityclosed}.  This mixes the first
10 NS--NS gauge fields among themselves, and the 16 R--R gauge fields among
themselves, and leaves the final NS--NS field invariant.
The $SO(5,5;\IZ)$ representations here correspond directly to the {\bf 10},
{\bf 16}, and {\bf 1} of $SO(10)$.

The low energy supergravity theory for this compactification has a
continuous symmetry $E_{6(6)}$ which is a non--compact version of
$E_6$.~\cite{julia}  This is one of those supergravity properties that was
ignored for some time, because there is no sign of it in (perturbative)
string theory. But now we know better:\,\cite{hullt} a discrete subgroup
$E_{6(6)}(\IZ)$ is supposed to be a good symmetry of the full theory.

The gauge bosons are in the {\bf 27} of $E_{6(6)}(\IZ)$, which is the
same as the {\bf 27} of $E_{6(6)}$.  The decomposition under $SO(10)
\sim SO(5,5;\IZ)$ is familiar from grand unified model building,
\begin{equation}
{\bf 27}\ \to\ {\bf 10}\ +\ {\bf 16}\ +\ {\bf 1}\ .
\end{equation}
The particle excitations carrying the {\bf 10} charges are just the
Kaluza--Klein and winding strings.  The U--duality requires also
states in the {\bf 16}.  These are just the various ways of wrapping
Dp--branes to give D--partiles (10 for D2, 5 for D4 and 1 for D0).
Finally, the state carrying the {\bf 1} charge is the NS5--brane,
wrapped entirely on the $T^5$.

\subsection{U--Duality and Bound States}

It is interesting to see how some of the bound state results from the
previous lecture fit the predictions of U--duality.  We will generate
U-transformations as a combination of T$_{mn\cdots p}$, which is a
T--duality in the indicated directions, and S, the IIB weak/strong
transformation.  The former switches between N and D boundary
conditions and between momentum and winding number in the indicated
directions.  The latter interchanges the NS and R two-forms but leaves
the R four-form invariant, and acts correspondingly on the solitons
carrying these charges.  We denote by ${\rm D}_{mn\cdots p}$ a
D--brane extended in the indicated directions, and similarly for ${\rm
  F}_m$ a fundamental string and $p_m$ a momentum-carrying BPS state.

The first duality chain is
\begin{equation}
({\rm D}_9,{\rm F}_9)\ \stackrel{{\rm T}_{78}}{\to}\ ({\rm D}_{789},{\rm F}_9)
\ \stackrel{\rm S}{\to}\ ({\rm D}_{789},{\rm D}_9)
\ \stackrel{{\rm T}_{9}}{\to}\ ({\rm D}_{78},{\rm D}_{\emptyset})\ .
\end{equation}
(The last symbol denotes a D0--brane, which is of course not extended
anywhere.) Thus the D--string--F--string bound state is U--dual to the
0--2 bound state.

The second chain is
\begin{equation}
({\rm D}_{6789},{\rm D}_{\emptyset})\ \stackrel{{\rm T}_{6}}{\to}\ 
({\rm D}_{789},{\rm D}_6)
\ \stackrel{\rm S}{\to}\ ({\rm D}_{789},{\rm F}_6)
\ \stackrel{{\rm T}_{6789}}{\to}\ ({\rm D}_{6},p_6)
\ \stackrel{\rm S}{\to}\ ({\rm F}_{6},p_6)
\end{equation}
The bound states of $n$ D0--branes and $m$ D4--branes are thus
U--dual to fundamental string states with momentum $n$ and winding
number $m$.  The bound state degeneracy~(\ref{degen}) for $m=1$
precisely matches the fundamental string
degeneracy.~\cite{vafwit,senbound2,vafa2}
For $m>1$ the same form~(\ref{degen}) should hold but with $n \to mn$.
This is believed to be the case, but the analysis (which requires the
instanton picture described in the next section) does not seem to be
complete.~\cite{vafa2}

A related issue is the question of branes ending on other
branes~\cite{andyopen}, and we shall see more of this later. An
F--string can of course end on a D--string, so from the first duality
chain it follows that a D$p$--brane can end on a D$(p+2)$--brane.
The key issue is whether the coupling between spacetime forms and
world--brane fields allows the source to be conserved, as with the
NS--NS two--form source in figure~\ref{boundfd}.  Similar arguments can
 be applied to the extended objects in
M--theory.~\cite{andyopen,towndf}

\section{D--Branes and Geometry I}

\subsection{D--Branes as a Probe of ALE Spaces}
\label{aleprobe}
One of the beautiful results which we uncovered soon after
constructing the type~II strings was that we can ``blow up'' the 16
fixed points of the $T^4/\IZ_2$ ``orbifold compactification'' to
recover string propagation on the smooth hyperK\"ahler manifold K3.
(See section~\ref{k3orbifold}.)  Strictly speaking, we only recovered
the algebraic data of the K3 this way, and it seemed plausible that
the full metric geometry of the surface is recovered, but how can we
see this directly?

We can recover this metric data by using a brane as a short distance
``probe'' of the geometry. This is a powerful technique, which has
many useful applications as we shall see in numerous examples as we
proceed.

Let us focus on a single fixed point, and the type~IIB theory.  The
full string theory is propagating on $\IR^6 \times (\IR^4/\IZ_2)$,
which arises from imposing a symmetry under the reflection ${\bf R}:
\,\, (x^6,x^7,x^8,x^9) \to (-x^6,-x^7,-x^8,-x^9)$. Now we can place a
D1--brane (a ``D--string'') in this plane at $x^{2},\ldots, x^{9} = 0$.
Here is a table to help keep track of where everything is:

\bigskip
\begin{center}
\begin{tabular}{|c|c|c|c|c|c|c|c|c|c|c|}
\hline
&$x^0$&$x^1$&$x^2$&$x^3$&$x^4$&$x^5$&$x^6$&$x^7$&$x^8$&$x^9$\\\hline
D1&$-$&$-$&$\bullet$&$\bullet$&$\bullet$&$\bullet$&$\bullet$
&$\bullet$&$\bullet$&$\bullet$\\\hline
ALE&$-$&$-$&$-$&$-$&$-$&$-$&$\bullet$&$\bullet$&$\bullet$&$\bullet$
\\
\hline
\end{tabular}
\end{center}
\bigskip
(We have put the $\IR^4/\IZ_2$ (ALE) space in as a ten dimensional extended
object too, since it only has structure in the directions
$x^6,x^7,x^8,x^9$.)

The D1--brane can quite trivially sit at the origin and respect the
symmetry ${\bf R}$, but if it moves off the fixed point, it will break
the $\IZ_2$ symmetry.  In order for it to be able to move off the
fixed point there needs to be an image brane moving to the mirror
image point also.  We therefore need two Chan--Paton indices: one for
the D--string and the other for its $\IZ_2$ image. So (to begin with)
the gauge group carried by our D--string system living at the origin
is $U(2)$, but this will be modified by the following considerations.
Since ${\bf R}$ exchanges the D--string with its image, it can be
chosen to act on an open string state as the exchange
$\gamma=\sigma_1\equiv\pmatrix{0&1\cr1&0}$.  So we can write the
representation of the action of ${\bf R}$ as:
\begin{eqnarray}
{\bf R}|{\psi,ij}\rangle &=&\gamma^{\phantom{-1}}_{ii'}
 |{{\bf R}\psi,i'j'}\rangle
\gamma^{-1}_{j'j}\ ,\quad\mbox{that is,} \nonumber\\
{\bf R} |{\psi,ij}\rangle &=& \sigma_{ii'} |{{\bf R}\psi,i'j'}\rangle
\sigma^1_{j'j}\ .  \end{eqnarray} 

So it acts on the oscillators in the usual way but also switches the
Chan--Paton factors for the brane and its image.  The idea~\cite{GP} is
that we must choose an action of the string theory orbifold symmetry
on the Chan--Paton factors when there are branes present and make sure
that the string theory is consistent in that sector too. Note that the
action on the Chan--Paton factors is again chosen to respect the
manner in which they appear in amplitudes, just as in
section~\ref{chanpat}.

\begin{figure}[ht]
\centerline{\psfig{figure=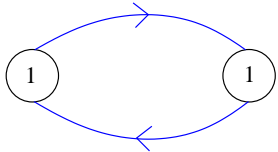,height=0.8in}}
\caption{A diagram showing the content of the probe gauge theory.  
The nodes give information about the gauge groups, while the links 
give the amount and charges of the mattter hypermultiplets.}
\label{quiver}
\end{figure}

We can therefore compute what happens: In the NS sector, the massless
$\bf R$--invariant states are, in terms of vertex operators:
\begin{eqnarray}
&&\partial_t X^\mu \sigma^{0,1}, \qquad \mu = 0,1 \nonumber\\
&&\partial_n X^i \sigma^{0,1}, \qquad i = 2,3,4,5 \nonumber\\ 
&&\partial_n X^m \sigma^{2,3}, \qquad m = 6,7,8,9.
\end{eqnarray}
The first row is the vertex operator describing a gauge field with
$U(1)\times U(1)$ as the gauge symmetry.  The next row constitutes
four scalars in the adjoint of the gauge group, parametrising the
position of the string within the six--plane $\IR^6$, and the last row
is four scalars in the ``bifundamental'' charges $(\pm1,\mp1)$ of the
gauge group the transverse position on $x^{6},x^7,x^8,x^{9}$.  Let us
denote the corresponding D--string fields $A^\mu, X^i, X^m$, all
$2\times2$ matrices. We may draw a ``quiver
diagram''~\cite{douglasmoore} displaying this gauge and matter
content. (see figure \ref{quiver}.)

Such diagrams have in general an integer $m$ inside each node,
representing a factor $U(m)$ in the gauge group. An arrowed edge of
the diagram represents hypermultiplet transforming as the fundamental
(for the sharp end) and anti--fundamental (for the blunt end) of the
two gauge groups corresponding to the connected nodes. The diagram is
simply a decorated version of the extended Dynkin diagram associated
to $A_1$. This will make even more sense shortly, since there is
geometric meaning to this. finally, note that one of the $U(1)$'s,
(the $\sigma_0$ one) is trivial: nothing transforms under it, and it
simply represents the overall centre of mass of the brane system.

Their bosonic action is the $d = 10$ $U(2)$ Yang--Mills action,
dimensionally reduced and $R$--projected (which breaks the gauge
symmetry to $U(1) \times U(1))$.  This dimensional reduction is easy
to do. There are kinetic terms:
\begin{equation}
T=-{1\over4 g_{\rm YM}^2}\left(F^{\mu\nu}F_{\mu\nu}+ \sum_i{\cal D}_\mu
 X^i {\cal D}^\mu X^i
+\sum_m{\cal D}_\mu X^m {\cal D}^\mu X^m\right)\ ,
\labell{stringkinetic}
\end{equation}
and potential terms:
\begin{equation}U = -{1\over4 g_{\rm YM}^2}\left(2\sum_{i,m} 
{\rm Tr}\,[X^i,X^m]^2 + \sum_{m,n} {\rm
Tr}\,[X^m,X^n]^2\right)\ . 
\labell{stringpotential} 
\end{equation} 
where using \reef{yangmillscoupling}, we have $g_{\rm
  YM}^{2}=(2\pi)^{-1}{\alpha^\prime}^{-1/2}g_s$. (Another potentially
non--trivial term disappears since the gauge group is Abelian.)  The
important thing to realize is that there are large {\sl families} of
vacua (here, $U=0$) of the theory. The space of such vacua is called
the ``moduli space'' of vacua, and they shall have an interesting
interpretation.  The moduli space has two branches:

On one, the ``Coulomb Branch'', $X^m = 0$ and $X^i = u^i \sigma^0 +
v^i \sigma^1$.  This corresponds to two D--strings moving
independently in the $\IR^6$, with positions $u^i \pm v^i$.  The gauge
symmetry is unbroken, giving independent $U(1)$'s on each D--string.

On the other, the ``Higgs Branch'', $X^m$ is nonzero and $X^i = u^i
\sigma^0$.  The $\sigma^1$ gauge invariance is broken and so we can
make the gauge choice $X^m = w^m \sigma^3$.  This corresponds to the
D--string moving off the fixed plane, the string and its image being
at $(u^i, \pm w^m)$. We see that this branch has the geometry of the
$\IR^6\times\IR^4/\IZ_2$ which we built in.

Now let us turn on twisted--sector fields which we uncovered in
section~\ref{k3orbifold}, where we learned that they give the blowup of the
geometry. They will appear as parameters in our D--brane gauge theory.
Define complex $q^m$ by $X^m = \sigma^3 {\rm Re}(q^m) + \sigma^2 {\rm
  Im}(q^m)$, and define two doublets,
\begin{equation}
\Phi_0 = \left( \begin{array}{c} q^6 + i q^7 \\ q^8 + i q^9
\end{array} \right), \qquad
\Phi_1 = \left( \begin{array}{c} \bar q^6 + i \bar q^7 \\ \bar q^8 + i
\bar q^9
\end{array} \right).
\end{equation}
These have charges $\pm 1$ respectively under the $\sigma^1$ $U(1)$.
The three NS--NS moduli can be written as a vector ${\bf D}$,
and the potential is proportional to
\begin{equation}
({\bf D}-\mbox{\boldmath$\mu$})^2\equiv
(\Phi_0^\dagger \mbox{\boldmath$\tau$} \Phi_0 - \Phi_1^\dagger
\mbox{\boldmath$\tau$} \Phi_1 + {\bf D})^2\ ,
\labell{moment}
\end{equation}
where the Pauli matrices are now denoted $\tau^I$ to emphasise that
they act in a different space. (The notation using vector
$\mbox{\boldmath$\mu$}$ has a significance which we shall discuss
later.) This reduces to the second term of the earlier potential
\reef{stringpotential} when ${\bf D} = 0$.  Its form is determined by
supersymmetry.

For ${\bf D} \neq 0$ the orbifold point is blown up.  The moduli space
of the gauge theory is simply the set of possible locations of the
probe {\it i.e.,} the blown up ALE space. (Note that the branch of the
moduli space with $v^i \neq 0$ is no longer present.)  

Let us count parameters and constants: The $X^m$ contain eight scalar
fields.  Three of them are removed by the ${\bf D}$--flatness condition that
the potential vanish, and a fourth is a gauge degree of freedom,
leaving the expected four moduli.  In terms of supermultiplets, the
system has the equivalent of $d=6$ $N=1$ supersymmetry.  The D-string
has two hypermultiplets and two vector multiplets, which are Higgsed
down to one hypermultiplet and one vector multiplet.  

The idea~\cite{douglasii} is that the metric on this moduli space, as
seen in the kinetic term for the D--string fields, should be the
smoothed ALE metric. Given the fact that we have eight supercharges,
it should be a hyperK\"ahler manifold,~\cite{deezee} and the ALE space
has this property. Let us explore this.~\cite{tensors} Three
coordinates on our moduli space are conveniently defined as (there are
dimensionful constants missing from this normalisation which we shall
ignore for now):
\begin{equation}
{\bf y} = \Phi_0^\dagger \mbox{\boldmath$\tau$} \Phi_0.
\label{hopf}
\end{equation}
The fourth coordinate, $z$, can be defined
\begin{equation}
z = 2\arg(\Phi_{0,1} \Phi_{1,1}).
\end{equation}
The ${\bf D}$--flatness condition
implies that
\begin{equation}
\Phi_1^\dagger \mbox{\boldmath$\tau$} \Phi_1 = {\bf y}
+ {\bf D},
\end{equation}
and $\Phi_0$ and $\Phi_1$ are determined in terms of ${\bf y}$ and
$z$, up to gauge choice.

The original metric on the space of hypermultiplet vevs is just the
flat metric $ds^2=d\Phi_0^\dagger d\Phi_0 + d\Phi_1^\dagger d\Phi_1$.
We must project this onto the space orthogonal to the $U(1)$ gauge
transformation. This is performed (for example) by coupling the
$\Phi_0, \Phi_1$ for two dimensional gauge fields according to their
charges, and integrating out the gauge field. (This whole
construction, imposing the {\bf D}--flatness conditions and making the
gauge identification, is known as the hyperK\"ahler
quotient.~\cite{hyper,kronheimer})  The result is
\begin{equation}
ds^2 = d\Phi_0^\dagger d\Phi_0 + d\Phi_1^\dagger
d\Phi_1 - \frac{(\omega_0 + \omega_1)^2}{4 (\Phi_0^\dagger \Phi_0 + 
\Phi_1^\dagger \Phi_1)}
\end{equation}
with
\begin{equation}
\omega_i = i(\Phi_i^\dagger d\Phi_i - d\Phi_i^\dagger \Phi_i)\ .
\end{equation}
It is straightforward~\cite{tensors} to express the
metric in terms of ${\bf y}$ and $t$ using the identity
$(\alpha^\dagger \tau^a \beta)(\gamma^\dagger \tau^a \delta) =
2(\alpha^\dagger \delta)(\gamma^\dagger \beta) - (\alpha^\dagger
\beta)(\gamma^\dagger \delta)$ for $SU(2)$ arbitrary doublets $\alpha,
\beta, \gamma, \delta$.  This gives:
\begin{eqnarray}
&&\Phi_0^\dagger \Phi_0 = 
|{\bf y}|,\qquad \Phi_1^\dagger \Phi_1=|{\bf y} + {\bf D}|,
\nonumber\\ 
&&d{\bf y} \cdot d{\bf y}\ =\ |{\bf y}| d\Phi_0^\dagger d\Phi_0 - \omega_0^2
\ =\ |{\bf y} + {\bf D}| d\Phi_1^\dagger d\Phi_1 - \omega_1^2,
\end{eqnarray}
and we find that our metric can be written as the $N=2$ case of the
Gibbons--Hawking metric:
\begin{eqnarray}
&&ds^2 = V^{-1}(dz - {\bf A} 
\cdot d{\bf y})^2 + V
d{\bf y} \cdot d{\bf y} \nonumber\\
&&V = \sum_{i=0}^{N-1} \frac{1}{|{\bf y} - {\bf y}_i|},
\qquad \mbox{\boldmath$\nabla$}V = \mbox{\boldmath$\nabla$}
\times{\bf A}\ .
\labell{gibbhawk}
\end{eqnarray}
Up to an overall normalisation (which we will fix later), we have the
normalisation ${\bf y}_0 = 0$, ${\bf y}_1 = {\bf D}$, and the vector
potential is
\begin{equation}
{\bf A}({\bf y}) 
\cdot d{\bf y} = 
|{\bf y}|^{-1} \omega_0 + |{\bf y} + {\bf D}|^{-1} \omega_1 + dz,
\end{equation}
and the field strength is readily obtained by taking the exterior
derivative and using the identity $\epsilon^{abc} (\alpha^\dagger \tau^b
\beta)(\gamma^\dagger \tau^c\delta) = i (\alpha^\dagger \tau^a
\delta)(\gamma^\dagger \beta) - i(\alpha^\dagger \delta)(\gamma^\dagger
\tau^a \beta)$.

Under a change of variables,~\cite{prasad} this metric (for $N=2$)
becomes the Eguchi--Hanson metric, \ref{eguchi} which we first
identified as the blowup of the orbifold point. The three parameters
in the vector ${\bf y}_1$ are the NS--NS fields representing the size
and orientation of the blown up $\IP^1$.

\begin{figure}[ht]
  \centerline{\psfig{figure=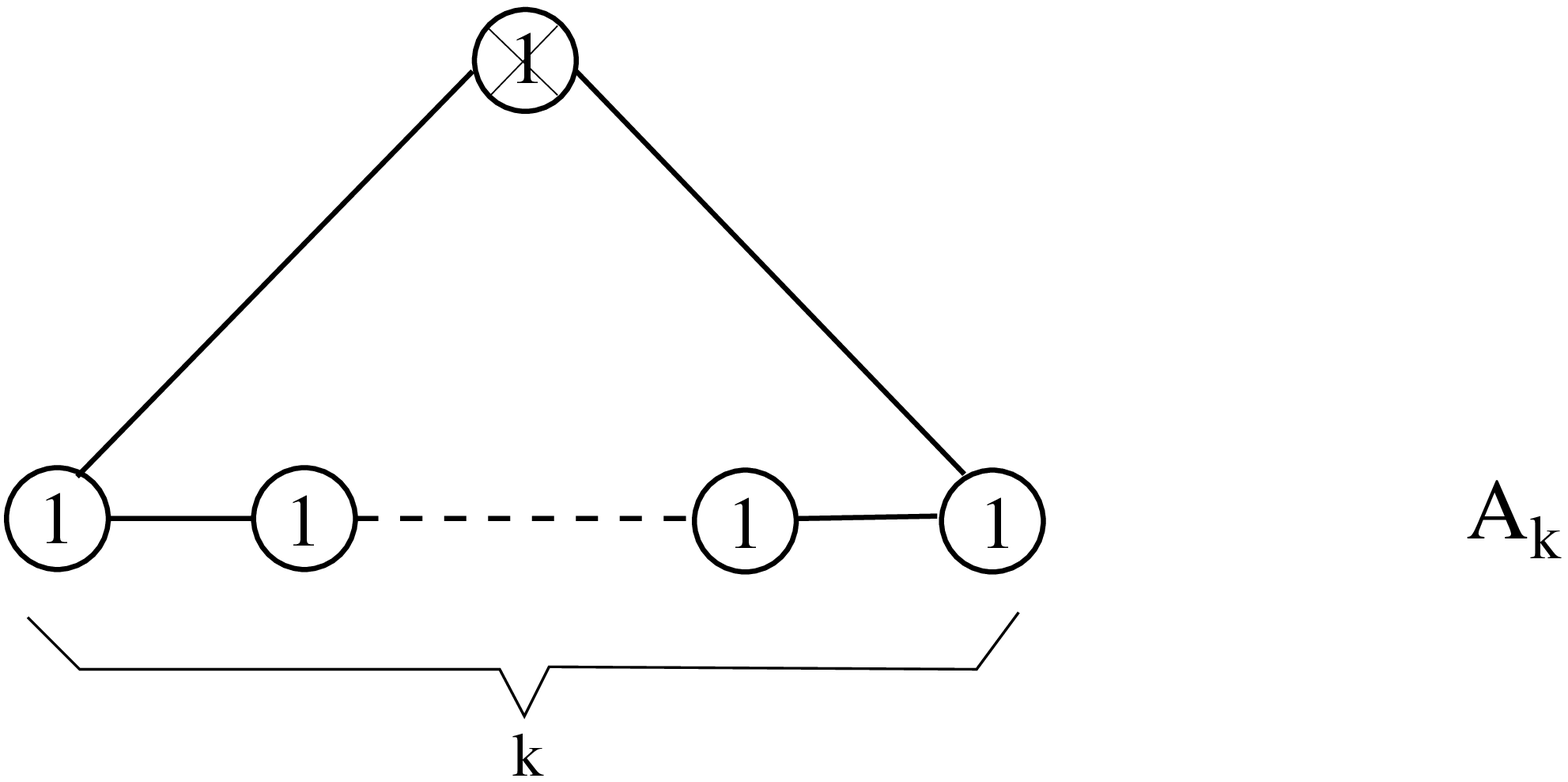,height=0.9in}}
\centerline{\psfig{figure=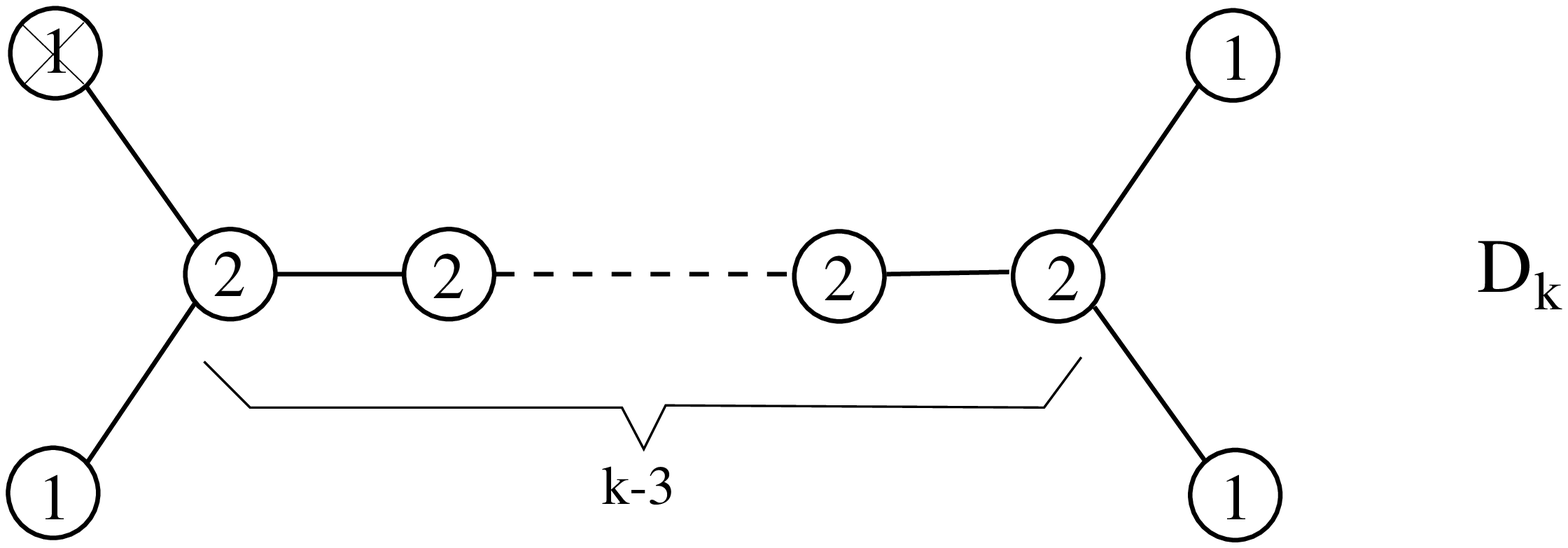,height=0.7in}}
\centerline{\psfig{figure=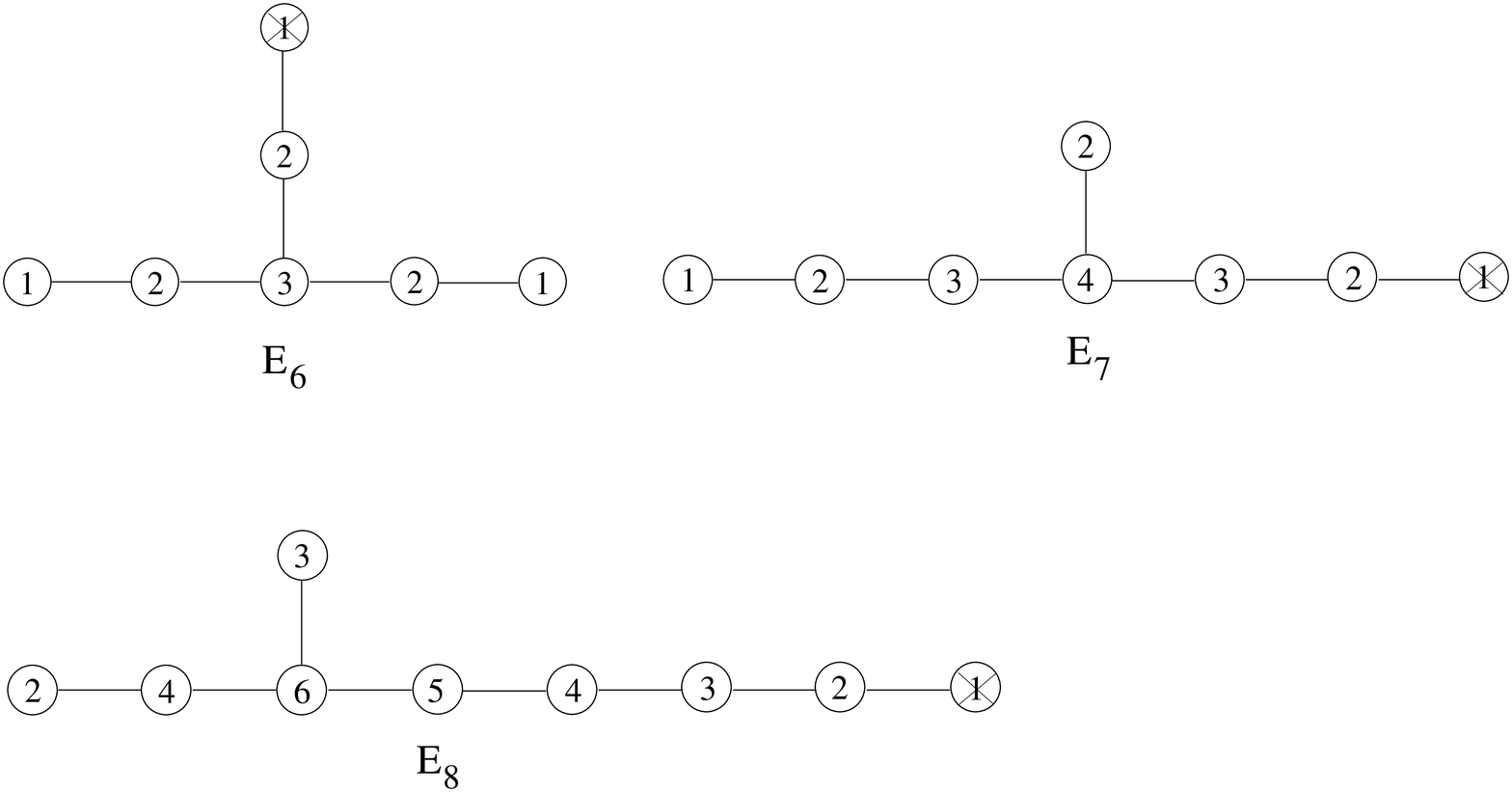,height=2.1in}}
\caption{The extended Dynkin diagrams for the A--D--E series. 
  As quiver diagrams, they give the gauge and matter content for the
  probe gauge theories which compute the resolved geometry of an ALE
  space. At the same time they also denote the actual underlying
  geometry of the ALE space, as each node denotes $\IP^1$, with the
  connecting edge representing a non--zero intersection.}
\label{extended}
\end{figure}

It is easy to carry out the generalisation to the full $A_{N-1}$
series, and get the metric \reef{gibbhawk} on the moduli space for a
D1--brane probing a $\IZ_N$ orbifold.  The gauge theory is just the
obvious generalisation derived from the extended Dynkin diagram:
$U(1)^N$, with $N+1$ bifundamental hypermultiplets with charges
$(1,-1)$ under the neighbouring $U(1)$'s.  (See
figure~\ref{extended}.)

There will be $3(N-1)$ NS--NS moduli which will become the $N-1$
differences ${\bf y}_i - {\bf y}_0$ in the resulting Gibbons--Hawking
metric \reef{gibbhawk}.  Geometrically, these correspond to the size
and orientation of $N-1$ separate $\IP^1$'s which can be blown up.  In
fact, we see that the there is another meaning to be ascribed to the
Dynkin diagram: Each node (except the trivial one) represents a
$\IP^1$ in the spacetime geometry that the probe sees on the Higgs
branch.

This entire construction which we have just described is a
``hyperK\"ahler quotient'', a powerful technique~\cite{hyper} for
describing hyperK\"ahler metrics of various types, and which has been
used to prove the existence of the full family of ALE
metrics.~\cite{kronheimer} It is remarkable that D--branes uncover the
spacetime using gauge theory variables and supersymmetry to construct
such a quotient, and that these are the same variables which appear in
the mathematical description of the construction. We shall see this
connection arising a number of other times in these notes.  A
reasonably elementary discussion, in this context, of the translation
between D--brane physics an the mathematics of hyperK\"ahler
quotients, can be found in the literature.~\cite{cvjmyers} For the
full A--D--E family of ALE spaces, there is a family of ``Kronheimer''
gauge theories which can be derived from the A--D--E extended Dynkin
diagram. (There is an excellent discussion,\cite{billo} with a stringy
flavour, in the bibliography.) This is the family of gauge theories
which arise on the world--volume of the D--brane
probes.~\cite{douglasmoore,cvjmyers} (See figure~\ref{extended}).
These families, and the correspondence to the A--D--E classification
arises as follows. We start with D--branes on $\IR^4/\Gamma$, where
$\Gamma$ is any discrete subgroup of $SU(2)$ (the cover of the $SO(3)$
which acts as rotations at fixed radii). It turns out that the
$\Gamma$ are classified in an ``A--D--E classification'', as shown by
McKay~\cite{mckay}. The $\IZ_{N}$ are the $A_{N-1}$ series. For the
$D_N$ and $E_{6,7,8}$ series, we have the binary dihedral
($\ID_{N-2}$), tetrahedral ($\cal T$), octahedral ($\cal O$) and
icosahedral ($\cal I$) groups.  In order to have the D--branes form a
faithful representation on the covering space of the quotient, we need
to start with a number equal to the order $|\Gamma|$ of the discrete
group.  This was two previously, and we started with $U(2)$.  So we
start with a gauge group $U(|\Gamma|)$, and then project, as before.

The resulting gauge groups are given by the extended Dynkin diagrams
suitably decorated.~\cite{kronheimer} (See figure~\ref{extended}.)  For
example, the simplest model in the D--series is D$_4$, which would
require 8 D1--branes on the covering space. The final probe gauge
theory after projecting is $U(2){\times}U(1)^4$, with four copies of a
hypermultipet in the ${\bf (2,1)}$. The families of hypermultiplets
(corresponding again to the links/edges in the diagram) and
D--flatness conditions, {\it etc.,} are precisely the variables and
algebraic condition which appear in Kronheimer's constructive proof of
the existence of the ALE metrics~\cite{kronheimer,cvjmyers}.
Unfortunately, it is a difficult and unsolved problem to obtain
explicit metrics for the resolved spaces in the D and E cases.

\subsection{Fractional D--Branes and Wrapped D--Branes}
\label{fractions}
\subby{Fractional Branes}

Let us pause to consider the following. In the previous section, we
noted that in order for the probe brane to move off the fixed point,
we needed to make sure that there were enough copies of it (on the
covering space) to furnish a representation of the discrete symmetry
$\Gamma$ that we were going to orbifold by. After the orbifold, we saw
that the Higgs branch corresponds to a {\sl single} D--brane moving
off the fixed point to non--zero $x^6,x^7,x^8,x^9$.  It is made up of the
$|\Gamma|$ D--branes we started with on the cover, which are now images
of each other under $\Gamma$. We can blow up the fixed point to a
smooth surface by setting the three NS--NS fields ${\bf D}$ non--zero.

When ${\bf D}=0$, there is a Coulomb branch. There, the brane is at
the fixed point $x^6,x^7,x^8,x^9=0$.  The $|\Gamma|$ D--branes are free
to move apart, independently, as they are no longer constrained by
$\Gamma$ projection.  So in fact, we have (as many as~\footnote{In the
  D and E cases, some of the branes are in clumps of size $n$
  (according to the nodes in figure \ref{extended}) and carry
  non--abelian $U(n)$}) $|\Gamma|$ independent branes, which therefore
have the interpretation as a fraction of the full brane.  None of
these individual fractional branes can move off.  They have charges
under the twisted sector R--R fields.  Twisted sector strings have no
zero mode, as we have seen, and so cannot propagate.

For an arbitrary number of these fractional branes (and there is no
reason not to consider any number that we want) a full $|\Gamma|$ of
them must come together to form a closed orbit of $\Gamma$, in order
for them to move off onto the Higgs branch as one single brane.  This
fits with the pattern of hypermultiplets and subsequent Higgs-ing
which can take place. There simply are not the hypermultiplets in the
model corresponding to the movement of an individual fractional brane
off the fixed point, and so they are ``frozen'' there, while they can
move within it,~\cite{GP,cvjmyers,ericme,douglasmoore} in the
$x^2,x^3,x^4,x^5$ directions.

\subby{Wrapped Branes}

There is  further understanding of what these individual fractional
branes mean. We see that when the ALE space is blown up, we fail to
get the fractional branes, suggesting that there is some geometrical
description to be found. The fancy language used at this point is that
the Coulomb branch is ``lifted'', which is to say it is no longer a
branch of degenerate vacua whose existence are protected by
supersymmetry.  While it is possible to blow up the point with the
separated fractional branes, it is not a supersymmetric operation. We
shall see why presently. First, let us set up the geometry.

Recall that each node (except for the extended one) in a Dynkin
diagram corresponds to a $\IP^1$ which can be blown up in the smooth
geometry. This is a cycle on which a D3--brane can be wrapped in order
to make a D1--brane on $\IR^6$.  For the A$_{N-1}$--series, where
things are simple, there are $N-1$ such cycles, giving that many
different species of D1--brane.

Where exactly is this cycle in the metric \reef{gibbhawk}? Notice that
the $4\pi$ periodic variable $z$ actually is a circle, but its radius
depends upon the prefactor $V^{-1}$, which varies with ${\bf y}$ in a
way which is set by the parameters (``centres'') ${\bf y}_i$.  When
${\bf y}={\bf y}_i$, the $z$--circle shrinks to zero size. There is a
$\IP^1$ between successive ${\bf y}_i$'s, which is the minimal surface
made up of the locus of $z$--circles which start out at zero size,
grow to some maximum value, and then shrink again to zero size, where
a $\IP^1$ then begins again as the neighbouring cycle, having
intersected with the previous one in a point.  The straight line
connecting this will give the smallest cycle, and so the area is
$4\pi|{\bf y}_i-{\bf y}_j|$ for the $\IP^1$ connecting centres ${\bf
  y}_{i,j}$.

If a brane is wrapped on a cycle, it cannot be pulled off (by
definition), even after the cycle has shrunken to zero size. Is this
perhaps responsible for the fractional brane description? If we can
get it to work for a single cycle~\cite{tensors,diacondouglas} (we need
to get rid of the total D3--brane charge), we can get it to work for
the entire A--D--E series of ALE spaces: the general Dynkin diagrams
telling us about the underlying geometry all have the interpretation
as the family of blown up cycles.

Here is one way to do it:~\cite{mukhiagain} Imagine a D3--brane with
some non--zero amount of $B+2\pi\alpha^\prime F$ on its world volume.
Recall that this corresponds to some D1--brane dissolved into the
worldvolume. We deduced this from T--duality in earlier sections. (We
did it with pure $F$, but we can always gauge in some $B$.) Since we
need a total D3--brane charge of zero in our final solution, let us
also consider a D3--brane with opposite charge, and with some
non--zero $B+2\pi\alpha^\prime {\widetilde F}$ on its worldvolume. We
write $\widetilde F$ to distinguish it from the $F$ on the other
brane's worldvolume, but the $B$'s are the same, since this is a
spacetime background field. So we have a worldvolume interaction:
\begin{equation}  
\mu_3\int C^{(2)}\wedge \left\{(B+2\pi\alpha^\prime F)-
(B+2\pi\alpha^\prime {\widetilde F})\right\}\ ,
\end{equation}
where we are keeping the terms separate for clarity. Our net D3--brane
charge is zero. Now let us choose $2\pi\alpha^\prime (\int_\Sigma
{F}-{\widetilde F})={\mu_1/\mu_3}$, and $\Phi_B\equiv
({\mu_3/\mu_1})\int_\Sigma {B}=1/2$ for some two dimensional spatial
subspace $\Sigma$ of the 3--volume. (Note that $\Phi_B\sim\Phi_B+1$.)
are only This gives a net D1--brane charge of $1/2+1/2=1$. The two
halves shall be our fractional branes. Right now, they are totally
delocalized in the world--volume of the D3--anti D3 system. We can
make the D1's more localised by identifying $\Sigma$ (the parts of the
3--volume where $B$ and $F$ are non--zero) with the $\IP^1$ of the ALE
space.  The smaller the $\IP^1$ is, the more localized the D1's are.
In the limit where it shrinks away we have the orbifold fixed point
geometry.  (Note that we still have $\Phi_{B}=1/2$ on the shrunken
cycle.  Happily, this is just the value needed to be present for a
sensible conformal field theory description of the orbifold
sector.~\cite{aspin})

Once the D1's are completely localized in $x^6,x^7,x^8,x^9$ from the
shrinking away of the $\IP^1$, then they are free to move
supersymmetrically in the $x^2,x^3,x^4,x^5$ directions. This should be
familiar as the general facts we uncovered about the D$p$--D$(p+2)$
bound state system: If the D$(p+2)$ is extended, the D$p$ cannot move
out of it and preserve supersymmetry. This is also T--dual to a single
brane at an angle and we shall see this next.

\subsection{Wrapped, Fractional and Stretched Branes}
\label{wrapfracstret}
There is yet another useful way of thinking of all the of the above
physics, and even more aspects of it will become manifest here. It
requires exploring a duality to another picture altogether. This
duality is morally a T--duality, although since it is a non--trivial
background that is involved, we should be careful. It is best trusted
at low energy, as we cannot be sure that the string theories are
completely dual at all mass levels. So we should probably claim only
that the backgrounds give the same low energy physics. Nevertheless,
once we arrive at our goal, we can forget about where it came from and
construct it directly in its own right.

Up to a change of variables, in the supergravity background
\reef{gibbhawk}, ${\bf y}$ can be taken to be the vector ${\bf
  y}=(x^7,x^8,x^9)$ while we will take $x^6$ to be our periodic
coordinate $z$. (There are some dimensionful parameters which were
left out of the derivation of \reef{gibbhawk}, for clarity, and we
shall put them in by hand, and try to fix the pure numbers with
T--duality.)

Then, using the T--duality rules \reef{backgroundT} we can arrive at
another background: (note that we have adjoined the flat transverse
spacetime $\IR^6$ to make a ten dimensional solution, and restored an
$\alpha^\prime$ for dimensions):
\begin{eqnarray}
ds^2&=&-dt^2+\sum_{m=1}^5dx^mdx^m+V(y)(dx^6dx^6+d{\bf y}\cdot d{\bf
y})\nonumber\\ e^{2\Phi}&=&V(y)=
\sum_{i=0}^{N-1}{\sqrt{\alpha^\prime}\over|{\bf y}-{\bf y}_i|}\ ,
\end{eqnarray}
which is also a ten dimensional solution if taken with a non--trivial
background field~\cite{vafatns,othertns}
$H_{mns}=\epsilon_{mns}^{\phantom{mns}r}\partial_r\Phi$ which defines
the potential $B_{6i}$ ($i=7,8,9$) as a vector $A_i$ which satisfies
$\nabla V=\nabla{\times}{\bf A}$.  Non--zero $B_{6i}$ arose because
the T--dual solution had non zero $G_{6i}$.

In fact, this is not quite the solution we are looking for. What we
have arrived at is a solution which is independent of the $x^6$
direction. This is neccessary if we are to use the operation
\reef{backgroundT}. In fact, we expect that the full solution we seek
has some structure in $x^6$, since translation invariance is certainly
broken there. This is because the $x^6$--circle of the ALE space has
$N$ places where winding number can change, since the circle shrinks
away there. So we expect that the same must be true for momentum in
the dual situation.~\cite{harveyruth} A simple guess for a solution
which is localised completely in the $x^6,x^7,x^8,x^9$ directions is
to simply as that it be harmonic there. We simply take ${\bf
  x}=(x^6,{\bf y})$ to mean a position in the full $\IR^4$, and
replace $V(y)$ by:
\begin{equation}
V(x)=1+
\sum_{i=0}^{N-1}{{\alpha^\prime}\over({\bf x}-{\bf x}_i)^2}
\end{equation}
We have done a bit more than just delocalized. By adding the 1 we have
endowed the solution with asymptotically flat behaviour. However,
adding the 1 is consistent with $V({\bf x})$ being harmonic in
$x^6,x^7,x^8,x^9$, and so it is still a solution.

The solution we have just uncovered is made up of a chain of $N$
objects which are pointlike in $\IR^4$ and magnetic sources of the
NS--NS potential $B_{\mu\nu}$. They are in fact the ``NS5--branes'' we
discovered by various arguments in previous sections, and a derivation
of the solution using S--duality transformations is presented in
insert 13 (p.\pageref{insert13}), with the result \reef{nsfive}. Here,
the NS5--branes are arranged in a circle on $x^6$, and distributed on
the rest of $\IR^4$ according to the centres ${\bf x}_i$,
$i=0,\cdots,N-1$.

\begin{figure}[ht]
  \centerline{\psfig{figure=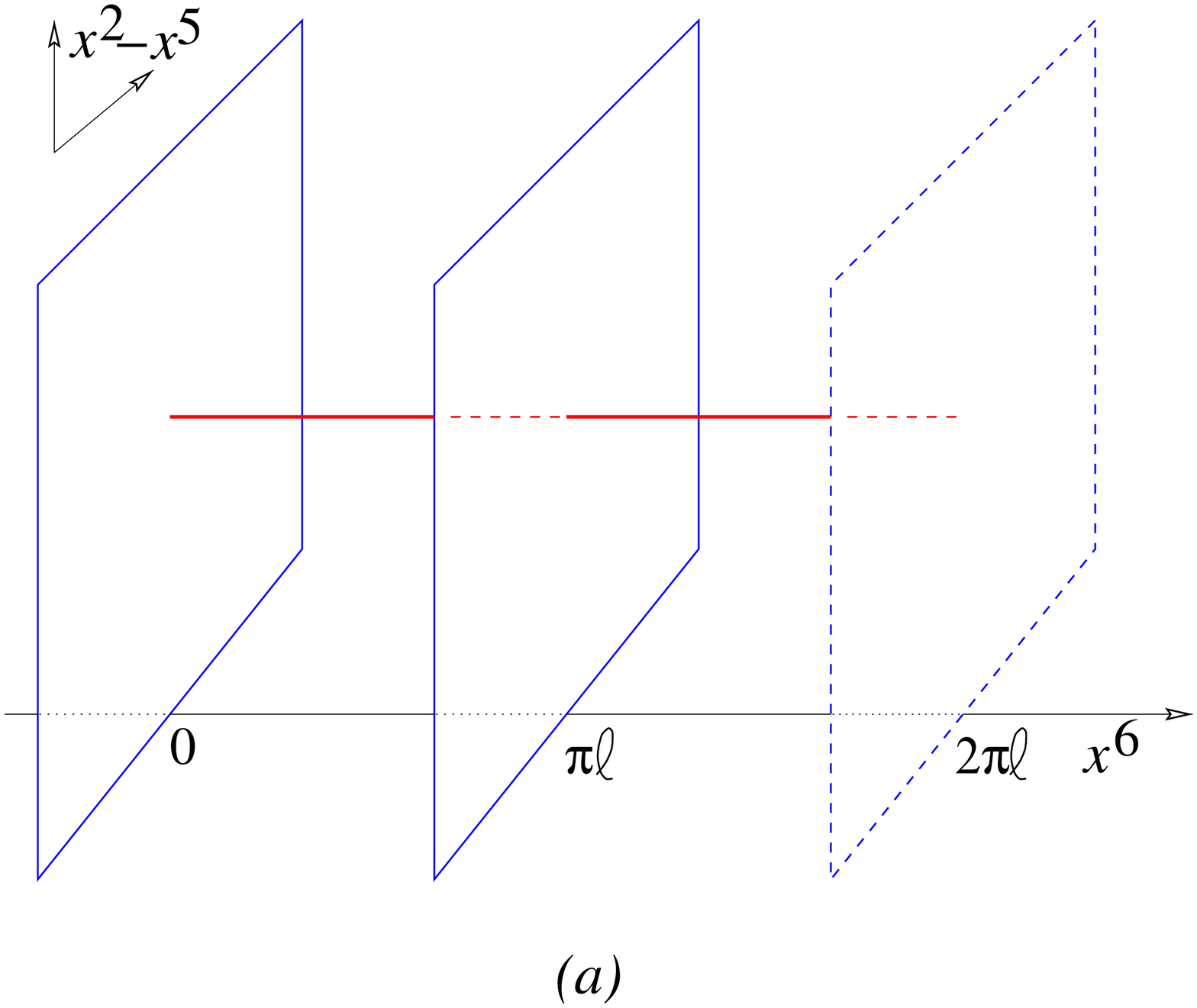,height=2.0in}
    \psfig{figure=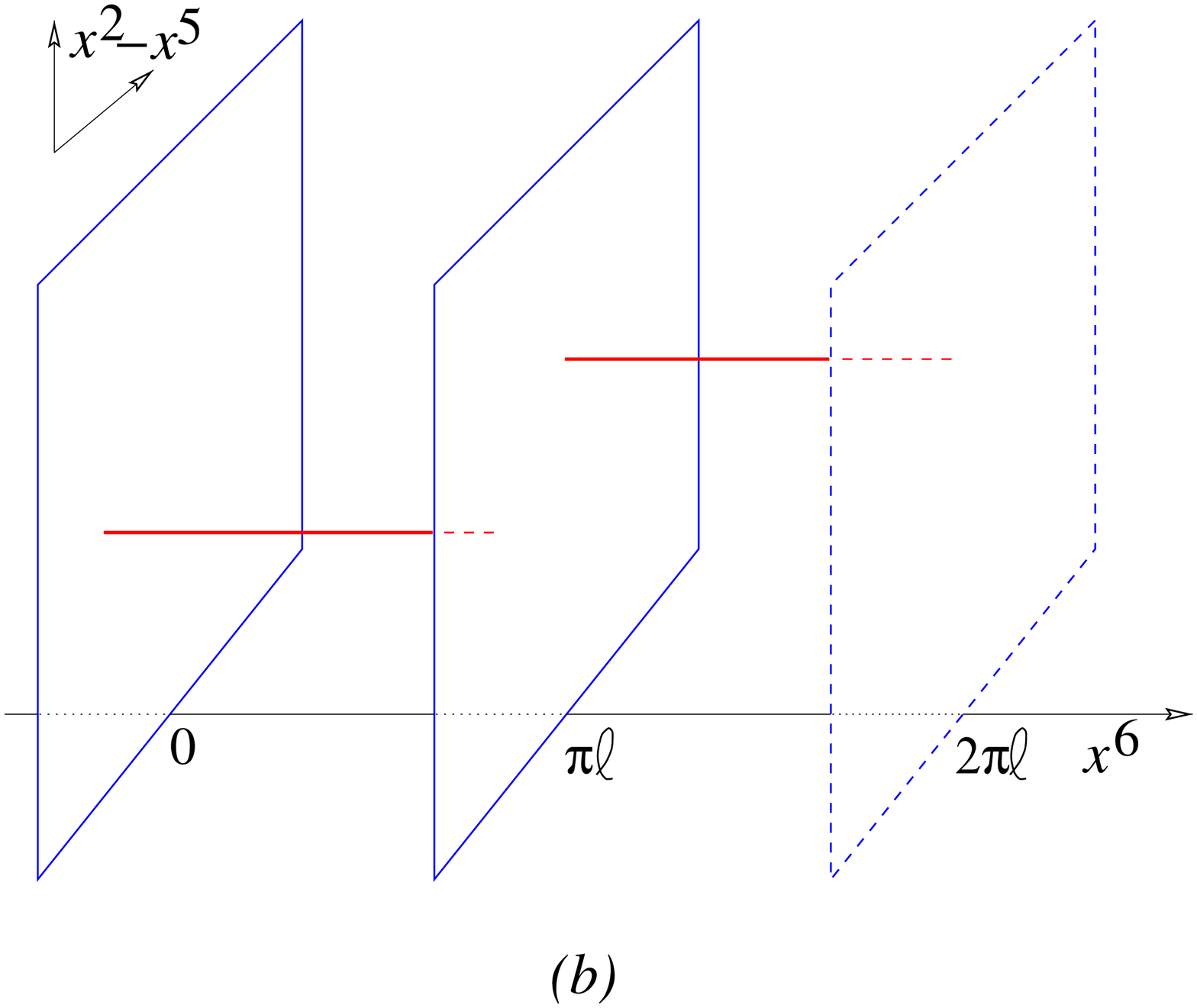,height=2.0in}}
\caption{{\it (a)} 
  This configuration of two NS5--branes on a circle with D2--branes
  streched between them is dual to a D1--brane probing an A$_1$ ALE
  space. {\it (b)} The Coulomb branch where the D2--brane splits into
  two ``fractional branes''.}
\label{stretched}
\end{figure}

Recall that we had a D1--brane lying along the $x^1$ direction,
probing the ALE space. By the rules of T--duality on a D--brane, it
becomes a D2--brane probing the space, with the extra leg of the
D2--brane extended along the compact $x^6$ direction.  The D2--brane
penetrates the two NS5--branes as it winds around once.  The point at
which it passes through an NS5--brane is given by four numbers ${\bf
  x}_i$ for the $i$th brane. The intersection point can be located
anywhere within the fivebrane's worldvolume in the directions
$x^2,x^3,x^4,x^5$. (See figure \ref{stretched}{\it (a)}.)

In the table below, we show the extension of the D2 in the $x^6$
direction as a $|-|$ to indicate that it may be of finite extent, if
it were ending on an NS5--brane.
\bigskip
\begin{center}
\begin{tabular}{|c|c|c|c|c|c|c|c|c|c|c|}
\hline
&$x^0$&$x^1$&$x^2$&$x^3$&$x^4$&$x^5$&$x^6$&$x^7$&$x^8$&$x^9$\\\hline
D2&$-$&$-$&$\bullet$&$\bullet$&$\bullet$&$\bullet$&$|-|$
&$\bullet$&$\bullet$&$\bullet$\\\hline
NS5&$-$&$-$&$-$&$-$&$-$&$-$&$\bullet$&$\bullet$&$\bullet$&$\bullet$
\\
\hline
\end{tabular}
\end{center}
\bigskip

This arrangement, with the branes lying in the directions which we
have described, preserves the same eight supercharges we discussed
before.  Starting with the 32 supercharges of the type~IIA
supersymmetry, the NS5--branes break a half, and the D2--brane breaks
half again. The infinite part of the probe, an effective one--brane
(string), has a $U(1)$ on its worldvolume, and its tension is
$\mu=2\pi\ell \mu_2$, where $\ell$ is the as yet unspecified length of
the new $x^6$ direction.  Note that if $\ell=\sqrt{\alpha^\prime}$, we
get the tension of a D1--brane, which apparently fixes all of our
parameters in the T--dual model in terms of the ALE
space.~\footnote{It might be useful to keep other values in mind,
  however.  Furthermore, the special value $\ell=\sqrt{\alpha^\prime}$
  coincides with the self--dual radius of simpler, toroidal
  compactifications, which is interesting.}

Let us focus on $N=2$. If the two fivebranes (with positions ${\bf
  x}_1,{\bf x}_2$; we can set ${\bf x}_0$ to zero) are located at the
same ${\bf y}=(x^7,x^8,x^9)$ position, then the D2--brane can break
into two segments, giving a $U(1)\times U(1)$ (one from each segment)
on the 1--brane part stretched in the infinite $x^1$ direction.  The
two segments can move independently within the NS5--brane worldvolume,
while still remaining parallel, preserving supersymmetry.~\footnote{It
  makes sense that the D2--brane can end on an NS--fivebrane. There is
  a two--form potential in the world--volume for which the
  string--like end can act as an electric source.}

\begin{figure}[ht]
  \centerline{\psfig{figure=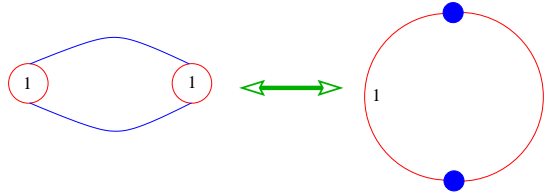,height=1.0in}}
\caption{There is a duality between the extended 
Dynkin diagram which gives the probe
  gauge theory and the diagram representing D--branes stretched
  between NS5--branes. The nodes in one are replaced by links in the
  other. In particular, the number inside the Dynkin nodes become the
  number of D--branes in the links in the dual diagram. The
  hypermultiplets associated with links in the Dynkin diagram arise
  from strings connecting the D--brane fragment ending on one side of
  an NS5--brane with the fragment on the other.}
\label{aleduals}
\end{figure}

This is the precise analogue of the Coulomb branch of the D1--brane
probing the ALE space that we saw earlier! The hypermultiplets of the
$U(1)\times U(1)$ theory are made here by stretching fundamental
strings across the NS5--branes in $x^6$ to make a connection between
the D--brane segments.~\cite{hanany} The three differences ${\bf
  y}_1-{\bf y}_2$ are the T--dual of the NS--NS parameters
representing the size and orientation of the ALE space's $\IP^1$. The
$x^6$ separation of the NS5--branes is dual to the flux
$2\pi\ell\Phi_B$. This is the length of one segment while
$2\pi\ell(1-\Phi_B)$ is the length of the other.  (This fits with the
fact that $\Phi_B\sim\Phi_B+1$.) Notice also that there is an
interesting duality between the quiver diagram and the arrangement of
branes in the dual picture. (See figure \ref{aleduals}.)

The original setup had the lengths equal, but we can change them at
will, and this is dual to changing $\Phi_B$. Note the possibility of
one of the lengths becoming zero. The NS--branes become coincident,
and at the same time a fractional brane becomes a tensionless string,
and we get an $A_1$ enhancement of the gauge symmetry carried by the
two--form potential which lives on the type~IIA
NS5--brane.~\cite{edcomm} If we had D1--branes stretched between NS5's
in type~IIB instead, we would get massless particles, and an enhanced
$SU(2)$ gauge symmetry. (See insert 11 (p.\pageref{insert11}))

\begin{figure}[ht]
  \centerline{\psfig{figure=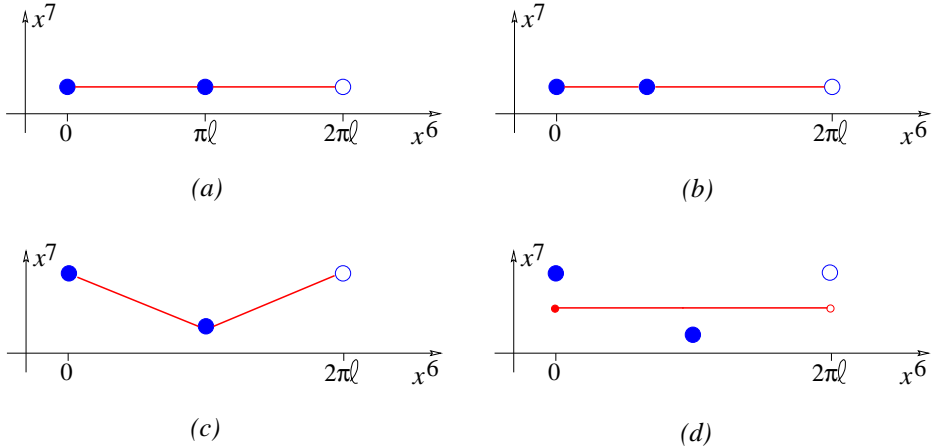,height=2.3in}}
\caption{Possible deformations of the brane arrangements, 
  and their gauge theory interpretation: {\it (a)} The configuration
  dual to the standard orbifold limit with the traditional ``half unit''
  of B--flux; {\it (b)} Varying the distribution of B--flux between
  segments. Sending it to zero will make the NS5--brane coincide and
  give an enhanced gauge symmetry; {\it (c)} Switching on a
  deformation parameter (an FI term in gauge theory) ``lifts'' the
  Coulomb branch: if there are separated D--brane fragments,
  supersymmetry cannot be retained; {\it (d)} First Higgs-ing to make a
  complete brane allows smooth movement onto the supersymmetric Higgs
  branch.}
\label{coulomb}
\end{figure}

If the segments are separated, and thus attached to the NS5--branes,
then when we move the NS5--branes out to different $x^{789}$
positions, the segments must tilt in order to remain stretched between
the two branes. They will therefore be oriented differently from each
other and will break supersymmetry.  This is how the Coulomb branch is
``lifted'' in this language. (See figure \ref{coulomb}{\it (c)}) A
segment at orientation gives a contribution
$\sqrt{(2\pi\ell\Phi_B)^2+({\bf y}_1-{\bf y}_2)^2}$ to the D1--brane's
tension. This formula should be familiar: it is of the form for the
more general formula for a bound state of a D1--D3 bound state, to
which this tilted D2--brane segment is dual.

For supersymmetric vacua to be recovered when the NS--fivebrane are
moved to different positions (the dual of smoothing the ALE space) the
branes segments must first rejoin with the other (Higgs--ing),
giving the single D--brane. Then it need not move with the NS5--branes
as they separate in ${\bf y}$, and can preserve supersymmetry by
remaining stretched as a single component. (See figure \ref{coulomb}{\it
  (d)}) Its ${\bf y}$ position and an $x^6$ Wilson line constitute the
Higgs branch parameters. Evidently the metric on these Higgs branch
parameters is that of an ALE space, since the 1+1 dimensional gauge
theory is the same as the discussion in section \ref{aleprobe}, and
hence the moduli spaces match. It is worth sharpening this into a
field theory proof of the low energy validity of the T--duality, but
we will not do that here.

It is worth noting here that once we have uncovered the existence of
fractional D--branes with a modulus for their separation, there is no
reason why we cannot separate them infinitely far from each other and
consider them in their own right. We also have the right to take a
limit where we focus on just one segment with a finite separation
between two NS5--branes, but with a {\sl non--compact} $x^6$
direction. This is achievable from what we started with here by
sending $\Phi_B\to0$, but changing to scaled variables in which there
is still a finite separation, and hence a finite gauge coupling on the
brane segment in question. (U--duality will then give us various
species of branes ending on branes which we will discuss later.)

Fractional branes, and their duals the stretched brane segments, are
useful objects since they are less mobile than a complete D--brane, in
that they cannot move in some directions.  One use of this is of
course the study of gauge theory on branes with a reduced number of
supersymmetries and a reduced number of charged
hypermultiplets.~\cite{hanany,elitzur} This has a lot of applications,
(there are reviews available~\cite{giveon,myreviewA,braneboxes}), some
of which we will consider later.

\subsection{D--Branes as Instantons}
\label{instantfun}
Consider a D0--brane and $N$ coincident D4--branes. There is a $U(1)$ on
the D0 and $U(N)$ on the D4's, which we shall take to be extended in
the $x^6,x^7,x^8,x^9$ directions.  The potential terms in the action
are
\begin{equation}
\frac{ \chi_i^{\dagger} \chi_i }{(2\pi{\alpha^\prime})^2 }
\sum_{a=1}^{5} ( X_a - Y_a )^2 + \frac{1}{4g_0^2}
\sum_{I=1}^3 (\chi_i^{\dagger}\tau^I\chi_i)^2 \ .
\label{potl}
\end{equation}
Here $a$ runs over the dimensions transverse to the D4--brane, and
$X_a$ and $Y_a$ are respectively the D0--brane and D4--brane
positions, and for now we ignore the position of the D0--brane within
the D4--branes' worldvolume.  This is the same action as in the
earlier case~(\ref{xyact}), but here the D4--branes have infinite
volume and so their $D$--term drops out. We have also written the 0--4
hypermultiplet field $\chi$ with a D4--brane index $i$.  (The
$SU(2)_R$ index is suppressed).  The potential~(\ref{potl}) is exact
on grounds of ${\cal N}{=}2$ supersymmetry.  The first term is the
${\cal N}{=}2$ coupling between the hypermultiplets $\chi$ and the
vector multiplet scalars $X$, $Y$.  The second is the $U(1)$
$D$--term.

For $N > 1$ there are two branches of moduli space, in direct analogy
with the ALE case.  The Coulomb branch is $(X \neq Y, \quad \chi =
0)$, which is simply the position of the D0--brane transverse to the
D4--branes. There is a mass for $\chi$ and so its vev is zero. The
Higgs\ branch $(X = Y, \quad \chi \neq 0)$ represents the physics of
the D0--brane being stuck on the world--volume of the D4--branes.  The
non--zero vev of $\chi$ Higgses away the $U(1)$ and some of the
$U(N)$. 

Let us count the dimension of moduli space.  There are $4N$ real
degrees of freedom in $\chi$.  The vanishing of the $U(1)$ $D$--term
imposes three constraints, and modding by the (broken) $U(1)$ removes
another degree of freedom leaving $4N - 4$. There are 4 moduli for the
position of the D0--brane inside the the D4--branes, giving a total of
$4N$ moduli.  This is in fact the correct dimension of moduli space
for an $SU(N)$ instanton when we do not mod out also the $SU(N)$
identifications.  For $k$ instantons this dimension becomes $4Nk$.

\begin{figure}[ht]
\centerline{\psfig{figure=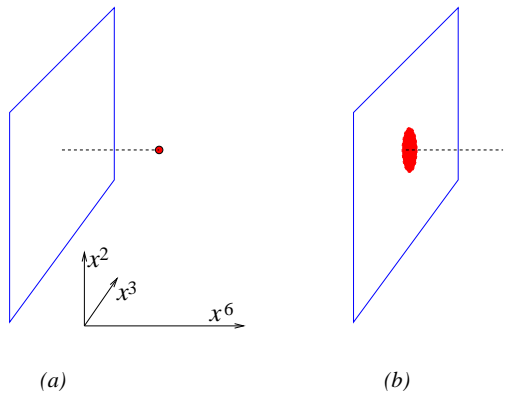,height=2.0in}}
\caption{ Instantons and the D$p$--D$(p+4)$ system. 
  {\it (a)} The Coulomb branch of the D$p$--brane theory represents a
  pointlike brane away from the D$(p+4)$--brane.  {\it (b)} The Higgs
  branch corresponds to it being stuck inside the D$(p+4)$--brane as a
  finite sized instanton of the D$(p+4)$--brane's gauge theory.}
\label{instanton}
\end{figure}

Another clue that the Higgs branch describes the D0--brane as a
D4--brane gauge theory instanton is the fact that the Ramond--Ramond
couplings include a term $\mu_4 C_{(1)}\wedge {\rm Tr}(F{\wedge}F)$.
As shown in section \ref{anomalousF}, when there is an instanton on
the D4--brane it carries D0--brane charge. The position of the
instanton is given by the 0--0 fields, while the 0--4 should give the
size and shape. (See figure \ref{instanton}).

The connection between D--branes and instantons was found first in the
case $(p,p') = (9,5)$ by Witten.~\cite{edsmall} This situation is
T--dual to the case we are discussing here, but does not have the
Coulomb branch, since the D9--branes fill spacetime.  The realization
that an instanton of D$p$--brane gauge theory can shrink to zero size
and move off as a D$(p-4)$--brane was noted by Douglas.~\cite{douglasii}

\subsection{Seeing the Instanton with a Probe}
Actually, we can really see the resulting instanton gauge fields by
using a D1--brane as a probe of the D9--D5 system.~\cite{douglasii} It
breaks half of the supersymmetries left over from the 9--5 system,
leaving four supercharges overall.  The effective 1+1 dimensional
theory is $(0,4)$ supersymmetric and is made of  1--1 fields, which
has two classes of hypermulitplets. One represents the motions of the
probe transverse to the D5, and the other parallel. The 1--5 and 1--9
fields are also hypermultiplets, while the 9--5 and 5--5 fields are
parameters in the model.

Let us place the D5--branes such that they are pointlike in
the $x^6,x^7,x^8,x^9$ directions. The D1--brane probe will lie along
the $x^1$ direction, as usual. 
\bigskip
\begin{center}
\begin{tabular}{|c|c|c|c|c|c|c|c|c|c|c|}
\hline
&$x^0$&$x^1$&$x^2$&$x^3$&$x^4$&$x^5$&$x^6$&$x^7$&$x^8$&$x^9$\\\hline
D1&$-$&$-$&$\bullet$&$\bullet$&$\bullet$&$\bullet$&$\bullet$
&$\bullet$&$\bullet$&$\bullet$\\\hline
D5&$-$&$-$&$-$&$-$&$-$&$-$&$\bullet$&$\bullet$&$\bullet$&$\bullet$
\\
\hline
\end{tabular}
\end{center}
\bigskip

This arrangement
of branes breaks the Lorentz group up as follows:

\begin{equation} 
 SO(1,9) \supset SO(1,1)_{01}\times SO(4)_{2345}\times SO(4)_{6789}\ ,
\labell{lorentz}
\end{equation}
where the superscripts denote the sub--spacetimes in which the
surviving factors act. We may label~\cite{wittenadhm,douglasii} the
worldsheet fields according to how they transform under the covering
group:
\begin{equation}
[SU(2)^\prime\times \widetilde{SU(2)}^\prime]_{2345}\times
[SU(2)_R\times SU(2)_L]_{6789}\ ,
\labell{cover}
\end{equation}
with doublet indices $(A^\prime,\tilde{A}^\prime,A,Y)$, respectively.

The analysis that we did for the D1--brane probe in the type~I string
theory in section \ref{hetdual} still applies, but there are some new
details. Now $\xi_-$ is further decomposed into $\xi^1_-$ and
$\xi^2_-$, where superscripts~1 and 2 denote the decomposition into
the (2345) sector and the (6789) sector, respectively.  So we have
that the  fermion $\xi^1_-$ (hereafter called
$\psi_-^{A\tilde{A}^\prime}$) is the right--moving superpartner of the
four component scalar field $b^{A^\prime\tilde{A}^\prime}$, while
$\xi^2_-$ (called $\psi_-^{A^\prime Y}$) is the right--moving
superpartner of $b^{AY}$.  The supersymmetry transformations are:
\begin{eqnarray}
&&\delta b^{A^\prime\tilde{A}^\prime}=i\epsilon_{AB}\eta^{A^\prime A}_{+} 
\psi_-^{B\tilde{A}^\prime}\nonumber\\
&&\delta b^{AY}=i\epsilon_{A^\prime B^\prime}\eta^{AA^\prime}_+
\psi_-^{B^\prime Y}\ .
\labell{susie}
\end{eqnarray}

In the 1--5 sector, there are four DN coordinates, and four DD
coordinates giving the NS sector a zero point energy of 0, with
excitations coming from integer modes in the $2345$ directions, giving
a four component boson.  The R sector also has zero point energy of
zero, with excitations coming from the 6789 directions, giving a four
component fermion $\chi$.

The GSO projections in either sector 
% are:
% \begin{eqnarray}
% &&(-1)^F_1=\Gamma^0\Gamma^1\Gamma^2\Gamma^3\Gamma^4\Gamma^5\nonumber\\
% &&(-1)^F_2=\Gamma^0\Gamma^1\Gamma^6\Gamma^7\Gamma^8\Gamma^9, 
% \labell{gsoagain}
% \end{eqnarray} which,
% upon application, 
reduce us to two bosonic states $\phi^{A^\prime}$ in
and decomposes the spinor $\chi$ into left and right moving two
component spinors, $\chi_-^A$ and $\chi_+^Y$, respectively. We see
that $\chi_-^A$ is the right--moving superpartner of
$\phi^{A^\prime}$.  Taking into account the fact that there is a
D5--brane index for these fields, we can display the
components~$(\phi^{A^\prime m},\chi_-^{Am})$ which are related by
supersymmetry:
\begin{equation}
{\delta\phi^{A^\prime m}=i\epsilon_{AB}\eta^{A^\prime A}_{+}
\chi_-^{Bm}.} 
\labell{susiei}
\end{equation}
and the $(0,4)$ supersymmetry parameter is denoted by
$\eta_+^{A^\prime A}$. Here, $m$ is a D5--brane group theory index.
Also, $\chi_+^{Y}$ has components $\chi_+^{Ym}$.

The supersymmetry transformation relating them to the
left moving fields are:
\begin{eqnarray}
&&\delta\lambda_+^M=\eta_+^{AA^\prime} C^M_{AA^\prime}\nonumber\\
&&\delta\chi_+^{Ym}=\eta_+^{AA^\prime} C^{Ym}_{AA^\prime}\ ,
\labell{susieii}
\end{eqnarray}
where $C^M_{AA^\prime}$ and $C^{Ym}_{AA^\prime}$ shall be determined
shortly. They will be made of the bosonic 1--1 fields and other
background couplings built out of the 5--5 and 5--9 fields.

The 5--5 and 5--9 couplings descend from the fields in the D9--D5
sector. There are some details of those fields which are peculiarities
of the fact that we are in type~I string theory. First, the gauge
symmetry on the D9--branes is $SO(32)$. Also, for $k$ coincident
D5--branes, there is a gauge symmetry $USp(2k)$,~\cite{edsmall} since
there is an extra $-1$ in the action of $\Omega$ on D5--brane fields.
~\cite{GP} The 5--5 sector hypermultiplet scalars (fluctuations in the
transverse $x^{6,7,8,9}$ directions) transform in the antisymmetric of
$USp(2k)$, which we call $X^{AY}_{mn}$, matching the notation in the
literature. ~\cite{douglasii} Meanwhile, the 5--9 sector produces a
$\bf{(2k,32)}$, denoted $h^{Am}_M$, with $m$ and $M$ as in D5-- and
D9--brane labels.

Using the form of the transformations \reef{susieii} allows us to
write the non--trivial part of the $(0,4)$ supersymmetric 1+1
dimensional Lagrangian containing the Yukawa couplings and the
potential of the $(0,4)$ model:
\begin{eqnarray}
&&{\cal L}_{\rm tot}={\cal L}_{\rm kinetic}
-{i\over4}\int d^2\!\sigma\biggl[  \lambda_+^M 
\left(
\epsilon^{BD}{\partial C^M_{BB^\prime} \over\partial b^{DY}}
\psi_-^{B^\prime Y}+
\epsilon^{B^\prime D^\prime}{\partial C^M_{B B^\prime}\over 
\partial\phi^{D^\prime m}}
\chi_-^{Bm}
\right)\biggr.\nonumber\\
\biggl.&&\hskip3cm+\chi_+^{Ym} 
\left(
\epsilon^{BD}{\partial C^{Ym}_{BB^\prime} \over\partial b^{DY}}
\psi_-^{B^\prime Y}+
\epsilon^{B^\prime D^\prime}{\partial C^{Ym}_{B B^\prime}\over 
\partial\phi^{D^\prime m}}
\chi_-^{Bm}
\right)\biggr.\nonumber\\
\biggl.&&\hskip3cm+{1\over2}
\epsilon^{AB}\epsilon^{A^\prime B^\prime}\left(C^M_{AA^\prime}
C^M_{BB^\prime}+C^{Ym}_{AA^\prime}C^{Ym}_{BB^\prime}\right)\biggr].
\labell{lagrange}
\end{eqnarray}

This is the most general~\cite{wittenadhm} $(0,4)$ supersymmetric
Lagrangian with these types of multiplets, providing that the $C$
satisfy the condition:
\begin{equation}
{C^M_{AA^\prime}C^M_{BB^\prime}
+C^{Ym}_{AA^\prime}C^{Ym}_{BB^\prime}+ C^M_{BA^\prime}
C^M_{AB^\prime}+C^{Ym}_{BA^\prime}C^{Ym}_{AB^\prime}=0\ ,}
\labell{hyperkahler}
\end{equation}
where ${\cal L}_{\rm kinetic}$ contains the usual kinetic terms for
all of the fields. Notice that the fields $b^{A^\prime {\tilde
    A}^\prime}$ and $\psi_-^{AA^\prime}$ are {\sl free}.

Now equation \reef{lagrange} might appear somewhat daunting, but is in
fact mostly notation. The trick is to note that general considerations
can allow us to fix what sort of things can appear in the matrices
$C_{AA^\prime}$. The distance between the D1--brane and the D5--branes
should set the mass of the 1--5 fields, $\phi^{A^\prime m}$ and its
fermionic partners $\chi^{Am}_-, \chi^{Ym}_+$.
 So there should be  terms of the form:
\begin{equation}
\phi_{A^\prime}^m\phi^{A^\prime n}(X^{AY}_{mn}-b^{AY}\delta_{mn})^2\ ,\quad
\chi^{Am}_-\chi^{Yn}_+(X^{AY}_{mn}-b^{AY}\delta_{mn})\ ,
\end{equation}
where the term in brackets is the unique translation invariant
combination of the appropriate 1--1 and 5--5 fields. There are also
1--5--9 couplings, which would be induced by couplings between 1--9,
1--5 and 5--9 fields, in the form $ \lambda^M_+\chi^A_{m-}h^m_{AM}$.

In fact, the required $C$'s which satisfy the requirements
\reef{hyperkahler} and give us the coupling which we expect
are:~\cite{douglasii}
\begin{eqnarray}
&&C^M_{AA^\prime}=h_{A}^{Mm}\phi_{A^\prime m}\nonumber\\
&&C^{Ym}_{AA^\prime}=\phi_{A^\prime}^n(X^{Ym}_{An}-b_A^{Y}\delta_{n}^m)\ .
\labell{cform}
\end{eqnarray}

The $(0,4)$ conditions \reef{hyperkahler} translate directly into a
series of equations for the D5--brane hypermultiplets to act as data
specifying an instanton {\it via} the ``ADHM description''.~\cite{adhm}
The crucial point is~\cite{wittenadhm} that the vacua of the sigma
model gives a space of solutions which is isomorphic to those of ADHM.

One can see that one has the right number of parameters as follows:
The potential is of the form $V=\phi^2((X-b)^2+h^2)$. So the term in
brackets acts as a mass term for $\phi$. The potential vanishes for
$\phi=0$, leaving this space of vacua to be parametrized by $X$ and
$h$, with $b$ giving the position of the D1--brane in the four
transverse directions. Let us write ${\widehat
  X}^{AY}{=}(X^{AY}{-}b^{AY})$ as the centre of mass field.

Notice that for these vacua ($\phi=0$), the Yukawa couplings are of
the form $\sum_a\lambda^a_+B^a_{Am}\chi_-^{Am}$ where
$B^a_{Am}=\partial C^a_{AB^\prime}/\partial\phi_{B^\prime m}$, and the
index $a$ is the set $(M,Y,m)$. There are 4$k$ fermions in $\chi_-$
and so this pairs with 4$k$ fermions in the set
$\lambda^a_+=(\chi_+^{Ym},\lambda^M_+)$, leaving a subspace of 32
massless modes describing the non--trivial gauge bundle.

The idea is to write the low energy sigma model action for these
massless fields.  This is done as follows: a basis of massless
components is given by $v_i^a$ ($i=1,\cdots,32$) defined by $\sum_a
v_v^a B^a_{Am}=0$, and we choose it to be orthonormal: $\sum_a v_i^a
v_j^a=\delta_{ij}$. The basis $v^a_i$ depends on ${\widehat X}$. So
substituting $\lambda^a_+=\sum_i v_i^a \lambda^i_+$ into the kinetic
energy gives:~\cite{wittenadhm}
\begin{equation}\lambda^a_+\partial_-\lambda^a_+=
\sum_{i,j}\biggl\{
\lambda_{+i}\left(\delta_{ij}\partial_-+\partial_-
{\widehat X}^{\mu}A_{\mu, ij}\right)\lambda_{+j}
\biggr\}\ ,
\labell{newkinetic}
\end{equation}
where 
\begin{equation}
A_{\mu, ij}\equiv A_{BY, ij}=\sum_a v^a_i{\partial v_j^a\over\partial 
{\widehat X}^{BY}}
\labell{instant}
\end{equation}
we have used the $x^6,x^7,x^8,x^9$ spacetime index $\mu$ on our 1--1
field ${\widehat X}^{BY}$ instead of the indices $(B,Y)$, for clarity.

So we see that the second term in \reef{newkinetic} shows the sigma
model couplings of the fermions to a background gauge field $A_\mu$.
Since we have generically
\begin{equation}
B^a_{Am}:\quad \left( {\widehat X}^{AY}, h^{Mm}_A\right)\ ,
\end{equation}
the orthonormal basis $v^a_i$ is 
\begin{equation}
v^a:\quad\left({h^{Mm}_A\over\sqrt{{\widehat X}^2+h^2}},
{-{\widehat X}^{AY}\over\sqrt{{\widehat X}^2+h^2}}\right)\ ,
\end{equation}
and from \reef{instant}, it is clear that the background gauge field
is indeed of the form of an instanton: The 5--9 field $h$ indeed sets
the scale size of the instanton, and the 5--5 field $X$ sets its
position.  Notice that this model gives a meaning to the instanton
even when its size drops to zero, well below any field theory or
string theory scale in the problem. This is another sign that
D--branes are able to see small ``substringy'' scales where new
physics is to be found.~\cite{shenk2,short,shorty} In the D$p$--D$(p+4)$
description, zero scale size is the place where the Higgs branch joins
onto the Coulomb branch representing the D$p$--brane becoming
pointlike (getting an enhanced gauge symmetry on its worldvolume), and
moves out of the worldvolume of the brane. (For $p=5$ this branch is
not present.)

\insertion{12}{The Heterotic NS5--brane\label{insert12}}{Recall that
  in insert 11 (p.\pageref{insert11}) we deduced that there must be a
  solitonic brane, the NS5--brane, which lives in $SO(32)$ heterotic
  string theory. This followed from the fact the D5--brane of type¬I
  had to map to such an object. This heterotic version of the
  NS5--brane inherits a number of properties from the D5--brane, the
  prinicipal one being that it must be an instanton of the $SO(32)$
  gauge theory of the heteroic string. In fact, it is the instanton
  property which led to its discovery early on. As a solution, it
  looks like the following (to leading order in
  $\ap$):~\cite{fivebranes,chstwo}
\begin{eqnarray}
ds^2&=&\eta_{\mu\nu}dx^\mu dx^\nu+e^{2\Phi}
\left(dr^r+r^2d\Omega_3^2\right)\nonumber\\
e^{2\Phi}&=&g_s^2\left(1+\alpha^\prime{(r^2+2\rho^2)
\over(r^2+\rho^2)^2}+O(\alpha^{\prime 2})\right)\ ,\quad
H_{\mu\nu\lambda}=
-\epsilon_{\mu\nu\lambda}^{\phantom{\mu\nu\lambda}\sigma}\partial_\sigma \Phi
\nonumber\\
A_\mu&=&\left({r^2\over r^2+\rho^2}\right)
g^{-1}\partial_\mu g\ , \quad g={1\over r}\left(\matrix{x^6+ix^7&x^8+ix^9\cr
x^8-ix^9&x^6-ix^7}\right) \ ,
\labell{heteroticfive}
\end{eqnarray}
showing its structure as an $SU(2)$ instanton localized in
$x^6,x^7,x^8,x^9$, with scale size $\rho$. $r^2$ is the radial
coordinate, and $d\Omega_3^2$ is a metric on a round $S^3$.}

\subsection{D--Branes as Monopoles}
Consider the case of a pair of parallel D3--branes, extended in the
directions $x^1, x^2, x^3$, and separated by a distance $L$ in the
$x^6$ direction.  Let us now stretch a family of $k$ parallel
D1--branes along the $x^6$ direction, and have them end on the
D3--branes. (This is U--dual to the case of D2--branes ending on
NS5--branes, as stated earlier in section~\ref{wrapfracstret}.) Let us
call the $x^6$ direction $s$, and place the D3--branes symmetrically
about the origin, choosing our units such that they are at $s=\pm1$.

\bigskip
\begin{center}
\begin{tabular}{|c|c|c|c|c|c|c|c|c|c|c|}
\hline
&$x^0$&$x^1$&$x^2$&$x^3$&$x^4$&$x^5$&$x^6$&$x^7$&$x^8$&$x^9$\\\hline
D1&$-$&$\bullet$&$\bullet$&$\bullet$&$\bullet$&$\bullet$&$|-|$
&$\bullet$&$\bullet$&$\bullet$\\\hline
D3&$-$&$-$&$-$&$-$&$\bullet$&$\bullet$&$\bullet$&$\bullet$&$\bullet$&$\bullet$
\\
\hline
\end{tabular}
\end{center}
\bigskip

This configuration preserves eight supercharges, as can be seen from
our previous discussion of fractional branes.  Also, a T$_6$--duality
yields a pair of D4--branes (with a Wilson line) in $x^1,x^2,x^3,x^6$
with $k$ (fractional) D0--branes.  This arrangement was shown to
preserve eight supercharges. (Also, we naively expect that this
construction should be related to our previous discussion of
instantons, but instead of on $\IR^4$, they are on $\IR^3\times S^1$.)
We can see it directly from the fact that the presence
of the D3-- and D1--branes world--volumes place the constraints:
\begin{equation}
\epsilon_L=\Gamma^0\Gamma^1\Gamma^2\Gamma^3\epsilon_R\ 
;\quad \epsilon_L=\Gamma^0\Gamma^6 \epsilon_R\ ,
\end{equation}
which taken together give eight supercharges, satisfying the condition
\begin{equation}
\epsilon_L=\Gamma^1\Gamma^2\Gamma^3\Gamma^6\epsilon_L\ .
\end{equation}

The 1--1 massless fields are simply the (1+1)--dimensional gauge field
$A^\mu(t,s)$ and eight scalars $\Phi^m(t,s)$ in the adjoint of $U(k)$,
the latter representing the transverse fluctuations of the branes.
There are fluctuations in $x^1,x^2,x^3$ and others in
$x^4,x^5,x^7,x^8,x^9$. We shall really only be interested in the
motions of the D1--brane within the D3--brane's directions
$x^1,x^2,x^3$, which is the ``Coulomb branch'' of the D1--brane moduli
space. So of the $\Phi^m$, we keep only the three for $m=1,2,3.$ There
are additionally 1--3 fields transforming in the $(\pm1,k)$. They form
a complex doublet of $SU(2)_R$ and are $1\times k$ matrices.
Crucially, these flavour fields are massless only at $s=\pm1$, the
locations where the D1--branes touch the D3--branes.  If we were to
write a Lagrangian for the massless fields, there will be a delta
function $\delta(s\mp1)$ in front of terms containing those.  The
structure of the Lagrangian is very similar to the one written for the
$p-(p+4)$ system, with the additional features of $U(k)$ non--abelian
structure.  Asking that the D--terms vanish, for a supersymmetric
vacuum, we get:~\cite{diaconescu}
\begin{equation}
{d \Phi^i\over ds}-[A_s,\Phi^i]+{1\over2}\epsilon^{ijk}[\Phi^j,\Phi^k]=0\ ,
\labell{nahm}
\end{equation}
where we have ignored possible terms on the right hand side supported
only at $s=\pm1$. These would arise from the interactions induced by
massless 1--3 fields there.~\cite{tsimpis}
We shall derive those effects in another
way by carefully considering the boundary conditions in a short while.

If we choose the gauge in which $A_s=0$, our equation \reef{nahm} can
be recognised as the Nahm equations,~\cite{wnahm} known to construct
the moduli space~\cite{modulispace} of $N$ SU(2) monopoles, {\it via}
an adaptation of the ADHM construction.~\cite{adhm} The covariant form
$A_s\neq0$, is useful for actually solving for the metric on the
moduli space of monopole solutions and for the spacetime monopole
fields themselves, as we shall show.~\cite{donaldson}

If our $k$ D1--branes were reasonably well separated, we would imagine
that the boundary condition at $s=\pm1$ is clearly
$2\pi\alpha^\prime\Phi^i(s=1))={\rm
  diag}\{x^i_1,x^i_2,\cdots,x^i_k\}$, where $x^i_n$, $i=1,2,3$ are the
three coordinates of the end of the $n$th D1--brane (similarly for the
other end). In other words, the off--diagonal fields corresponding to
the 1--1 strings stretching between the individual D1--branes are
heavy, and therefore lie outside the description of the massless
fields.  However, this is not quite right.  In fact, it is very badly
wrong. To see this, note that the D1--branes have tension, and
therefore must be pulling on the D3--brane, deforming its shape
somewhat. In fact, the shape must be given, to a good approximation, by
the following description. The function $s({\bf x})$ describing the
position of the D3--brane along the $x^6$ direction as a function of
the three coordinates $x^i$ should satisfy the equation $\nabla^2
s(x)=0$, where $\nabla^2$ is the three dimensional Laplacian. A
solution to this is
\begin{equation}
s=1+{c\over|{\bf x}-{\bf x}_0|}\ ,
\end{equation}
where $1$ is the position along the $s$ direction and $c$ and
${\bf x}_0$ are constants. So, far away from ${\bf x}_0$, we see that the
solution is $s=1$, telling us that we have a description of a flat
D3--brane. Nearer to ${\bf x}_0$, we see that $s$ increases away from 0, and
eventually blows up at ${\bf x}_0$. 

\begin{figure}[ht]
\centerline{\psfig{figure=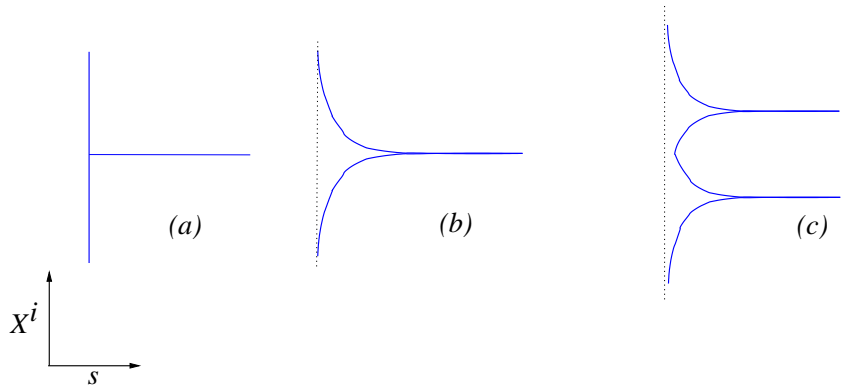,height=2.0in}}
\caption{{\it (a)}: A D3--brane (vertical)  with a D1--brane ending on it 
  (horizontal) is actually pulled {\it (b)} into a smooth
  interpolating shape. {\it (c)}: Finitely separated D1--branes can
  only be described with non--commutative coordinates (see text)}
\label{pulling}
\end{figure}

We sketch this shape in figure \ref{pulling}{\it (a)}. It is again our
BIon--type solution, described before in section \ref{BIonsection}.
The D3--brane smoothly interpolates between a pure D1--brane geometry
far away and a spiked shape resembling D1--brane behaviour at the
centre.  A multi--centred solution is easy to construct as a
superposition of harmonic solutions of the above type. Considering two
of them, we see that in fact for any finite separation of the
D1--branes (as measured far enough along the $s$--direction), by time
we get to $s=1$, they will be arbitrarily close to each other (see
\ref{pulling}{\it (b)}).  We therefore cannot forget~\cite{giveon}
about the off--diagonal parts of $\Phi^m$ corresponding to 1--1
strings stretching between the branes, and in fact we are forced to
describe the geometry of the branes' endpoints on the D3--brane using
non--abelian $X^m$.  This is another example of the ``natural''
occurrence of a non--commutativity arising in what we would have
naively interpreted as ordinary spacetime coordinates.

We can see precisely what the boundary conditions must be, since we
are simply asking that there be a pole in $\Phi^i(s)$ as $s\to\pm1$:
\begin{equation}
\Phi^i(s)\to{\Sigma^i\over s\mp1}\ ,
\end{equation}
and placing this into \reef{nahm}, we see that the $k{\times}k$
residues must satisfy \begin{equation}
[\Sigma^i,\Sigma^j]=2{\rm i}\epsilon_{ijk}\Sigma^k\ .
\end{equation}
In other words, they must form an $k$--dimensional representations of
$SU(2)$! This representation {\sl must} be irreducible, as we have
seen. Otherwise it necessarily captures only the physics of $m$
infinitely separated clumps of D1--branes, for the case where the
representation is reducible into $m$ smaller irreducible
representations.

The problem we have constructed is that of
monopoles~\cite{thooftpolyakov,poles} of $SU(2)$ spontaneously broken
to $U(1)$ via an adjoint Higgs field ${\bf H}$.~\cite{monoreview}
Ignoring the centre of mass of the D3--brane pair, this $SU(2)$ is on
their world volume, and the separation is given by the vev, $H$ of the
Higgs field.  The first order ``Bogomol'nyi'' equations~\cite{bogo}
are:
\begin{eqnarray}
&&B_i\equiv{1\over2}\epsilon_{ijk}F_{jk}=D_i\H\ ,\quad{\rm with}\nonumber\\
&&\quad F_{ij}=\partial_iA_j-\partial_jA_i+[A_i,A_j]; 
\quad \quad D_i\H=\partial_i\H+[A_i,\H]\ ,
\label{bogomolnyi}
\end{eqnarray}
with gauge invariance $(g({\bf x})\in SU(2))$:
\begin{equation}
A_i\to g^{-1}A_ig+g^{-1}\partial_ig;\quad \H\to g^{-1}\H g\ .
\end{equation}
Static, finite energy monopole solutions satisfy
\begin{equation}
\|\H({\bf x})\|\equiv{1\over2}\Tr \left[\H^*\H\right]\to H\  
\quad {\rm as}\quad r\to\infty\ ,
\end{equation}
where ${\bf x}=(x_1,x_2,x_3)$, $r^2=x_1^2+x_2^2+x_3^2$, and
$(2\pi\alpha^\prime)H=L/2$, where $L$ is the separation of our
D3--branes. The topological magnetic charge the monopoles carry comes
from the fact that the vacuum manifold, which is $SU(2)/U(1)\sim S^2$,
can wind an integer number of times around the $S^2$ at infinity,
giving a stable solution whose charge is a fixed number times that integer.

In fact, we can construct the Higgs field and gauge field of monopole
solution of the 3+1 dimensional gauge theory as follows. Given
$k{\times}k$ Nahm data
$(\Phi^1,\Phi^2,\Phi^3)=2\pi\alpha^\prime(T_1,T_2,T_3)$ solving the
equation \reef{nahm}, there is an associated differential equation for
a $2k$ component vector ${\bf v}(s)$:
$$
\left\{{\bf 1}_{2N}{d\over ds}+ \left({x^a\over 2}{\bf 1}_k
  +iT_a\right)\otimes\sigma^a\right\}{\bf v}=0\ .
$$
There is a unique solution normalisable with respect to the
inner product
$$
<{\bf v}_1,{\bf  v}_2>=\int_{-1}^1 {\bf v}^\dagger_1{\bf v}_2 ds\ .
$$
In fact, the space of normalisable solutions to the equation is
four dimensional, or complex dimension 2. Picking an orthonormal basis
${\widehat {\bf v}}_1,{\widehat {\bf v}}_2$, we construct the Higgs
and gauge potential as:
\begin{eqnarray}
{\bf H}&=&i\left[\matrix{<s{\widehat {\bf v}}_1,{\widehat {\bf v}}_1>
&<s{\widehat {\bf v}}_1,{\widehat {\bf v}}_2>\cr
<s{\widehat {\bf v}}_2,{\widehat {\bf v}}_1>
&<s{\widehat {\bf v}}_2,{\widehat {\bf v}}_2>
}\right]\ ,
\nonumber\\
 A_i&=&\left[\matrix{<{\widehat {\bf v}}_1,\partial_i{\widehat {\bf v}}_1>
&<{\widehat {\bf v}}_1,\partial_i{\widehat {\bf v}}_2>\cr
<{\widehat {\bf v}}_2,\partial_i{\widehat {\bf v}}_1>
&<{\widehat {\bf v}}_2,\partial_i{\widehat {\bf v}}_2>
}\right]\ 
\end{eqnarray}
The reader may notice a similarity between this means of extracting
the gauge and Higgs fields, and the extraction
\reef{newkinetic}\reef{instant} of the instanton gauge fields in the
previous section.  This is not an accident. The Nahm construction is
in fact a hyperK\"ahler quotient which modifies the ADHM procedure.
The fact that this arrangement of branes is T--dual to that of the
$p$--$(p+4)$ system is the physical realisation of this fact, showing
that the basic families of hypermultiplet fields upon which the
construction is based (in the brane context) are present here too.

\begin{figure}[h]
\centerline{\psfig{figure=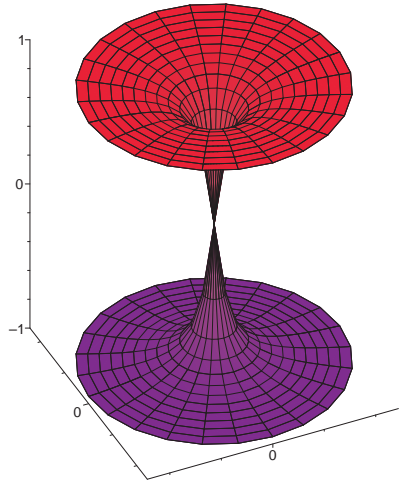,height=2.5in}}
\caption{{\it (a)}: A slice through part of 
  two (horizontal) D3--branes with a (vertical) D1--brane acting as a
  single BPS monopole. This is made by plotting the exact BPS
  solution.}
\label{monoshape}
\end{figure}

It is worth studying the case $k=1$, for orientation. In this case,
the solutions $T_i$ are simply real constants
$(2\pi\alpha^\prime)\Phi_i=-ia_i/2$, having the meaning of the
position of the monopole at ${\bf x}=(a_1,a_2,a_3)$.  Let us place it
at the origin. Furthermore, as this situation is spherically
symmetric, we can write ${\bf x}=(0,0,r)$. Writing components ${\bf
  v}=(w_1,w_2)$, we get a pair of simple differential equations with
solution
\begin{equation}
w_1=c_1e^{-rs/2}\ ,\quad w_2=c_2e^{rs/2}\ .
\end{equation}
An orthonormal basis is given by
\begin{eqnarray}
{\widehat {\bf v}}_1\ :\left(c_1=0\ ,c_2=\sqrt{{r\over e^{2r}-1}}\,\right)\,\,
\ ;{\widehat{\bf v}}_2\ :\left(c_2=0\ , c_1=\sqrt{{r\over 1-e^{-2r}}}\,\right) 
\end{eqnarray}
and the Higgs field is simply: 
\begin{eqnarray}
&&{\bf H}(r)={\hat x}_i\sigma_i{\varphi(r)\over r}\ , 
\quad{\rm  with}\nonumber \\
&&\varphi(r)={r\over(e^{2r}-1)}\int_{-1}^1 se^{rs} ds={r\coth r-1}\ .
\end{eqnarray}
(here ${\hat x}=(0,0,1)$) while the gauge field is:
\begin{equation}
A_i(r)=\epsilon_{ijk}\sigma_j {\hat x}_k{\sinh r-r\over r^2\sinh r}\ .
\end{equation}
This is the standard one--monopole solution of Bogomol'nyi, Prasad and
Sommerfield, the prototypical ``BPS monopole''.~\cite{bogo,onemonopole}
We can insert the required dimensionful quantities:
\begin{equation}
% {\bf H}\to {{\bf H}\over g^2_{\bf YM}}\ ;\quad g^{-2}_{\rm YM}=\mu_3
% g^{-1}_s(2\pi\alpha^\prime)^2\ ; \quad 
\varphi(r)\to
\varphi\left(Lr/4\pi\alpha^\prime\right)\ ,
\end{equation}
to get the Higgs field:
\begin{equation}
{\bf H}={\sigma_3\over r}
\varphi\left({Lr\over4\pi\alpha^\prime}\right)
\longrightarrow {L\over 4\pi\alpha^\prime}\sigma_3\ ,
 \quad {\rm as}\quad r\to\infty\ ,
\end{equation}
showing the asymptotic positions of the D3--branes to be $\pm L/2$,
after multiplying by $2\pi\alpha^\prime$ to convert the Higgs field
(which has dimensions of a gauge field) to a distance in $x^6$. A
picture of the resulting shape~\cite{roberto,akibend} of the D3--brane is
shown in figure \ref{monoshape}.

There is also a simple generalisation of the purely magnetic solution
which makes a ``dyon'', a monopole with an additional $n$ units of
electric charge.  It interpolates between the magnetic monopole
behaviour we see here and the spike electric solution we found in
section \ref{BIonsection}. It is amusing to note~\cite{dominic} that an
evaluation of the mass of the solution gives the correct formula for
the bound state mass of a D1--string bound to $n$ fundamental strings,
as it should, since an electric point source is in fact the
fundamental string.

\section{D--Branes and Geometry II}
\subsection{The Geometry produced by D--Branes}
By studying the supergravities arising in the low energy limit of the
superstring theory, it was shown that there exist extended solutions
resembling generalisations of charged black holes. The $p$ dimensional
extended solution carries charges under the R--R form $C^{(p+1)}$.
The extremal cases are BPS solutions, and they differ from
Reissner--Nordstrom black holes in that their horizons at extremality
have zero area~\footnote{This latter fact is interesting, but we will
  leave it to the reader to consult the lectures of Amanda
  Peet\cite{peettasi} and Mike Duff\cite{dufftasi} to see how this
  relates to the understanding of black hole entropy {\it via}
  D--branes,~\cite{sv,peet} and the AdS/CFT correspondence.}.  The BPS
(extremal) solution is:~\cite{blackp,ramzireview}
\begin{eqnarray}
ds^2 &=&  Z_p^{-1/2} \eta_{\mu\nu} dx^\mu dx^\nu +
 Z_p^{1/2} dx^i dx^i \ ,\nonumber
\\
e^{2\Phi } &=& g^2_s { Z_p}^{(3-p)\over2}\ , \nonumber\\
C_{ (p+1)} &=& ({Z_p}^{-1} -1)g_s^{-1} dx^0 \wedge \cdots \wedge dx^p
\ , \labell{branes}
\end{eqnarray}
where $\mu=0,\ldots,p$, and $i=p+1,\ldots, 9$, and the harmonic
function $Z_p$ is
\begin{equation}
Z_p=1+{d_p(2\pi)^{p-2}g_sN\alpha^{\prime (7-p)/2}\over r^{7-p}}\ ;
 \quad d_p= 2^{7-2p}\pi^{9-3p\over 2}
\Gamma\left({7-p\over 2}\right) \ .
\labell{harmbranes}
\end{equation}
More complicated supergravity solutions preserving fewer
supersymmetries (in the extremal case) can be made by combining these
simple solutions in various ways, by intersecting them with each
other, boosting them to finite momentum, and by wrapping, and/or
warping them on compact geometries.  This allows for the construction
of finite area horizon solutions, corresponding to R--R charged
Reissner--Nordstrom black holes, and generalisations thereof.

These solutions are R--R charged, but we have already established to
all orders in string perturbation theory that D$p$--brane actually are
the {\it basic sources} of these R--R fields. In fact, the solutions
\reef{branes} are normalised such that they carry $N$ units of the
basic D--brane charge $\mu_p$.

It is natural to suppose that there is a connection between these two
families of objects: Perhaps the solution \reef{branes} is ``made of
D--branes'' in the sense that it is actually the field due to $N$
D$p$--branes, all located at $r=0$. This is precisely how we are to
make sense of this solution as a supergravity soliton solution. We
{\it must} do so, since (except for $p=3$) the solution is actually
singular at $r=0$, and so one might have simply discarded them as
pathological, since solitons ``ought to be smooth'', like the
NS5--brane solution~\footnote{At $r=0$, the NS5--brane geometry (see
  \reef{nsfive}) opens up into an infinite throat geometry, which is
  smooth, being $\IR^7{\times}S^3$, with a dilaton which is linear in
  the distance down one of the $\IR^7$ directions. The $p=3$ version
  of the geometry in \reef{branes} also has a smooth throat, but the
  geometry is AdS$_5{\times}S^5$, with constant dilaton. String theory
  propagating on these throat backgrounds is, in each case, believed
  to be dual to a non--gravitational
  theory.\cite{mythroat,oferetal,juan,agmoo,myreviewB}}.  However, string
duality {\it forces} us to consider them, since smooth NS--NS solitons
of various extended sizes (which can be made by wrapping or warping
NS5--branes in an arbitrary compactification) are
mapped~\cite{isstrings} into these R--R solitons under it,
generalizing what we have already seen in ten dimensions (see {\it
  e.g.} insert~11, (p.\pageref{insert11})). With the understanding
that there are D--branes ``at their core'', which fits with the fact
that they are R--R charged, they make sense of the whole spectrum of
extended solitons in string theory.

\insertion{13}{The Type~II NS5--brane\label{insert13}}{In insert 11
  (p.\pageref{insert11}) we deduced that there must be a solitonic
  brane, the NS5--brane, in type~II string theory. We can deduce its
  supergravity fields by using the ten dimensional S--duality
  transformations to convert the case $p=5$ of equations
  \reef{branes}, \reef{harmbranes}, to give:~\cite{fivebranes,chstwo}
\begin{eqnarray}
ds^2&=&-dt^2+(dx^1)^2+\cdots+(dx^5)^2+{\tilde Z}_5
\left(dr^r+r^2d\Omega_3^2\right)\nonumber\\
e^{2\Phi}&=&g_s^2{\tilde Z}_5 =g_s^2\left(1+
{\alpha^\prime N\over r^2}\right)\ ,\nonumber\\
B_{ (6)} &=& ({{\tilde Z}_5}^{-1} -1)g_s 
dx^0 \wedge \cdots \wedge dx^5
\ .
\labell{nsfive}
\end{eqnarray}
This solution has $N$ units of the basic magnetic charge of $B_{(2)}$,
and is a point in $x^6,x^7,x^8,x^9$. Here, $r^2$ is the radial
coordinate, and $d\Omega_3^2$ is a metric on a round $S^3$. The
tension of this BPS object was deduced in insert 11
(p.\pageref{insert11}) to be: $\tau^{\rm
  F}_5=(2\pi)^{-5}\alpha^{\prime -3}g_s^{-2}$. (Note that the same
transformation will give a solution for the fields around a
fundamental IIB string, by starting with the $p=1$ case of
\reef{branes}.~\cite{fundstring,fstring}) Recall also that we deduced
the structure of this solution already using (a cavalier) T--duality
to an ALE space in section \ref{wrapfracstret}. Here, we have used
S--duality to the precise D--brane computations to see that our
normalisations in those sections were correct.}

Let us build up the logic of how they can be related to D--branes.
Recall that the form of the action of the ten dimensional supergravity
with NS--NS and R--R field strengths $H$ and $G$ respectively is,
roughly:
\begin{equation}
S = \int d^{10}\!x\, \left( e^{-2\Phi} R -  e^{-2\Phi} H^2 -  G^2
\right) \ .  
\labell{roughact}
\end{equation}
There is a balance between the dilaton dependence of the NS--NS and
gravitational parts, and so the mass of a soliton
solution~\cite{ramzireview} carrying NS--NS charge (like the
NS5--brane) scales like the action: $T_{\rm NS}\sim e^{-2\Phi}\sim
g_s^{-2}$. A R--R charged soliton has, on the other hand, a mass which
goes like the geometric mean of the dilaton dependence of the R--R and
gravitational parts: $T_{\rm R}\sim e^{-\Phi}\sim g_s^{-1}$. This is
just the behaviour we saw for the tension of the D$p$--brane, computed
in string perturbation theory, treating them as boundary conditions.

D$p$--branes have been treated so far largely as point--like (in their
transverse dimensions) in an otherwise flat spacetime, and we were
able to study an arbitrary number of them by placing the appropriate
Chan--Paton factors into amplitudes. However, the solutions
\reef{branes} have non--trivial spacetime curvature, and is only
asymptotically flat. How are these two descriptions related?

The point is as follows: For every D$p$--brane which is added to a
situation, another boundary is added to the problem, and so a typical
string diagram has a factor $g_sN$ since every boundary brings in a
factor $g_s$ and there is the trace over the $N$ Chan--Paton factors. So
perturbation theory is good as long as $g_sN<1$. Notice that this is the
regime where the supergravity solution \reef{branes} fails to be valid,
since the curvatures are high.  On the other hand, for $g_sN>1$, the
supergravity solution has its curvature weakened, and can be
considered as a workable solution.  This regime is where the
D$p$--brane perturbation theory, on the other hand, breaks down.

So we have a fruitful complementarity between the two descriptions. In
particular, since we are only really good at string perturbation
theory, {\it i.e.} $g_s<1$, for most computations, we can work with
the supergravity solution with the interpretation that $N$ is very
large, such that the curvatures are small. Alternatively, if one
restricts oneself to studying only the BPS sector, then one can work
with arbitrary $N$, and extrapolate results ----computed with the
D--brane description for small $g_s$--- to the large $g_s$ regime,
(since there are often non--renormalisation theorems which apply)
where they can be related to properties of the non--trivial curved
solutions. This is the basis of the successful statistical enumeration
of the entropy of black holes, for cases where the solutions
\reef{branes} are used to construct R--R charged black
holes.~\cite{sv,peet} This exciting subject will be described in the
notes of Amanda Peet.~\cite{peettasi}

In summary, for a large enough number of coincident D--branes or for
strong enough string coupling, one cannot consider them as points in
flat space: they deform the spacetime according to the geometry given
in eqn.~\reef{branes}.  Given that D--branes are also described very
well at low energy by gauge theories, this gives plenty of scope for
finding a complementarity between descriptions of non--trivially
curved geometry and of gauge theory. This is the basis of what might
be called ``gauge theory/geometry'' correspondences.  In some cases,
when certain conditions are satisfied, there is a complete decoupling
of the supergravity description from that of the gauge theory,
signalling a complete {\it duality} between the two.  This is the
basis of the AdS/CFT correspondence,~\cite{juan,agmoo} aspects of which are
described in the lectures of Mike Duff and others.

\subsection{Probing D--Branes' Geometry with D--Branes: $p$ with D$p$}

%\subby{probe like with like} 
In the last section, we argued that the spacetime geometry given by
equations \reef{branes} represents the spacetime fields produced by
$N$ D$p$--branes.  We noted that as a reliable (or ``trustworthy'')
solution to supergravity, the product $g_sN$ ought be be large enough
that the curvatures are small. This corresponds to either having $N$
small and $g_s$ large, or {\it vice--versa}. Since we are good at
studying situations with $g_s$ small, we can safely try to see if it
makes sense to make $N$ large.

One way to imagine that this spacetime solution came about at weak
coupling was that we built it by bringing in $N$ D$p$--branes, one by
one, from infinity. If this is to be a sensible process, we must study
whether it is really possible to do this. Imagine that we have been
building the geometry for a while, bringing up one brane at a time
from $r=\infty$ to $r=0$. Let us now imagine bringing the next brane
up, in the background fields created by all the other $N$ branes.
Since the branes share $p$ common directions where there is no
structure to the background fields, we can ignore those directions and
see that the problem reduces to the motion of a test particle in the
transverse $9-p$ spatial directions. What is the mass of this
particle, and what is the effective potential that it moves in?

This sort of question is answered by the still--developing toolbox
which combines the fact that we have a gauge theory on D--branes with
the fact that the probe brane is a heavy object which can examine many
distance scales, and has seen many applications in our understanding
of spacetime geometry in various
situations.~\cite{shenk2,short,shorty,probebranes,probeholes,seibergprobe}

We can derive the answers to all of the present questions by deriving
an effective Lagrangian for the problem which results from the
world--volume action of the brane. We can exploit the fact that we
have spacetime Lorentz transformations and world--volume
reparametrisations at our disposal to choose the work in the ``static
gauge''. In this gauge, we align the world--volume coordinates,
$\xi^a$, of the brane with the spacetime coordinates such that:
\begin{eqnarray}
&&\xi^0=x^0=t\ ;\nonumber\\
&&\xi^i=x^i\ ;\quad i=1\cdots p\ ,\nonumber\\
&&\xi^m=\xi^m(t)\ ;\quad m=p+1\cdots 9\ .
\label{staticguage}
\end{eqnarray}
The Dirac--Born--Infeld part of the action \reef{diracborninfeld}
requires the insertion of the induced metric derived from the metric
in question.  In static gauge, it is easy to see that the induced
metric becomes:
\begin{equation}
[G]_{ab}=\pmatrix{G_{00}+\sum_{mn} G_{mn}v_mv_n&0&0&\cdots&0\cr
0&G_{11}&0&\cdots&\vdots\cr
\vdots&\vdots&\ddots&\vdots&\vdots\cr
\vdots&\vdots&\vdots&\ddots&\vdots\cr
0&0&0&\cdots&G_{pp}}\ ,
\label{induce}
\end{equation}
where $v_m\equiv dx^m/d\xi^0={\dot x}^m$.

In our particular case of a simple diagonal metric, the determinant
turns out as
\begin{equation}
\det[-G_{ab}]=Z_p^{-{(p+1)\over2}}\left(1-Z_p\sum_{m=p+1}^9 v_m^2\right)
=Z_p^{-{(p+1)\over2}}\left(1-Z_pv^2\right)\ .
\end{equation}
The Wess--Zumino term representing the electric coupling of the brane
is, in this gauge:
\begin{eqnarray}
\mu_p\int C_{(p+1)}&=&\mu_p\int d^{p+1}\xi \,\,%\varepsilon^{a_0a_1\cdots a_p}
[C_{(p+1)}]_{\mu_0\mu_1...\mu_p}{\partial x^{\mu_0}\over\partial\xi^{a_0}}
{\partial x^{\mu_1}\over\partial\xi^{a_1}}
\cdots {\partial x^{\mu_p}\over\partial\xi^{a_p}}\nonumber\\
&=&\mu_pV_p\int dt \left[Z_p^{-1}-1\right]g^{-1}\ ,
\end{eqnarray}
where $V_p=\int d^p\!x$, the spatial world--volume of the brane.  Now,
we are going to work in the approximation that we bring the branes {\sl
  slowly} up the the main stack of branes so we keep the velocity $v$
small enough such that only terms up to quadratic order in $v$ are
kept in our computation. We can therefore the expand the square root
of our determinant, and putting it all together (not forgetting the
crucial insertion of the background functional dependence of the
dilaton from \reef{branes}) we get that the action is:
\begin{eqnarray}
S&=&\mu_pV_p\int dt \left(-g_s^{-1}Z_p^{-1}+
{1\over2g_s}v^2+ g_s^{-1}Z_p^{-1}-g_s^{-1}\right)\nonumber\\
&=&\int dt {\cal L}=\int dt\left( {1\over2}m_pv^2-m_p\right)\ ,
\label{particlemotion}
\end{eqnarray} 
which is just a Lagrangian for a free particle moving in a constant
potential, (which we can set to zero) where $m_p=\tau_p V_p$ is the
mass of the particle.
 
This result has a number of interesting interpretations. The first is
simply that we have successfully demonstrated that our procedure of
``building'' our geometry \reef{branes} by successively bringing branes
up from infinity to it, one at a time, makes sense: There is no
non--trivial potential in the effective Lagrangian for this process,
so there is no force required to do this; correspondingly there is no
binding energy needed to make this system.

That there is no force is simply a restatement of the fact that these
branes are BPS states, all of the same species. This manifests itself here
as the fact that the R--R charge is equal to the tension (with a
factor of $1/g_s$), saturating the BPS bound. It is this fact which
ensured the cancellation between the $r$--dependent parts in
\reef{particlemotion} which would have otherwise resulted in a
non--trivial potential $U(r)$. (Note that the cancellation that we saw
only happens at order $v^2$ ---the slow probe limit. Beyond that
order, the BPS condition is violated, since it really only applies to
statics.)

\subsection{The Metric on Moduli Space}

All of this has pertinent meaning from the point of view of field
theory as well. Recall that there is a $U(N)$ $(p+1)$--dimensional
gauge theory on a family of $N$ D$p$--branes. Recall furthermore that
there is a sector of the theory which consists of a family of $(9-p)$
scalars, $\Phi^m$, in the adjoint.  Geometrically, these are the
collective coordinates for motions of the branes transverse to their
world--volumes. Classical background values for the fields, (defining
vacua about which we would then do perturbation theory) are equivalent
to data about how the branes are distributed in this transverse space.
Well, we have just confirmed that there is in fact a ``moduli space''
of inequivalent vacua of the theory corresponding to the fact that one
can give a vacuum expectation value to a component of an $\Phi^m$
representing, representing a brane moving away from the clump of $N$
branes. That there is no potential translates into that fact that we
can place the brane anywhere in this transverse clump, and it will
stay there.

It is also worth noting that this metric on the moduli space is {\sl
  flat}; treating the fields $\Phi^m$ as coordinates on the space
$\IR^{9-p}$, we see (from the fact that the velocity squared term in
\reef{particlemotion} appears as $v^2=\delta_{mn}v^mv^n$) that the
metric seen by the probe is simply
\begin{equation}
ds^2\sim\delta_{mn}d\Phi^m d\Phi^n\ .
\labell{flatmetric}
\end{equation} 
This flatness is a consequence of the high amount of supersymmetry (16
supercharges). For the case of D3--branes (whether or not they are in
the AdS$_5\times S^5$ limit), this result translates into the fact
there that there is no running of the gauge coupling $g^2_{\rm YM}$ of
the superconformal gauge theory on the brane.  This is read off from
the prefactor $g_{\rm
  YM}^{-2}=\tau_3(2\pi\alpha^\prime)^2=(2\pi g_s)^{-1}$ in the metric. The
supersymmetry ensures that any corrections which could have been
generated are zero. We shall now see a less trivial version, where we
have a nontrivial metric in the case of eight supercharges.

\subsection{Probing D--Branes' Geometry with D--Branes: $p$ with  D$(p-4)$.} 
Let us probe the geometry of the $p$--branes with a D$(p-4)$--brane.
{}From our analysis of section \ref{peepee}, we know that this system is
supersymmetric. Therefore, we expect that there should still be a trivial
potential for the result of the probe computation, but there is not
enough supersymmetry to force the metric to be flat. There are
actually two sectors within which the probe brane can move
transversely. Let us choose static gauge again, with the probe aligned
so that its $p-4$ spatial directions $\xi^1-\xi^{p-4}$ are aligned
with the directions $x^1-x^{p-4}$. Then there are four transverse
directions {\sl within} the $p$--brane background, labelled
$x^{p-3}-x^{p}$, and which we can call $x_\parallel^i$ for short.
There are $9-p$ remaining transverse directions which are transverse
to the $p$--brane as well, labelled $x^{p+1}-x^9$ which we'll
abbreviate to $x_\perp^m$. The 6--2 case is tabulated as a visual
guide:
\bigskip
\begin{center}
\begin{tabular}{|c|c|c|c|c|c|c|c|c|c|c|}
\hline
&$x^0$&$x^1$&$x^2$&$x^3$&$x^4$&$x^5$&$x^6$&$x^7$&$x^8$&$x^9$\\\hline
D2--brane&$-$&$-$&$-$&$\bullet$&$\bullet$&$\bullet$&$\bullet$
&$\bullet$&$\bullet$&$\bullet$\\\hline
6--brane&$-$&$-$&$-$&$-$
&$-$&$-$&$-$&$\bullet$&$\bullet$&$\bullet$\\
\hline
\end{tabular}
\end{center}
\bigskip

Following the same lines of reasoning as above, the determinant which
shall go into our Dirac--Born--Infeld Lagrangian is:
\begin{equation}
\det[-G_{ab}]=Z_p^{-{(p-3)\over2}}\left(1-v_\parallel^2-Z_pv_\perp^2\right)\ ,
\end{equation}
where the velocities come from the time ($\xi^0$) derivatives of
$x_\parallel$ and $x_\perp$. This is nice, since in forming the action
by multiplying by the exponentiated dilaton factor and expanding in
small velocities, we get the Lagrangian
\begin{equation}
{\cal L}={1\over2}m_{p-4}\left(v_\parallel^2+Z_pv_\perp^2-2\right)\ ,
\end{equation}
which again has a constant potential which we can discard, and pure
kinetic terms. We see that there is a purely flat metric on the moduli
space for the motion inside the four dimensions of the $p$--brane
geometry, while there is a metric
\begin{equation}
ds^2=Z_p(r)\delta_{mn}dx^mdx^n\ ,
\end{equation}
for the transverse motion. This is the Coulomb branch, in gauge theory
terms, and the flat metric was on the Higgs branch. (In fact, the
Higgs result does not display all of the richness of this system that we
have seen. In addition to the flat metric geometry inside the brane
that we see here, there is additional geometry describing the
D$p$--D$(p-4)$ fields corresponding to the full instanton geometry. This
``Yang--Mills geometry'' comes from the fact that the D$(p-4)$--brane
behaves as an instanton of the non--abelian gauge theory on the
world--volume of the coincident D$p$--branes.)

Notice that for the fields we have studied, we obtained a trivial
potential for free without having to appeal to a cancellation due to
the coupling of the charge $\mu_{p-4}$ of the probe. This is good,
since there is no electric source of this in the background for it to
couple to. Instead, the form of the solution for the background makes
it force--free automatically.

\subsection{D2--branes and 6--branes: Kaluza--Klein Monopoles
 and M--Theory}
\label{d2probemagic}
Actually, when $p\geq5$, something interesting happens. The electric
source of $C_{(p+1)}$ potential in the background produces a {\sl
  magnetic} source of $C_{(7-p)}$. The rank of this is low enough for
there to be a chance for the D$(p-4)$--probe brane to couple to it
even in the Abelian theory. For example, for $p=5$ there is a magnetic
source of $C_2$ to which the D1--brane probe can couple. Meanwhile for
$p=6$, there is a magnetic source of $C_1$.  The D2--brane probes see
this in an interesting way. Let us linger here to study this case a
bit more closely. Since there is always a trivial $U(1)$ gauge field
on the world volume of a D2--brane probe, corresponding to the centre
of mass of the brane, we should include the coupling of the
world--volume gauge potential $A_a$ (with field strength $F_{ab}$) to
any of the fields coming from the background geometry.

In fact, as we saw before in section \ref{anomalousF} there is a
coupling
\begin{equation}
2\pi\alpha^\prime\mu_2 \int_M C_1 \wedge F\ ,
\end{equation}
where $C_1=C_\phi d\phi$ is the magnetic potential produced by the
6--brane background geometry, which is easily computed to be:
$C_\phi={-}(r_6/g_s)\cos\theta$, where $r_6=gN\alpha^{\prime 1/2}/2$.

This extra degree of freedom on the world volume is equivalent to one
scalar, since it comes from a gauge field in three dimensions. In our
computations we may exchange $A_a$ for a scalar $s$, by Hodge duality
in the (2+1)--dimensional world-volume. (This is of course a feature
specific to the $p{=}2$ case.)

To get the coupling for this extra scalar correct, we should augment
the probe computation. As we have seen, the Dirac--Born--Infeld action
is modified by an extra term in the determinant:
\begin{equation}
-{\rm det}g_{ab}\to-{\rm det}(g_{ab}+2\pi\alpha^\prime F_{ab})\ .
\end{equation}
We can~\cite{towndf,schmidhuber} introduce an auxiliary vector field
$v_a$, replacing $2\pi\alpha^\prime F_{ab}$ by
$e^{2\phi}\mu_2^{-2}v_av_b$ in the Dirac action, and adding the term
$2\pi\alpha^\prime\int_M F\wedge v$ overall. Treating $v_a$ as a
Lagrange multiplier, the path integral over $v_a$ will give the action
involving $F$ as before.  Alternatively, we may treat $F_{ab}$ as a
Lagrange multiplier, and integrating it out enforces
\begin{equation}
\epsilon^{abc}\partial_b(-\mu_2 {\hat C}_c+v_c)=0\ .
\end{equation}
Here, ${\hat C}_c$ are the components of the pullback of $C_1$ to the
probe's world--volume.  The solution to the constraint above is
 \begin{equation}
-\mu_2{\hat C}_a+v_a=\partial_a s\ ,
\end{equation} where $s$ is our dual scalar.
We may now replace $v_a$ by $\partial_a
s+\mu_2{\hat C}_a$ in the action.

The static gauge computation picks out only ${\dot s}+\mu_2C_\phi{\dot
  \phi}$, and recomputing the determinant gives
\begin{equation}
\det=Z_6^{-{3\over2}}
\left(1-v_\parallel^2-Z_6v_\perp^2-{Z_6^{1\over2}e^{2\Phi}\over\mu_2^2}
\left[{\dot s}+\mu_2C_\phi{\dot \phi}\right]^2\right)\ .
\end{equation}
Again, in the full Dirac--Born--Infeld action, the dilaton factor
cancels the prefactor exactly, and including the factor of $-\mu_2$
and the trivial integral over the worldvolume directions to give a
factor $V_2$, the resulting Lagrangian is
\begin{equation}
{\cal L}={1\over2}m_2(v_\parallel^2-2)+{1\over2}V_2
\left({\mu_2Z_6\over g}v_\perp^2
+{g_s\over \mu_2 Z_6}\left({\dot s}+\mu_2C_\phi{\dot\phi}\right)^2\right)\ ,
\end{equation}
which is (after ignoring the constant potential) again a purely
kinetic lagrangian for motion in {\sl eight} directions. There is a
non--trivial metric in the part transverse to both branes:
\begin{eqnarray}
&&ds^2= V(r)\left({d r}^2
 +r^2{d\Omega}^2 \right)
+V(r)^{-1}\left({d s}+A_\phi d\phi\right)^2\ ,\nonumber\\
\mbox{with}&& V(r)={\mu_2Z_6\over g_s}\quad \mbox{and}\quad 
 A={\mu_2r_6\over g_s}\cos\theta d\phi\ , 
\label{taubnut}
\end{eqnarray}
where ${d\Omega}^2={d\theta}^2+\sin^2\!\theta\,{d\phi}^2$.  There is a
number of fascinating interpretations of this result. In pure
geometry, the most striking feature is that there are now {\sl eleven}
dimensions for our spacetime geometry. The D2--brane probe computation
has uncovered, in a very natural way, an extra transverse dimension.
This extra dimension is compact, since $s$ is periodic, which is
inherited from the gauge invariance of the dual world--volume gauge
field.  The radius of the extra dimension is proportional to the
string coupling, which is also interesting.  This eleventh dimension
is of course the M--direction we saw earlier. The D2--brane has
revealed that the six--brane is a Kaluza--Klein monopole~\cite{kkmono}
of eleven dimensional supergravity on a circle,~\cite{town} which is
constructed out of a Taub--NUT geometry \reef{taubnut}. This fits very
well with the fact that the D6 is the Hodge dual of the D0--brane,
which we already saw is a Kaluza--Klein electric particle.

\subsection{The Metric on Moduli Space}
\label{muchado}

As before, the result also has a field theory interpretation. The
$(2+1)$--dimensional $U(1)$ gauge theory (with eight supercharges) on
the worldvolume of the D2--brane has $N_f=N$ extra hypermultiplets
coming from light strings connecting it to the $N_f=N$ D6--branes. The
$SU(N_f)$ symmetry on the worldvolume of the D6--branes is a global
``flavour'' symmetry of the $U(1)$ gauge theory on the D2--brane. A
hypermultiplet $\Psi$ has four components $\Psi_i$ corresponding to
the 4 scalar degrees of freedom given by the four positions
$\Psi^i\equiv(2\pi\alpha^\prime)^{-1}x_\parallel^i$. The vector
multiplet contains the vector $A_a$ and three scalars $\Phi^m\equiv
(2\pi\alpha^\prime)^{-1}x_\perp^m$.  The Yang--Mills coupling is
$g^2_{\rm YM}=g_s{\alpha^\prime}^{-1/2}$.

The branch of vacua of the theory with $\Psi\neq0$ is called the
``Higgs'' branch of vacua while that with $\Phi\neq0$ constitutes the
``Coulomb'' branch, since there is generically a $U(1)$ left unbroken.
There is a non--trivial four dimensional metric on the Coulomb branch.
This is made of the three $\Phi^m$, and the dual scalar of the $U(1)$'s
photon. Let us focus on the quantities which survive in the low energy
limit $\alpha^\prime\to0$ and hold fixed any sensible gauge theory
quantities which appear in our expressions.  (Such procedures will be
studied a lot in other lecture courses in this school).  The metric
which appears in \reef{taubnut} survives the limit as
\begin{eqnarray}
&&ds^2=V(U)(dU^2+U^2d\Omega^2_2)+V(U)^{-1}(d\sigma+A_\phi d\phi)^2\nonumber\\
&&\mbox{where}\quad
V(U)={1\over 4\pi^2g^2_{\rm YM}}\left(1+{g^2_{\rm YM}N_f\over 2U}\right)
\ ;\quad A_\phi={N_f\over8\pi^2}\cos\theta\ ,
\labell{tauby}
\end{eqnarray}
where $U=r/\alpha^\prime$ has the dimensions of an energy scale in the
gauge theory.  Also, $\sigma=\alpha^\prime s$, and we will
fix its period shortly.

In fact, the naive tree level metric on the moduli space is that on
$\IR^3\times S^1$, of form $ds^2=g^{-2}_{\rm YM}dx_\perp^2+g^2_{\rm
  YM}d\sigma^2$. Here, we have the tree level and one loop result:
$V(U)$ has the interpretation as the sum of the tree level and
one--loop correction to the gauge coupling of the 2+1 dimensional
gauge theory.~\cite{seibergprobe} Note the factor $N_f$ in the one loop
correction. This multiplicity comes from the number of hypermultiplets
which can run around the loop.  Similarly, the cross term from the
second part of the metric has the interpretation as a one--loop
correction to the naive four dimensional topology, changing it to the
(Hopf) fibred structure above.

Actually, the moduli space's dimension had to be a multiple of four,
as it generally has to be hyperK\"ahler for $D{=}2+1$ supersymmetry
with eight supercharges.~\cite{deezee} Our metric is indeed
hyperK\"ahler since it is the Taub--NUT metric: The hyperK\"ahler
condition on the metric in the form it is written is the by--now
familiar equation: $\nabla\times {\bf A}={ \nabla} V(U)$, which is
satisfied.

In fact, we are not quite done yet. With some more care we can
establish some important facts quite neatly. We have not been careful
about the period of $\sigma$, the dual to the gauge field, which is
not surprising given all of the factors of 2, $\pi$ and
$\alpha^\prime$. To get it right is an important %(but thankless)
task, which will yield interesting physics.  We can work it out in a
number of ways, but the following is quite instructive. If we perform
the rescaling $U=\rho /4g^2_{\rm YM}$ and $\psi=8\pi^2\sigma/N_f$, our
metric is:
\begin{eqnarray}
&&ds^2={g_{\rm YM}^2\over64\pi^2}\, ds^2_{\rm TN}\ ,\quad {\rm
  where}\nonumber\\
&&
ds^2_{\rm TN}=\left(1+{2N_f\over\rho}\right)(d\rho^2+\rho^2d\Omega^2_2)+
4N^2\left(1+{2N_f\over\rho}\right)^{-1}
(d\psi+\cos\theta d\phi)^2\ ,\nonumber\\
\labell{taubytwo}
\end{eqnarray}
which is a standard form for the Taub--NUT metric, with mass $N_f$,
equal to the ``nut parameter'' for this self--dual case.~\cite{hawk}
This metric is apparently singular at $\rho=0$, and in fact, for the
correct choice of periodicity for $\psi$, this pointlike structure,
called a ``nut'', is removable, just like the case of the bolt
singularity encountered for the Eguchi--Hanson space. (See insert 10, 
p.\pageref{insert10}.)  Just for fun, insert 14
(p.\pageref{insert14}) carries out the analysis and finds that $\psi$
should have period~$4\pi$, and so in fact the full $SU(2)$ isometry of
the metric is preserved.

\insertion{14}{Removing the ``Nut'' Singularity from
  Taub--NUT\label{insert14}}{The metric \reef{taubytwo} will be
  singular at at the point $\rho=0$, for arbitrary periodicity of $\psi$.
  This will be a pointlike singularity which is called a
  ``nut'',~\cite{nuts,egh} in contrast to the ``bolt'' we encountered
  for the Eguchi--Hanson space in insert 10 (p.\pageref{insert10}),
  which was an $S^2$.  In this case, near $\rho=0$, if we make the
  space look like the {\it origin} of $\IR^4$, we can make this
  pointlike structure into nothing but a coordinate singularity. Near
  $\rho=0$, we have:
$$
ds^2_{\rm TN}={2N_f\over\rho}\left(d\rho^2+\rho^2d\Omega^2_2+\rho^2
(d\psi+\cos\theta d\phi)^2\right)\ ,
$$
which is just the right metric for $\IR^4$ if $\Delta\psi=4\pi$,
the standard choice for the Euler coordinate. (This may have seemed
somewhat heavy--handed for a result one would perhaps have guessed
anyway, but it is worthwhile seeing it, in preparation for more
complicated examples.)}

What does this all have to do with gauge theory? Let us consider the
case of $N_f=1$, which means one six brane. This is 2+1 dimensional
$U(1)$ gauge theory with one hypermultiplet, a rather simple theory.
We see that after restoring the physical scales to our parameterers,
our original field $\sigma$ has period $1/2\pi$, and so we see that
the dual to the photon is more sensibly defined as
${\widetilde\sigma}=4\pi^2\sigma$, which would have period $2\pi$,
which is a more reasonable choice for a scalar dual to a photon. We
shall use this choice later.  With this choice, the metric on the
Coulomb branch of moduli space is completely non--singular, as should
be expected for such a simple theory.

Let us now return to arbitrary $N_f$. This means that we have $N_f$
hypermultiplets, but still a $U(1)$ 2+1 dimensional gauge theory with
a global ``flavour'' symmetry of $SU(N_f)$ coming from the
six--branes.  There is no reason for the addition of hypermultiplets
to change the periodicity of our dual scalar and so we keep it fixed
and accept the consequences when we return to physical coordinates
$(U,{\widetilde\sigma})$: {\it The metric on the Coulomb branch is
  singular at $U=0\,!$} This is so because insert 14 told us to give
$\widetilde\sigma$ a periodicity of~$2\pi N_f$, but we are keeping it
as $2\pi$. So our metric in physical units has $\widetilde\sigma$ with
period~$2\pi$ appearing in the combination $(2d{\widetilde\sigma}
+N_f\cos\theta d\phi)^2 $. This means that the metric is no longer has
an $SU(2)$ acting, since the round $S^3$ has been deformed into a
``squashed'' $S^3$, where the squashing is controlled by $N_f$. In fact,
there is a deficit angle at the origin corresponding to an $A_{N_f-1}$
singularity.

How are we to make sense of this singularity? Well, happily, this all
fits rather nicely with the fact that for $N_f>1$ there is an $SU(N_f)$
gauge theory on the sixbranes, and so { there is a Higgs
branch, corresponding to the D2-brane becoming an $SU(N_f)$ instanton!}
The singularity of the Coulomb branch is indeed a signal that we are
at the origin of the Higgs branch, and it also fits that there is no
singularity for $N_f=1$.

It is worthwhile carrying out this computation for the case of $N_f$
D6--branes in the presence of a negative orientifold 6--plane oriented
in the same way. In that case we deduce from facts we learned before
that the presence of the O6--plane gives global flavour group
$SO(2N_f)$ for $N_f$ D6--branes. The D2--brane however carries an
$SU(2)$ gauge group. This is T--dual to the earlier statement made in
section \ref{instantfun} about D9--branes in type~I string theory
carrying $SO(N_f)$ groups while D5's carry $USp(2M)$
groups.~\cite{edsmall,GP}: The orientifold forces a pair of D2--branes
to travel as one, with a $USp(2)=SU(2)$ group.

So the story now involves 2+1 dimensional $SU(2)$ gauge theory with
$N_f$ hypermultiplets. The Coulomb branch for $N_f=0$
must be completely non--singular, since again there is no Higgs branch to
join to. This fits with the fact that there are no D6--branes; just
the O6--plane. The result for the metric on moduli space can be
deduced from a study of the gauge theory (with the intuition gained
from this stringy situation), and has been proven to be the
Atiyah--Hitchin manifold.~\cite{atiyah,seibergprobe,SWtwo,valya} Some
of this will be discussed in more detail in
subsection~\ref{muchmoreado}. For the case of $N_f=1$, the result is
also non--singular (there is again no Higgs branch for 1 D6--brane)
and the result is a certain cover of the Atiyah--Hitchin
manifold.~\cite{atiyah,SWtwo}. The case of general $N_f$ gives certain
generalisations of the Atiyah--Hitchin manifold.~\cite{SWtwo,dancer}
The manifolds have $D_{N_f}$ singularities, consistent with the fact
that there is a Higgs branch to connect to. Note also that a sringy
interpretation of this result is that the strong coupling limit of
these O6--planes is in fact M--theory on the Atiyah--Hitchin manifold,
just like it is Taub--NUT for the D6--brane.~\footnote{It is amusing
  to note ---and the reader may bave already spotted it--- that the
  story above seems to be describing local pieces of K3, which has ADE
  singularities of just the right type, with the associated $SU(N)$
  and $SO(2N)$ enhanced gauge symmetries appearing also (global
  flavour groups for the 2+1 dimensional theory here). (The existence
  of three new exceptional theories, for $E_6, E_7, E_8$, is then
  immediate.~\cite{seibergprobe}.) What we are actually recovering is
  the fact~\cite{wit} that there is a strong/weak coupling duality
  between type~I (or $SO(32)$ heterotic) string theory on $T^3$ and
  M--theory on K3!!}

\subsection{When Supergravity Lies: Repulson Vs. Enhan\c con}
Despite the successes we have achieved in the previous section with
interpretation of supergravity solutions in terms of constituent
D--branes, we should be careful, even in the case when we have
supersymmetry to steer us away from potential pathologies.  It is not
always case that if someone presents us with a solution of
supergravity with R--R charges that we should believe that it has an
interpretation as being ``made of D--branes''.

Consider again the case of eight supercharges. We studied brane
configurations with this amount of supersymmetry by probing the
geometry of $N$ (large) D$p$--branes with a single D$(p{-}4)$--brane.
As described in previous sections, another simple way to achieve a
geometry with eight supercharges from D--branes is to simply wrap
branes on a manifold which already breaks half of the supersymmetry.
The example mentioned was the four dimensional case of K3. In this
case, we learned that if we wrap a D$(p{+}4)$--brane (say) on K3, we
induce precisely one unit of negative D$p$--brane charge~\cite{bsv}
supported on the unwrapped part of the worldvolume (see
eqn.\reef{wrapcharge}). At large $N$ therefore, we might expect that
there is a simple supergravity geometry which might be obtained by
taking the  solution for the D$(p+4)$--D$p$ system, and modifying
the asymptotic charges to suit this situation.  The resulting geometry
naively should have the interpretation as that due to a large number
$N$ of wrapped D$(p{+}4)$ branes:
\begin{eqnarray}
ds^2 &=& Z_2^{-1/2} Z_6^{-1/2} \eta_{\mu\nu} dx^\mu dx^\nu +
Z_2^{1/2} Z_6^{1/2} dx^i dx^i + V^{1/2} Z_2^{1/2} Z_6^{-1/2} ds^2_{\rm K3} \
,\nonumber
\\
e^{2\Phi } &=& g_s^2 {Z_p}^{(3-p)/2}{ Z_{p+4}}^{-(p+1)/2}\ , \nonumber\\
C_{\it (p+1)} 
&=& ({Z_p}^{-1}-1)g_s^{-1} 
dx^0 \wedge dx^1\wedge\cdots \wedge dx^{p+1} \nonumber\\
C_{\it (p+5)} &=& 
(Z_{p+4}^{-1}-1)g_s^{-1} dx^0 \wedge dx^1 \wedge \cdots\wedge dx^{p+5}
\ . \labell{peeplus}
\end{eqnarray}
Here, $\mu,\nu$ run over the $x^0-x^{p+1}$ directions, which are
tangent to all the branes. Also $i$ runs over the directions
transverse to all branes, $x^{p+2}-x^5$, and in the remaining
directions, transverse to the induced brane but inside the large
brane, $ds^2_{\rm K3}$ is the metric of a K3 surface of unit volume.
$V$ is the volume of the K3 as measured at infinity, but the
supergravity solution adjusts itself such that $V(r){=}VZ_p/Z_{p+4}$ is
the measured volume of the K3 at radius $r$.

Let us focus on the case $p=2$, where we wrap D6--branes to get
induced D2--branes.~\footnote{This will also teach us a lot about the
  pure $SU(N)$ gauge theory on the remaining 2+1 dimensional
  world--volume. Wrapping D7--branes ($p=3$) teaches us~\cite{jpp}
  about pure $SU(N)$ gauge theory in 3+1 dimensions, where we should
  make a connection to Seiberg--Witten theory at large
  $N$.~\cite{seibwitt,dougshenk}}
\bigskip
\begin{center}
\begin{tabular}{|c|c|c|c|c|c|c|c|c|c|c|}
\hline
&$x^0$&$x^1$&$x^2$&$x^3$&$x^4$&$x^5$&$x^6$&$x^7$&$x^8$&$x^9$\\\hline
D2&$-$&$-$&$-$&$\bullet$&$\bullet$&$\bullet$&$\bullet$
&$\bullet$&$\bullet$&$\bullet$\\\hline
D6&$-$&$-$&$-$&$\bullet$&$\bullet$&$\bullet$&$-$
&$-$&$-$&$-$\\\hline
K3&$-$&$-$&$-$&$-$&$-$&$-$&$\bullet$&$\bullet$&$\bullet$&$\bullet$
\\
\hline
\end{tabular}
\end{center}
\bigskip
The harmonic functions are
\begin{eqnarray}
Z_2 &=& 1+\frac{r_2}{r}\ ,\quad  r_2 =
-\frac{ (2\pi)^4 g_s N \alpha'^{5/2} }{ 2V } \ , \nonumber\\
Z_6 &=& 1+\frac{r_6}{r}\ ,\quad  r_6 = \frac{g_sN\alpha'^{1/2}}{2} \ ,
\end{eqnarray}
normalised such that the D2-- and D6--brane charges are
$Q_2{=}{-}Q_6=-N$. Note that the smaller brane is delocalised in the
K3 directions, as it should be, since the same is true of K3's
curvature.

We worked out the spectrum of type~IIA supergravity theory
compactified to six dimensions on K3 in subsection \ref{k3orbifold}. The six
dimensional supergravity theory has as an additional sector
twenty--four $U(1)$'s in the R--R sector. Of these, twenty--two come
from wrapping the ten dimensional two--form on the 19+3 two--cycles of
K3.  The remaining two are special $U(1)$'s for our purposes: One of
them arises from wrapping IIA's five--form entirely on K3, while the
final one is simply the plain one--form already present in the
uncompactified theory.

It is easy to see that there is something wrong with the geometry
which we have just written down, representing the wrapping of the
D6--branes on the K3.  There is a naked singularity at $r=|r_2|$,
known as the ``repulson'', since it represents a repulsive
gravitational potential,~\cite{repulson} as can be seen by scattering
test particles in to small enough $r$. The curvature diverges there
which is related to the fact that the volume of the K3 goes to zero
there, and the geometry stops making sense. Let us look carefully to
see if this is really the geometry produced by the branes.~\cite{jpp}

The object we have made should be a BPS membrane made of $N$ identical
objects. These objects feel no force due to each other's presence, and
therefore the BPS formula for the total mass is simply (see
eqn.\reef{wraptension})
\begin{equation}
\tau_N = \frac{N}{g_s} (  \mu_6 V -\mu_2 )  
\labell{massformula}
\end{equation}
with $\mu_6 = (2\pi)^{-6} \alpha'^{-7/2}$ and $\mu_2 = (2\pi)^{-2}
\alpha'^{-3/2}$. In fact, the BPS membrane is actually a monopole of
one of the six dimensional $U(1)$'s. It is obvious which $U(1)$ this
is; the diagonal combination of the two special ones we mentioned
above. The D6--brane component is already a monopole of the IIA R--R
one--form, and the D2 is a monopole of the five--form, which gets
wrapped. 

\ennbee{As we shall see, the final combination is a non--singular
BPS monopole, having been appropriately dressed~\cite{melaur} by the
Higgs field associated to the volume of K3. Also, it
maps~\cite{isstrings} (under the strong/weak coupling duality of the
type~IIA string on K3 to the heterotic string on
$T^4$)~\cite{hullt,wit,aspinwall} to a bound state of a Kaluza--Klein
monopole~\cite{kkmono} and an H--monopole~\cite{hmono}, made by
wrapping the heterotic NS5--fivebrane.~\cite{jpp,morten,ami}}

 If we are to interpret our geometry as having been
made by bringing together $N$ copies of our membrane, we ought to be
able to carry out the procedure we described in the previous sections.
We should see that the geometry as seen by the probe is
potential--free and well--behaved, allowing us the interpretation of
being able to bring the probe in from infinity.

The effective action for a D6--brane probe (wrapped on the K3) is:
\begin{equation}
S = - \int_{M}d^3\xi\, e^{-\Phi(r)} (\mu_6 V(r) - \mu_2)
(-\det{g_{ab}})^{1/2} + \mu_6 \int_{M \times\rm K3}\! C_{\it 7} -  \mu_2
\int_{M} C_{\it 3}\ . 
\labell{probeaction}
\end{equation}
Here $M$ is the projection of the world-volume onto the three non--compact
dimensions.  As discussed previously
(see eqn.\reef{wrapped} and surrounding discussion), the first term is
the Dirac--Born--Infeld action with the position dependence of the
tension~\reef{massformula} taken into account; in particular, $V(r)=V
Z_2(r)/Z_6(r)$.  The second and third terms are the couplings of the
probe charges $(\mu_6,-\mu_2)$ to the background R--R potentials,
following from eqn \reef{wrapcharge}, and surrounding discussion.

Having derived the action, the calculation proceeds very much as we
outlined in the previous sections, with the result:
\begin{eqnarray}
{\cal L} &=& - \frac{\mu_6 V Z_2 - \mu_2 Z_6}{Z_6 Z_2 g_s}  + \frac{\mu_6
V}{g_s}(Z_6^{-1} - 1) - \frac{\mu_2}{g_s} (Z_2^{-1} - 1)  \nonumber\\
&&\qquad\qquad\qquad\qquad{}+\frac{1}{2g_s}
(\mu_6 V Z_2 -
\mu_2 Z_6) v^2 + O(v^4)\ . \label{lag}
\end{eqnarray}
The position--dependent potential terms cancel as expected for a
supersymmetric system, leaving the constant potential $(\mu_6 V -
\mu_2)/g$ and a nontrivial metric on moduli space (the $O(v^2)$ part)
as expected with eight supersymmetries.  The metric is proportional to
\begin{equation}
ds^2={1\over g_s}\left(\mu_6 V Z_2 - \mu_2
  Z_6\right)dx_\perp^2 =  {\alpha'^{-3/2}\over (2\pi)^2 g} \Biggl(
\frac{V}{V_*} - 1 - \frac{g_s N 
\alpha'^{1/2}}{r}\Biggr)(dr^2+r^2d\Omega^2_2)\ .\label{kinetic}
\end{equation}
We assume that $V > V_* \equiv (2\pi)^4 \alpha'^2$, so that the metric
at infinity (and the membrane tension) are positive.  However, as $r$
decreases the metric eventually becomes negative, and this occurs at a
radius
\begin{equation}
r = \frac{2V}{{V} - V_* } |r_2| \equiv r_{\rm e}
\end{equation}
which is  greater than the radius $r_{\rm r} = |r_2|$ of the
repulson singularity.

In fact, our BPS monopole is becoming massless as we approach the
special radius. This should mean that the $U(1)$ under which it is
charged is becoming enhanced to a non--abelian group. This is the
case. There is a purely stringy phenomenon which lies outside the
W--bosons are wrapped D4--branes, which enhance the $U(1)$ to an
$SU(2)$. The masses of wrapped D4--branes is just like that of the
membrane, and so becomes zero when the K3's volume
reaches the value $V_*{\equiv}(2\pi\sqrt{\alpha^\prime})^4$.

The point is that the repulson geometry represents supergravity's best
attempt to construct a solution with the correct asymptotic charges.
In the solution, the volume of the K3 decreases from its asymptotic
value $V$ as one approaches the core of the configuration.  At the
centre, the K3 radius is zero, and this is the singularity.  This
ignores rather interesting physics, however. At a finite distance from
the putative singularity (where $V_{\rm K3}=0$), the volume of the K3
gets to $V{=}V_*$, so the stringy phenomena ---including new massless
fields--- giving the enhanced $SU(2)$ should have played a
role.~\footnote{Actually, this enhancement of $SU(2)$ is even less
  mysterious in the heterotic--on--$T^4$ dual picture mentioned two
  pages ago.~\cite{jpp} It is just the $SU(2)$ of a self--dual circle
  in this picture, which we studied extensively in
  section~\ref{selfdual}} So the aspects of the supergravity solution
near and inside the special radius, called the ``enhan\c con radius'',
should not be taken seriously at all, since it ignored this stringy
physics.

To a first approximation, the supergravity solution should only be
taken as physical down to the enhan\c con radius $r_{\rm e}$. That
locus of points, a two--sphere $S^{2}$, is itself called an ``enhan\c
con''.~\cite{jpp}

Note also that the {\it size} of the monopole is inverse to the mass
of the $W$ bosons, and so in fact by time our probe gets to the
enhan\c con radius, it has smeared out considerably, and in fact
merges into the geometry, forming a ``shell'' with the other monopoles
at that radius.  Since by this argument we cannot place sharp sources
inside the enhan\c con radius, evidently, and so the geometry on the
inside must be very different from that of the repulson. In fact, to a
first approximation, it must simply be flat, forming a smooth junction
with the outside geometry at $r=r_{\rm e}$.

In general, one expects the same sort of reasoning to apply for all
$p$, and so the enhan\c{c}on locus resulting from  wrapping a
D$(p+4)$--brane on K3 is $S^{4-p} \times R^{p+1}$, whose
interior is $(5+1)$--dimensional.  For even $p$ the theory in
the interior has an $SU(2)$ gauge symmetry, while for odd $p$ there is
the $A_1$ two--form gauge theory.  The details of the smoothing will be very
case dependent, and it should be interesting to work out those
details.

One can also study $SO(2N)$, $SO(2N{+}1)$ and $USp(2N)$ gauge theories
with eight supercharges in various dimensions using similar
techniques, placing an orientifold O6--plane into the system parallel
to the D6--branes. The enhan\c con then becomes an
$\mathbf{RP}^2$.~\cite{jj}

Note that the Lagrangian (\ref{lag}) depends only on three moduli
space coordinates, $(x^3,x^4,x^5)$, or $(r,\theta,\phi)$ in polar
coordinates.  As mentioned before, a (2+1) dimensional theory with
eight supercharges, should have a moduli space metric which is
hyperK\"ahler.~\cite{deezee} So we need at least one extra modulus,
$s$. A similar procedure to that used in section \ref{d2probemagic}
can be used to introduce the gauge field's correct couplings and
dualize to introduce the scalar $s$. A crucial difference is that one
must replace $2\pi\alpha^\prime F_{ab}$ by
$e^{2\phi}(\mu_6V(r)-\mu_2)^{-2}v_av_b$ in the Dirac--Born--Infeld
action, the extra complication being due to the $r$ dependent nature
of the tension. The static gauge computation gives for the kinetic
term:
\begin{equation}
{\cal L}=
F(r)
 \left({\dot r}^2
 +r^2{\dot\Omega}^2 \right)
+F(r)^{-1}\left({\dot s}/2-\mu_2C_\phi{\dot\phi}/2\right)^2\ ,
\labell{fourth}
\end{equation}
where
\begin{equation}
F(r)={Z_6\over 2g_s}\left(\mu_6V(r)-\mu_2\right)\ ,
\end{equation}
and ${\dot\Omega}^2={\dot\theta}^2+\sin^2\!\theta\,{\dot\phi}^2.$

\subsection{The Metric on Moduli Space}
\label{muchmoreado}
Again, there is gauge theory information to be extracted here. We have
pure gauge $SU(N)$ theory with no hypermultiplets, and eight
supercharges. We should be able to cleanly separate the gauge theory
data from everything else by taking the decoupling limit
$\alpha^\prime\to0$ while holding the gauge theory coupling $g^2_{\rm
  YM} = g^2_{{\rm YM},p}V^{-1}=(2\pi)^4g_s\alpha^{\prime 3/2} V^{-1}$
and the energy scale $U=r/\alpha^\prime$ (proportional to $M_W$)
fixed.  
In doing this, we get
the metric:
\begin{eqnarray}
&&ds^2= f(U) \left({\dot U}^2 +U^2{d\Omega}^2\right) +f(U)^{-1}
 \left({d\sigma} -{N\over{4\pi^2}}A_\phi{d\phi}\right)^2\ ,\nonumber\\
\mbox{where}&&
f(U)={1\over 4\pi^2 g^2_{\rm YM}}
\left(1-{ g^2_{\rm YM}N \over U}\right)\ ,
\labell{probenear}
\end{eqnarray}
the $U(1)$ monopole potential is $A_\phi=\pm1-\cos\theta,$ and
$\sigma=s{\alpha^\prime}$, and the metric is meaningful only for
$U{>}U_{\rm e} = \lambda$. This metric, which should be contrasted
with equation \reef{tauby}, is the hyperK\"ahler Taub--NUT metric, but
this time with a negative mass. This metric is singular, but the full
metric, obtained by instanton corrections to this one--loop result,
should be smooth, as we will discuss. 
The details of this smoothing will teach us more
about this $p=2$ case of the enhan\c con geometry and the
interpolation between the exterior supergravity solution and the
interior region, which is flat to leading order.

{}From the point of view of the monopole description, this manifold
should be related to the metric on the moduli space of monopoles. This
fits with the fact that the moduli space of the gauge theory and that
of the monopole problem are known to be
identified.~\cite{SWtwo,CH,hanany} It is clearly a submanifold of the
full $4N{-}4$ dimensional metric on the {\it relative} moduli
space~\cite{modulispace} of $N$ BPS monopoles which is known to be
smooth.~\cite{nakajima} For the problem of two monopoles, that moduli
space manifold\cite{manton} is the Atiyah--Hitchin
manifold,~\cite{atiyah} while for general $N$ it is more complicated.
This should remind the reader of our study in
subsection~\ref{muchado}. Recalling that this is also a study of
$SU(N)$ gauge theory with no hypermultiplets, we know the result for
$N=2$: The metric on the moduli space must be smooth, as there is no
Higgs branch to connect to {\it via} the singularity. This is true for
all $SU(N)$, and matches the monopole result. For $N=2$, we saw that
the metric on the moduli space is actually the Atiyah--Hitchin
manifold.

The structure of our particular four dimensional submanifold of the
general moduli space is very similar to that of an Atiyah--Hitchin
manifold, in fact!  To see this,~\cite{justme} change variables in our
probe metric (\ref{probenear}) by absorbing a factor of
$\lambda/2=g^2_{\rm YM}N/2$ into the radial variable $U$, defining
$\rho=2U/\lambda$.  Further absorb $\psi=\sigma 8\pi^2/N$ and gauge
transform to $A_\phi={-}\cos\theta$. Then we get:
\begin{eqnarray}
&&ds^2= {g^2_{\rm YM}N^2\over 32\pi^2}ds^2_{\rm TN-}\ 
,\quad{\rm with}\label{taubnutty}\\
&&ds^2_{\rm TN-}=\left(1-{2\over\rho}\right)
 \left({d \rho}^2 +\rho^2{d\Omega}^2\right) 
+4\left(1-{2\over\rho}\right)^{-1}
\left({d\psi}+\cos\theta{d\phi}\right)^2\ .
\nonumber
\labell{scaledTN}
\end{eqnarray}
The latter is precisely the form of the Taub--NUT metric that one gets
by expanding the Atiyah--Hitchin metric in large $\rho$ and neglecting
exponential corrections.\footnote{The reader should compare this
  result to that in equation \reef{taubytwo} to see that it is the
  case of $N=-1$, using the meaning that $N$ has in
  subsection~\ref{muchado}.}

Now for the same reasons as in subsection~\ref{muchado}, the
periodicity of $\sigma$ is $1/2\pi$, and we will use
$\widetilde\sigma=4\pi^2\sigma$ as our $2\pi$ periodic scalar dual to
the photon on the probe's world--volume. Looking at the choices we
made above, this implies that for the $SU(2)$ case, the
coordinate $\psi$ has period~$2\pi$~!  This is surprising (perhaps),
but does not lead to a ``nut'' singularity (see insert 14,
p.\pageref{insert14}) for the following reason: The nut would be at
$\rho=0$, but there is a more dangerous singularity already at
$\rho=2$. This new singularity is an artifact of a large $\rho$
expansion, however. There is a {\it unique} and completely
non--singular manifold whose metric is as asymptotically close to
$ds^2_{\rm TN-}$ up to exponential corrections, which is determined as
follows:

In this case of $N=2$, there is an $SO(3)=SU(2)/\IZ_2$ isometry in the
problem, and not the naive $SU(2)$ of the Taub--NUT space, since
$\psi$ has period $2\pi$ and not $4\pi$. This isometry, smoothness,
and the condition of hyperK\"ahlerity pick out {\it uniquely} the
Atiyah--Hitchin manifold as the completion of the negative mass
Taub--NUT and completes the story for the $SU(2)$ gauge theory moduli
space problem.~\cite{SWtwo} The Atiyah--Hitchin manifold can be
written in the following manifestly $SO(3)$ invariant
manner:~\cite{atiyah,gibbonsmanton}
\begin{eqnarray}
&&ds^2_{\rm AH}=f^2d\rho^2+
a^2\sigma_1^2+b^2\sigma_2^2+c^2\sigma_3^2\ ;\nonumber\\
{2bc\over f}{da\over d\rho}&=&(b-c)^2-a^2\ ,\mbox{ and cyclic perms.;}\quad 
\rho =2K\left(\sin{\beta\over2}\right),
\labell{atiyahhitchin}
\end{eqnarray} 
where the choice $f=-b/r$ can be made, the $\sigma_i$ are defined in
\reef{oneforms}, and $K(k)$ is the elliptic integral of the first
kind:
\begin{equation}
K(k)=\int_0^{\pi\over2}(1-k^2\sin^2\tau)^{1\over2}d\tau\ .
\labell{elliptic}
\end{equation}
Also, $k{=}\sin(\beta/2)$, the ``modulus'', runs from $0$ to $1$, so
$\pi\leq\rho\leq\infty$.

In fact, the solution for $a,b,c$ can be written out in terms of
elliptic functions, but we shall not do that here. It is enough to
note that when $\rho$ is large, the difference between this metric and
$ds^2_{\rm TN-}$ is exponentially small in~$\rho$. These exponential
corrections for smaller $\rho$ remove the singularity: $\rho=2$ is
just an artefact of the large $\rho$ metric in the above form
\reef{scaledTN}. 

The exponential corrections have the expected interpretation in the
gauge theory as the instanton corrections.~\cite{valya} Translating
back to physical variables, we see that these corrections go as
$\exp{(-U/g^2_{\rm YM})}$, which has the correct form of action for a
gauge theory instanton.  (We have just described a {\it cover} of the
Atiyah--Hitchin manifold needed for the $SU(2)$ case. There is an
additional identification to be discussed below.)

Can we learn anything from this for our case of general $N$, especially
for large $N$, to teach us about the enhan\c con geometry? We have to
be careful. Now, fixing our period of $\widetilde\sigma$ to be~$2\pi$
as before, for general $N$ the reulting period of $\psi$ in the scaled
variables is $\Delta\psi=4\pi/N$. Therefore our isometry is not
$SO(3)$ but $SU(2)/\IZ_N$. (So the boundary at infinity is the
squashed $S^3$, given by $S^3/\IZ_N$).

So the manifold we need is not quite the Atiyah--Hitchin manifold, but
probably a close cousin; as the Atiyah--Hitchin manifold goes once
around its $\psi$--circle, the manifold we need goes around $N/2$
times, and it is tempting to wonder if the manifold we seek is simply
a smooth quotient of it. It would be interesting to find this manifold
using requirements of uniqueness and smoothness. This manifod
certainly exists, given the data that we have presented from the point
of view of the gauge theory and the monopole physics.

Once we have found this manifold in scaled coordinates, we can then
rescale everything back to the original physical variables.  The rescaled
exponential corrections should be the gauge theory instanton
corrections which we expect, although for large $N$ they will be quite
small, and the dominant geometry will be that of the negative mass
Taub--NUT for a wide range of validity.~\cite{smallcorrect}

Even without precise knowledge of the manifold we seek, we can learn
much about our problem at large $N$: We are working on a very
symmetric subspace of the full $4N-4$ dimensional relative moduli
space of monopoles. The problem is of a large charge $N$ monopole
being approached by a small charge 1 monopole probe. The
Atiyah--Hitchin manifold ($N=2$) in standard variables used in
\reef{scaledTN} and \reef{atiyahhitchin} represents two charge 1
monopoles approaching one another from asymptotic large relative
separation $\rho$.  We can borrows some of the intuitive behaviour of
the two monopole case, some interpretation: For the two charge 1 case,
for small $\rho$ they begin to merge into a charge 2 monopole, and
$\rho$ no longer has distinct meaning as a separation.  The
singularity at $\rho=2$ is never reached, as it is an artifact of the
large $\rho$ expansion; instead $\rho=\pi$ is the case where the
monopoles are coincident. It is a removable ``bolt singularity'' in
the full Atiyah--Hitchin geometry, of exactly the type we saw in the
case of the Eguchi--Hanson space in insert 10 (p.\pageref{insert10}).

Actually, we have described a trivial cover of the true
Atiyah--Hitchin space. The two monopole problem has an obvious $\IZ_2$
symmetry coming from the fact that the monopoles are identical. Some
field configurations described by the manifold as described up to now
are overcounted, and so we must divide by this $\IZ_2$. The result is
that the bolt is an $\mathbf{RP}^2$ instead of an $S^2$. We will not
have such an identification for $N>2$.

Note that when we scale $\rho$ back to $U$, our coordinate $U$ (for
large $U$) is truly a radial coordinate, as one extremely heavy
monopole is at the centre, being probed by a charge 1 monopole.  The
generalisation of the Atiyah--Hitchin bolt then represents the place
of closest approach of the probe, where it has smoothed out.  This is
the smoothed, ``nonperturbative'' enhan\c con locus.

\section{D--Branes and Geometry III: Non--Commutativity}
\subsection{Open Strings with a Background B--Field}
Let us return briefly to where we started out. Writing down the open
string sigma model (in conformal gauge). Gathering together the
various pieces from the early chapters, we have:
\begin{equation}
  S={1\over 4\pi\alpha^\prime}\int_\Sigma
\! d^2\!\sigma \left\{\left(g^{ab}
G_{\mu\nu}(X)+\epsilon^{ab}B_{\mu\nu}(X)\right)
\partial_a X^\mu\partial_b X^\nu\right\}
+\int_{\partial\Sigma}
\! d\tau A_i(X)\partial_\tau X^i\ .
\labell{sigmasigma}
\end{equation}
We are going to focus on the case where we have some gauge field on
the world volume of a D$p$--brane, which has world--volume coordinates
$X^i$, for $i=0,\ldots,p$. Transverse coordinates are $X^m$, for
$m=p+1,\ldots,9$. We shall also have, as usual a trivial background
$G_{\mu\nu}=\eta_{\mu\nu};\,\,\Phi={\rm constant}$, and a constant
background $B$ field.  We can go and vary the action as we did before,
and we will find that again our $X^\mu$'s satisfy the 2D wave
equation, but we have slightly different boundary conditions at
$\sigma=0,\pi$:
\begin{equation}
  \partial_\sigma X^i+\partial_\tau X^j{\cal F}^i_j=0\ ,\quad %i,j=0,\ldots,p
  X^m=x_0^m\ ,%\quad m=p+1,\ldots,9
\labell{condits}
\end{equation}
where we have written the gauge invariant combination ${\cal
  F}=B+2\pi\alpha^\prime F$. The second part is the Dirichlet boundary
condition, fixing $x_0^m$ as the positions of the D$p$--brane.

\begin{figure}[ht]
\centerline{\psfig{figure=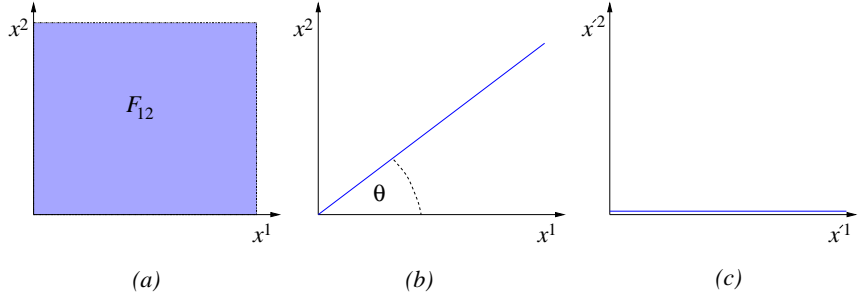,height=1.5in}}
\caption{A 
  brane in the 1--2 plane with a background field, {\it (a)}, is dual
  to a tilted brane of one extended dimension fewer {\it (b)}. It may
  then be rotated {\it (c)} to lie along a coordinate direction.  }
\label{tilted}
\end{figure}

Before going any further, it is worth trying to interpret the
modification to the Neumann boundary condition, in the light of what we
already know. Let us choose two directions in which there are non
trivial components of ${\cal F}$, let us say $X^1$ and $X^2$. So we
have either non--zero $B_{12}$ or $F_{12}$, or both. Then writing out
the condition, we have:
\begin{eqnarray}
&& \partial_\sigma X^1+\partial_\tau X^2 {\cal F}_2^1=0\ ;\\
&& \partial_\sigma X^2-\partial_\tau X^1 {\cal F}_2^1=0\ ,
\labell{mixed}
\end{eqnarray}
where we have used the fact that ${\cal F}$ is antisymmetric. Now, if we write 
${\cal F}_2^1=\cot\theta$, then we have 
\begin{eqnarray}
&&\phantom{-} \cos\theta \partial_\sigma X^1
+\sin\theta\partial_\tau X^2 =0\ ;\\
&& -\sin\theta\partial_\tau X^1+\cos\theta\partial_\sigma X^2 =0\ .
\labell{mixedtwo}
\end{eqnarray}
Now if we do a T--duality in the 2 direction, we exchange
$\partial_\sigma$ and $\partial_\tau$'s action on $X^2$.  Then we see
that we can rotate by an angle $\theta$ in the 1--2 plane, to new axes
$X^{\prime1},X^{\prime2}$ to get:
\begin{eqnarray}
 \partial_\sigma X^{\prime 1}=0\ ,\quad \partial_\tau X^{\prime 2}=0\ .
\labell{unmixed}
\end{eqnarray}
Now, $\partial_\tau X^a=0$ is not quite a Dirichlet condition in the
direction $X^a$, but nearly. Instead of saying that there is a
definite position $X^a=x_0^a$ that the string endpoint must be on, it
is in fact a definite statement about the conjugate momentum. So we
interpret this to mean that there is a Dirichlet condition, but that
the associated position has not been specified, and so it can be
anywhere in the direction $X^a$. So in fact, we have gone from a
D2--brane filling the $X^1,X^2$ directions to a D--brane lying along
the $X^{\prime1}$ direction (see figure \ref{tilted}{\it (c)}). Also,
before rotation, we see that $\partial_\sigma X^1
+\tan\theta\partial_\sigma X^2 =0$ is simply specifying that there be
a D1--brane lying at an angle $\theta$ in the 1--2 plane (See figure
\ref{tilted}{\it (b)}). We saw this in previous sections, but it is
worth repeating here. Furthermore, we can now look at the original
mixed condition \reef{mixed} and see that it is simply the
specification of a D2--brane lying in the 1--2 plane, but the presence
of ${\cal F}$ mixes in a D0--brane, but it is in fact completely
delocalized in the plane. We know that this must be true, since it is
only in that case that a D$p$--D$(p-2)$ combination can be
supersymmetric, and it also must be so in order to be T--dual to a
D1--brane.

Further consequences come when we try to carry out the line of
reasoning that we did in the early stages, in order to quantize the
theory.~\cite{chu} We can solve
the 2D wave equation with the boundary conditions \reef{mixed} to get
the general solution:
\begin{eqnarray}
&&X^i=x^i+2\alpha^\prime(p^i_0\tau-p^j_0{\cal F}^i_j 
\sigma)\nonumber\\
&&\hskip3cm +(2\alpha^\prime)^{1/2}
\sum_{n\neq0}{1\over n}e^{-in\tau}\left(i\alpha^i_n\cos n\sigma
-\alpha^j_n{\cal F}^i_j\sin n\sigma\right)\ ,\nonumber\\
&&
  X^m=x^m+Y^m 
\sigma+(2\alpha^\prime)^{1/2}
\sum_{n\neq0}{1\over n}e^{-in\tau}\alpha^i_n\sin n\sigma
\ ,
\labell{newmodes}  
\end{eqnarray}
We have included the possibility that there is more than one D--brane,
so that looking at $X^m$, it is clear that $Y^m$ is the separation
between the brane that the ends of the string rest on.  We will
henceforth assume that both ends of the string lie on the same brane,
and so $Y^m=0$. Also, in this case, we can rewrite our boundary term
in the action as a bulk term $(1/2)\int_\Sigma d^2\!\sigma\, 
\epsilon^{\alpha\beta} F_{ij}\partial_\alpha X^i\partial_\beta X^j$,
so that we see the appearance of $\cal F$ explicitly in the sigma
model action.

It is interesting to follow the route further. The canonical momenta
to $X^\mu$'s are
\begin{eqnarray}
\Pi^m={1\over 2\pi\alpha^\prime}\partial_\tau X^i\ ;\qquad
\Pi^i={1\over 2\pi\alpha^\prime }\left(\partial_\tau X^i
+\partial_\sigma X^j{\cal F}_j^i\right)
\ ,\labell{canons}
\end{eqnarray}
from which we can compute the total conserved momentum:
\begin{equation}
P^m_{\rm tot}=0\ ;\qquad 
P^i_{\rm tot}=\int_0^\pi d\sigma \Pi^i(\tau,\sigma)=p^j_0M_j^i\ ,  
\label{momentagain}
\end{equation}
where 
\begin{equation}
  M_{ij}=\eta_{ij}-{\cal F}_i^k{\cal F}_{kj}\ .
\label{theem}  
\end{equation}
The Hamiltonian is then, using equations (\ref{destiny},\ref{hammy}):
\begin{equation}
  H={1\over2}\left(M_{ij}p^i_0p^j_0+
\sum_{n\neq0}(M_{ij}\alpha_n^i\alpha_{-n}^j+\alpha^m_n\alpha^m_{-n})\right)\ ,
\labell{hamiltonagain}
  \end{equation}
  where we can see the non--trivial modification in the directions
  parallel to the brane, and nowhere else.
  
\subsection{Non--Commutative Geometry and D--branes}
Now the fun comes when we try to quantize.~\cite{chu} The first thing
to notice is that if we use the expression for the canonical momentum,
and the boundary condition, we can derive that:
\begin{equation}
2\pi\alpha^\prime \Pi^j(\tau,0){\cal F}_j^i+\partial_\sigma X^j(\tau,0)M_j^i=0
\ ,
  \labell{derive}
\end{equation}
so that, in particular
\begin{equation}
2\pi\alpha^\prime [\Pi^j(\tau,0),\Pi^k(\tau,\sigma)]
{\cal F}_j^i=-[\partial_\sigma X^j(\tau,0),,\Pi^k(\tau,\sigma)]M_j^i\ .
  \labell{cannot}
\end{equation}
But this is completely incompatible with our next step,~\cite{chu} which is to
try to impose the canonical commutation relations \reef{commute}.

This is a case where our naive quantisation procedures break down, as
happens in gauge theory when the gauge fixing condition is
incompatible with the canonical approach. Like that situation, one has
to use more careful methods, such as the constrained quantisation
techniques of Dirac. We will not that here, but state the result, and
refer the reader to the literature~\cite{chu} for the details. The equal
time commutators for the modes may be derived quite straightforwardly,
and then used to infer the relations on the spacetime fields
$X(\sigma,\tau),\Pi(\sigma,\tau)$:
\begin{eqnarray}
&&[\Pi^i(\tau,\sigma),\Pi^j(\tau,\sigma^\prime)]=0\ ,\quad 
[X^i(\tau,\sigma),X^j(\tau,\sigma^\prime)]=0\ ; {\sigma\neq\sigma^\prime}
\nonumber\\
&&[X^i(\tau,\sigma),\Pi^j(\tau,\sigma^\prime)]
=i\eta^{ij}{1\over\pi}\left(1+\sum_{n\neq0}
\cos n\sigma\cos n\sigma^\prime\right)\ ; 
\nonumber\\
&&[X^i(\tau,\sigma),X^j(\tau,\sigma^\prime)]=
e^{i\sigma}2i\pi\alpha^\prime(M^{-1}{\cal F})^{ij}\ ;\quad 
\sigma=\sigma^\prime=0,\pi
\labell{newrelations}
\end{eqnarray}

Now the remarkable thing is that the coordinates in the interior of
the string ({\it i.e.,} away from $\sigma=0,\pi$) satisfy the usual
canonical commutation relations, but at the ends of the string, we see
that there is actually some non--commutativity. For definiteness, let
us look at our case of just the $X^1,X^2$ plane again, and we see
that, at the string endpoints, if we set only the spatial parts of
${\cal F}$ to be non--zero:
\begin{equation}
[X^1,X^2]=2\pi i\alpha^\prime {{\cal F}\over1+{\cal F}^2}\ ,
  \labell{noncommute}
\end{equation}
which is quite remarkable. Notice that we can have non--commutativity in
time as well, if we turn on components of ${\cal F}$ in the time
direction. Note that these ``electric'' components correspond to a
boosted D--brane in the T--dual picture.

It is worth remarking that although this seems a bit strange, it is
again just ordinary string theory looked at in a different way. Note
that the rest of the studies we did in early sections go through. For
example, the imposition of diffeomorphism invariance will still allow
us to derive Virasoro generators:
\begin{equation}
L_m={1\over 2}\sum_{n=-\infty}^\infty\left(M_{ij}
\alpha^i_{m-n}\alpha^j_n+\alpha_{m-n}\cdot\alpha_{n}\right)\ ,
  \labell{virasoroagain}
\end{equation}
where $\alpha_{l}\cdot\alpha_{k}\equiv \alpha^m_l\alpha_k^m$, the dot
product in the transverse directions. After the standard normal
ordering passing to the quantum story, it can be shown that they
satisfy the usual Virasoro algebra.~\cite{chu}

One last thing to notice is the fact that the mass spectrum may appear
rather puzzling now. If we again follow the standard route, asking
$L_0-1$ to vanish, we will get a formula
\begin{equation}
M_{ij}p_0^ip_0^j+\sum_{n=1}^\infty\left(M_{ij}
\alpha^i_{m-n}\alpha^j_n+\alpha_{m-n}\cdot\alpha_{n}\right)-1=0\ .
  \labell{massform}
\end{equation}
The question is what to take as the definition of the mass. If we use
the usual definition $M^2=-p^ip_i$, then we will have not only
discrete contributions to the mass spectrum coming from the
oscillators, but we will have continuous pats as well, coming from
non--zero parts ${\cal F}$ in $M$. So we have a choice. We can either
interpret this as a new feature of the string, or we can take the
simpler approach and interpret all continuous parts as coming from the
string having been streched. In other words, defining
$M^2=-M_{ij}p^ip^j$ measures correctly the length of the string such
that all other contributions to the mass spectrum are from the
discrete oscillation energies. So to measure the length of our vector
$p_0^i$, we used a natural metric associated to the open string in the
presence of non--zero ${\cal F}$ which is different from $\eta_{ij}$,
the metric that closed strings see. $M_{ij}$ is often called the
``open string metric'' in this context, and this is the reason why.

So we see that the spacetime coordinates on a D--brane in the presence
of non--zero $\cal F$ are actually non--commutative.~\cite{cds,chu}
This makes a lot of sense, given our picture which we built up in the
previous section: When ${\cal F}=0$, the endpoints of the string are
instructed to simply end on the D$p$--brane, but for non--zero ${\cal
  F}$ there are D$(p-2)$--branes in the world--volume, but totally
delocalised, since its presence is specified by a definite condition
on momentum and not position. So the location of the string endpoints,
in as much as they now make any definite sense, necessarily inherit an
admixture of this delocalisation, taking on some of the
characteristics of momentum, resulting in non--commutativity.

Many of the pieces of physics which we have investigated so far are
worth revisiting in this light, and it might be worth keeping this
picture in mind when the non--commutativity seems hard to accept. In
particular, this means that the $\alpha^\prime\to0$ limit of the open
string sector should give Yang--Mills theory on non--commutative
spacetimes.  This has the amusing and sometimes confusing feature of
endowing even Abelian Yang--Mills theory with non--commutative
features.  Gauge theories on non--commutative backgrounds is a subject
of intense research at the time of writing. \footnote{That comment
  serves as a signal to the reader to be prepared to encounter the
  subject. I will not attempt to give any citations for this rapidly
  developing area, as I will not be able to truly representative since
  that subject is beyond the scope of these notes.}

\subsection{Yang--Mills Geometry I: D--branes and the Fuzzy Sphere}

In our many studies of the geometry seen by D--branes throughout these
lectures, we kept using the idea that the spacetime coordinates
transverse to the brane appear as scalar fields in a gauge theory on
the brane's world--volume. The vevs of these fields give the allowed
positions of the brane in the spacetime, and so on. This allows for a
remarkably rich dialogue between geometrical techniques and those of
gauge theory.

When there are many branes, however, we know that the gauge theory
becomes non--Abelian. This immediately leads to the idea~\cite{edbound}
that this description forces us to consider non--commutative geometry
in our spacetime, since the fields which we wish to interpret as
coordinates have failed to commute. 

This leads to non--commutative geometry of a naively different type
from that which we encountered in the previous section, and there is
potential for confusion. There really should not be. As we proceed
with this process of blurring the distinction between descriptions of
spacetime and other structures like string theory and gauge theory,
the idea of non--commutative geometry as a natural language will arise
again and again. One envisions it as something like the concept of the
derivative: Differential calculus arises in many different situations,
some of which are connected and some not. We do not search for deep
connections for too long, but just see it as a tool and move on,
knowing that being too philosophical about it is not necessarily very
useful as a pursuit in itself; one expects that the same will be true
of how we will regard the various situations where ``geometry'' has
some degree of non--commutativity.

For the purposes of these notes, however, and because some readers
might be trying to sort out the similarities and differences between
these situations at a learning stage~\footnote{I certainly am.}, I
will call the non--commutative geometry in this section ``Yang--Mills
Geometry'' and hope that this term is not too confusing.

The most familiar non--Abelian term which shows that there is
something interesting to occur is of course the familiar scalar
potential of the Yang--Mills theory.  This of course appears in the
Yang--Mills theory in the usual way, and can be thought of as
resulting from the reduction of the ten dimensional Yang--Mills
theory. It also arises as the leading part of the expansion of the
$\det(Q^i_j)$ term in the non--abelian Born--Infeld action, in the
case when the brane is embedded in the trivial flat background
$G_{\mu\nu}=\eta_{\mu\nu}$, as discussed in section
\ref{yangmillsstuff}:
\begin{equation}
V=\tau_p\,\Tr\sqrt{\det(Q^i{}_j)}= N\tau_p+{\tau_p(2\pi\alpha^\prime)^2\over4}
\Tr([\Phi^i,\Phi^j]\,[\Phi^j,\Phi^i])+\ldots\ ,
\labell{commpot}
\end{equation}
where $i=p+1,\ldots,9$.  As we have discussed in a number of cases
before, the simplest solution extremising $V$ is that the $\Phi^i$ all
commute, in which case we can write them as diagonal matrices
$\Phi^i=(2\pi\alpha^\prime)^{-1} X^i$, where $X^i={\rm
  diag}(x^i_1,x^i_2,\ldots,x^i_N)$. The interpretation is that $x^i_n$
is the coordinate of the $n$th D$p$--brane in the $X^i$ direction; we
have $N$ parallel flat D$p$--branes, identically oriented, at
arbitrary positions in a flat background, $\IR^{9-p}$. The centre of
mass of the D$p$--branes is at $x^i_0=\Tr(X^i)/N$.  The potential is
$N\tau_p$, which is simply the sum of all of the rest energies of the
branes. We shall discard it in much of what follows.

When we look  for situations with non--zero
commutators, things become more complicated in interesting ways,
giving us the possibility of new interesting extrema of the potential
in the presence of non--trivial backgrounds. This is because the
commutators appear in many parts of the worldvolume action, and in
particular appear in couplings to the R--R fields, as we have seen in
section¬\ref{nonablestuff}. Furthermore, the background fields
themselves depend upon the transverse coordinates $X^i$ even in the
abelian case, and so will depend upon the full $\Phi^i$ in the
non--abelian generalisation.

In general, this is all rather complicated, but we shall focus on one
of the simpler cases as an illustration of the rich set of physical
phenomena waiting to be uncovered.~\cite{robdielectric} Imagine that we
have $N$ D$p$--branes in a constant background R--R $(p+4)$--form
field strength $G_{(p+4)}=dC_{(p+3)}$, with non--trivial components:
\begin{equation}
G_{01...pijk}\equiv G_{tijk}=\matrix{-2f \varepsilon_{ijk}& i,j,k\in
 \{1,2,3\}}
\labell{RRbackgrnd}
\end{equation}
(We have suppressed the indices $1...p$, as there is no structure
there, and will continue to do so in what follows.)  Let the
D$p$--brane be pointlike in the directions $x^i$, $(i=1,2,3)$, and
extended in $p$ other directions.  None of these D$p$--branes in
isolation is an electric source of this R--R field strength.  Recall
however, that there is a coupling of the D$p$--branes to the R--R
$(p+3)$--form potential in the non--Abelian case, as shown in
\reef{nonabeliancs}.  We will assume a static configuration, choose
static gauge
\begin{equation}
\zeta^0=t\ ,\quad\zeta^\mu=X^\mu\ ,\quad \mbox{ for  } \mu=1,\ldots,p\ ,
\end{equation}
and get (see \reef{nonabeliancs}):
\begin{eqnarray}
&&(2\pi\alpha^\prime)\mu_p\int \Tr\, P\left[{\rm i}_\Phi {\rm i}_\Phi
C\right]=\nonumber\\ &&\quad =(2\pi\alpha^\prime)\mu_p\int dt
\Tr\left[\Phi^j\Phi^i\left( C_{ijt}(\Phi,t)+(2\pi\alpha^\prime)
C_{ijk}(\Phi,t)\,D_t\Phi^k\right)\right]\ .
\end{eqnarray}
We can now do a ``non--Abelian Taylor
expansion''~\cite{myersscatter,kabattaylor} of the background field
about $\Phi^i$. Generally, this is defined as:
\begin{equation}
F(\Phi^i)=\sum_{n=0}^\infty {(2\pi\alpha^\prime)^n\over n!}
\Phi^{i_1}\cdots\Phi^{i_n}\partial_{x^{i_1}}\cdots
\partial_{x^{i_1}}F(x^i)|_{x=0}\
.
\end{equation}
and so:
\begin{equation}
C_{ijk}(\Phi,t)=C_{ijk}(t)+(2\pi\alpha^\prime)
\Phi^k\partial_kC_{ijk}(t)+{(2\pi\alpha^\prime)^2\over2}
\Phi^l\Phi^k\partial_l\partial_kC_{ijk}(t)+\ldots
\labell{interinter}
\end{equation}
Now since $C_{ijt}(t)$ does not depend on $\Phi^i$, the quadratic
term containing it vanishes, since it is antisymmetric in $(ij)$ and
we are taking the trace. This leaves the cubic parts:
\begin{eqnarray}
&&(2\pi\alpha^\prime)^2\mu_p\int dt\,\Tr\left(\Phi^j\Phi^i\left[
\Phi^k\partial_kC_{ijt}(t)+C_{ijk}(t)\,D_t\Phi^k\right]\right)
\nonumber\\ &&\qquad={1\over3}(2\pi\alpha^\prime)^2\mu_p\int
dt\,\Tr\left(\Phi^i\Phi^j\Phi^k\right) G_{tijk}(t)\ ,
\labell{interaction}\end{eqnarray} after an integration by
parts. Note that the final expression only depends on the gauge
invariant field strength, $G_{(p+4)}$. Since we have chosen it to be
constant, this interaction \reef{interaction} is the only term that
need be considered, since of the higher order terms implicit in
equation~\reef{interinter} will give rise to terms depending on derivatives
of $G$.

Combining equation~\reef{interaction} with the part arising in the
Dirac--Born-Infeld potential \reef{commpot} yields our effective
Lagrangian in the form $S=-\int dt {\cal L}$. This is a static
configuration, so there are no kinetic terms and so ${\cal L}=-V(\Phi)$,
with
\begin{equation}
V(\Phi)=-{(2\pi\alpha^\prime)^2\tau_p\over4}\Tr([\Phi^i,\Phi^j]^2)
-{1\over3}(2\pi\alpha^\prime)^2\mu_p\Tr\left(\Phi^i\Phi^j\Phi^k\right)
G_{tijk}(t)\ .
\labell{thepotential}
\end{equation}
Let us substitute our choice of background field \reef{RRbackgrnd}. The
Euler--Lagrange equations $\delta V(\Phi)/\delta\Phi^i=0$ yield
\begin{equation}
[[\Phi^i,\Phi^j],\Phi^j]+f\varepsilon_{ijk}[\Phi^j,\Phi^k]=0\ .
\labell{euler}
\end{equation}
Now of course, the situation of $N$ parallel static branes,
$[\Phi^i,\Phi^j]=0$ is still a solution, but there is a far more
interesting one.~\cite{robdielectric} In fact, the non--zero
commutator:
\begin{equation}
[\Phi^i,\Phi^j]=f\,\varepsilon_{ijk}\Phi^k\ ,
\labell{noncomm}
\end{equation}
is a solution.  In other words we can choose
\begin{equation}
\Phi^i=-{\rm i}\,{f\over2}\,\Sigma^i
\labell{fuzzy}
\end{equation}
where $\Sigma^i$ are any $N\times N$ matrix representation of
the $SU(2)$ algebra
\begin{equation}
[\Sigma^i,\Sigma^j]=2{\rm i}\,\epsilon_{ijk}\,\Sigma^k\ .
\labell{algebra}
\end{equation}
The $N\times N$ irreducible representation of $SU(2)$ has
\begin{equation}
(\Sigma^i)^2={1\over3}(N^2-1) {\bf I}_{N\times N} \quad{\rm for}\ i=1,2,3.
\labell{trace}
\end{equation}
where ${\bf I}_{N\times N}$ is the identity.
Now the value of the potential \reef{thepotential}
for this  solution is
\begin{equation}
V_{\rm N}=-{\tau_p(2\pi\alpha^\prime)^2f^2\over6}\sum_{i=1}^3\Tr[(\Phi^i)^2]
=-{(2\pi)^{-p+2}\alpha^{\prime{3-p\over2}}f^4\over12g}N(N^2-1)\ .
\labell{potvalue}
\end{equation}
So our noncommutative solution solution has {\it lower energy} than
the commuting solution, which has $V=0$ (since we threw away the
constant rest energy). This means that the configuration of separated
D$p$--branes is unstable to collapse to the new configuration.

What is the geometry of this new configuration?  Well, the $\Phi$'s
are the transverse coordinates, and so we should try to understand
their geometry, despite the fact that they do not commute. In fact,
the choice \reef{fuzzy} with the algebra \reef{algebra} is that
corresponding to the non--commutative or ``fuzzy''
two--sphere~\cite{madore}.  The radius of this sphere is given by
\begin{equation}
R_N^2=(2\pi\alpha^\prime)^2{1\over N}\sum_{i=1}^3\Tr[(\Phi^i)^2]=
\pi^2\alpha^{\prime2} f^2(N^2-1)\ ,
\labell{radius}
\end{equation}
and so at large $N$: $R_N\simeq\pi\alpha^\prime f N$. The fuzzy sphere
construction may be unfamiliar, and we refer the reader to the
references for the details.~\cite{madore}  It suffices to say that as
$N$ gets large, the approximation to a smooth sphere improves.

Note that the {\it irreducible} $N\times N$ representation is not the
only solution.  A reducible $N\times N$ representation can be made by
direct product of $k$ smaller irreducible representations. Such a
representation gives a $\Tr[(\Sigma^i)^2]$ which is less than that for
the irreducible representation \reef{noncomm}, and therefore yields
higher values for their corresponding potential.  Therefore, these
smaller representations representations, corresponding geometrically to
smaller spheres, are unstable extrema of the potential which again
would collapse into the single large sphere of radius $R_N$.  It is
amusing to note that we can adjust the solution representing an sphere
of size $n$ by
\begin{equation}
\Phi^i=-{\rm i}{f\over2}\,\Sigma_{n}^i + x^i \,{\bf I}_{n\times n}\ .
\end{equation}
This has the interpretation of shifting the position of its centre of
mass by $x^i$.

What we have constructed is a D$(p+2)$--brane with topology
$\IR^p\times S^2$. The $\IR^p$ part is where the $N$ D$p$--branes are
extended and the $S^2$ is the fuzzy sphere. There is no net
D$(p+2)$--brane charge, as each infinitesimal element of the
spherical brane which would act as a source of $C_{(p+3)}$ potential
has an identical oppositely oriented (and hence oppositely charged)
partner. There is therefore a ``dipole'' coupling due to the
separation of these oppositely oriented surface elements.  This type
of construction is useful in matrix theory, where one can construct
for example, spherical D2--brane backgrounds in terms of $N$
D0--branes variables.~\cite{kabattaylor,rey,dewit}

One way~\cite{roberto,robdielectric} to confirm that we have made a spherical
brane at large $N$, is to {\it start} with a spherical
D$(p+2)$--brane, (topology $\IR^p\times S^2$) and bind $N$
D$p$--branes to it, aligned along an $\IR^p$. We can then place it in
the background R--R field we first thought of and see if the system
will find a static configuration keeping the topology $\IR^p\times
S^2$, with radius $R_N$. Failure to find a non--zero radius as a
solution of this probe problem would be a sign that we have not
interpreted our physics correctly.

Let us write the ten dimensional flat space metric with spherical
polar coordinates on the part where the sphere is to be located
($x^1,x^2,x^3$):
\begin{equation}
ds^2=-dt^2+dr^2+r^2\left(d\theta^2+\sin^2\theta\,d\phi^2\right)
+\sum_{i=4}^9(dx^i)^2\ .
\labell{ourmetric}
\end{equation}
Our constant background fields  in these coordinates is (again,
suppresing the $1,\ldots,p$ indices):
\begin{equation}
G_{tr\theta\phi}=-2fr^2\,\sin\theta\ \quad\mbox{and so   }\ 
C_{t\theta\phi}={2\over3}fr^3\,\sin\theta\ .
\labell{background}
\end{equation}
As we have seen many times before, $N$ bound D$p$--branes in the
D$(p+2)$--brane's worldvolume corresponds to a flux due to the
coupling:
\begin{equation}
(2\pi\alpha^\prime)\mu_{p+2}\int_{{\cal M}^3} C_{(p+1)}\wedge F 
= {\mu_p\over 2\pi}
\int\! dt\,C_{(p+1)}\wedge F \ ,
\end{equation}
where $C_{(p+1)}$ is the R--R potential to which the D$p$--branes
couple, and is not to be confused with the $C_{(p+3)}$ we are using in our
background, in \reef{background}.  We need exactly $N$ D$p$--branes, so
let us determine what $F$--flux we need to achieve this. If we work
again in static gauge, with the D$(p+2)$--brane's world--volume
coordinates in the interesting directions being:
\begin{equation}
\zeta^0=t\ ,\quad\zeta^1=\theta\ ,\quad\zeta^2=\phi\ ,
\end{equation}
then
\begin{equation}
F_{\theta\phi}={N\over 2}\sin\theta\ ,
\labell{flux}
\end{equation}
is correctly normalised magnetic field to give our desired flux.

We now have our background, and our $N$ bound D$p$--branes, so let us seek
a static solution of the form
\begin{equation}
r=R\quad \mbox{and $x^i=0$}\ , \,\,\mbox{for $i=4,\ldots,9$}\ .
\labell{sphere}
\end{equation}
The world volume action for our D$(p+2)$--brane is:
\begin{equation}
S=-\tau_{p+2}\int \!dt\,d\theta\, d\phi
\,e^{-\Phi}{\det{}^{1\over2}(-G_{ab}+2\pi\alpha^\prime F_{ab})} +\mu_{p+2}\int
C_{(p+3)}\ .
\end{equation}
Assuming that we have the static trial solution \reef{sphere},
inserting the fields \reef{background}, a trivial dilaton, and the
metric from \reef{ourmetric}, the potential energy is:
\begin{eqnarray}
V(R)&=&-\int\! d\theta\, d\phi\, {\cal L} \nonumber\\ &=&4\pi
\tau_{p+2}\left(\left[R^4+{(2\pi\alpha^\prime)^2 N^2\over
4}\right]^{\ha}-{2f\over3}R^3\right) \nonumber\\ &=&{ N} \tau_p
+{2\tau_p\over(2\pi\alpha^\prime)^2{N}}R^4
-{4\tau_p\over3(2\pi\alpha^\prime)}fR^3+\ldots\ . 
\labell{potty}
\end{eqnarray}
In the above we expanded the square root assuming that 
$2R^2/(2\pi\alpha^\prime){\rm N}<<1$, and kept the first two terms
in the expansion.  As usual we have substituted
$\tau_p=4\pi^2\alpha^\prime \tau_{p+2}$.

The constant term in the potential energy corresponds to the rest
energy of $N$ D$p$--branes, and we discard that as before in order to
make our comparison. The case $V=0$ corresponds to $R=0$, the solution
representing flat D$p$--branes.  Happily, there is another extremum:
\begin{eqnarray}
&&R=R_N=\pi\alpha^\prime f N\quad{\rm with}\ 
V=-{(2\pi)^{-p+2}\alpha^{\prime{3-p\over2}}f^4\over12g_s}
N^3\ .
\nonumber
\end{eqnarray}
To leading order in $1/N$, we see that we have recovered the radius
(and potential energy) of the non--commutative sphere configuration
which we found in equations \reef{radius} and \reef{potvalue}.

As noted before, this spherical D$(p+2)$--brane configuration carries
no net D$(p+2)$--brane charge, since each surface element of it has an
antipodal part of opposite orientation and hence opposite charge. However,
as the sphere is at a finite radius, there is a finite dipole
coupling.

This is the D--brane analogue~\cite{robdielectric} of the dielectric
effect in  electromagnetism.  If we place D$p$--branes in a
background R--R field under which the D$p$--branes would normally be
regarded as neutral, the external field ``polarises'' the
D$p$--branes, making them puff out into a (higher dimensional)
non--commutative world--volume geometry.  Just as in electromagnetism,
where an external field may induce a separation of charges in neutral
materials, the D--branes respond through the production of electric
dipole and possibly higher multipole couplings {\it via} the non--zero
commutators of the world--volume scalars.\footnote{Note that this very
  fuzzy sphere geometry arises for branes in the background NS--NS
  field $B_{(2)}$,\cite{alekseev} further illustrating the already
  noted artificiality of distinguishing the two types of
  non--commutative geometry discussed in this and the previous
  subsection!}

There is clearly a rich set of physical phenomena to be uncovered by
considering non--commuting $\Phi$'s. Already there have been
applications of this mechanism to the understanding of a number of
systems, such as large $N$ gauge theory in the AdS/CFT
correspondence.~\cite{adsdielectric}

\subsection{Yang--Mills Geometry II: Enhan\c cons and Monopoles}

As a final example of how ``Yang--Mills'' non--commutative geometry
naturally arises, let us return to our study of the enhan\c con.
There, we we probed the metric geometry of the $N$ D6 branes wrapped
on K3, and found that the true geometry deviates from the naive
geometry due to stringy effects invisible in supergravity. The
deviations were consistent with the fact that the probe was actually a
magnetic monopole of one of the $U(1)$'s of the six dimensional
theory. At a special radius the $U(1)$ gets enhanced to $SU(2)$ and
the monopole becomes massless. Crucially for our concerns here, the
monopole also stops being pointlike, and begins to spread out.

If this is the case, then in the light of what we have learned, it
should mean that the world--volume fields describing the transverse
coordinates of the wrapped brane must have become non--commutative
describing their smearing.  Does there exist a useful description of
this?  Luckily, the answer to this is in the affirmative.

Recall that the wrapped D6--brane is actually a charge $N$ monopole of
the spontaneously broken $SU(2)$ six dimensional gauge theory. There
is already a description of the $N$ monopole solution in terms of
$N\times N$ matrices of $SU(N)$. It is Nahm's equations shown in
equation \reef{nahm}. While we derived them for D1--branes stretched
between D3--branes, the monopole aspect of the description is
essentially the same. This can also be seen by the following chain of
dualities: K3 shrinking to volume $V_*=(2\pi)^4\alpha^{\prime2}$ is in
fact T--dual to K3 at a collapsed $A_1$ singularity, where the B--flux
is going to zero. The wrapped D6--branes become D4--branes wrapping
the collapsed singularity~\cite{jpp} We are on the Coulomb branch
where the resulting D2--branes have split into two fractional ones,
each carrying an $SU(N)$. We are focusing on one of them, and so send
the other one off to infinity.~\footnote{For wrapped D7--branes on K3,
  the dual situation is a D5--brane wrapped on the collapsed cycle
  giving fractional D3--branes, and the large $N$ gauge theory study
  of such systems {\it via} supergravity is
  underway.~\cite{mukhiagain,igoretal}}

As we learned in subsection~\ref{wrapfracstret} this situation is in
turn T--dual to fractional D3--branes stretched between NS5--branes,
where we focus on just one segment, and send the other to infinity. A
D3--brane stretched between NS5--branes in this way is a monopole of
the $U(1)$ gauge theory on the fivebrane worldvolume. The B--field is
the distance between the NS5--branes, and when it goes to zero they
coincide and there is an enhanced $SU(2)$.
\bigskip
\begin{center}
\begin{tabular}{|c|c|c|c|c|c|c|c|c|c|c|}
\hline
&$x^0$&$x^1$&$x^2$&$x^3$&$x^4$&$x^5$&$x^6$&$x^7$&$x^8$&$x^9$\\\hline
D3&$-$&$-$&$-$&$\bullet$&$\bullet$&$\bullet$&$|-|$
&$\bullet$&$\bullet$&$\bullet$\\\hline
NS5&$-$&$-$&$-$&$-$&$-$&$-$&$\bullet$&$\bullet$&$\bullet$&$\bullet$
\\
\hline
\end{tabular}
\end{center}
\bigskip

\begin{figure}[ht]
\centerline{\psfig{figure=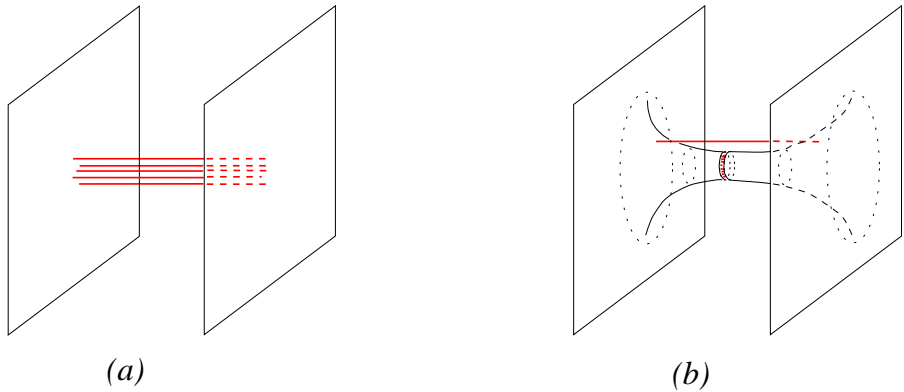,height=2.0in}}
\caption{\small
  $N$ D$(p+1)$--branes ending on NS5--branes: {\it (a)} The naive
  picture {\it (b)} The resulting bending of the NS5--branes cannot be
  neglected for large $g_sN$. The separated brane is the probe which
  becomes massless at the enhan\c con locus, an $S^{4-p}$ (a circle in
  the figure).}
\label{bent}
\end{figure}

The enhan\c con phenomenon, where the $SU(2)$ is enhanced on a sphere
in spacetime, is the result of the bending of the NS5--branes as the
D3--branes pull on them.~\cite{jpp} (This is shown for all $p$ in figure
\ref{bent}. The case we are discussing here is $p=2$.)
The $N$ D3--brane configuration has a description as an $N$--monopole
in the NS5--brane worldvolume. The earlier appearance of the Nahm
equations is therefore manifestly connected to the geometry of the
arrangement in the figure. The distance between the NS5--branes is the
Higgs vev.

\ennbee{Using type~IIB's S--duality converts the NS5--branes to
  D5--branes, and leaves alone the D3--branes stretched between them.
  A T--duality in the two spatial directions common to all the D--branes will
  complete the journey to the system of the D1's stretched between two
  D3's.}

The $N{\times}N$ fields $\Phi^i(s)$ which appear in Nahm's equations
\reef{nahm} represent the transverse coordinates of the $N$
D3--branes, in directions $x^3,x^4,x^5$. However as we have already
seen in the discussion of monopoles, they are {\sl necessarily}
non--commutative.  At the ends of the interval they must form an
irreducible $N$ dimensional representation of $SU(2)$. These are
precisely the same data which built the fuzzy sphere in our previous
example.~\cite{justme}

It is clear from this that at large $N$, a cross section of the
$N$--monopole depicted in figure \ref{bent} has a description as a
fuzzy sphere. The enhan\c con, which is the surface of the central
slice is therefore describable as a fuzzy sphere. (Other points in the
full monopole moduli space will describe other fuzzy geometries.)  As
$N$ is large, this is spherical to a good approximation and matches
onto the spherically symmetric supergravity geometry in
\reef{peeplus}.

Unfortunately, it is has not been possible to write down the spacetime
gauge and Higgs fields for multi--monopole solutions, as it is clearly
interesting to study them more in detail. The construction of the full
solutions are rather implicit, using algrebraic, and other methods
from scattering theory, {\it
  etc}.~\cite{ward,hitchin,atiyahward,backlund,nmonopoles,manymono,forgacs}
It would be an interesting problem to study how those fields match
onto the asymptotic spherically symmetric supergravity fields of the
solution \reef{peeplus}. The explicit solutions, if we had them, might
tell us much about both the supergravity geometry and possibly the
large $N$ gauge theory on the wrapped brane, perhaps deepening the
already known correspondence between their moduli
spaces,~\cite{SWtwo,CH,hanany} as discussed in detail in
section~\ref{muchmoreado}.

Is short, we see that the intuitive reason for non--commutativity in
this and the previous subsection is simply the fact that branes, for
one reason or another, cease to be pointlike and/or lose their
identity among other branes, becoming ``smeared'' or ``dissolved''.
This process is controlled/described by non--commutativity in some
choice of variables.  Since the endpoints of the strings are meant to
be located on the branes, the smearing results in a natural departure
from commutativity for our spacetime coordinates if we insist on
continuing to use the variables we used when the branes were
pointlike.

%\newpage

\section{Closing Remarks}
I think that it is high time that I stopped, since these notes have
begun to become unwieldy. While there has been some repetition of
ideas and phrases in various places, it was worth doing since it is by
understanding something in as many different ways as one can that one
can move beyond it.  Particularly repetitious were the continued
T--duality demonstrations (most things seemed to be explained by
tilting a brane!), for which I make no apology.

As a means of making up for the rather large size of the notes, I
collect towards the end a page or two of some of the most useful
formulae that people like to have to hand (and their number in the
text so that you can find where they are discussed/derived). Also, I
have listed the titles and locations of the various inserts, which I
hope are useful.

\bigskip

\rightline{{\it ---cvj}}

%\bigskip

\bigskip \hskip3.5cm\psfig{figure=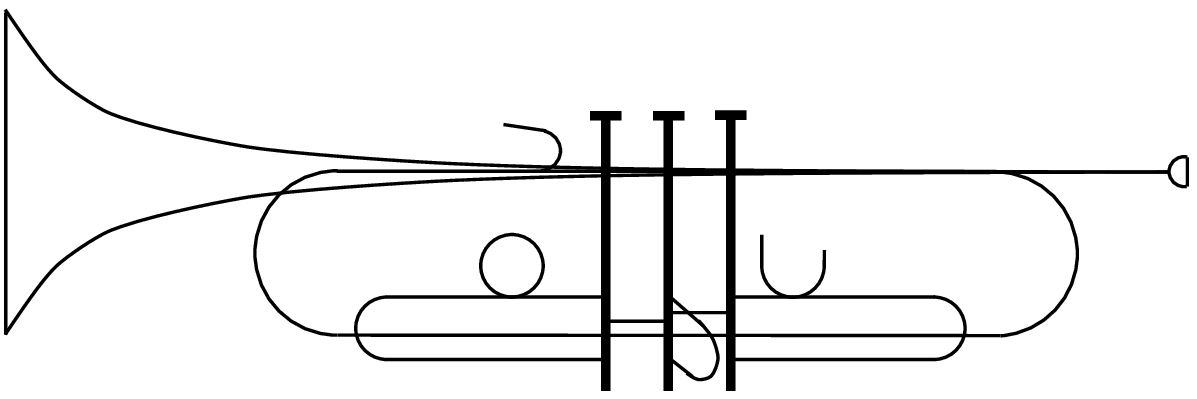,height=0.5in}

%\vfill\eject

\bigskip

\section*{Acknowledgements}
The first two weeks or so of this work was supported by an NSF CAREER
grant, \#9733173. I am grateful to the organisers of the 1998 Trieste
(ICTP) Spring School, the organisers of the 1999 Theoretical Advanced
Summer Institute (TASI), and the organisers of the 1999 British
Universities Summer School in Theoretical Elementary Particle Physics
(BUSSTEPP) for the invitations to give these lectures, and to their
associated staff for helping to make my time at Trieste, Boulder and
Southampton (respectively) so pleasant. Many thanks to Andreas
Recknagel, Marco Bill\'o, Daniel Bundzik, Laur J\"arv, Ken Lovis,
David Page, Volker Schomerus and Arkady Tseytlin for some comments on
this manuscript.

\newpage

%\bigskip
\appendix

\section{Collection of (Hopefully) Useful Formulae}

\bigskip

\centerline{\bf Charges and tensions}
\medskip

\begin{itemize}
\item
The fundamental string tension: $$\tau^{\rm F}_1\equiv
T=(2\pi\alpha^\prime)^{-1}=\nu_1\ .$$

($\nu_1$ is its $B_{(2)}$--charge.)
\item
The tension and charge of a D$p$--brane in superstring theory \reef{dcharge}:
$$\tau_p=\mu_p g_s^{-1}=(2\pi)^{-p}{\alpha^\prime}^{-(p+1)/2}g_s^{-1}\ .$$

\item
A recursion relation \reef{trec}:
$$
\tau_p=\tau_{p'}(2\pi\sqrt{\alpha^\prime})^{p'-p}\ . 
$$
\item
The tension of the NS5--brane (see insert 11 (p.\pageref{insert11}))
$$\tau^{\rm F}_5=(2\pi)^{-5}\alpha^{\prime{-3}}g_s^{-2}\ .$$

\item The Yang Mills coupling for the field theory on a brane
  \reef{yangmillscoupling}:
$$
g^2_{{\rm YM},p}=\tau_p^{-1}(2\pi\alpha^\prime)^{-2}=(2\pi)^{p-2}
\alpha^{\prime(p-3)/2}\ .$$

\item
Orientifold charge and tension \reef{orrycharge}: 
$$
\mu'_p = \mp 2^{p - 5} \mu_p, \qquad \tau'_p = \mp 2^{p - 5} \tau_p \ .
$$
(The minus sign is correlated with $SO$ and the plus with $USp$.)

\item The  product of the dual D--branes' tensions 
$$\tau_p\tau_{6-p}=2\pi(2\pi)^{-7}\alpha^{\prime -4}g_s^{-2}
={2\pi\over 2\kappa^2}$$ is the
minimum allowed by the quantum theory, with the following formula:

\item
The 10 dimensional Newton's constant \reef{nice}
$$
2\kappa^2\equiv 2\kappa_0^2 g_s^2
=(16\pi G_N)=(2\pi)^{-1}(4\pi^2\alpha^\prime)^4g_s^2=
 (2\pi)^7\alpha^{\prime 4}g_s^2\ .
$$

\item The tensions of the M2-- and M5--branes of 11 dimensional supergravity:
$$
\tau^{\rm M}_2=(2\pi)^{-2}\ell_p^{-3}\ ;\qquad\tau^{\rm
  M}_5=(2\pi)^{-5}\ell_p^{-6}\ ,
$$
can be deduced from the fact that the low energy type~IIA string
theory is 11 dimensional supergravity at strong coupling, that the
D2--branes and NS5--branes directly lift to become the M--branes, and the
following:

\item The 11 dimensional Planck length $\ell_p$ \reef{elevenrelation}:
$$\ell_p= g_s^{1/3}\sqrt{\alpha^{\prime}}.$$

\item The  product of the M--branes' tensions $$\tau^{\rm M}_2\tau^{\rm
  M}_5=2\pi(2\pi)^{-8}\ell_p^{-9}={2\pi\over 2\kappa_{11}^2}$$ is the
minimum allowed by the quantum theory, with:

\item The  11 dimensional Newton constant  \reef{elevenrelation}:
$$16\pi G_N^{11}=2\kappa_{11}^2\ ; \qquad \kappa_{11}^2=2^7\pi^8\ell_p^9.$$

\item
D--brane action \reef{diracborninfeld}, \reef{RRpart}:
$$
S_p=-T_p\int d^{p+1}\xi\,e^{-\Phi}\det\!^{1/2}(G_{ab}+B_{ab}
+2\pi\alpha^\prime
F_{ab})+\mu_p\int_{{\cal M}_{p+1}} C_{(p+1)}\ .  
$$

\item
Some curvature couplings \reef{curvy}:
$$
\mu_p\int_{{\cal M}_{p+1}}\sum_i C_{(i)}\left[ e^{2\pi\alpha^\prime
 F+B} \right] \sqrt{{\hat {\cal A}}(4\pi^2\alpha^\prime R)}\ .
$$
where the ``A--roof'' or ``Dirac'' genus has its square root defined as:
$$
\sqrt{{\hat {\cal
A}}(R)}=1-{p_1(R)\over48}+p_1^2(R){7\over11520}-{p_2(R)\over2880}+\cdots
$$
The $p_i(R)$'s are the $i$th Pontryagin class. For example, 
$$
p_1(R)=-{1\over8\pi^2}{\rm Tr} R\wedge R\ .
$$

\end{itemize}

\bigskip

\centerline{\bf Bosonic Effective Actions}
\medskip

\begin{itemize}

\item
The Dirac--Born--Infeld--Wess--Zumino Action \reef{diracborninfeld}

$$
S= -\tau_p\int d^{p+1}X\,\det\!^{1/2}(\eta_{ab}+ \partial_aX^m\partial_bX_m
+2\pi\alpha^\prime F_{ab})+\mu_p\int C_{(p+1)}\ ,\nonumber
$$

\item
Type~IIA string frame effective actions \reef{Aaction},\reef{Baction}
\begin{eqnarray}
&&S_{\rm IIA} = {1\over2\kappa_0^2}\int\!d^{10}\!x (-G)^{1/2} \left\{
e^{-2 \Phi}\left[R + 4 ( \nabla \phi)^2 -{1 \over
12}(H^{(3)})^2\right] \right.\nonumber\\ &&\qquad\quad\left.- {1 \over
4} (G^{(2)})^2- {1 \over
48}(G^{(4)})^2\right\}-{1\over4\kappa_0^2}\int B^{(2)}
dC^{(3)}dC^{(3)}\ . \nonumber
\end{eqnarray}
$H^{(3)}=dB^{(2)}$, $G^{(2)}=dC^{(1)}$ and $G^{(4)}=dC^{(3)}+H^{(3)}\wedge C^{(1)}$.

\begin{eqnarray}
&&S_{\rm IIB} = {1\over2\kappa_0^2}\int\!d^{10}\!x (-G)^{1/2}
\left\{e^{-2 \phi}\left[R + 4( \nabla \phi)^2
-{1 \over 12} (H^{(3)})^2\right] 
\right.
\nonumber\\
&&\left.\qquad\qquad\qquad
- {1 \over 12} (G^{(3)} + C^{(0)} H^{(3)})^2 
-{1\over2}(d C^{(0)})^2-{1\over 480}(G^{(5)})^2\right\}\nonumber\\
&&\qquad\qquad\qquad
+{1\over4\kappa_0^2}\int \left(C^{(4)}+{1\over2}B^{(2)}\,C^{(2)}\right)
\,G^{(3)}\,H^{(3)}\ .\nonumber
\end{eqnarray}
$G^{(3)}=dC^{(2)}$ and $G^{(5)}=dC^{(4)}+H^{(3)}C^{(2)}$,  $C^{(0)}$. Impose
self--duality of  $C^{(4)}$ {\it via} 
$F^{(5)}=\,^*\!F^{(5)}$ by hand in the equations of
motion.

\item Use
$${\widetilde G}_{\mu\nu}=e^{-\Phi/2}G_{\mu\nu}\ ,$$
to go to the 
Einstein frame.

\item
Type~I Bosonic Action \reef{Iaction}
\begin{eqnarray}
&&S_{\rm I} = {1\over2\kappa_0^2}\int\!d^{10}\!x (-G)^{1/2}
\Biggl\{e^{-2 \Phi}\left[R + 4( \nabla \phi)^2\right] \nonumber\\
&&\hskip4cm - {1 \over 12} ({\widetilde
G}^{(3)})^2-{\alpha^\prime\over8}e^{-\Phi}\Tr (F^{(2)})^2\Biggr\}\ .\nonumber
\end{eqnarray}
Here
$$
{\widetilde G}^{(3)}=dC^{(2)}-{\alpha^\prime\over4}\left[\omega_{\rm
3Y}(A)-\omega_{\rm 3L}(\Omega)\right]\ ,
$$

\item
Heterotic actions \reef{Haction}
\begin{eqnarray}
S_{\rm H} = {1\over2\kappa_0^2}\int\!d^{10}\!x (-G)^{1/2}
e^{-2 \Phi}\Biggl\{ R + 4( \nabla \phi)^2
%\nonumber\\&&\hskip4cm
- {1 \over 12} ({\widetilde H}^{(3)})^2-{\alpha^\prime\over 8}\Tr
(F^{(2)})^2\Biggr\}\ , \nonumber
\end{eqnarray}

$$
{\widetilde H}^{(3)}=dB^{(2)}-{\alpha^\prime\over4}\left[\omega_{\rm
3Y}(A)-\omega_{\rm 3L}(\Omega)\right]\ .
$$

\item Chern--Simons three--form:
$$
\omega_{\rm 3Y}(A)\equiv \Tr\left(A\wedge dA+{2\over3}A\wedge A\wedge
A\right)\ ,\, \,\mbox{with}\,\,\, d\omega_{\rm 3Y}=\Tr F\wedge F\ .\nonumber
$$
with a similar expression for the spin connection $\Omega$, to make
$\omega_{\rm 3L}$.

\end{itemize}

\bigskip

\centerline{\bf Supergravity Brane (and other) Solutions}
\medskip

\begin{itemize}

\item The 10 dimensional $p$--brane solutions \reef{branes}:
\begin{eqnarray}
ds^2 &=&  Z_p^{-1/2} \eta_{\mu\nu} dx^\mu dx^\nu +
 Z_p^{1/2} dx^i dx^i \ ,\nonumber
\\
e^{2\Phi } &=& g^2_s { Z_p}^{(3-p)\over2}\ , \nonumber\\
C_{ (p+1)} &=& ({Z_p}^{-1} -1)g_s^{-1} dx^0 \wedge \cdots \wedge dx^p
\ ,\nonumber
\end{eqnarray}
where $\mu=0,\ldots,p$, and $i=p+1,\ldots, 9$, and the harmonic
function $Z_p$ is
$$
Z_p=1+{d_p(2\pi)^{p-2}g_sN\alpha^{\prime (7-p)/2}\over r^{7-p}}\ ;
 \quad d_p= 2^{7-2p}\pi^{9-3p\over 2}
\Gamma\left({7-p\over 2}\right) \ .
$$

\item The 10 dimensional type~IIA NS5--brane \reef{nsfive}:

\begin{eqnarray}
ds^2&=&-dt^2+(dx^1)^2+\cdots+(dx^5)^2+{\tilde Z}_5
\left(dr^r+r^2d\Omega_3^2\right)\nonumber\\
e^{2\Phi}&=&g_s^2{\tilde Z}_5 =g_s^2\left(1+
{\alpha^\prime N\over r^2}\right)\ ,\nonumber\\
B_{ (6)} &=& ({{\tilde Z}_5}^{-1} -1)g_s 
dx^0 \wedge \cdots \wedge dx^5
\ .
\nonumber
\end{eqnarray}

\item The 11 dimensional M2--brane:
\begin{eqnarray}
ds^2=f_3^{-2/3}\left(-dt^2+(dx^1)^2+(dx^2)^2\right)
+f_3^{1/3}(dr^2+r^2d\Omega_7^2)\nonumber\\
f_3=\left(1+{\pi N\ell_p^3\over r^3}\right)\ ,\quad
A_{(3)}=f_3^{-1}dt\wedge dx^1\wedge dx^2\ .\nonumber
\end{eqnarray}

\item The 11 dimensional M5--brane:
\begin{eqnarray}
ds^2=f_5^{-1/3}\left(-dt^2+(dx^1)^2+\cdots+(dx^5)^2\right)
+f_5^{2/3}(dr^2+r^2d\Omega_4^2)\nonumber\\
f_5=\left(1+{32\pi^2 N\ell_p^6\over r^6}\right)\ ,\quad
A_{(6)}=f_5^{-1}dt\wedge dx^1\wedge\cdots \wedge dx^5\ .\nonumber
\end{eqnarray}

\item Sometimes useful are the $SU(2)_L$ invariant one--forms:
\begin{eqnarray}
&&\sigma_1=-\sin\psi d\theta+\cos\psi\sin\theta d\phi\ ;\nonumber\\
&&\sigma_2=\cos\psi d\theta+\sin\psi\sin\theta d\phi\ ;\nonumber\\
&&\sigma_3=d\psi +\cos\theta d\phi\ ,\nonumber
\end{eqnarray}
($0<\theta<\pi$, $0<\phi<2\pi$, $0<\psi<4\pi$ are the $S^3$ Euler
angles). 

\noindent Note:
$d\sigma_i={1\over2}\epsilon_{ijk}\sigma_j\wedge\sigma_k $ (The
$SU(2)_R$ invariants come from $\psi\leftrightarrow\phi$.)

\item $\sigma_1^2+\sigma_2^2=d\theta^2+\sin^2\theta d\phi^2 \equiv
  d\Omega_2^2$
\item The Eguchi--Hanson metric \reef{eguchi}:
  $$
  ds^2=\left(1-\left({a\over r}\right)^4\right)^{-1}dr^2 +
  r^2\left(1-\left({a\over
        r}\right)^4\right)\sigma_3^2+r^2(\sigma_1^2+\sigma_2^2)\ ,
$$
\noindent Note: period of $\psi$ is $2\pi.$ There is an $SO(3)$ isometry.

\item The A--series ALE spaces \reef{gibbhawk}:
\begin{eqnarray}
&&ds^2 = V^{-1}(dz - {\bf A} 
\cdot d{\bf y})^2 + V
d{\bf y} \cdot d{\bf y} \nonumber\\
&&V = \sum_{i=0}^{N-1} \frac{\sqrt{\alpha^\prime}}{|{\bf y} - {\bf y}_i|},
\qquad \mbox{\boldmath$\nabla$}V = \mbox{\boldmath$\nabla$}
\times{\bf A}\ .\nonumber
\end{eqnarray}
\noindent Note: case $N=2$ is equivalent to Eguchi--Hanson.~\cite{prasad}

\item The Self--Dual Taub--NUT metric \reef{taubytwo}:
  $$
  ds^2_{\rm
    TN}=\left(1+{2N\over\rho}\right)(d\rho^2+\rho^2(\sigma_1^2+\sigma_2^2))+
  4N^2\left(1+{2N\over\rho}\right)^{-1} \sigma_3^2\ .
$$

\noindent Note: period of $\psi$ is $4\pi.$  There is an $SU(2)$ isometry.

\noindent The singular case $N=-1$ results from  taking the large $\rho$ limit
of the smooth Atiyah--Hitchin manifold \reef{atiyahhitchin}, and in
that case the period of $\psi$ is $2\pi$. There is an $SO(3)$
isometry.

\item The A--series ALF (multi--Taub--NUT) spaces:~\cite{hawk}
\begin{eqnarray}
&&ds^2 = V^{-1}(dz - {\bf A} 
\cdot d{\bf y})^2 + V
d{\bf y} \cdot d{\bf y} \nonumber\\
&&V = 1+\sum_{i=0}^{N-1} \frac{2n_i\sqrt{\alpha^\prime}}
{|{\bf y} - {\bf y}_i|},
\qquad \mbox{\boldmath$\nabla$}V = \mbox{\boldmath$\nabla$}
\times{\bf A}\ .\nonumber
\end{eqnarray}
\noindent Note: case $N=2$ is equivalent to self--dual Taub--NUT.

\end{itemize}

\bigskip

\section{List of Inserts}

\noindent
Insert 1: $T$ is for Tension\dotfill\pageref{insert1}

\noindent
Insert 2: A Rotating Open String\dotfill\pageref{insert2}

\noindent
Insert 3: Cylinders, Strips and the Complex Plane\dotfill\pageref{insert3}

\noindent
Insert 4: Partition Functions\dotfill\pageref{insert4}

\noindent
Insert 5: World Sheet Perturbation Theory: Diagrammatics
\dotfill\pageref{insert5}

\noindent
Insert 6: Particles and Wilson Lines\dotfill\pageref{insert6}
 
\noindent
Insert 7: Vacuum Energy\dotfill\pageref{insert7}

\noindent
Insert 8: Translating Closed to Open\dotfill\pageref{insert8}

\noindent
Insert 9: Forms and Branes \dotfill\pageref{insert9}

\noindent
Insert 10: A Closer Look at the Eguchi--Hanson Space and its ``Bolt''
\dotfill\pageref{insert10}

\noindent
Insert 11: Dual Branes from 10D String--String
Duality\dotfill\pageref{insert11}

\noindent
Insert 12: The Heterotic NS5--brane \dotfill\pageref{insert12}

\noindent
Insert 13: The Type~II NS5--brane\dotfill\pageref{insert13}

\noindent
Insert 14: Removing the ``Nut'' Singularity from
  Taub--NUT \dotfill\pageref{insert14}

\end{document}